\documentclass[aps,prd,          groupedaddress,amsmath,amssymb]{revtex4}

\usepackage{epsfig}
\usepackage{graphicx}
\usepackage{dcolumn}
\usepackage{bm}
\begin{document}

\def\d{{\rm d}}
\def\eps{\varepsilon}
\def\lp{\left. }
\def\rp{\right. }
\def\lr{\left( }
\def\rr{\right) }
\def\le{\left[ }
\def\re{\right] }
\def\lg{\left\{ }
\def\rg{\right\} }
\def\lb{\left| }
\def\rb{\right| }
\def\beq{\begin{equation}}
\def\eeq{\end{equation}}
\def\bea{\begin{eqnarray}}
\def\eea{\end{eqnarray}}

\preprint{KA-TP-07-2007}
\preprint{LPSC 07-023}
\preprint{SFB/CPP-07-09}
\title{Squark and Gaugino Hadroproduction and Decays in Non-Minimal \\
 Flavour Violating Supersymmetry}
\author{Giuseppe Bozzi}
\affiliation{Institut f\"ur Theoretische Physik, Universit\"at
  Karlsruhe, Postfach 6980, D-76128 Karlsruhe, Germany}
\author{Benjamin Fuks}
\author{Bj\"orn Herrmann}
\author{Michael Klasen}
\email[]{klasen@lpsc.in2p3.fr}
\affiliation{Laboratoire de Physique Subatomique et de Cosmologie,
 Universit\'e Joseph Fourier/CNRS-IN2P3,
 53 Avenue des Martyrs, F-38026 Grenoble, France}
\date{\today}
\begin{abstract}
 We present an extensive analysis of squark and gaugino hadroproduction and
 decays in non-minimal flavour violating supersymmetry. We employ the
 so-called super-CKM basis to define the possible misalignment of quark
 and squark rotations, and we use generalized (possibly complex) charges to
 define the mutual couplings of (s)quarks and gauge bosons/gauginos. The
 cross sections for all squark-(anti-)squark/gaugino pair and squark-gaugino
 associated production processes as well as their decay widths are then
 given in compact analytic form. For four different constrained
 supersymmetry breaking models with non-minimal flavour violation in the
 second/third generation squark sector only, we establish the parameter
 space regions allowed/favoured by low-energy, electroweak precision, and
 cosmological constraints and display the chirality and flavour
 decomposition of all up- and down-type squark mass eigenstates. Finally,
 we compute numerically the dependence of a representative sample of
 production cross sections at the LHC on the off-diagonal mass matrix
 elements in the experimentally allowed/favoured ranges.
\end{abstract}
\maketitle


\section{Introduction}
\label{sec:1}

\vspace*{-125mm}
\noindent KA-TP-07-2007\\
\noindent LPSC 07-023\\
\noindent SFB/CPP-07-09
\vspace*{110.5mm}

Weak scale supersymmetry (SUSY) remains a both theoretically and
phenomenologically attractive extension of the Standard Model (SM) of
particle physics \cite{Nilles:1983ge,Haber:1984rc}. Apart from linking
bosons with fermions and unifying internal and external (space-time)
symmetries, SUSY allows for a stabilization of the gap between the Planck
and the electroweak scale and for gauge coupling unification at high
energies. It appears naturally in string theories, includes gravity, and
contains a stable lightest SUSY particle (LSP) as a dark matter candidate.
Spin partners of the SM particles have not yet been observed, and in order
to remain a viable solution to the hierarchy problem, SUSY must be broken at
low energy via soft mass terms in the Lagrangian. As a consequence, the SUSY
particles must be massive in comparison to their SM counterparts, and the
Tevatron and the LHC will perform a conclusive search covering a wide range
of masses up to the TeV scale.

If SUSY particles exist, they should also appear in virtual
particle loops and affect low-energy and electroweak precision
observables. In particular, flavour-changing neutral currents
(FCNC), which appear only at the one-loop level even in the SM,
put severe constraints on new physics contributions appearing at
the same perturbative order. Extended technicolour and many
composite models have thus been ruled out, while the Minimal
Supersymmetric Standard Model (MSSM) has passed these crucial
tests. This is largely due to the assumption of constrained
Minimal Flavour Violation (cMFV) \cite{Ciuchini:1998xy,
Buras:2000dm} or Minimal Flavour Violation (MFV)
\cite{Hall:1990ac, D'Ambrosio:2002ex, Altmannshofer:2007cs}, where
heavy SUSY particles may appear in the loops, but flavour changes
are either neglected or completely dictated by the structure of
the Yukawa couplings and thus the CKM-matrix \cite{Cabibbo:1963yz,
Kobayashi:1973fv}.

The squark mass matrices $M^2_{\tilde{Q}}$, $M^2_{\tilde{U}}$, and
$M^2_{\tilde{D}}$ are usually expressed in the super-CKM flavour
basis \cite{Hall:1985dx}. In MFV SUSY scenarios, their flavour
violating non-diagonal entries $\Delta_{ij}$, where $i,j=L,R$
refer to the helicity of the (SM partner of the) squark, stem from
the trilinear Yukawa couplings of the fermion and Higgs
supermultiplets and the resulting different renormalizations of
the quark and squark mass matrices, which induce additional
flavour violation at the weak scale through renormalization group
running \cite{Donoghue:1983mx, Duncan:1983iq, Bouquet:1984pp,
Borzumati:1986qx}, while in cMFV scenarios, these flavour
violating off-diagonal entries are simply neglected at both the
SUSY-breaking and the weak scale.

When SUSY is embedded in larger structures such as grand unified
theories (GUTs), new sources of flavour violation can appear
\cite{Gabbiani:1988rb}. For example, local gauge symmetry allows
for $R$-parity violating terms in the SUSY Lagrangian, but these
terms are today severely constrained by proton decay and collider
searches. In non-minimal flavour violating (NMFV) SUSY, additional
sources of flavour violation are included in the mass matrices at
the weak scale, and their flavour violating off-diagonal terms
cannot be simply deduced from the CKM matrix alone. NMFV is then
conveniently parameterized in the super-CKM basis by considering
them as free parameters. The scaling of these entries with the
SUSY-breaking scale $M_{\rm SUSY}$ implies a hierarchy
$\Delta_{\rm LL}\gg\Delta_{\rm LR,RL}\gg\Delta_{\rm RR}$
\cite{Gabbiani:1988rb}.

Squark mixing is expected to be largest for the second and third
generations due to the large Yukawa couplings involved
\cite{Brax:1995up}. In addition, stringent experimental
constraints for the first generation are imposed by precise
measurements of $K^0-\bar{K}^0$ mixing and first evidence of
$D^0-\bar{D}^0$ mixing
\cite{Hagelin:1992tc,Gabbiani:1996hi,Ciuchini:2007}. Furthermore,
direct searches of flavour violation depend on the possibility of
flavour tagging, established experimentally only for heavy
flavours. We therefore consider here only flavour mixings of
second- and third-generation squarks.

The direct search for SUSY particles constitutes a major physics goal of
present (Tevatron) and future (LHC) hadron colliders. SUSY particle
hadroproduction and decay has therefore been studied in detail
theoretically. Next-to-leading order (NLO) SUSY-QCD calculations exist for
the production of squarks and gluinos \cite{Beenakker:1996ch}, sleptons
\cite{Baer:1997nh}, and gauginos \cite{Beenakker:1999xh} as well as for
their associated production \cite{Berger:1999mc}. The production of top
\cite{Beenakker:1997ut} and bottom \cite{Berger:2000mp} squarks with large
helicity mixing has received particular attention. Recently, both QCD
one-loop and electroweak tree-level contributions have been calculated for
non-diagonal, diagonal and mixed top and bottom squark pair production
\cite{Bozzi:2005sy}. However, flavour violation has never been considered in
the context of collider searches apart from the CKM-matrix appearing in the
electroweak stop-sbottom production channel \cite{Bozzi:2005sy}.

It is the aim of this paper to investigate for the first time the possible
effects of non-minimal flavour violation at hadron colliders. To this end,
we re-calculate all squark and gaugino production and decay helicity
amplitudes, keeping at the same time the CKM-matrix and the quark masses to
account for non-diagonal charged-current gaugino and Higgsino Yukawa
interactions, and generalizing the two-dimensional helicity mixing matrices,
often assumed to be real, to generally complex six-dimensional helicity and
generational mixing matrices. We keep the notation compact by presenting all
analytical expressions in terms of generalized couplings. In order to obtain
numerical predictions for hadron colliders, we have implemented all our
results in a flexible computer program. In our phenomenological analysis of
NMFV squark and gaugino production, we concentrate on the LHC due to its
larger centre-of-mass energy and luminosity. We pay particular attention to
the interesting interplay of parton density functions (PDFs), which are
dominated by light quarks, strong gluino contributions, which are generally
larger than electroweak contributions and need not be flavour-diagonal, and
the appearance of third-generation squarks in the final state, which are
easily identified experimentally and generally lighter than first- and
second-generation squarks.

After reviewing the MSSM with NMFV and setting up our notation in Sec.\
\ref{sec:2}, we define in Sec.\ \ref{sec:3} generalized couplings of
quarks, squarks, gauge bosons, and gauginos. We then use these couplings to
present our analytical calculations in concise form. In particular, we have
computed partonic cross sections for NMFV squark-antisquark and
squark-squark pair production, squark and gaugino associated and gaugino
pair production as well as NMFV two-body decay widths of all squarks and
gauginos. Section \ref{sec:4} is devoted to a precise numerical analysis of
the experimentally allowed NMFV SUSY parameter space, an investigation of
the corresponding helicity and flavour decomposition of the up- and
down-type squarks, and the definition of four collider-friendly benchmark
points. These points are then investigated in detail in Sec.\ \ref{sec:5} so
as to determine the possible sensitivity of the LHC experiments on the
allowed NMFV parameter regions in the above-mentioned production channels.
Our conclusions are presented in Sec.\ \ref{sec:6}.

\section{Non-Minimal Flavour Violation in the MSSM}
\label{sec:2}

Within the SM, the only source of flavour violation arises through the
rotation of the up-type (down-type) quark interaction eigenstates $u_{L,R}'$
($d_{L,R}'$) to the basis of physical mass eigenstates $u_{L,R}$
($d_{L,R}$), such that
\bea
 d_{L,R} ~=~ V^d_{L,R}\ d_{L,R}' & ~~~{\rm and}~~~ &
 u_{L,R} ~=~ V^u_{L,R}\ u_{L,R}'.
\eea
The four bi-unitary matrices $V_{L,R}^{u,d}$ diagonalize the quark Yukawa
matrices and render the charged-current interactions proportional to the
unitary CKM-matrix \cite{Cabibbo:1963yz,Kobayashi:1973fv}
\bea
 \label{eq:ckm}
 V~=~V^u_L\,V^{d\dag}_L=\left( \begin{array}{ccc} V_{ud} & V_{us} & V_{ub}\\
 V_{cd} & V_{cs} & V_{cb} \\ V_{td} & V_{ts} & V_{tb} \end{array}\right).
\eea

In the super-CKM basis \cite{Hall:1985dx}, the squark interaction
eigenstates undergo the same rotations at high energy scale as their quark
counterparts, so that their charged-current interactions are also
proportional to the SM CKM-matrix. However, different renormalizations of
quarks and squarks introduce a mismatch of quark and squark field rotations
at low energies, so that the squark mass matrices effectively become
non-diagonal \cite{Donoghue:1983mx,Duncan:1983iq,Bouquet:1984pp,%
Borzumati:1986qx}. NMFV is then conveniently parameterized by non-diagonal
entries $\Delta_{ij}^{qq'}$ with $i,j=L,R$ in the squared squark mass
matrices
\bea
 M_{\tilde{u}}^2 &=& \left(\begin{array}{ccc|ccc}
 M_{\tilde L_u}^2 & \Delta_{LL}^{uc} & \Delta_{LL}^{ut} & m_u X_u  & \Delta_{LR}^{uc} & \Delta_{LR}^{ut} \\
 \Delta_{LL}^{cu} & M_{\tilde L_c}^2 & \Delta_{LL}^{ct} & \Delta_{RL}^{cu} & m_c X_c  & \Delta_{LR}^{ct} \\
 \Delta_{LL}^{tu} & \Delta_{LL}^{tc} & M_{\tilde L_t}^2 & \Delta_{RL}^{tu} &\Delta_{RL}^{tc} &m_t X_t \\
 \hline
 m_u X_u & \Delta_{RL}^{uc} & \Delta_{RL}^{ut} & M_{\tilde R_u}^2 & \Delta_{RR}^{uc} & \Delta_{RR}^{ut} \\
 \Delta_{LR}^{cu} & m_c X_c & \Delta_{RL}^{ct} & \Delta_{RR}^{cu} & M_{\tilde R_c}^2 & \Delta_{RR}^{ct} \\
 \Delta_{LR}^{tu} & \Delta_{LR}^{tc} & m_t X_t & \Delta_{RR}^{tu} & \Delta_{RR}^{tc} & M_{\tilde R_t}^2
 \end{array}\right)
\eea
and
\bea
 M_{\tilde{d}}^2 &=& \left( \begin{array}{ccc|ccc}
 M_{\tilde L_d}^2 & \Delta_{LL}^{ds} & \Delta_{LL}^{db} & m_d X_d &  \Delta_{LR}^{ds} &  \Delta_{LR}^{db} \\
 \Delta_{LL}^{sd} & M_{\tilde L_s}^2 & \Delta_{LL}^{sb} & \Delta_{RL}^{sd} & m_s X_s &  \Delta_{LR}^{sb} \\
 \Delta_{LL}^{bd} & \Delta_{LL}^{bs} & M_{\tilde L_b}^2 & \Delta_{RL}^{bd} & \Delta_{RL}^{bs} & m_b X_b  \\
 \hline
 m_d X_d & \Delta_{RL}^{ds} & \Delta_{RL}^{db} & M_{\tilde R_d}^2 & \Delta_{RR}^{ds} & \Delta_{RR}^{db} \\
 \Delta_{LR}^{sd} & m_s X_s & \Delta_{RL}^{sb} & \Delta_{RR}^{sd} & M_{\tilde R_s}^2 & \Delta_{RR}^{sb} \\
 \Delta_{LR}^{bd} & \Delta_{LR}^{bs} & m_b X_b & \Delta_{RR}^{bd} & \Delta_{RR}^{bs} & M_{\tilde R_b}^2 \end{array} \right),
\eea
where the diagonal elements are given by
\bea
\label{eq:5}
 M_{\tilde L_q}^2 &=& M_{\tilde Q_q}^2 + m_q^2 +
                     \cos 2\beta M_Z^2 (T_q^3 - e_q s_W^2),  \\
\label{eq:6}
 M_{\tilde R_q}^2 &=& M_{\tilde U_q}^2 + m_q^2 +
                      \cos 2\beta M_Z^2 e_q s_W^2 ~~~{\rm for~up-type~squarks,}  \\
\label{eq:7}
 M_{\tilde R_q}^2 &=& M_{\tilde D_q}^2 + m_q^2 +
                      \cos 2\beta M_Z^2 e_q s_W^2 ~~~{\rm for~down-type~squarks},
\eea
while the well-known squark helicity mixing is generated by the elements
\bea
 X_q &=& A_q - \mu \left\{\begin{array}{l}
 \cot\beta\hspace*{3.mm} {\rm for~up-type~squarks,}\\
 \tan\beta\hspace*{2.8mm}{\rm for~down-type~squarks.}
 \end{array}\right.
\eea
Here, $m_q$, $T_q^3$, and $e_q$ denote the mass, weak isospin quantum
number, and electric charge of the quark $q$. $m_Z$ is the $Z$-boson mass,
and $s_W$ ($c_W$) is the sine (cosine) of the electroweak mixing angle
$\theta_W$. The soft SUSY-breaking mass terms are $M_{\tilde Q_q}$ and
$M_{\tilde U_q,\tilde D_q}$ for the left- and right-handed squarks. $A_q$
and $\mu$ are the trilinear coupling and off-diagonal Higgs mass parameter,
respectively, and $\tan\beta = v_u / v_d$ is the ratio of vacuum expectation
values of the two Higgs doublets. The scaling of the flavour violating
entries $\Delta_{ij}^{qq'}$ with the SUSY-breaking scale $M_{\rm SUSY}$
implies a hierarchy $\Delta_{\rm LL}^{qq'}\gg\Delta_{\rm LR,RL}^{qq'}\gg
\Delta_{\rm RR}^{qq'}$ among them \cite{Gabbiani:1988rb}. They are usually
normalized to the diagonal entries \cite{Gabbiani:1996hi}, so that
\bea
 \Delta_{ij}^{qq'} &=& \lambda^{qq'}_{ij} M_{\tilde i_q} M_{\tilde j_{q'}}.
 \label{eq:delta}
\eea
Note also that $SU(2)$ gauge invariance relates the (numerically largest)
$\Delta_{LL}^{qq'}$ of up- and down-type quarks through the CKM-matrix,
implying that a large difference between them is not allowed.

The diagonalization of the mass matrices $M_{\tilde{u}}^2$ and
$M_{\tilde{d}}^2$ requires the introduction of two additional $6 \times 6$
matrices $R^u$ and $R^d$ with
\bea
 {\rm diag}\,(m_{\tilde u_1}^2, \ldots, m_{\tilde u_6}^2) ~=~ R^u\,
 M_{\tilde{u}}^2\,R^{u\dag} &{\rm ~~~~and~~~~}&
 {\rm diag}\,(m_{\tilde d_1}^2, \ldots, m_{\tilde d_6}^2) ~=~ R^d\,
 M_{\tilde{d}}^2\, R^{d\dag}.
\eea
By convention, the masses are ordered according to $m_{\tilde q_1} < \ldots
< m_{\tilde q_6}$. The physical mass eigenstates are given by
\bea
 \begin{pmatrix} \tilde{u}_1 \\ \tilde{u}_2 \\
 \tilde{u}_3 \\ \tilde{u}_4 \\ \tilde{u}_5 \\ \tilde{u}_6 \\
 \end{pmatrix} = R^{u} \begin{pmatrix} \tilde{u}_L \\ \tilde{c}_L
 \\ \tilde{t}_L \\ \tilde{u}_R \\ \tilde{c}_R \\ \tilde{t}_R \\
 \end{pmatrix} &{\rm ~~~~and~~~~} \begin{pmatrix} \tilde{d}_1 \\
 \tilde{d}_2 \\ \tilde{d}_3 \\ \tilde{d}_4 \\ \tilde{d}_5 \\
 \tilde{d}_6 \\ \end{pmatrix} = R^{d} \begin{pmatrix} \tilde{d}_L
 \\ \tilde{s}_L \\  \tilde{b}_L \\ \tilde{d}_R \\ \tilde{s}_R \\
 \tilde{b}_R \\ \end{pmatrix}.
\eea In the limit of vanishing off-diagonal parameters
$\Delta_{ij}^{qq'}$, the matrices $R^q$ become flavour-diagonal,
leaving only the well-known helicity mixing already present in
cMFV.

\section{Analytical Results for Production Cross Sections and Decay Widths}
\label{sec:3}

In this section, we introduce concise definitions of generalized strong and
electroweak couplings in NMFV SUSY and compute analytically the
corresponding partonic cross sections for squark and gaugino production as
well as their decay widths. The cross sections of the production processes
\bea
 a_{h_a}(p_a)\, b_{h_b}(p_b) &\to& \left\{\begin{array}{l}
 \tilde{q}^{(\ast)}_i(p_1)\, \tilde{q}^{\prime(\ast)}_j(p_2),~\\
 \tilde{\chi}^\pm_j(p_1)\, \tilde{q}^{(\ast)}_i(p_2),~\\
 \tilde{\chi}^{\pm(0)}_i(p_1)\, \tilde{\chi}^{\pm(0)}_j(p_2)
 \end{array}\right.
\eea
are presented for definite helicities $h_{a,b}$ of the initial partons
$a,b=q,\bar{q},g$ and expressed in terms of the squark, chargino,
neutralino, and gluino masses $m_{\tilde{q}_k}$, $m_{\tilde{\chi}^\pm_k}$,
$m_{\tilde{\chi}^0_k}$, and $m_{\tilde{g}}$, the conventional
Mandelstam variables,
\bea
 s=(p_a+p_b)^2,~ t=(p_a-p_1)^2 \mbox{,~and~} u=(p_a-p_2)^2,
\eea
and the masses of the neutral and charged electroweak gauge bosons $m_Z$ and
$m_W$. Propagators appear as mass-subtracted Mandelstam variables,
\bea
 \begin{array}{l c l c l c l c}
 s_w &=& s-m_W^2&,~& s_z &=& s-m_Z^2 &,\\
 t_{\tilde{\chi}^0_k} &=& t - m_{\tilde{\chi}^0_k}^2&,~&
 u_{\tilde{\chi}^0_k} &=& u-m_{\tilde{\chi}^0_k}^2&, \\
 t_{\tilde{\chi}_j} &=& t-m_{\tilde{\chi}^\pm_j}^2 &,~&
 u_{\tilde{\chi}_j} &=& u-m_{\tilde{\chi}^\pm_j}^2 &, \\
 t_{\tilde{g}} &=& t - m^2_{\tilde{g}} &,~& u_{\tilde{g}} &=& u -
 m^2_{\tilde{g}} &, \\
 t_{\tilde{q}_i} &=& t-m_{\tilde{q}_i}^2 &,~& u_{\tilde{q}_i} &=&
 u-m_{\tilde{q}_i}^2 &. \end{array}&&
\eea
Unpolarized cross sections, averaged over initial spins, can easily be
derived from the expression
\bea
 \d\hat{\sigma}=\frac{\d\hat{\sigma}_{ 1, 1} + \d\hat{\sigma}_{
 1,-1} + \d\hat{\sigma}_{-1, 1} + \d\hat{\sigma}_{-1,-1}}{4},
\eea
while single- and double-polarized cross sections, including the same
average factor for initial spins, are given by
\bea
 \d\Delta\hat{\sigma}_L=\frac{\d\hat{\sigma}_{ 1, 1} +
 \d\hat{\sigma}_{1,-1} - \d\hat{\sigma}_{-1, 1} -
 \d\hat{\sigma}_{-1,-1}}{4} & ~~~{\rm and}~~~ &
 \d\Delta\hat{\sigma}_{LL}=\frac{\d\hat{\sigma}_{ 1, 1} -
 \d\hat{\sigma}_{ 1,-1} - \d\hat{\sigma}_{-1, 1} +
 \d\hat{\sigma}_{-1,-1}}{4},
\eea
so that the single- and double-spin asymmetries become
\bea
 A_L = \frac{\d\Delta\hat{\sigma}_L}{\d\hat{\sigma}}& {\rm ~~~and~~~}&
 A_{LL} = \frac{\d\Delta\hat{\sigma}_{LL}}{\d\hat{\sigma}}.
\eea

\subsection{Generalized Strong and Electroweak Couplings in NMFV SUSY}

Considering the strong interaction first, it is well known that the
interaction of quarks, squarks, and gluinos, whose coupling is normally just
given by $g_s=\sqrt{4\pi\alpha_s}$, can in general lead to flavour violation
in the left- and right-handed sectors through non-diagonal entries in the
matrices
$R^q$,
\bea
 \left\{L_{\tilde{q}_j q_k \tilde{g}}, R_{\tilde{q}_j q_k
 \tilde{g}} \right\} &=& \left\{R^q_{jk}, -
 R^q_{j(k+3)}\right\}.
 \label{eq:coup2}
\eea
Of course, the involved quark and squark both have to be up- or down-type,
since the gluino is electrically neutral.

For the electroweak interaction, we define the square of the weak coupling
constant $g_W^2=e^2/\sin^2\theta_W$ in terms of the electromagnetic fine
structure constant $\alpha=e^2/(4\pi)$ and the squared sine of the
electroweak mixing angle $x_W=\sin^2\theta_W=s_W^2 = 1-\cos^2\theta_W =
1-c_W^2$. Following the standard notation, the $W^\pm-\tilde{\chi}^0_i-
\tilde{\chi}^\pm_j$, $Z-\tilde{\chi}^+_i-\tilde{\chi}^-_j$, and
$Z-\tilde{\chi}^0_i- \tilde{\chi}^0_j$ interaction vertices are proportional
to \cite{Haber:1984rc}
\bea
 O^L_{ij} =
 -\frac{1}{\sqrt{2}} N_{i4} V^\ast_{j2} + N_{i2} V^\ast_{j1} &{\rm
 ~~~~and~~~~}& O^R_{ij} = \frac{1}{\sqrt{2}} N_{i3}^\ast U_{j2} +
 N_{i2}^\ast U_{j1},~\nonumber \\
 O^{\prime L}_{ij} = -V_{i1} V_{j1}^\ast -
 \frac{1}{2} V_{i2} V_{j2}^\ast + \delta_{ij} x_W &{\rm
 ~~~~and~~~~}& O^{\prime R}_{ij} = -U_{i1}^\ast U_{j1} -
 \frac{1}{2} U_{i2}^\ast U_{j2} + \delta_{ij} x_W,~\nonumber \\
 O^{\prime\prime L}_{ij} = -\frac{1}{2} N_{i3} N_{j3}^\ast +
 \frac{1}{2} N_{i4}N_{j4}^\ast &{\rm ~~~~and~~~~}& O^{\prime\prime
 R}_{ij} = \frac{1}{2}
 N_{i3}^\ast N_{j3} - \frac{1}{2} N_{i4}^\ast N_{j4}.
\eea

In NMFV, the coupling strengths of left- and right-handed (s)quarks to the
electroweak gauge bosons are given by
\bea
 \{ L_{q q^\prime Z},R_{qq^\prime Z} \}&=& (2\,T^{3}_q -
 2\,e_q\,x_W) \times \delta_{q q^\prime},~ \nonumber\\
 \{ L_{\tilde{q}_i \tilde{q}_j Z},R_{\tilde{q}_i \tilde{q}_j Z} \}
 &=& (2\,T^{3}_{\tilde{q}}-2\,e_{\tilde{q}}\,x_W) \times
 \sum_{k=1}^3 \{R^u_{ik}\, R^{u\ast}_{jk}, R^u_{i(3+k)}\,
 R^{u\ast}_{j(3+k)} \},~\nonumber\\
 \{L_{qq^{\prime}W},R_{qq^{\prime}W}\} &=&
 \{\sqrt{2}\,c_W\,V_{qq^{\prime}}, 0\},~\nonumber\\
 \{L_{\tilde{u}_i \tilde{d}_j W}, R_{\tilde{u}_i \tilde{d}_j W}\}
 &=& \sum_{k,l=1}^3\{\sqrt{2}\,c_W\, V_{u_kd_l}\, R^u_{ik}\,
 R^{d\ast}_{jl},\, 0\},
\eea
where the weak isospin quantum numbers are $T_{\{q,\tilde{q}\}}^3 = \pm1/2$
for left-handed and $T_{\{q,\tilde{q}\}}^3=0$ for right-handed up- and
down-type (s)quarks, their fractional electromagnetic charges are denoted by
$e_{\{q,\tilde{q}\}}$, and $V_{kl}$ are the elements of the CKM-matrix
defined in Eq.\ (\ref{eq:ckm}). To simplify the notation, we have introduced
flavour indices in the latter, $d_1=d$, $d_2=s$, $d_3=b$, $u_1=u$, $u_2=c$,
and $u_3=t$.

The SUSY counterparts of these vertices correspond to the
quark-squark-gaugino couplings,
\bea
 L_{\tilde{d}_j d_k\tilde{\chi}^0_i} &=& \bigg[ (e_q - T^3_q)\, s_W\, N_{i1}
 +T^3_q\, c_W\, N_{i2} \bigg] R^{d\ast}_{jk}+ \frac{m_{d_k}\, c_W\,N_{i3}\,
 R^{d\ast}_{j(k+3)}}{2\, m_W\, \cos\beta},~  \nonumber\\
 -R_{\tilde{d}_j d_k \tilde{\chi}_i^0}^\ast &=& e_q\, s_W\,
 N_{i1}\, R^d_{j(k+3)} - \frac{m_{d_k}\, c_W\, N_{i3}\,
 R^d_{jk}}{2\, m_W\, \cos\beta} ,~\nonumber\\
 L_{\tilde{u}_j u_k\tilde{\chi}^0_i} &=& \bigg[ (e_q - T^3_q)\, s_W\, N_{i1}
 +T^3_q\, c_W\, N_{i2} \bigg] R^{u\ast}_{jk}+ \frac{m_{u_k}\, c_W\,
 N_{i4}\, R^{u\ast}_{j(k+3)}}{2\, m_W\, \sin\beta}  ,~\nonumber\\
 -R_{\tilde{u}_j u_k \tilde{\chi}_i^0}^\ast &=& e_q\, s_W\,
 N_{i1}\, R^u_{j(k+3)} - \frac{m_{u_k}\, c_W\, N_{i4}\,
 R^u_{jk}}{2\, m_W\, \sin\beta} ,~\nonumber\\ L_{\tilde{d}_j u_l
 \tilde{\chi}_i^\pm}&=& \sum_{k=1}^3\bigg[  U_{i1}\, R^{d\ast}_{jk}
 - \frac{m_{d_k}\, U_{i2}\, R^{d\ast}_{j(k+3)}}{\sqrt{2}\, m_W\,
 \cos\beta} \bigg] V_{u_l d_k} ,~\nonumber\\ -R^\ast_{\tilde{d}_j
 u_l \tilde{\chi}_i^\pm} &=& \sum_{k=1}^3 \frac{m_{u_l}\, V_{i2}\,
 V_{u_l d_k}^\ast\, R^d_{jk}}{\sqrt{2}\, m_W\, \sin\beta} ,~\nonumber\\
 L_{\tilde{u}_j d_l \tilde{\chi}_i^\pm}&=& \sum_{k=1}^3 \bigg[
 V_{i1}^\ast\, R^u_{jk} - \frac{m_{u_k}\, V_{i2}^\ast\,
 R^u_{j(k+3)}}{\sqrt{2}\, m_W\, \sin\beta}  \bigg] V_{u_k d_l}
 ,~\nonumber \\ -R^\ast_{\tilde{u}_j d_l \tilde{\chi}_i^\pm} &=&
 \sum_{k=1}^3 \frac{m_{d_l}\, U_{i2}^\ast V_{u_k d_l}^\ast\,
 R^{u\ast}_{jk}}{\sqrt{2}\, m_W\, \cos\beta}~,\label{eq:coup1}
\eea
where the matrices $N$, $U$ and $V$ relate to the gaugino/Higgsino mixing
(see App.\ \ref{sec:a}). All other couplings vanish due to (electromagnetic)
charge
conservation (e.g.\ $L_{\tilde{u}_j u_l \tilde{\chi}_i^\pm}$). These general
expressions can be simplified by neglecting the Yukawa couplings except for
the one of the top quark, whose mass is not small compared to $m_W$.
For the sake of simplicity, we use the generic notation
\bea
 \left\{\mathcal{C}^1_{a b c}, \mathcal{C}^2_{a b c} \right\} = \left\{
 L_{a b c},R_{a b c} \right\}
\eea
in the following.

\subsection{Squark-Antisquark Pair Production}

\begin{figure}
 \centering
 \includegraphics[width=0.75\columnwidth]{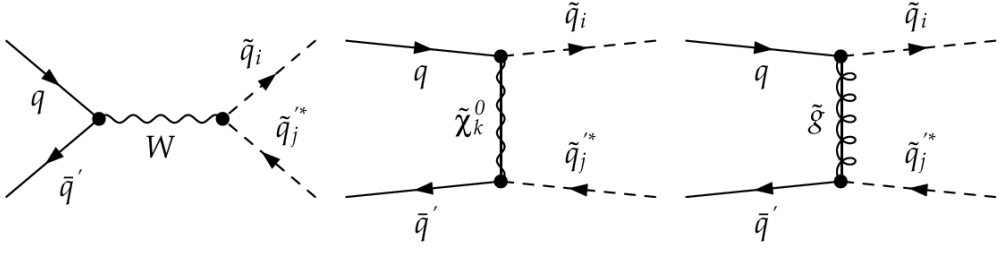}
 \caption{\label{fig:1}Tree-level Feynman diagrams for the production of
          charged squark-antisquark pairs in quark-antiquark collisions.}
\end{figure}

The production of charged squark-antisquark pairs
\bea
 \label{proc:sqSQ}
 q(h_a, p_a)\, \bar{q}^\prime(h_b, p_b) \to \tilde{u}_i(p_1)\,
 \tilde{d}^\ast_j(p_2),
\eea
where $i,j=1,...,6$
label up- and down-type squark mass eigenstates, $h_{a,b}$
helicities, and $p_{a,b,1,2}$ four-momenta, proceeds from an equally charged
quark-antiquark initial state through the tree-level Feynman diagrams shown
in Fig.\ \ref{fig:1}.
%
%
The corresponding cross section can be written in a compact way as
\bea \frac{{\rm d}
 \hat{\sigma}^{q\bar{q}'}_{h_a, h_b}}{dt} &=& (1-h_a) (1+h_b) \Bigg[
 \frac{\mathcal{W}}{s_w^2} + \bigg( \sum_{k,l=1,...,4}
 \frac{\mathcal{N}^{kl}_{11}}{t_{\tilde{\chi}^0_k}\,
 t_{\tilde{\chi}^0_l}}\bigg) +
 \frac{\mathcal{G}_{11}}{t_{\tilde{g}}^2} + \bigg( \sum_{k=1,...,4}
 \frac{\mathcal{[NW]}^k}{t_{\tilde{\chi}^0_k}\, s_w} \bigg) +
 \frac{\mathcal{[GW]}}{t_{\tilde{g}}\, s_w} \Bigg] \nonumber\\ &+&
 (1-h_a) (1-h_b) \Bigg[ \bigg( \sum_{k,l=1,...,4}
 \frac{\mathcal{N}^{kl}_{12}}{t_{\tilde{\chi}^0_k}\,
 t_{\tilde{\chi}^0_l}}\bigg) +
 \frac{\mathcal{G}_{12}}{t_{\tilde{g}}^2}\Bigg] + (1+h_a) (1+h_b)
 \Bigg[ \bigg( \sum_{k,l=1,...,4}
 \frac{\mathcal{N}^{kl}_{21}}{t_{\tilde{\chi}^0_k}\,
 t_{\tilde{\chi}^0_l}}\bigg) +
 \frac{\mathcal{G}_{21}}{t_{\tilde{g}}^2}\Bigg] \nonumber\\ &+&
 (1+h_a)(1-h_b) \Bigg[ \bigg( \sum_{k,l=1,...,4}
 \frac{\mathcal{N}^{kl}_{22}}{t_{\tilde{\chi}^0_k}\,
 t_{\tilde{\chi}^0_l}}\bigg) +
 \frac{\mathcal{G}_{22}}{t_{\tilde{g}}^2}\Bigg]
\eea
thanks to the form factors
\bea
 \mathcal{W} &=& \frac{\pi\, \alpha^2}{16\, x_W^2\, (1-x_W)^2\,
 s^2} \left| L^\ast_{q q^\prime W}\, L_{\tilde{u}_i \tilde{d}_j
 W}\right|^2 \left( u\, t - m^2_{\tilde{u}_i}\,
 m^2_{\tilde{d}_j}\right),~\nonumber\\ \mathcal{N}_{mn}^{kl} &=&
 \frac{\pi\, \alpha^2}{x_W^2\, (1 - x_W)^2\, s^2}
 \mathcal{C}^n_{\tilde{d}_j q^\prime \tilde{\chi}_k^0}\,
 \mathcal{C}^{m\ast}_{\tilde{u}_i q \tilde{\chi}_k^0}\,
 \mathcal{C}^{n\ast}_{\tilde{d}_j q^\prime \tilde{\chi}_l^0}\,
 \mathcal{C}^m_{\tilde{u}_i q \tilde{\chi}_l^0}\, \Bigg[ \left( u\,
 t - m^2_{\tilde{u}_i}\, m^2_{\tilde{d}_j}\right) \delta_{mn} +
 \left( m_{\tilde{\chi}^0_k}\, m_{\tilde{\chi}^0_l}\, s \right)
 \left(1-\delta_{mn} \right) \Bigg],~\nonumber\\
 \mathcal{G}_{mn} &=& \frac{2\, \pi\, \alpha_s^2}{9\, s^2} \left|
 \mathcal{C}^{n\ast}_{\tilde{d}_j q^\prime \tilde{g}}\,
 \mathcal{C}^m_{\tilde{u}_i q \tilde{g}}\right|^2 \Bigg[ \left( u\,
 t - m^2_{\tilde{u}_i}\, m^2_{\tilde{d}_j}\right) \delta_{mn} +
 \left(m_{\tilde{g}}^2\, s \right) \left(1-\delta_{mn} \right)
 \Bigg],~\nonumber \\ \mathcal{[NW]}^k &=& \frac{\pi\,
 \alpha^2}{6\, x_W^2\, (1-x_W)^2\, s^2}\, {\rm Re} \left[ L^\ast_{q
 q^{\prime} W}\, L_{\tilde{u}_i \tilde{d}_j W}\, L_{\tilde{u}_i q
 \tilde{\chi}_k^0}\, L^\ast_{\tilde{q}_j q^\prime \tilde{\chi}_k^0}
 \right] \left( u\, t - m^2_{\tilde{u}_i}\,
 m^2_{\tilde{d}_j}\right) ,~\nonumber\\ \mathcal{[GW]} &=&
 \frac{4\, \pi\, \alpha_s\, \alpha}{18\, x_W\, (1 - x_W)\, s^2}\,
 {\rm Re} \left[ L^\ast_{\tilde{u}_i q \tilde{g}}\, L_{\tilde{d}_j
 q^\prime \tilde{g}}\, L^\ast_{q q^\prime W}\, L_{\tilde{u}_i
 \tilde{d}_j W} \right] \left( u\, t - m^2_{\tilde{u}_i}\,
 m^2_{\tilde{d}_j}\right),
\eea which combine coupling constants and Dirac traces of the
squared and interference diagrams. In cMFV, superpartners of heavy
flavours can only be produced through the purely left-handed
$s$-channel $W$-exchange, since the $t$-channel diagrams are
suppressed by the small bottom and negligible top quark densities
in the proton, and one recovers the result in Ref.\
\cite{Bozzi:2005sy}. In NMFV, $t$-channel exchanges can, however,
contribute to heavy-flavour final state production from
light-flavour initial states and even become dominant, due to the
strong gluino coupling.

\begin{figure}
 \centering
 \includegraphics[width=\columnwidth]{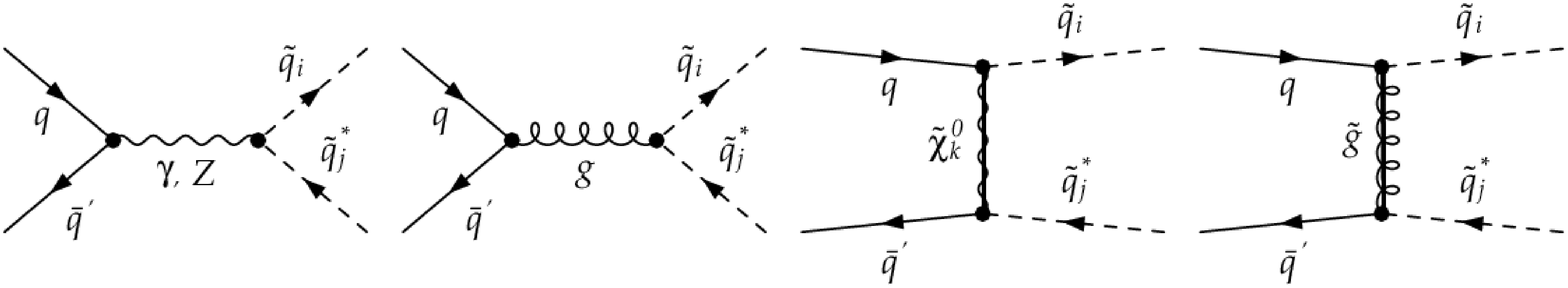}
 \includegraphics[width=\columnwidth]{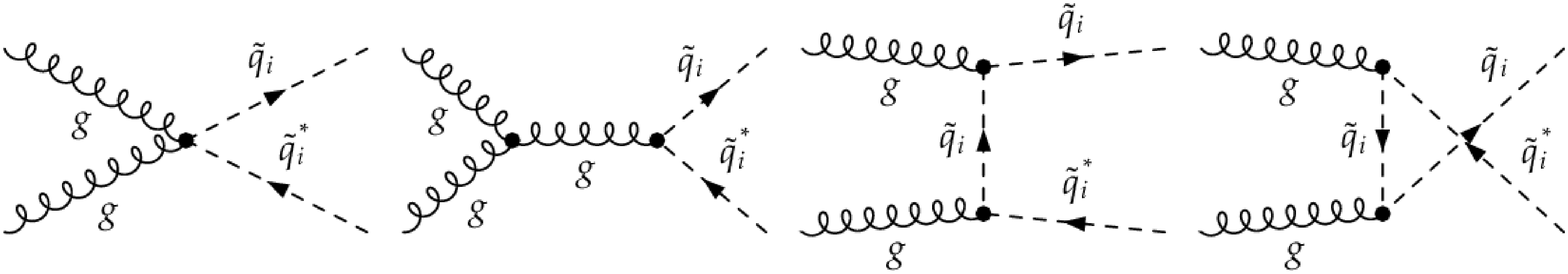}
 \caption{\label{fig:2}Tree-level Feynman diagrams for the production of
          neutral squark-antisquark pairs in quark-antiquark (top) and
          gluon-gluon collisions (bottom).}
\end{figure}

Neutral squark-antisquark pair production proceeds either from equally
neutral quark-antiquark initial states
\bea
 q(h_a, p_a)\, \bar{q}^\prime(h_b, p_b) \to \tilde{q}_i(p_1)\,
 \tilde{q}^\ast_j(p_2),
\eea
through the five different gauge-boson/gaugino exchanges shown in Fig.\
\ref{fig:2} (top) or from gluon-gluon initial states
%
%
\bea
 g(h_a,p_a)\, g(h_b, p_b) \to \tilde{q}_i(p_1)\, \tilde{q}^\ast_i(p_2)
\eea
through the purely strong couplings shown in Fig.\ \ref{fig:2} (bottom). The
differential cross section for quark-antiquark scattering
\bea
 \frac{{\rm d}
 \hat{\sigma}^{q\bar{q}'}_{h_a, h_b}}{dt} &=& (1-h_a) (1+h_b) \Bigg[
 \frac{\mathcal{Y}}{s^2} + \frac{\mathcal{Z}_1}{s_z^2} +
 \frac{\mathcal{G}}{s^2} +
 \frac{\widetilde{\mathcal{G}}_{11}}{t_{\tilde{g}}^2} +
 \frac{\mathcal{[YZ]}_1}{s\, s_z} +
 \frac{[\widetilde{\mathcal{G}}\mathcal{Y}]_1}{t_{\tilde{g}}\, s} +
 \frac{[\widetilde{\mathcal{G}}\mathcal{Z}]_1}{t_{\tilde{g}}\, s_z}
 + \frac{[\widetilde{\mathcal{G}}\mathcal{G}]_1}{t_{\tilde{g}}\, s}
 \nonumber\\ &+& \sum_{k,l=1,...,4}
 \bigg(\frac{\mathcal{N}^{kl}_{11}}{t_{\tilde{\chi}^0_k}\,
 t_{\tilde{\chi}^0_l}}\bigg) + \sum_{k=1,...,4} \bigg(
 \frac{\mathcal{[NY]}^k_1}{t_{\tilde{\chi}^0_k}\, s} +
 \frac{\mathcal{[NZ]}^k_1}{t_{\tilde{\chi}^0_k}\, s_z} +
 \frac{\mathcal{[NG]}^k_1}{t_{\tilde{\chi}^0_k}\, s} \bigg)
 \Bigg]\nonumber \\
 &+& (1+h_a) (1-h_b) \Bigg[ \frac{\mathcal{Y}}{s^2} +
 \frac{\mathcal{Z}_2}{s_z^2} + \frac{\mathcal{G}}{s^2} +
 \frac{\widetilde{\mathcal{G}}_{22}}{t_{\tilde{g}}^2} +
 \frac{\mathcal{[YZ]}_2}{s\, s_z} +
 \frac{[\widetilde{\mathcal{G}}\mathcal{Y}]_2}{t_{\tilde{g}}\, s} +
 \frac{[\widetilde{\mathcal{G}}\mathcal{Z}]_2}{t_{\tilde{g}}\, s_z}
 + \frac{[\widetilde{\mathcal{G}}\mathcal{G}]_2}{t_{\tilde{g}}\, s}
 \nonumber\\ &+& \sum_{k,l=1,...,4}
 \bigg(\frac{\mathcal{N}^{kl}_{22}}{t_{\tilde{\chi}^0_k}\,
 t_{\tilde{\chi}^0_l}}\bigg) + \sum_{k=1,...,4} \bigg(
 \frac{\mathcal{[NY]}^k_2}{t_{\tilde{\chi}^0_k}\, s} +
 \frac{\mathcal{[NZ]}^k_2}{t_{\tilde{\chi}^0_k}\, s_z} +
 \frac{\mathcal{[NG]}^k_2}{t_{\tilde{\chi}^0_k}\, s} \bigg)
 \Bigg]\nonumber\\
 &+& (1-h_a) (1-h_b) \Bigg[
 \frac{\widetilde{\mathcal{G}}_{12}}{t_{\tilde{g}}^2} +
 \sum_{k,l=1,...,4}
 \bigg(\frac{\mathcal{N}^{kl}_{12}}{t_{\tilde{\chi}^0_k}\,
 t_{\tilde{\chi}^0_l}}\bigg) \Bigg] + (1+h_a) (1+h_b) \Bigg[
 \frac{\widetilde{\mathcal{G}}_{21}}{t_{\tilde{g}}^2} +
 \sum_{k,l=1,...,4}
 \bigg(\frac{\mathcal{N}^{kl}_{21}}{t_{\tilde{\chi}^0_k}\,
 t_{\tilde{\chi}^0_l}}\bigg) \Bigg]
\eea
involves many different form factors,
\bea
 \mathcal{Y} &=&  \frac{\pi\, \alpha^2\, e_q^2\, e_{\tilde{q}}^2\,
 \delta_{ij}\, \delta_{qq^\prime}}{s^2} \left( u\, t -
 m^2_{\tilde{q}_i}\, m^2_{\tilde{q}_j^\prime}\right),~ \nonumber \\
 \mathcal{Z}_m &=&  \frac{\pi\, \alpha^2}{16\, s^2\, x_W^2
 (1-x_W)^2} \left| L_{\tilde{q}_i \tilde{q}_j Z} + R_{\tilde{q}_i
 \tilde{q}_j Z} \right|^2\, \left( C^m_{q q^\prime Z}\right)^2
 \left( u\, t - m^2_{\tilde{q}_i}\, m^2_{\tilde{q}_j^\prime}
 \right) ,~\nonumber \\ \mathcal{G} &=& \frac{2\, \pi\,
 \alpha_s^2\, \delta_{ij}\, \delta_{qq^\prime}}{9\, s^2} \left( u\,
 t - m^2_{\tilde{q}_i}\, m^2_{\tilde{q}_j^\prime}\right) ,~
 \nonumber \\ \mathcal{N}_{mn}^{kl} &=& \frac{\pi\,
 \alpha^2}{x_W^2\, (1 - x_W)^2\, s^2}\,
 \mathcal{C}^{m\ast}_{\tilde{q}_i q \tilde{\chi}_k^0}\,
 \mathcal{C}^m_{\tilde{q}_i q \tilde{\chi}_l^0}\,
 \mathcal{C}^n_{\tilde{q}_j q^\prime \tilde{\chi}_k^0}
 \mathcal{C}^{n\ast}_{\tilde{q}_j q^\prime \tilde{\chi}_l^0} \Bigg[
 \left( u\, t - m^2_{\tilde{q}_i}\, m^2_{\tilde{q}_j}\right)\,
 \delta_{mn} + \left( m_{\tilde{\chi}^0_k}\, m_{\tilde{\chi}^0_l}\,
 s \right)\, \left( 1-\delta_{mn} \right) \Bigg] ,~ \nonumber \\
 \widetilde{\mathcal{G}}_{mn} &=& \frac{2\, \pi\, \alpha_s^2}{9\,
 s^2}\, \left| \mathcal{C}^m_{\tilde{q}_i q \tilde{g}}\,
 \mathcal{C}^{n\ast}_{\tilde{q}_j q^\prime \tilde{g}}\right|^2
 \Bigg[ \left( u\, t - m^2_{\tilde{q}_i}\,
 m^2_{\tilde{q}_j}\right)\, \delta_{mn} + \left( m_{\tilde{g}}^2\,
 s \right)\, \left(1-\delta_{mn} \right) \Bigg],~ \nonumber\\
 \mathcal{[YZ]}_m &=&  \frac{\pi\, \alpha^2\, e_q\, e_{\tilde{q}}\,
 \delta_{ij}\, \delta_{qq^\prime}}{2\, s^2\, x_W (1-x_W)}\, {\rm
 Re} \left[ L_{\tilde{q}_i \tilde{q}_j Z} + R_{\tilde{q}_i
 \tilde{q}_j Z} \right]\, C^m_{q q^\prime Z} \left( u\, t -
 m^2_{\tilde{q}_i}\, m^2_{\tilde{q}_j^\prime}\right) ,~ \nonumber
 \\ \mathcal{[NY]}_m^k &=& \frac{2\, \pi\, \alpha^2\, e_q\,
 e_{\tilde{q}}\, \delta_{ij}\, \delta_{qq^\prime}}{3\, x_W\, (1 -
 x_W)\, s^2}\, {\rm Re} \left[\mathcal{C}^m_{\tilde{q}_i q
 \tilde{\chi}_k^0}\, \mathcal{C}^{m\ast}_{\tilde{q}_j q^\prime
 \tilde{\chi}_k^0} \right] \left( u\, t - m^2_{\tilde{q}_i}\,
 m^2_{\tilde{q}_j}\right),~ \nonumber
\eea
\bea
 \mathcal{[NZ]}_m^k &=&
 \frac{\pi\, \alpha^2}{6\, x_W^2\, (1 - x_W)^2\, s^2}\, {\rm Re}
 \left[\mathcal{C}^m_{\tilde{q}_i q \tilde{\chi}_k^0}\,
 \mathcal{C}^{m\ast}_{\tilde{q}_j q^\prime \tilde{\chi}_k^0} \left(
 L_{\tilde{q}_i \tilde{q}_j Z} + R_{\tilde{q}_i \tilde{q}_j Z}
 \right)\right]\, \mathcal{C}^m_{q q^\prime Z}\, \left( u\, t -
 m^2_{\tilde{q}_i}\, m^2_{\tilde{q}_j}\right),~ \nonumber\\
 \mathcal{[NG]}_m^k &=& \frac{8\, \pi\, \alpha\, \alpha_s\,
 \delta_{ij}\, \delta_{qq^\prime}}{9\, x_W\, (1 - x_W)\, s^2}\,
 {\rm Re} \left[\mathcal{C}^m_{\tilde{q}_i q \tilde{\chi}_k^0}\,
 \mathcal{C}^{m\ast}_{\tilde{q}_j q^\prime \tilde{\chi}_k^0}
 \right] \left( u\, t - m^2_{\tilde{q}_i}\,
 m^2_{\tilde{q}_j}\right),~ \nonumber\\
 \big[\widetilde{\mathcal{G}} \mathcal{G}\big]_m &=& - \frac{4\,
 \pi\, \alpha_s^2\, \delta_{ij}\, \delta_{qq^\prime}}{27\, s^2}\,
 {\rm Re} \left[\mathcal{C}^{m\ast}_{\tilde{q}_i q \tilde{g}}\,
 \mathcal{C}^m_{\tilde{q}_j q^\prime \tilde{g}} \right] \left( u\,
 t - m^2_{\tilde{q}_i}\, m^2_{\tilde{q}_j}\right),~ \nonumber\\
 \mathcal{[\widetilde{\mathcal{G}} Y]}_m &=& \frac{8\, \pi\,
 \alpha\, \alpha_s\, e_q\, e_{\tilde{q}}\, \delta_{ij}\,
 \delta_{qq^\prime}}{9\, s^2}\, {\rm Re}
 \left[\mathcal{C}^{m\ast}_{\tilde{q}_i q \tilde{g}}\,
 \mathcal{C}^m_{\tilde{q}_j q^\prime \tilde{g}} \right] \left( u\,
 t - m^2_{\tilde{q}_i}\, m^2_{\tilde{q}_j}\right),~ \nonumber\\
 \mathcal{[\widetilde{\mathcal{G}} Z]}_m &=& \frac{2\, \pi\,
 \alpha\, \alpha_s}{9\, x_W\, (1 - x_W)\, s^2}\, {\rm Re}
 \left[\mathcal{C}^{m\ast}_{\tilde{q}_i q \tilde{g}}\,
 \mathcal{C}^m_{\tilde{q}_j q^\prime \tilde{g}} \left(
 L_{\tilde{q}_i \tilde{q}_j Z} + R_{\tilde{q}_i \tilde{q}_j Z}
 \right)\right]\, \mathcal{C}^m_{q q^\prime Z}\, \left( u\, t -
 m^2_{\tilde{q}_i}\, m^2_{\tilde{q}_j}\right),
\eea
since only very few interferences (those between strong and electroweak
channels of the same propagator type) are eliminated due to colour
conservation. On the other hand, the gluon-initiated cross section
\bea
 \frac{{\rm d} \hat{\sigma}^{gg}_{h_a,
 h_b}}{dt} &=& \frac{\pi\alpha_s^2}{128 s^2} \left[ 24 \left(
 1-2\frac{t_{\tilde{q}_i} u_{\tilde{q}_i}}{s^2}\right) -
 \frac{8}{3}\right] \left[ (1-h_a h_b)-2 \frac{s
 m_{\tilde{q}_i}^2}{t_{\tilde{q}_i} u_{\tilde{q}_i}} \left( (1-h_a
 h_b) - \frac{s m_{\tilde{q}_i}^2}{t_{\tilde{q}_i} u_{\tilde{q}_i}}
 \right)\right]
\eea involves only the strong coupling constant and is thus quite
compact. In the case of cMFV, but diagonal or non-diagonal squark
helicity, our results agree with those in Ref.\
\cite{Bozzi:2005sy}. Diagonal production of identical
squark-antisquark mass eigenstates is, of course, dominated by the
strong quark-antiquark and gluon-gluon channels. Their relative
importance depends on the partonic luminosity and thus on the type
and energy of the hadron collider under consideration.
Non-diagonal production of squarks of different helicity or
flavour involves only electroweak and gluino-mediated
quark-antiquark scattering, and the relative importance of these
processes depends largely on the gluino mass.

\subsection{Squark Pair Production}

\begin{figure}
 \centering
 \includegraphics[width=0.75\columnwidth]{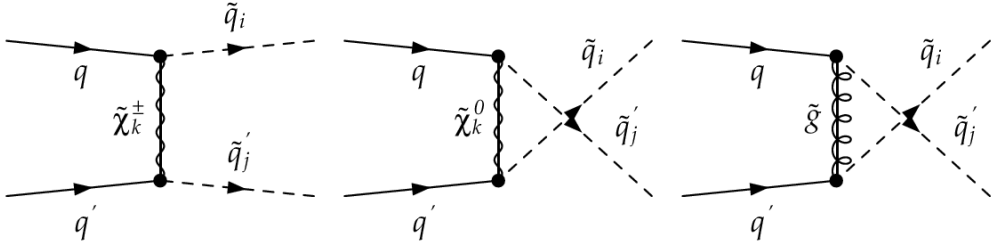}
 \caption{\label{fig:3}Tree-level Feynman diagrams for the production of
          one down-type squark ($\tilde{q}_i$) and one up-type squark
          ($\tilde{q}_j^\prime$) in the collision of an up-type quark ($q$)
          and a down-type quark ($q^\prime$).}
\end{figure}

While squark-antisquark pairs are readily produced in $p\bar{p}$ collisions,
e.g.\ at the Tevatron, from valence quarks and antiquarks, $pp$ colliders
have a larger quark-quark luminosity and will thus more easily lead to
squark pair production. The production of one down-type and one up-type
squark
\bea
 q(h_a,p_a)\,q'(h_b,p_b) \to \tilde{d}_i(p_1)\,\tilde{u}_j(p_2),
\eea
in the collision of an up-type quark $q$ and a down-type quark $q'$ proceeds
through the $t$-channel chargino or $u$-channel neutralino and gluino
exchanges shown in Fig.\ \ref{fig:3}. The corresponding cross section
%
%
\bea
 \frac{d\hat{\sigma}^{qq'}_{h_a, h_b}}{dt} &=& (1\! - \!h_a) (1\! - \!h_b)
 \Bigg[\bigg( \sum_{k=1,2 \atop l=1,2}\!
 \frac{\mathcal{C}^{kl}_{11}}{t_{\tilde{\chi}_k}\,
 t_{\tilde{\chi}_l}}\bigg) \! + \! \bigg( \sum_{k=1,\ldots,4\atop
 l=1,\ldots,4}\! \frac{\mathcal{N}_{11}^{kl}}
 {u_{\tilde{\chi}_k^0}\, u_{\tilde{\chi}_l^0}}\bigg) \! + \!
 \frac{\mathcal{G}_{11}}{u_{\tilde{g}}^2} \! + \!  \bigg(
 \sum_{k=1,2\atop l=1,\ldots,4}\!
 \frac{\mathcal{[CN]}^{kl}_{11}}{t_{\tilde{\chi}_k}\,
 u_{\tilde{\chi}_l^0}}\bigg) \! + \! \bigg( \sum_{k=1,2}\!
 \frac{\mathcal{[CG]}^k_{11}}{t_{\tilde{\chi}_k}\,
 u_{\tilde{g}}}\bigg) \Bigg] \nonumber \\ & \! + \!& (1\! + \!h_a)
 (1\! + \!h_b) \Bigg[ \bigg( \sum_{k=1,2 \atop l=1,2} \!
 \frac{\mathcal{C}^{kl}_{22}}{t_{\tilde{\chi}_k}\,
 t_{\tilde{\chi}_l}}\bigg) \! + \! \bigg(\sum_{k=1,\ldots,4 \atop
 l=1,\ldots,4} \!
 \frac{\mathcal{N}_{22}^{kl}}{u_{\tilde{\chi}_k^0}\,
 u_{\tilde{\chi}_l^0}}\bigg) \! + \!
 \frac{\mathcal{G}_{22}}{u_{\tilde{g}}^2} \! + \!  \bigg(
 \sum_{k=1,2\atop l=1,\ldots,4}\!
 \frac{\mathcal{[CN]}^{kl}_{22}}{t_{\tilde{\chi}_k}\,
 u_{\tilde{\chi}_l^0}}\bigg) \! + \! \bigg( \sum_{k=1,2}\!
 \frac{\mathcal{[CG]}^k_{22}}{t_{\tilde{\chi}_k}\,
 u_{\tilde{g}}}\bigg) \Bigg]\nonumber \\ &\! + \!& (1\! - \!h_a)
 (1\! + \!h_b) \Bigg[\bigg( \sum_{k=1,2 \atop l=1,2} \!
 \frac{\mathcal{C}^{kl}_{12}}{t_{\tilde{\chi}_k}\,
 t_{\tilde{\chi}_l}}\bigg) \! + \! \bigg(\sum_{k=1,\ldots,4 \atop
 l=1,\ldots,4}\!
 \frac{\mathcal{N}_{12}^{kl}}{u_{\tilde{\chi}_k^0}\,
 u_{\tilde{\chi}_l^0}}\bigg) \! + \!
 \frac{\mathcal{G}_{12}}{u_{\tilde{g}}^2} \! + \! \bigg(
 \sum_{k=1,2\atop l=1,\ldots,4}\!
 \frac{\mathcal{[CN]}^{kl}_{12}}{t_{\tilde{\chi}_k}\,
 u_{\tilde{\chi}_l^0}}\bigg) \! + \! \bigg( \sum_{k=1,2}\!
 \frac{\mathcal{[CG]}^k_{12}}{t_{\tilde{\chi}_k}\,
 u_{\tilde{g}}}\bigg)\Bigg] \nonumber \\ & \! + \!& (1\! + \!h_a)
 (1\! - \!h_b) \Bigg[\bigg( \sum_{k=1,2 \atop l=1,2} \!
 \frac{\mathcal{C}^{kl}_{21}}{t_{\tilde{\chi}_k}\,
 t_{\tilde{\chi}_l}}\bigg) \! + \! \bigg(\sum_{k=1,\ldots,4\atop
 l=1,\ldots,4}\!
 \frac{\mathcal{N}_{21}^{kl}}{u_{\tilde{\chi}_k^0}\,
 u_{\tilde{\chi}_l^0}}\bigg) \! + \!
 \frac{\mathcal{G}_{21}}{u_{\tilde{g}}^2} \! + \! \bigg(
 \sum_{k=1,2\atop l=1,\ldots,4}\!
 \frac{\mathcal{[CN]}^{kl}_{21}}{t_{\tilde{\chi}_k}\,
 u_{\tilde{\chi}_l^0}}\bigg) \! + \! \bigg( \sum_{k=1,2}\!
 \frac{\mathcal{[CG]}^k_{21}}{t_{\tilde{\chi}_k}\,
 u_{\tilde{g}}}\bigg) \Bigg]~~
\eea
involves the form factors
\bea
 \mathcal{C}_{mn}^{kl} &=& \frac{\pi\, \alpha^2}{4\, x_W^2\, s^2}
 \mathcal{C}^n_{\tilde{u}_j q^\prime \tilde{\chi}_k^\pm}\,
 \mathcal{C}^{m\ast}_{\tilde{d}_i q \tilde{\chi}_k^\pm}\,
 \mathcal{C}^{n\ast}_{\tilde{u}_j q^\prime \tilde{\chi}_l^\pm}\,
 \mathcal{C}^m_{\tilde{d}_i q \tilde{\chi}_l^\pm}\, \Bigg[ \left(
 u\, t - m^2_{\tilde{d}_i}\, m^2_{\tilde{u}_j}\right)
 \left(1-\delta_{mn} \right)  + m_{\tilde{\chi}^\pm_k}\,
 \,m_{\tilde{\chi}^\pm_l}\, s\, \delta_{mn} \Bigg],~ \nonumber\\
 \mathcal{N}_{mn}^{kl} &=& \frac{\pi\, \alpha^2}{x_W^2\, (1 -
 x_W)^2\, s^2}\mathcal{C}^{m\ast}_{\tilde{u}_j q
 \tilde{\chi}_k^0}\, \mathcal{C}^{n\ast}_{\tilde{d}_i q'
 \tilde{\chi}_k^0}\, \mathcal{C}^m_{\tilde{u}_j q
 \tilde{\chi}_l^0}\, \mathcal{C}^n_{\tilde{d}_i q'
 \tilde{\chi}_l^0}\, \Bigg[ \left( u\, t - m^2_{\tilde{d}_i}\,
 m^2_{\tilde{u}_j}\right) \left( 1-\delta_{mn} \right) +
 m_{\tilde{\chi}^0_k}\, m_{\tilde{\chi}^0_l}\, s\, \delta_{mn}
 \Bigg],~ \nonumber\\
 \mathcal{G}_{mn} &=& \frac{2\, \pi\,
 \alpha_s^2 }{9\, s^2} \left| \mathcal{C}^m_{\tilde{u}_j q
 \tilde{g}}\, \mathcal{C}^n_{\tilde{d}_i q' \tilde{g}}\right|^2
 \Bigg[ \left( u\, t - m^2_{\tilde{d}_i}\, m^2_{\tilde{u}_j}\right)
 \left( 1-\delta_{mn} \right) + m_{\tilde{g}}^2\, s\,\delta_{mn}
 \Bigg],~ \nonumber \\
 \mathcal{[CN]}^{kl}_{mn} &=& \frac{\pi\,
 \alpha^2}{3\, x_W^2\, (1 - x_W)\, s^2} {\rm Re} \left[
 \mathcal{C}^n_{\tilde{u}_j q^\prime \tilde{\chi}_k^\pm}\,
 \mathcal{C}^{m\ast}_{\tilde{d}_i q \tilde{\chi}_k^\pm}\,
 \mathcal{C}^m_{\tilde{u}_j q \tilde{\chi}_l^0}\,
 \mathcal{C}^n_{\tilde{d}_i q' \tilde{\chi}_l^0} \right] \Bigg[
 \left( u\, t - m^2_{\tilde{d}_i}\, m^2_{\tilde{u}_j}\right) \left(
 \delta_{mn}-1 \right) + m_{\tilde{\chi}^\pm_k}\,
 m_{\tilde{\chi}^0_l}\, s\, \delta_{mn} \Bigg],~~  \nonumber \\
 \mathcal{[CG]}^k_{mn} &=& \frac{4\, \pi\, \alpha\, \alpha_s}{9\,
 s^2\, x_W} {\rm Re} \left[ \mathcal{C}^n_{\tilde{u}_j q^\prime
 \tilde{\chi}_k^\pm}\, \mathcal{C}^{m\ast}_{\tilde{d}_i q
 \tilde{\chi}_k^\pm}\, \mathcal{C}^{m\ast}_{\tilde{u}_j q
 \tilde{g}}\, \mathcal{C}^{n\ast}_{\tilde{d}_i q' \tilde{g}} \right]
 \Bigg[ \left( u\, t - m^2_{\tilde{d}_i}\, m^2_{\tilde{u}_j}\right)
 \left( \delta_{mn} - 1 \right) + m_{\tilde{\chi}^\pm_k}\,
 m_{\tilde{g}}\, s\, \delta_{mn} \Bigg],
\eea where the neutralino-gluino interference term is absent due
to colour conservation. The cross section for the charge-conjugate
production of antisquarks from antiquarks can be obtained from the
equations above by replacing $h_{a,b}\to-h_{a,b}$. Heavy-flavour
final states are completely absent in cMFV due to the negligible
top quark and small bottom quark densities in the proton and can
thus only be obtained in NMFV.

\begin{figure}
 \centering
 \includegraphics[width=\columnwidth]{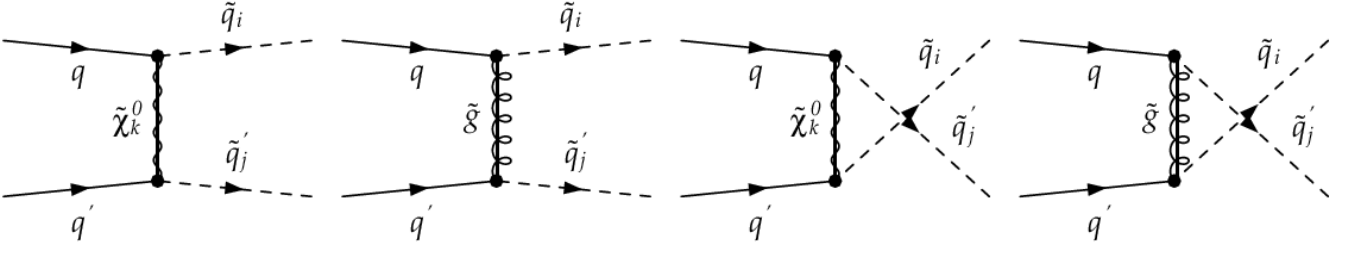}
 \caption{\label{fig:4}Tree-level Feynman diagrams for the production of
          two up-type or down-type squarks.}
\end{figure}

The Feynman diagrams for pair production of two up- or down-type squarks
\bea
 q(h_a,p_a)\, q'(h_b,p_b) &\to&\tilde{q}_i(p_1)\, \tilde{q}_j(p_2)
\eea
are shown in Fig.\ \ref{fig:4}. In NMFV, neutralino and gluino exchanges can
%
%
lead to identical squark flavours for different quark initial states, so
that both $t$- and $u$-channels contribute and may interfere. The cross
section
\bea
 \frac{d\hat{\sigma}^{qq'}_{h_a, h_b}}{dt} &=& (1-h_a) (1-h_b) \Bigg[\bigg(
 \sum_{k=1,\ldots,4\atop l=1,\ldots,4}
 \frac{[\mathcal{NT}]_{11}^{kl}} {t_{\tilde{\chi}_k^0}\,
 t_{\tilde{\chi}_l^0}} + \frac{[\mathcal{NU}]_{11}^{kl}}
 {u_{\tilde{\chi}_k^0}\, u_{\tilde{\chi}_l^0}} +
 \frac{[\mathcal{NTU}]_{11}^{kl}} {t_{\tilde{\chi}_k^0}\,
 u_{\tilde{\chi}_l^0}} \bigg) +
 \frac{[\mathcal{GT}]_{11}}{t_{\tilde{g}}^2} +
 \frac{[\mathcal{GU}]_{11}}{u_{\tilde{g}}^2}+
 \frac{[\mathcal{GTU]}_{11}}{u_{\tilde{g}}t_{\tilde{g}}}\nonumber
 \\ &+& \bigg( \sum_{k=1,\ldots,4} \frac{[\mathcal{NGA}]_{11}^{k}}
 {t_{\tilde{\chi}_k^0}\, u_{\tilde{g}}} +
 \frac{[\mathcal{NGB}]_{11}^{k}} {u_{\tilde{\chi}_k^0}\,
 t_{\tilde{g}}} \bigg)\Bigg]\frac{1}{1+\delta_{ij}} \nonumber \\
 &+& (1+h_a) (1+h_b) \Bigg[\bigg( \sum_{k=1,\ldots,4\atop
 l=1,\ldots,4} \frac{[\mathcal{NT}]_{22}^{kl}}
 {t_{\tilde{\chi}_k^0}\, t_{\tilde{\chi}_l^0}} +
 \frac{[\mathcal{NU}]_{22}^{kl}} {u_{\tilde{\chi}_k^0}\,
 u_{\tilde{\chi}_l^0}} + \frac{[\mathcal{NTU}]_{22}^{kl}}
 {t_{\tilde{\chi}_k^0}\, u_{\tilde{\chi}_l^0}}\bigg) +
 \frac{[\mathcal{GT]}_{22}}{t_{\tilde{g}}^2} +
 \frac{[\mathcal{GU]}_{22}}{u_{\tilde{g}}^2}+
 \frac{[\mathcal{GTU]}_{22}}{u_{\tilde{g}}t_{\tilde{g}}}\nonumber
 \\ &+& \bigg( \sum_{k=1,\ldots,4} \frac{[\mathcal{NGA}]_{22}^{k}}
 {t_{\tilde{\chi}_k^0}\, u_{\tilde{g}}} +
 \frac{[\mathcal{NGB}]_{22}^{k}} {u_{\tilde{\chi}_k^0}\,
 t_{\tilde{g}}} \bigg)\Bigg]\frac{1}{1+\delta_{ij}} \nonumber \\
 &+& (1-h_a) (1+h_b) \Bigg[\bigg( \sum_{k=1,\ldots,4\atop
 l=1,\ldots,4} \frac{[\mathcal{NT}]_{12}^{kl}}
 {t_{\tilde{\chi}_k^0}\, t_{\tilde{\chi}_l^0}} +
 \frac{[\mathcal{NU}]_{12}^{kl}} {u_{\tilde{\chi}_k^0}\,
 u_{\tilde{\chi}_l^0}} + \frac{[\mathcal{NTU}]_{12}^{kl}}
 {t_{\tilde{\chi}_k^0}\, u_{\tilde{\chi}_l^0}} \bigg) +
 \frac{[\mathcal{GT]}_{12}}{t_{\tilde{g}}^2} +
 \frac{[\mathcal{GU]}_{12}}{u_{\tilde{g}}^2}+
 \frac{[\mathcal{GTU]}_{12}}{u_{\tilde{g}}t_{\tilde{g}}}\nonumber
 \\ &+& \bigg( \sum_{k=1,\ldots,4} \frac{[\mathcal{NGA}]_{12}^{k}}
 {t_{\tilde{\chi}_k^0}\, u_{\tilde{g}}} +
 \frac{[\mathcal{NGB}]_{12}^{k}} {u_{\tilde{\chi}_k^0}\,
 t_{\tilde{g}}} \bigg)\Bigg]\frac{1}{1+\delta_{ij}} \nonumber \\
 &+& (1+h_a) (1-h_b) \Bigg[\bigg( \sum_{k=1,\ldots,4\atop
 l=1,\ldots,4} \frac{[\mathcal{NT}]_{21}^{kl}}
 {t_{\tilde{\chi}_k^0}\, t_{\tilde{\chi}_l^0}} +
 \frac{[\mathcal{NU}]_{21}^{kl}} {u_{\tilde{\chi}_k^0}\,
 u_{\tilde{\chi}_l^0}} + \frac{[\mathcal{NTU}]_{21}^{kl}}
 {t_{\tilde{\chi}_k^0}\, u_{\tilde{\chi}_l^0}} \bigg) +
 \frac{[\mathcal{GT]}_{21}}{t_{\tilde{g}}^2} +
 \frac{[\mathcal{GU]}_{21}}{u_{\tilde{g}}^2} +
 \frac{[\mathcal{GTU]}_{21}}{u_{\tilde{g}}t_{\tilde{g}}}\nonumber
 \\ &+& \bigg( \sum_{k=1,\ldots,4} \frac{[\mathcal{NGA}]_{21}^{k}}
 {t_{\tilde{\chi}_k^0}\, u_{\tilde{g}}} +
 \frac{[\mathcal{NGB}]_{21}^{k}} {u_{\tilde{\chi}_k^0}\,
 t_{\tilde{g}}} \bigg)\Bigg] \frac{1}{1+\delta_{ij}}
\eea
depends therefore on the form factors
\bea [\mathcal{NT}]_{mn}^{kl} &=& \frac{\pi\, \alpha^2}{x_W^2\, (1
 - x_W)^2\, s^2}\mathcal{C}^{n\ast}_{\tilde{q}_j q^\prime
 \tilde{\chi}_k^0}\, \mathcal{C}^{m\ast}_{\tilde{q}_i q
 \tilde{\chi}_k^0}\, \mathcal{C}^n_{\tilde{q}_j q^\prime
 \tilde{\chi}_l^0}\, \mathcal{C}^m_{\tilde{q}_i q
 \tilde{\chi}_l^0}\, \Bigg[ \left( u\, t - m^2_{\tilde{q}_i}\,
 m^2_{\tilde{q}_j}\right) \left( 1-\delta_{mn} \right) +
 m_{\tilde{\chi}^0_k}\, m_{\tilde{\chi}^0_l}\, s\, \delta_{mn}
 \Bigg],~\nonumber\\ \large[\mathcal{NU}\large]_{mn}^{kl} &=&
 \frac{\pi\, \alpha^2}{x_W^2\, (1 - x_W)^2\,
 s^2}\mathcal{C}^{n\ast}_{\tilde{q}_i q^\prime \tilde{\chi}_k^0}\,
 \mathcal{C}^{m\ast}_{\tilde{q}_j q \tilde{\chi}_k^0}\,
 \mathcal{C}^n_{\tilde{q}_i q^\prime \tilde{\chi}_l^0}\,
 \mathcal{C}^m_{\tilde{q}_j q \tilde{\chi}_l^0}\, \Bigg[ \left( u\,
 t - m^2_{\tilde{q}_i}\, m^2_{\tilde{q}_j}\right) \left(
 1-\delta_{mn} \right) + m_{\tilde{\chi}^0_k}\,
 m_{\tilde{\chi}^0_l}\, s\, \delta_{mn} \Bigg],~ \nonumber\\
 \large[\mathcal{NTU}\large]_{mn}^{kl} &=& \frac{2\,\pi\,
 \alpha^2}{3\, x_W^2\, (1 - x_W)^2\, s^2} {\rm Re} \left[
 \mathcal{C}^{m\ast}_{\tilde{q}_i q \tilde{\chi}_k^0}\,
 \mathcal{C}^{n\ast}_{\tilde{q}_j q^\prime \tilde{\chi}_k^0}\,
 \mathcal{C}^n_{\tilde{q}_i q^\prime \tilde{\chi}_l^0}\,
 \mathcal{C}^m_{\tilde{q}_j q \tilde{\chi}_l^0}\right]\, \Bigg[
 \left( u\, t - m^2_{\tilde{q}_i}\, m^2_{\tilde{q}_j}\right) \left(
 1-\delta_{mn} \right) + m_{\tilde{\chi}^0_k}\,
 m_{\tilde{\chi}^0_l}\, s\, \delta_{mn} \Bigg],~ \nonumber\\
 \large[\mathcal{GT}\large]_{mn} &=& \frac{2\, \pi\, \alpha_s^2
 }{9\, s^2} \left| \mathcal{C}^n_{\tilde{q}_j q^\prime \tilde{g}}\,
 \mathcal{C}^m_{\tilde{q}_i q \tilde{g}}\right|^2 \Bigg[ \left( u\,
 t - m^2_{\tilde{q}_i}\, m^2_{\tilde{q}_j}\right) \left(
 1-\delta_{mn} \right) + m_{\tilde{g}}^2\, s\,\delta_{mn}  \Bigg],~
 \nonumber \\ \large[\mathcal{GU}\large]_{mn} &=& \frac{2\, \pi\,
 \alpha_s^2 }{9\, s^2} \left| \mathcal{C}^m_{\tilde{q}_i q^\prime
 \tilde{g}}\, \mathcal{C}^n_{\tilde{q}_j q \tilde{g}}\right|^2
 \Bigg[ \left( u\, t - m^2_{\tilde{q}_i}\, m^2_{\tilde{q}_j}\right)
 \left( 1-\delta_{mn} \right) + m_{\tilde{g}}^2\, s\,\delta_{mn}
 \Bigg],~ \nonumber \\ \large[\mathcal{GTU}\large]_{mn} &=&
 \frac{-4\, \pi\, \alpha_s^2 }{27\, s^2} {\rm Re} \left[
 \mathcal{C}^m_{\tilde{q}_i q \tilde{g}}\,
 \mathcal{C}^n_{\tilde{q}_j q^\prime \tilde{g}} \mathcal{C}^{m
 \ast}_{\tilde{q}_i q^\prime \tilde{g}}\, \mathcal{C}^{n
 \ast}_{\tilde{q}_j q \tilde{g}} \right] \Bigg[ \left( u\, t -
 m^2_{\tilde{q}_i}\, m^2_{\tilde{q}_j}\right) \left( 1-\delta_{mn}
 \right) + m_{\tilde{g}}^2\, s\,\delta_{mn}  \Bigg],~ \nonumber \\
 \large[\mathcal{NGA}\large]_{mn}^k &=& \frac{8\, \pi\, \alpha
 \alpha_s}{9\, s^2\, x_W\, (1 - x_W)} {\rm Re} \left[
 \mathcal{C}^{n\ast}_{\tilde{q}_j q^\prime \tilde{\chi}_k^0}\,
 \mathcal{C}^{m\ast}_{\tilde{q}_i q \tilde{\chi}_k^0}\,
 \mathcal{C}^{m \ast}_{\tilde{q}_i q^\prime \tilde{g}}\,
 \mathcal{C}^{n \ast}_{\tilde{q}_j q \tilde{g}} \right] \Bigg[
 \left( u\, t - m^2_{\tilde{q}_i}\, m^2_{\tilde{q}_j}\right) \left(
 1-\delta_{mn} \right) + m_{\tilde{\chi}^0_k}\, m_{\tilde{g}}\,
 s\,\delta_{mn}  \Bigg],~ \nonumber \\
 \large[\mathcal{NGB}\large]_{mn}^k &=& \frac{8\, \pi\, \alpha
 \alpha_s}{9\, s^2\, x_W\, (1 - x_W)} {\rm Re} \left[
 \mathcal{C}^{n\ast}_{\tilde{q}_i q^\prime \tilde{\chi}_k^0}\,
 \mathcal{C}^{m\ast}_{\tilde{q}_j q
 \tilde{\chi}_k^0}\,\mathcal{C}^{n \ast}_{\tilde{q}_j q^\prime
 \tilde{g}}\, \mathcal{C}^{m\ast}_{\tilde{q}_i q \tilde{g}} \right]
 \Bigg[ \left( u\, t - m^2_{\tilde{q}_i}\, m^2_{\tilde{q}_j}\right)
 \left( 1-\delta_{mn} \right) + m_{\tilde{\chi}^0_k}\,
 m_{\tilde{g}}\, s\,\delta_{mn} \Bigg].
\eea
Gluinos will dominate over neutralino exchanges due to their strong
coupling, and the two will only interfere in the mixed $t$- and $u$-channels
due to colour conservation. At the LHC, up-type squark pair production
should dominate over mixed up-/down-type squark production and down-type
squark pair production, since the proton contains two valence up-quarks and
only one valence down-quark. As before, the charge-conjugate production of
antisquark pairs is obtained by making the replacement $h_{a,b}\to-h_{a,b}$.
If we neglect electroweak contributions as well as squark flavour and
helicity mixing and sum over left- and right-handed squark states, our
results agree with those of Ref.\ \cite{Dawson:1983fw}.

\subsection{Associated Production of Squarks and Gauginos}

\begin{figure}
 \centering
 \includegraphics[width=0.5\columnwidth]{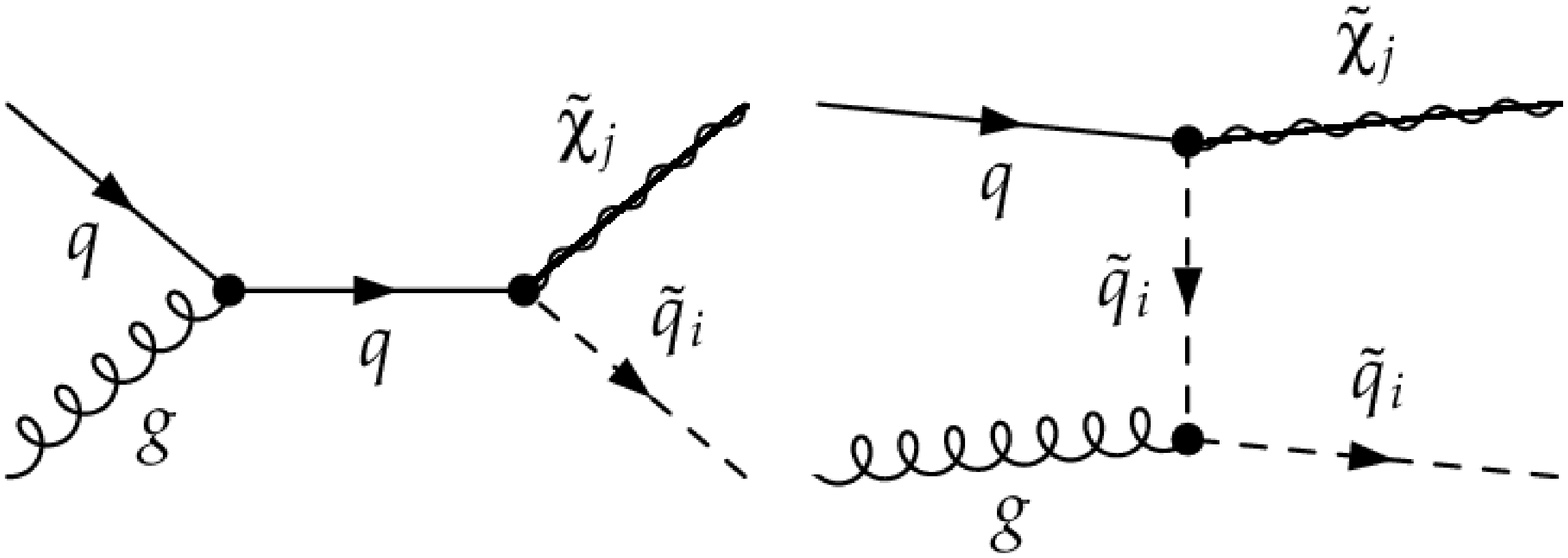}
 \caption{\label{fig:5}Tree-level Feynman diagrams for the associated
          production of squarks and gauginos.}
\end{figure}

The associated production of squarks and neutralinos or charginos
\bea
 q(h_a,p_a)\, g(h_b,p_b) \to\tilde{\chi}_j(p_1)\, \tilde{q}_i(p_2)
\eea
is a semi-weak process that originates from quark-gluon initial states
and has both an $s$-channel quark and a $t$-channel squark contribution.
They involve both a quark-squark-gaugino vertex that can in general be
flavour violating. The corresponding Feynman diagrams can be seen in Fig.\
\ref{fig:5}. The squark-gaugino cross section
%
%
\bea
 \frac{d \hat{\sigma}^{qg}_{h_a, h_b}}{dt} &=&
 \frac{\pi\, \alpha\, \alpha_s}{n_{\tilde\chi}\, s^2} \Bigg\{
 \frac{-u_{\tilde{\chi}_j}}{s} \bigg[(1-h_a)(1-h_b) \left|
 L_{\tilde{q}_i q \tilde{\chi}_j}\right|^2 + (1+h_a)(1+h_b)
 \left| R_{\tilde{q}_i q \tilde{\chi}_j}\right|^2\bigg] \nonumber
 \\ &+& \frac{t_{\tilde{\chi}_j}\left( t + m^2_{\tilde{q}_i}
 \right)}{t_{\tilde{q}_i}^2} \bigg[(1-h_a)\left| L_{\tilde{q}_i q
 \tilde{\chi}_j}\right|^2 + (1+h_a) \left| R_{\tilde{q}_i q
 \tilde{\chi}_j}\right|^2\bigg] \nonumber \\ & +& \frac{2\,(u\, t
 -m^2_{\tilde{q}_i}\, m_{\tilde{\chi}_j}^2)}{s\, t_{\tilde{q}_i}}
 \bigg[(1-h_a)(1-h_b) \left| L_{\tilde{q}_i q
 \tilde{\chi}_j}\right|^2 + (1+h_a)(1+ h_b) \left|
 R_{\tilde{q}_i q \tilde{\chi}_j}\right|^2\bigg] \nonumber \\
 &+& \frac{t_{\tilde{\chi}_j} (t_{\tilde{\chi}_j} -
 u_{\tilde{q}_i}) }{s\, t_{\tilde{q}_i}} \bigg[(1-h_a)\left|
 L_{\tilde{q}_i q \tilde{\chi}_j}\right|^2 + (1+h_a) \left|
 R_{\tilde{q}_i q \tilde{\chi}_j}\right|^2\bigg] \Bigg\},
\eea
where $n_{\tilde\chi}=6x_W(1-x_W)$ for neutralinos and $n_{\tilde\chi}=12
x_W$ for charginos, is sufficiently compact to be written without the
definition of form factors. Note that the $t$-channel diagram involves the
coupling of the gluon to scalars and does thus not depend on its helicity
$h_b$. The cross section of the charge-conjugate process can be obtained by
taking $h_a\to-h_a$. Third-generation squarks can only be produced in NMFV,
preferably through a light (valence) quark in the $s$-channel. For
non-mixing squarks and gauginos, we agree again with the results of Ref.\
\cite{Dawson:1983fw}.

\subsection{Gaugino Pair Production}

\begin{figure}
 \centering
 \includegraphics[width=0.75\columnwidth]{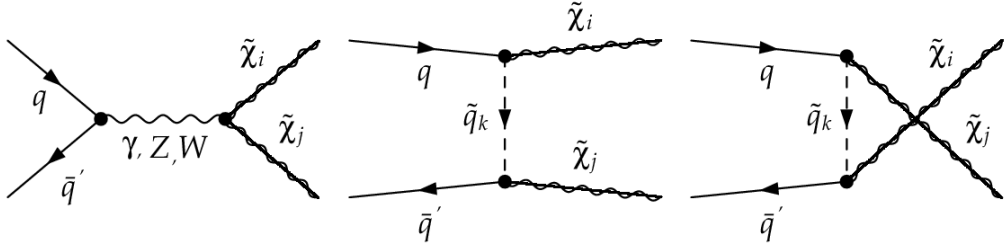}
 \caption{\label{fig:6}Tree-level Feynman diagrams for the production
          of gaugino pairs.}
\end{figure}

Finally, we consider the purely electroweak production of gaugino pairs
\bea
 q(h_a,p_a)\, \bar{q}^\prime(h_b,p_b) \to
 \tilde{\chi}_i(p_1)\, \tilde{\chi}_j(p_2)
\eea
from quark-antiquark initial states, where flavour violation can occur
via the quark-squark-gaugino vertices in the $t$- and $u$-channels (see
Fig.\ \ref{fig:6}). However, if it were not for different parton density
weights, summation over complete squark multiplet exchanges would make these
channels insensitive to the exchanged squark flavour. Furthermore there are
no final state squarks that could be experimentally tagged. The cross
section can be expressed generically as
%
%
\bea
 \frac{d \hat{\sigma}^{q\bar{q}'}_{h_a, h_b}}{dt} &=&
 \frac{\pi \alpha^2}{3 s^2}(1-h_a) (1+h_b) \Big[ \left| Q^u_{LL}
 \right|^2 u_{\tilde{\chi}_i} u_{\tilde{\chi}_j} + \left| Q_{LL}^t
 \right|^2 t_{\tilde{\chi}_i} t_{\tilde{\chi}_j} + 2 {\rm Re}
 [Q_{LL}^{u\ast} Q_{LL}^t] m_{\tilde{\chi}_{i}}
 m_{\tilde{\chi}_{j}} s \Big] \nonumber \\ & +& \frac{\pi
 \alpha^2}{3 s^2}(1+h_a) (1-h_b) \Big[ \left| Q_{RR}^u \right|^2
 u_{\tilde{\chi}_i} u_{\tilde{\chi}_j} + \left| Q_{RR}^t \right|^2
 t_{\tilde{\chi}_i} t_{\tilde{\chi}_j} + 2 {\rm Re} [Q_{RR}^{u\ast}
 Q_{RR}^t] m_{\tilde{\chi}_{i}} m_{\tilde{\chi}_{j}} s
 \Big]\nonumber \\& +& \frac{\pi \alpha^2}{3 s^2}(1+h_a) (1+h_b)
 \Big[ \left| Q_{RL}^u \right|^2 u_{\tilde{\chi}_i}
 u_{\tilde{\chi}_j} + \left| Q_{RL}^t \right|^2 t_{\tilde{\chi}_i}
 t_{\tilde{\chi}_j} + {\rm Re} [Q_{RL}^{u\ast} Q_{RL}^t] (u t -
 m^2_{\tilde{\chi}_{i}} m^2_{\tilde{\chi}_{j}}) \Big]\nonumber \\ &
 +& \frac{\pi \alpha^2}{3 s^2}(1-h_a) (1-h_b) \Big[ \left| Q_{LR}^u
 \right|^2 u_{\tilde{\chi}_i} u_{\tilde{\chi}_j} + \left| Q_{LR}^t
 \right|^2 t_{\tilde{\chi}_i} t_{\tilde{\chi}_j} + {\rm Re}
 [Q_{LR}^{u\ast} Q_{LR}^t] (u t - m^2_{\tilde{\chi}_{i}}
 m^2_{\tilde{\chi}_{j}}) \Big],
 \label{eq:xsecgg}
\eea
i.e.\ in terms of generalized charges.
For $\tilde{\chi}_i^-\tilde{\chi}_j^+$-production, these charges are given
by
\bea
 Q_{LL}^{u-+} &=& \Bigg(\frac{e_q \delta_{ij}
 \delta_{qq^\prime}}{s} - \frac{L_{q q^\prime Z} O^{\prime
 R\ast}_{ij}}{2 \, x_W \, (1-x_W) \,s_z} + \sum_{k=1}^6
 \frac{L_{\tilde{d}_k q^\prime \tilde{\chi}_i^\pm}
 L^\ast_{\tilde{d}_k q \tilde{\chi}_j^\pm}}{2\, x_W\,
 u_{\tilde{d}_k}}\Bigg) ,~\nonumber\\ Q_{LL}^{t-+} &=&
 \Bigg(\frac{e_q \delta_{ij} \delta_{qq^\prime}}{s} - \frac{L_{q
 q^\prime Z} O^{\prime L\ast}_{ij}}{2 \,x_W \,(1-x_W) \,s_z} -
 \sum_{k=1}^6 \frac{L^\ast_{\tilde{u}_k q^\prime
 \tilde{\chi}_j^\pm} L_{\tilde{u}_k q \tilde{\chi}_i^\pm}}{2
 \,x_W\, t_{\tilde{u}_k}}\Bigg),~ \nonumber\\ Q_{RR}^{u-+} &=&
 \Bigg(\frac{e_q \delta_{ij} \delta_{qq^\prime}}{s} - \frac{R_{q
 q^\prime Z} O^{\prime L\ast}_{ij}}{2 \, x_W \, (1-x_W) \,s_z}+
 \sum_{k=1}^6 \frac{R_{\tilde{d}_k q^\prime
 \tilde{\chi}_i^\pm} R^\ast_{\tilde{d}_k q \tilde{\chi}_j^\pm}}{2\,
 x_W\, u_{\tilde{d}_k}}\Bigg),~ \nonumber\\ Q_{RR}^{t-+} &=&
 \Bigg(\frac{e_q \delta_{ij} \delta_{qq^\prime}}{s} - \frac{R_{q
 q^\prime Z} O^{\prime R\ast}_{ij}}{2 \,x_W \,(1-x_W) \,s_z} -
 \sum_{k=1}^6 \frac{R^\ast_{\tilde{u}_k q^\prime
 \tilde{\chi}_j^\pm} R_{\tilde{u}_k q \tilde{\chi}_i^\pm}}{2
 \,x_W\, t_{\tilde{u}_k}}\Bigg),~ \nonumber\\ Q_{LR}^{u-+} &=&
 \sum_{k=1}^6 \frac{R_{\tilde{d}_k q^\prime
 \tilde{\chi}_i^\pm} L^\ast_{\tilde{d}_k q \tilde{\chi}_j^\pm}}{2\,
 x_W\, u_{\tilde{d}_k}} ,~ \nonumber\\ Q_{LR}^{t-+} &=&
 \sum_{k=1}^6 \frac{R^\ast_{\tilde{u}_k q^\prime
 \tilde{\chi}_j^\pm} L_{\tilde{u}_k q \tilde{\chi}_i^\pm}}{2
 \,x_W\, t_{\tilde{u}_k}},~ \nonumber\\ Q_{RL}^{u-+} &=&
 \sum_{k=1}^6 \frac{L_{\tilde{d}_k q^\prime
 \tilde{\chi}_i^\pm} R^\ast_{\tilde{d}_k q \tilde{\chi}_j^\pm}}{2\,x_W\,
 u_{\tilde{d}_k}} ,~ \nonumber\\ Q_{RL}^{t-+} &=&\sum_{k=1}^6
 \frac{L^\ast_{\tilde{u}_k q^\prime \tilde{\chi}_j^\pm}
 R_{\tilde{u}_k q \tilde{\chi}_i^\pm}}{2 \,x_W \,
 t_{\tilde{u}_k}}.
\eea
Note that there is no interference between $t$- and $u$-channel diagrams due
to (electromagnetic) charge conservation.
The cross section for chargino-pair production in $e^+e^-$-collisions can be
deduced by setting $e_q \to e_l = -1$, $L_{q q^\prime Z}\to L_{e e Z} =
(2\,T^{3}_l - 2\,e_l\,x_W)$ and $R_{q q^\prime Z}\to R_{e e Z} = - 2\,e_l\,
x_W$. Neglecting all Yukawa couplings, we can then reproduce the
calculations of Ref.\ \cite{Choi:1998ei}.

The charges of the chargino-neutralino associated production are given by
\bea
 Q_{LL}^{u+0} &=&
 \frac{1}{\sqrt{2\,(1-x_W)}\, x_W} \left[ \frac{O^{L\ast}_{ji}
 L^\ast_{qq^\prime W}}{\sqrt{2}\,s_w} + \sum_{k=1}^6
 \frac{L_{\tilde{u}_k q^\prime \tilde{\chi}_i^\pm}^\ast
 L_{\tilde{u}_k q \tilde{\chi}_j^0}^\ast}{u_{\tilde{u}_k}}
 \right],~\nonumber \\ Q_{LL}^{t+0} &=&
 \frac{1}{\sqrt{2\,(1-x_W)}\, x_W} \left[ \frac{O^{R\ast}_{ji}
 L^\ast_{qq^\prime W}}{\sqrt{2}\,s_w} - \sum_{k=1}^6
 \frac{L_{\tilde{d}_k q \tilde{\chi}_i^\pm}^\ast L_{\tilde{d}_k
 q^\prime \tilde{\chi}_j^0}}{t_{\tilde{d}_k}} \right],~\nonumber \\
 Q_{RR}^{u+0} &=& \frac{1}{\sqrt{2\,(1-x_W)}\, x_W} \sum_{k=1}^6
 \frac{R_{\tilde{u}_k q^\prime \tilde{\chi}_i^\pm}^\ast
 R_{\tilde{u}_k q \tilde{\chi}_j^0}^\ast}{u_{\tilde{u}_k}} ,~
 \nonumber \\ Q_{RR}^{t+0} &=& \frac{-1}{\sqrt{2\,(1-x_W)}\, x_W}
 \sum_{k=1}^6 \frac{R_{\tilde{d}_k q \tilde{\chi}_i^\pm}^\ast
 R_{\tilde{d}_k q^\prime \tilde{\chi}_j^0}}{t_{\tilde{d}_k}} ,~
 \nonumber\\
 Q_{LR}^{u+0} &=& \frac{1}{\sqrt{2\,(1-x_W)}\, x_W}
 \sum_{k=1}^6 \frac{R_{\tilde{u}_k q^\prime
 \tilde{\chi}_i^\pm}^\ast L_{\tilde{u}_k q
 \tilde{\chi}_j^0}^\ast}{u_{\tilde{u}_k}} ,~ \nonumber \\
 Q_{LR}^{t+0} &=& \frac{1}{\sqrt{2\,(1-x_W)}\, x_W} \sum_{k=1}^6
 \frac{L_{\tilde{d}_k q \tilde{\chi}_i^\pm}^\ast R_{\tilde{d}_k
 q^\prime \tilde{\chi}_j^0}}{t_{\tilde{d}_k}} ,~ \nonumber \\
 Q_{RL}^{u+0} &=& \frac{1}{\sqrt{2\,(1-x_W)}\, x_W} \sum_{k=1}^6
 \frac{L_{\tilde{u}_k q^\prime \tilde{\chi}_i^\pm}^\ast
 R_{\tilde{u}_k q \tilde{\chi}_j^0}^\ast}{u_{\tilde{u}_k}} ,~
 \nonumber \\ Q_{RL}^{t+0} &=& \frac{1}{\sqrt{2\,(1-x_W)}\, x_W}
 \sum_{k=1}^6 \frac{R_{\tilde{d}_k q \tilde{\chi}_i^\pm}^\ast
 L_{\tilde{d}_k q^\prime \tilde{\chi}_j^0}}{t_{\tilde{d}_k}}.
\eea
The charge-conjugate process is again obtained by making the replacement
$h_{a,b}\to -h_{a,b}$
in Eq.\ (\ref{eq:xsecgg}). In the case of non-mixing squarks with neglected
Yukawa couplings, we agree with the results of Ref.\
\cite{Beenakker:1999xh}, provided we correct a sign in their Eq.\ (2) as
described in Ref.\ \cite{Spira:2000vf}.

Finally, the charges for the neutralino pair production are given by
\bea
 Q_{LL}^{u00} &=&
 \frac{1}{x_W\,(1-x_W)\,\sqrt{1+ \delta_{ij}}} \left[ \frac{L_{q
 q^\prime Z} O^{\prime\prime L}_{ij} }{2 s_z} + \sum_{k=1}^6
 \frac{L_{\tilde{Q}_k q^\prime \tilde{\chi}_i^0} L_{\tilde{Q}_k q
 \tilde{\chi}_j^0}^\ast}{u_{\tilde{Q}_k}} \right] ,~ \nonumber \\
 Q_{LL}^{t00} &=& \frac{1}{x_W\,(1-x_W)\,\sqrt{1+ \delta_{ij}}}
 \left[ \frac{L_{q q^\prime Z} O^{\prime\prime R}_{ij} }{2 s_z} -
 \sum_{k=1}^6 \frac{L_{\tilde{Q}_k q \tilde{\chi}_i^0}^\ast
 L_{\tilde{Q}_k q^\prime \tilde{\chi}_j^0}}{t_{\tilde{Q}_k}}
 \right] ,~ \nonumber \\ Q_{RR}^{u00} &=&
 \frac{1}{x_W\,(1-x_W)\,\sqrt{1+ \delta_{ij}}} \left[ \frac{R_{q
 q^\prime Z} O^{\prime\prime R}_{ij} }{2 s_z} + \sum_{k=1}^6
 \frac{R_{\tilde{Q}_k q^\prime \tilde{\chi}_i^0} R_{\tilde{Q}_k q
 \tilde{\chi}_j^0}^\ast}{u_{\tilde{Q}_k}} \right] ,~ \nonumber \\
 Q_{RR}^{t00} &=& \frac{1}{x_W\,(1-x_W)\,\sqrt{1+ \delta_{ij}}}
 \left[ \frac{R_{q q^\prime Z} O^{\prime\prime L}_{ij} }{2 s_z} -
 \sum_{k=1}^6 \frac{R_{\tilde{Q}_k q \tilde{\chi}_i^0}^\ast
 R_{\tilde{Q}_k q^\prime \tilde{\chi}_j^0}}{t_{\tilde{Q}_k}}
 \right] ,~\nonumber \\ Q_{LR}^{u00} &=&
 \frac{1}{x_W\,(1-x_W)\,\sqrt{1+ \delta_{ij}}} \sum_{k=1}^6
 \frac{R_{\tilde{Q}_k q^\prime \tilde{\chi}_i^0} L_{\tilde{Q}_k q
 \tilde{\chi}_j^0}^\ast}{u_{\tilde{Q}_k}} ,~\nonumber \\
 Q_{LR}^{t00} &=& \frac{1}{x_W\,(1-x_W)\,\sqrt{1+ \delta_{ij}}}
 \sum_{k=1}^6 \frac{L_{\tilde{Q}_k q \tilde{\chi}_i^0}^\ast
 R_{\tilde{Q}_k q^\prime \tilde{\chi}_j^0}}{t_{\tilde{Q}_k}} ,~
 \nonumber \\ Q_{RL}^{u00} &=& \frac{1}{x_W\,(1-x_W)\,\sqrt{1+
 \delta_{ij}}} \sum_{k=1}^6 \frac{L_{\tilde{Q}_k q^\prime
 \tilde{\chi}_i^0} R_{\tilde{Q}_k q
 \tilde{\chi}_j^0}^\ast}{u_{\tilde{Q}_k}} ,~ \nonumber \\
 Q_{RL}^{t00} &=& \frac{1}{x_W\,(1-x_W)\,\sqrt{1+ \delta_{ij}}}
 \sum_{k=1}^6 \frac{R_{\tilde{Q}_k q \tilde{\chi}_i^0}^\ast
 L_{\tilde{Q}_k q^\prime \tilde{\chi}_j^0}}{t_{\tilde{Q}_k}},
\eea
which agrees with the results of Ref.\ \cite{Gounaris:2004fm} in
the case of non-mixing squarks.

\subsection{Squark Decays}

\begin{figure}
 \centering
 \includegraphics[width=0.75\columnwidth]{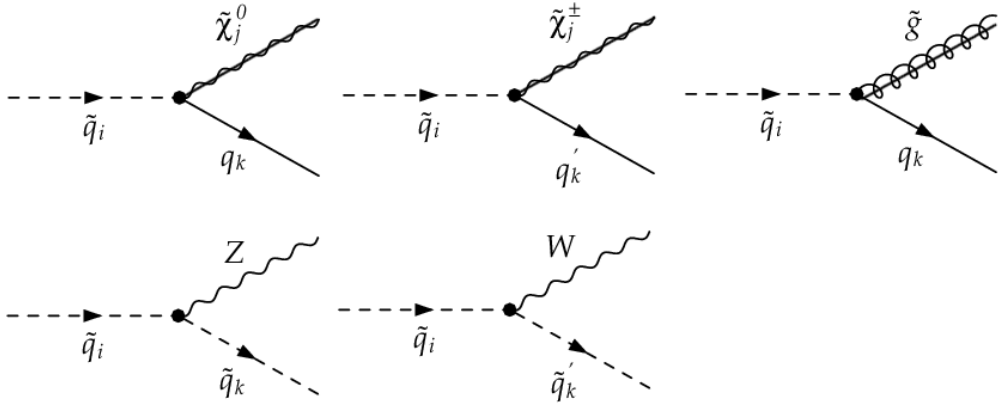}
 \caption{\label{fig:7a}Tree-level Feynman diagrams for squark decays
          into gauginos and quarks (top) and into electroweak gauge bosons
          and lighter squarks (bottom).}
\end{figure}

We turn now from SUSY particle production to decay processes and show in
Fig.\ \ref{fig:7a} the possible decays of squarks into gauginos
%
%
and quarks (top) as well as into electroweak gauge bosons and lighter
squarks (bottom). Both processes can in general induce flavour violation.
The decay widths of the former are given by
\bea
 \Gamma_{\tilde{q}_i \to \tilde{\chi}^0_j q_k} &=&
 \frac{\alpha}{2\, m^3_{\tilde{q}_i}\, x_W\, (1-x_W)}
 \Bigg(\bigg(m^2_{\tilde{q}_i} - m^2_{\tilde{\chi}^0_j} -
 m^2_{q_k}\bigg) \bigg(\left| L_{\tilde{q}_i q_k \tilde{\chi}_j^0}
 \right|^2 + \left|
 R_{\tilde{q}_i q_k \tilde{\chi}_j^0} \right|^2\bigg) \nonumber \\
 &-& 4\, m_{\tilde{\chi}^0_j}\, m_{q_k}\, {\rm
 Re}\left[L_{\tilde{q}_i q_k \tilde{\chi}_j^0} R^\ast_{\tilde{q}_i
 q_k \tilde{\chi}_j^0}\right]\Bigg)\,
 \lambda^{1/2}(m^2_{\tilde{q}_i}, m^2_{\tilde{\chi}^0_j},
 m^2_{q_k}),~  \label{eq:sqneq} \\
 \Gamma_{\tilde{q}_i \to \tilde{\chi}^\pm_j
 q^\prime_k} &=& \frac{\alpha}{4\, m^3_{\tilde{q}_i}\, x_W}
 \Bigg(\bigg(m^2_{\tilde{q}_i} - m^2_{\tilde{\chi}^\pm_j} -
 m^2_{q^\prime_k}\bigg) \bigg(\left| L_{\tilde{q}_i q^\prime_k
 \tilde{\chi}_j^\pm} \right|^2 + \left| R_{\tilde{q}_i q^\prime_k
 \tilde{\chi}_j^\pm} \right|^2\bigg)\nonumber \\ &-& 4\,
 m_{\tilde{\chi}^\pm_j}\, m_{q^\prime_k}\, {\rm
 Re}\left[L_{\tilde{q}_i q^\prime_k \tilde{\chi}_j^\pm}
 R^\ast_{\tilde{q}_i q^\prime_k \tilde{\chi}_j^\pm}\right]\Bigg)\,
 \lambda^{1/2}(m^2_{\tilde{q}_i}, m^2_{\tilde{\chi}^\pm_j},
 m^2_{q^\prime_k}) ,~ \label{eq:sqchq} \\
 \Gamma_{\tilde{q}_i \to \tilde{g} q_k} &=&
 \frac{2\, \alpha_s}{3\, m^3_{\tilde{q}_i}\, x_W}
 \Bigg(\bigg(m^2_{\tilde{q}_i} - m^2_{\tilde{g}} - m^2_{q_k}\bigg)
 \bigg(\left| L_{\tilde{q}_i q_k \tilde{g}} \right|^2 + \left|
 R_{\tilde{q}_i q_k \tilde{g}} \right|^2\bigg)- 4\, m_{\tilde{g}}\,
 m_{q_k}\, {\rm Re}\left[L_{\tilde{q}_i q_k \tilde{g}}
 R^\ast_{\tilde{q}_i q_k \tilde{g}}\right]\Bigg)\nonumber\\
 &\times& \lambda^{1/2}(m^2_{\tilde{q}_i}, m^2_{\tilde{g}},
 m^2_{q_k}),
\eea
while those of the latter are given by
\bea
 \Gamma_{\tilde{q}_i \to Z
 \tilde{q}_k} &=& \frac{\alpha}{16\, m^3_{\tilde{q}_i}\, m^2_Z\,
 x_W\, (1-x_W)} \left| L_{\tilde{q}_i \tilde{q}_k Z} +
 R_{\tilde{q}_i \tilde{q}_k Z}\right|^2\,
 \lambda^{3/2}(m^2_{\tilde{q}_i}, m^2_Z,
 m^2_{\tilde{q}_k}),\\
 \Gamma_{\tilde{q}_i \to W^\pm \tilde{q}^\prime_k}
 &=& \frac{\alpha}{16\, m^3_{\tilde{q}_i}\, m^2_W\, x_W\, (1-x_W)}
 \left| L_{\tilde{q}_i \tilde{q}^\prime_k W} \right|^2\,
 \lambda^{3/2}(m^2_{\tilde{q}_i}, m^2_W,
 m^2_{\tilde{q}^\prime_k}).
\eea
The usual K\"allen function is
\bea
 \lambda(x,y,z) = x^2 + y^2 + z^2 - 2 (x\,y + y\,z + z\,x).
\eea In cMFV, our results agree with those of Ref.\
\cite{Bartl:1994bu}.

\subsection{Gluino Decays}

\begin{figure}
 \centering
 \includegraphics[width=0.25\columnwidth]{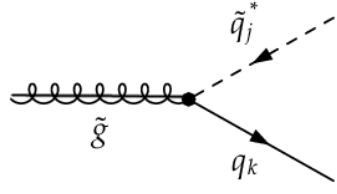}
 \caption{\label{fig:7b}Tree-level Feynman diagram for gluino decays
          into squarks and quarks.}
\end{figure}

Heavy gluinos can decay strongly into squarks and quarks as shown in Fig.\
\ref{fig:7b}. The corresponding decay width
%
%
\bea
 \Gamma_{\tilde{g} \to \tilde{q}^\ast_j
 q_k} &=& \frac{\alpha_s}{8\, m^3_{\tilde{g}}}
 \Bigg(\bigg(m^2_{\tilde{g}} - m^2_{\tilde{q}_j} + m^2_{q_k}\bigg)
 \bigg(\left| L_{\tilde{q}_j q_k \tilde{g}} \right|^2 + \left|
 R_{\tilde{q}_j q_k \tilde{g}} \right|^2\bigg) + 4\,
 m_{\tilde{g}}\, m_{q_k}\, {\rm Re}\left[L_{\tilde{q}_j q_k
 \tilde{g}} R^\ast_{\tilde{q}_j q_k
 \tilde{g}}\right]\Bigg)\nonumber\\ &\times&
 \lambda^{1/2}(m^2_{\tilde{g}}, m^2_{\tilde{q}_j},
 m^2_{q_k})
\eea can in general also induce flavour violation. In cMFV, our
result agrees again with the one of Ref.\ \cite{Bartl:1994bu}.

\subsection{Gaugino Decays}

\begin{figure}
 \centering
 \includegraphics[width=0.75\columnwidth]{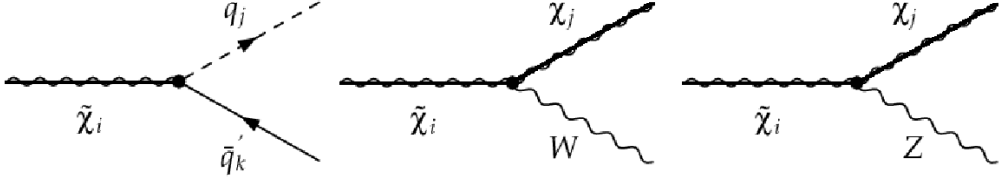}
 \caption{\label{fig:7c}Tree-level Feynman diagrams for gaugino decays
          into squarks and quarks (left) and into lighter gauginos and
          electroweak gauge bosons (centre and right).}
\end{figure}

Heavier gauginos can decay into squarks and quarks as shown in Fig.\
\ref{fig:7c} (left) or into lighter gauginos and electroweak gauge bosons
(Fig.\ \ref{fig:7c} centre and right). The analytical decay widths are
%
%
\bea
 \Gamma_{\tilde{\chi}_i^\pm \to \tilde{q}_j
 \bar{q}^\prime_k} &=& \frac{3\, \alpha}{8\, m^3_{\tilde{\chi}_i^\pm}\,
 x_W} \Bigg(\bigg( m^2_{\tilde{\chi}_i^\pm} - m^2_{\tilde{q}_j} +
 m^2_{q^\prime_k}\bigg) \bigg(\left| L_{\tilde{q}_j q^\prime_k
 \tilde{\chi}_i^\pm} \right|^2 + \left| R_{\tilde{q}_j q^\prime_k
 \tilde{\chi}_i^\pm} \right|^2\bigg) \nonumber \\&+& 4\,
 m_{\tilde{\chi}_i^\pm}\, m_{q^\prime_k}\, {\rm
 Re}\left[L_{\tilde{q}_j q^\prime_k \tilde{\chi}_i^\pm}
 R^\ast_{\tilde{q}_j q^\prime_k \tilde{\chi}_i^\pm}\right]\Bigg) \,
 \lambda^{1/2}(m^2_{\tilde{\chi}_i^\pm}, m^2_{\tilde{q}_j},
 m^2_{q^\prime_k})
\eea
and
\bea
 \Gamma_{\tilde{\chi}_i^\pm \to
 \tilde{\chi}_j^0 W^\pm} &=& \frac{\alpha}{8\,
 m^3_{\tilde{\chi}_i^\pm}\, m_W^2\, x_W} \Bigg(\bigg(
 m^4_{\tilde{\chi}_i^\pm} + m^4_{\tilde{\chi}_j^0} - 2\, m^4_W +
 m^2_{\tilde{\chi}_i^\pm}\, m^2_W + m^2_{\tilde{\chi}_j^0}\, m^2_W
 - 2\, m^2_{\tilde{\chi}_i^\pm}\, m^2_{\tilde{\chi}_j^0} \bigg)
 \nonumber \\ &\times & \bigg(\left| O^L_{ij} \right|^2 + \left|
 O^R_{ij} \right|^2\bigg) - 12\, m_{\tilde{\chi}_i^\pm}\, m_W^2\,
 m_{\tilde{\chi}_j^0}\, {\rm Re}\left[O^L_{ij}
 O^{R\ast}_{ij}\right]\Bigg)\,
 \lambda^{1/2}(m^2_{\tilde{\chi}_i^\pm}, m^2_{\tilde{\chi}_j^0},
 m^2_W) ,~ \\ \Gamma_{\tilde{\chi}_i^\pm \to \tilde{\chi}_j^\pm Z}
 &=& \frac{\alpha}{8\, m^3_{\tilde{\chi}_i^\pm}\, m_Z^2\, x_W\,
 (1-x_W)} \Bigg(\bigg( m^4_{\tilde{\chi}_i^\pm} +
 m^4_{\tilde{\chi}_j^\pm} - 2\, m^4_Z + m^2_{\tilde{\chi}_i^\pm}\,
 m^2_Z + m^2_{\tilde{\chi}_j^\pm}\, m^2_Z - 2\,
 m^2_{\tilde{\chi}_i^\pm}\, m^2_{\tilde{\chi}_j^\pm}
 \bigg)\nonumber \\ &\times & \bigg(\left| O^{\prime L}_{ij}
 \right|^2 + \left| O^{\prime R}_{ij} \right|^2\bigg) - 12\,
 m_{\tilde{\chi}_i^\pm}\, m_Z^2\, m_{\tilde{\chi}_j^\pm}\, {\rm
 Re}\left[O^{\prime L}_{ij} O^{\prime R\ast}_{ij}\right]\Bigg)\,
 \lambda^{1/2}(m^2_{\tilde{\chi}_i^\pm}, m^2_{\tilde{\chi}_j^\pm},
 m^2_Z)
\eea
for charginos and
\bea
 \Gamma_{\tilde{\chi}_i^0 \to \tilde{q}_j \bar{q}_k} &=&
 \frac{3\, \alpha}{4\, m^3_{\tilde{\chi}_i^0}\, x_W\, (1-x_W)}
 \Bigg(\bigg( m^2_{\tilde{\chi}_i^0} - m^2_{\tilde{q}_j} +
 m^2_{q_k}\bigg) \bigg(\left| L_{\tilde{q}_j q_k \tilde{\chi}_i^0}
 \right|^2 + \left| R_{\tilde{q}_j q_k \tilde{\chi}_i^0}
 \right|^2\bigg) \nonumber \\&+& 4\, m_{\tilde{\chi}_i^0}\,
 m_{q_k}\, {\rm Re}\left[L_{\tilde{q}_j q_k \tilde{\chi}_i^0}
 R^\ast_{\tilde{q}_j q_k \tilde{\chi}_i^0}\right]\Bigg) \,
 \lambda^{1/2}(m^2_{\tilde{\chi}_i^0}, m^2_{\tilde{q}_j},
 m^2_{q_k})
\eea
and
\bea
 \Gamma_{\tilde{\chi}_i^0 \to \tilde{\chi}_j^\pm
 W^\mp} &=& \frac{\alpha}{8\, m^3_{\tilde{\chi}_i^0}\, m_W^2\, x_W}
 \Bigg(\bigg( m^4_{\tilde{\chi}_i^0} + m^4_{\tilde{\chi}_j^\pm} -
 2\, m^4_W + m^2_{\tilde{\chi}_i^0}\, m^2_W +
 m^2_{\tilde{\chi}_j^\pm}\, m^2_W - 2\, m^2_{\tilde{\chi}_i^0}\,
 m^2_{\tilde{\chi}_j^\pm} \bigg) \nonumber \\ &\times &
 \bigg(\left| O^L_{ij} \right|^2 + \left| O^R_{ij} \right|^2\bigg)
 - 12\, m_{\tilde{\chi}_i^0}\, m_W^2\, m_{\tilde{\chi}_j^\pm}\,
 {\rm Re}\left[O^L_{ij} O^{R\ast}_{ij}\right]\Bigg)\,
 \lambda^{1/2}(m^2_{\tilde{\chi}_i^0}, m^2_{\tilde{\chi}_j^\pm},
 m^2_W) ,~ \\ \Gamma_{\tilde{\chi}_i^0 \to \tilde{\chi}_j^0 Z} &=&
 \frac{\alpha}{8\, m^3_{\tilde{\chi}_i^0}\, m_Z^2\, x_W\, (1-x_W)}
 \Bigg(\bigg( m^4_{\tilde{\chi}_i^0} + m^4_{\tilde{\chi}_j^0} - 2\,
 m^4_Z + m^2_{\tilde{\chi}_i^0}\, m^2_Z + m^2_{\tilde{\chi}_j^0}\,
 m^2_Z - 2\, m^2_{\tilde{\chi}_i^0}\, m^2_{\tilde{\chi}_j^0}
 \bigg)\nonumber \\ &\times & \bigg(\left| O^{\prime\prime L}_{ij}
 \right|^2 + \left| O^{\prime\prime R}_{ij} \right|^2\bigg) - 12\,
 m_{\tilde{\chi}_i^0}\, m_Z^2\, m_{\tilde{\chi}_j^0}\, {\rm
 Re}\left[O^{\prime\prime L}_{ij} O^{\prime\prime
 R\ast}_{ij}\right]\Bigg)\, \lambda^{1/2}(m^2_{\tilde{\chi}_i^0},
 m^2_{\tilde{\chi}_j^0}, m^2_Z)
\eea for neutralinos, respectively. Chargino decays into a slepton
and a neutrino (lepton and sneutrino) can be deduced from the
previous equations by taking the proper limits, i.e.\ by removing
colour factors and up-type masses in the coupling definitions. Our
results agree then with those of Ref.\ \cite{Obara:2005qi} in the
limit of non-mixing sneutrinos. Note that the same simplifications
also permit a verification of our results for squark decays into a
gaugino and a quark in Eqs.\ (\ref{eq:sqneq}) and (\ref{eq:sqchq})
when compared to their leptonic counterparts in Ref.\
\cite{Obara:2005qi}.

\section{Experimental Constraints, Scans and Benchmarks in NMFV SUSY}
\label{sec:4}

In the absence of experimental evidence for supersymmetry, a large variety
of data can be used to constrain the MSSM parameter space. For example,
sparticle mass limits can be obtained from searches of charginos
($m_{\tilde{\chi}^\pm_1}\geq85$ GeV for heavier sneutrinos at LEP2),
neutralinos ($m_{\tilde{\chi}^0_1}\geq59$ GeV in minimal supergravity
(mSUGRA) from the combination of LEP2 results), gluinos ($m_{\tilde g}\geq
195$ GeV from CDF), stops ($m_{\tilde{t}_1}\geq95\dots96$ GeV for neutral-
or charged-current decays from the combination of LEP2 results), and other
squarks ($m_{\tilde q}\geq300$ GeV for gluinos of equal mass from CDF) at
colliders \cite{Yao:2006px}. Note that all of these limits have been
obtained assuming minimal flavour violation. For non-minimal flavour
violation, rather strong constraints can be obtained from low-energy,
electroweak precision, and cosmological observables. These are discussed in
the next subsection, followed by several scans for experimentally
allowed/favoured regions of the constrained MSSM parameter space and the
definition of four NMFV benchmark points/slopes. Finally, we exhibit the
corresponding chirality and flavour decomposition of the various squark mass
eigenstates.

\subsection{Low-Energy, Electroweak Precision, and Cosmological Constraints}
\label{sec:const}

In a rather complete analysis of FCNC constraints more than ten years ago
\cite{Gabbiani:1996hi}, upper limits from the neutral kaon sector (on
$\Delta m_K$, $\eps$, $\eps'/\eps$), on $B$- ($\Delta m_B$) and $D$-meson
oscillations ($\Delta m_D$), various rare decays (BR($b\to s\gamma$),
BR($\mu\to e\gamma$), BR($\tau\to e \gamma$), and BR($\tau\to\mu\gamma$)),
and electric dipole moments ($d_n$ and $d_e$) were used to impose
constraints on non-minimal flavour mixing in the squark and slepton sectors.
The limit obtained for the absolute value in the left-handed, down-type
squark sector was rather weak ($|\lambda_{LL}^{sb}|< 4.4\dots26$ for varying
gluino-to-squark mass ratio), while the limits for the mixed/right-handed,
imaginary or sleptonic parts were already several orders of magnitude
smaller. In the meantime, many of the experimental bounds have been improved
or absolute values for the observables have been determined, so that an
updated analysis could be performed \cite{Ciuchini:2007ha}. The results for
the down-type squark sector are cited in Tab.\ \ref{tab:1}.
%
\begin{table}
 \caption{\label{tab:1}The 95\% probability bounds on $|\lambda_{ij}^{d_k
          d_l}|$ obtained in Ref.\ \cite{Ciuchini:2007ha}.}
 \begin{tabular}{c|cccc}
  \underline{ij} & LL & LR & RL & RR\\
  $kl$           &    &    &    &   \\
  \hline
  12 & 1.4$\times10^{-2}$ & 9.0$\times10^{-5}$ & 9.0$\times10^{-5}$ & 9.0$\times10^{-3}$ \\
  13 & 9.0$\times10^{-2}$ & 1.7$\times10^{-2}$ & 1.7$\times10^{-2}$ & 7.0$\times10^{-2}$ \\
  23 & 1.6$\times10^{-1}$ & 4.5$\times10^{-3}$ & 6.0$\times10^{-3}$ & 2.2$\times10^{-1}$ \\
 \end{tabular}
\end{table}
%
As can be seen and as has already been hinted at in the introduction, only
mixing between second- and third-generation squarks can be substantial, and
this only in the left-left or right-right chiral sectors, the latter being
disfavoured by its scaling with the soft SUSY-breaking mass.
Independent analyses focusing on this particular sector, i.e.\ on BR($b\to
s\gamma$), BR($b\to s\mu\mu$), and $\Delta m_{B_s}$, have been performed
recently by two other groups \cite{Foster:2006ze,Hahn:2005qi} with very
similar results.

In our own analysis, we take implicitly into account all of the previously
mentioned constraints by restricting ourselves to the case of only one real
NMFV
parameter, $\lambda\equiv\lambda_{LL}^{sb}=\lambda_{LL}^{ct}$. Allowed
regions for this parameter are then obtained by imposing explicitly a number
of low-energy, electroweak precision, and cosmological constraints.
We start by imposing the theoretically robust inclusive branching ratio
\bea
 {\rm BR}(b\to s \gamma) = (3.55\pm 0.26) \times 10^{-4},
\eea
obtained from the combined measurements of BaBar, Belle, and CLEO
\cite{Barberio:2006bi}, at the 2$\sigma$-level on the two-loop QCD/one-loop
SUSY calculation \cite{Hahn:2005qi,Kagan:1998bh}, which affects directly
the allowed squark mixing between the second and third generation.

A second important consequence of NMFV in the MSSM is the generation of
large splittings between squark-mass eigenvalues. The splitting within
isospin doublets influences the $Z$- and $W$-boson self-energies at
zero-momentum $\Sigma_{Z,W}(0)$ in the electroweak $\rho$-parameter
\bea
 \Delta\rho=\Sigma_Z(0)/M_Z^2-\Sigma_W(0)/M_W^2
\eea
and consequently the $W$-boson mass $M_W$ and the squared sine of the weak
mixing angle $\sin^2\theta_W$. The latest combined fits of the $Z$-boson
mass, width, pole asymmetry, $W$-boson and top-quark mass constrain new
physics contributions to $T=-0.13\pm0.11$ \cite{Yao:2006px}
or
\bea
 \Delta\rho=-\alpha T=0.00102\pm0.00086,
\eea where we have used $\alpha(M_Z)=1/127.918$. This value is
then imposed at the 2$\sigma$-level on the one-loop NMFV and
two-loop cMFV SUSY calculation \cite{Heinemeyer:2004by}.

A third observable sensitive to SUSY loop-contributions is the anomalous
magnetic moment $a_\mu=(g_\mu-2)/2$ of the muon, for which recent BNL data
and the SM prediction disagree by \cite{Yao:2006px}
\bea
 \Delta a_\mu=(22\pm10)\times 10^{-10}.
\eea
In our calculation, we take into account the SM and MSSM contributions up to
two loops \cite{Heinemeyer:2003dq,Heinemeyer:2004yq} and require them to
agree with the region above within two standard deviations.

For cosmological reasons, i.e.\ in order to have a suitable candidate
for non-baryonic cold dark matter \cite{Ellis:1983ew}, we require the
lightest SUSY particle (LSP) to be stable, electrically neutral, and a
colour singlet. The dark matter relic density is then calculated using
a modified version of DarkSUSY 4.1 \cite{Gondolo:2004sc}, that takes into
account the six-dimensional squark helicity and flavour mixing, and
constrained to the region
\bea
 0.094<\Omega_{CDM} h^2<0.136
\eea
at 95\% (2$\sigma$) confidence level. This limit has recently been obtained
from the three-year data of the WMAP satellite, combined with the SDSS and
SNLS survey and Baryon Acoustic Oscillation data and interpreted within an
eleven-parameter inflationary model \cite{Hamann:2006pf}, which is more
general than the usual six-parameter ``vanilla'' concordance model of
cosmology. Note
that this range is well compatible with the older, independently obtained
range of $0.094 <\Omega_{CDM} h^2<0.129$ \cite{Ellis:2003cw}.

\subsection{Scans of the Constrained NMFV MSSM Parameter Space}

The above experimental limits are now imposed on the constrained
MSSM (cMSSM), or minimal supergravity (mSUGRA), model with five
free parameters $m_0$, $m_{1/2}$, $\tan\beta$, $A_0$, and
sgn$(\mu)$ at the grand unification scale. Since our scans of the
cMSSM parameter space in the common scalar mass $m_0$ and the
common fermion mass $m_{1/2}$ depend very little on the trilinear
coupling $A_0$, we set it to zero in the following. Furthermore,
we fix a small (10), intermediate (30), and large (50) value for
the ratio of the Higgs vacuum expectation values $\tan\beta$. The
impact of the sign of the off-diagonal Higgs mass parameter $\mu$
is investigated for $\tan\beta=10$ only, before we set it to
$\mu>0$ for $\tan\beta=30$ and 50 (see below).

With these boundary conditions at the grand unification scale, we solve the
renormalization group equations numerically to two-loop order using the
computer program SPheno 2.2.3 \cite{Porod:2003um} and compute the soft
SUSY-breaking masses at the electroweak scale with the complete one-loop
formulas, supplemented by two-loop contributions in the case of the neutral
Higgs bosons and the $\mu$-parameter. At this point we generalize the squark
mass matrices as described in Sec.\ \ref{sec:2} in order to account for
flavour mixing in the left-chiral sector of the second- and third-generation
squarks, diagonalize these mass matrices, and compute the low-energy,
electroweak precision, and cosmological observables with the computer
programs FeynHiggs 2.5.1 \cite{Heinemeyer:1998yj} and DarkSUSY 4.1
\cite{Gondolo:2004sc}.

For the masses and widths of the electroweak gauge bosons and the mass of
the top quark, we use the current values of $m_Z=91.1876$ GeV, $m_W=80.403$
GeV, $m_t = 174.2$ GeV, $\Gamma_Z=2.4952$ GeV, and $\Gamma_W=2.141$ GeV. The
CKM-matrix elements are computed using the parameterization
\bea
 V = \left(\begin{array}{c c c} c_{12} c_{13} & s_{12}
 c_{13} & s_{13} e^{-i \delta}\\ -s_{12} c_{23} - c_{12} s_{23}
 s_{13} e^{i \delta}& c_{12} c_{23} - s_{12} s_{23} s_{13} e^{i
 \delta}& s_{23} c_{13}\\ s_{12} s_{23} - c_{12} c_{23} s_{13} e^{i
 \delta}& -c_{12} s_{23} - s_{12} c_{23} s_{13} e^{i \delta}&
 c_{23} c_{13} \end{array}\right),
\eea
where $s_{ij} = \sin\theta_{ij}$ and $c_{ij} = \cos\theta_{ij}$ relate to
the mixing of two specific generations $i$ and $j$ and $\delta$ is the SM
$CP$-violating complex phase. The numerical values are given by
\bea
 s_{12} = 0.2243,~s_{23} = 0.0413,~ s_{13} = 0.0037,~{\rm and}~
 \delta = 1.05.
\eea
The squared sine of the electroweak mixing angle $\sin^2\theta_W=1-
m_W^2/m_Z^2$ and the electromagnetic fine structure constant $\alpha=
\sqrt{2} G_F m_W^2\sin^2\theta_W/\pi$ are calculated in the improved
Born approximation using the world average value of $G_F=1.16637\cdot
10^{-5}$ GeV$^{-2}$ for Fermi's coupling constant \cite{Yao:2006px}.

\begin{figure}
 \centering
 \includegraphics[width=0.24\columnwidth]{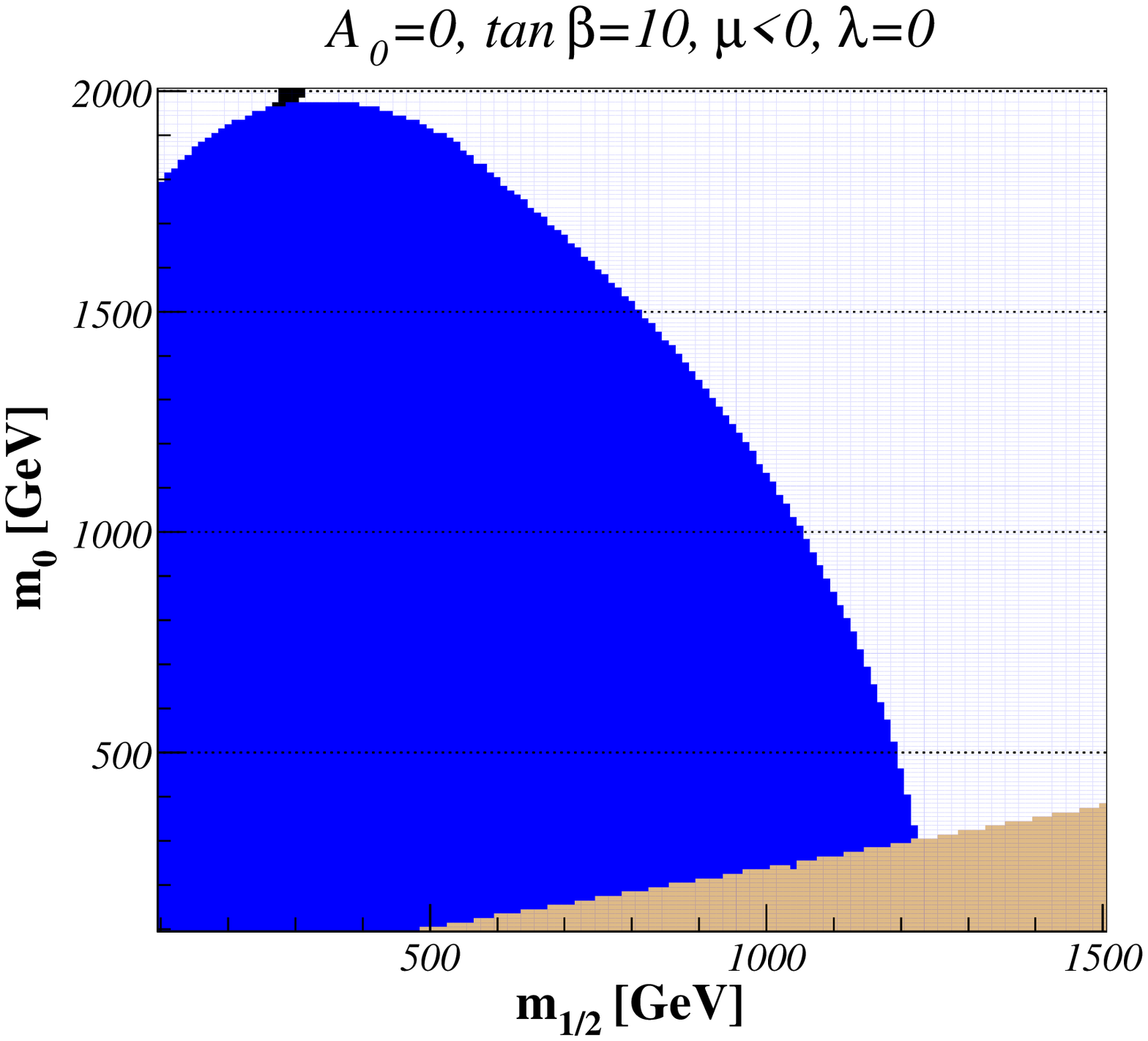}\hspace{1mm}
 \includegraphics[width=0.24\columnwidth]{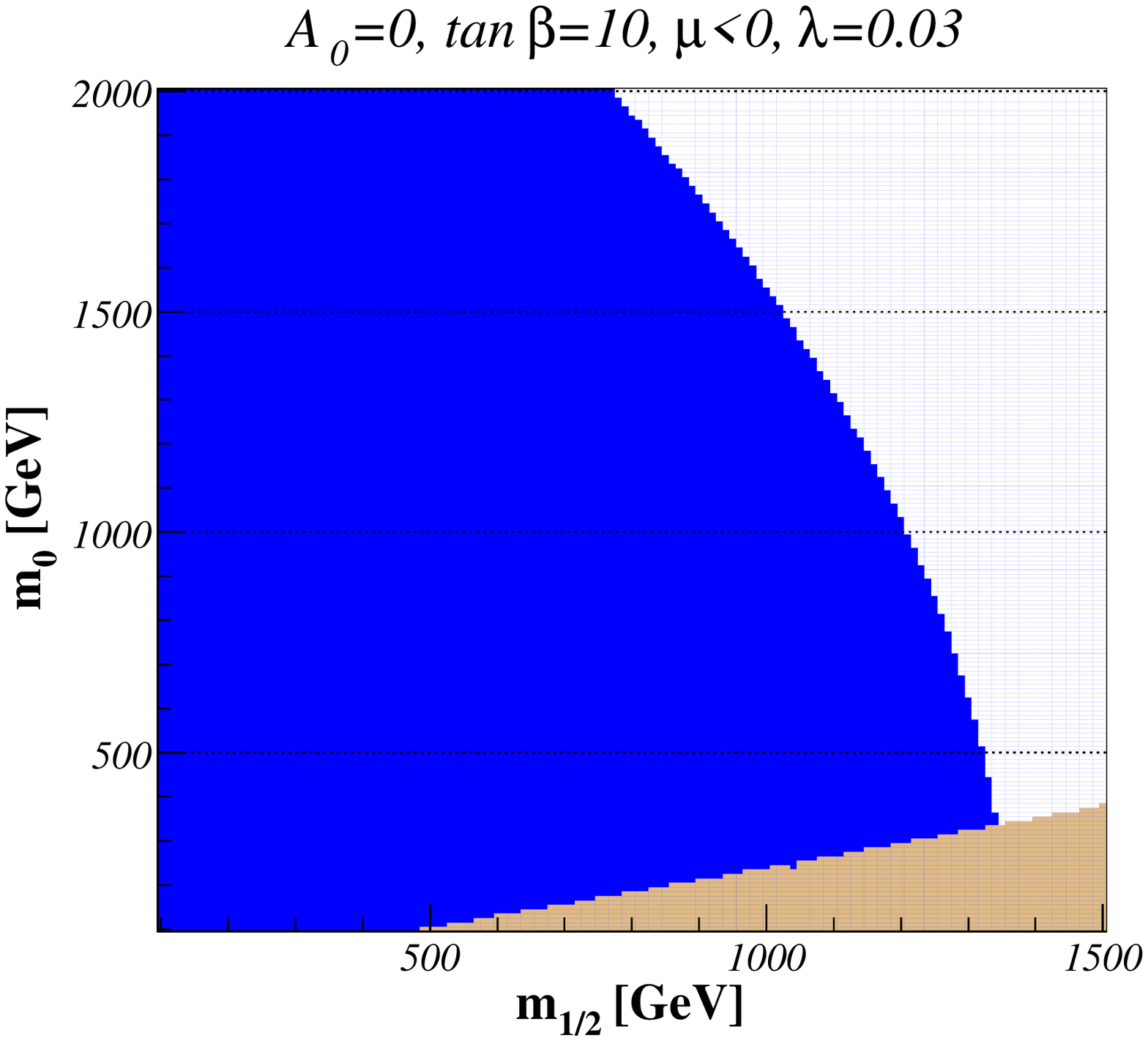}\hspace{1mm}
 \includegraphics[width=0.24\columnwidth]{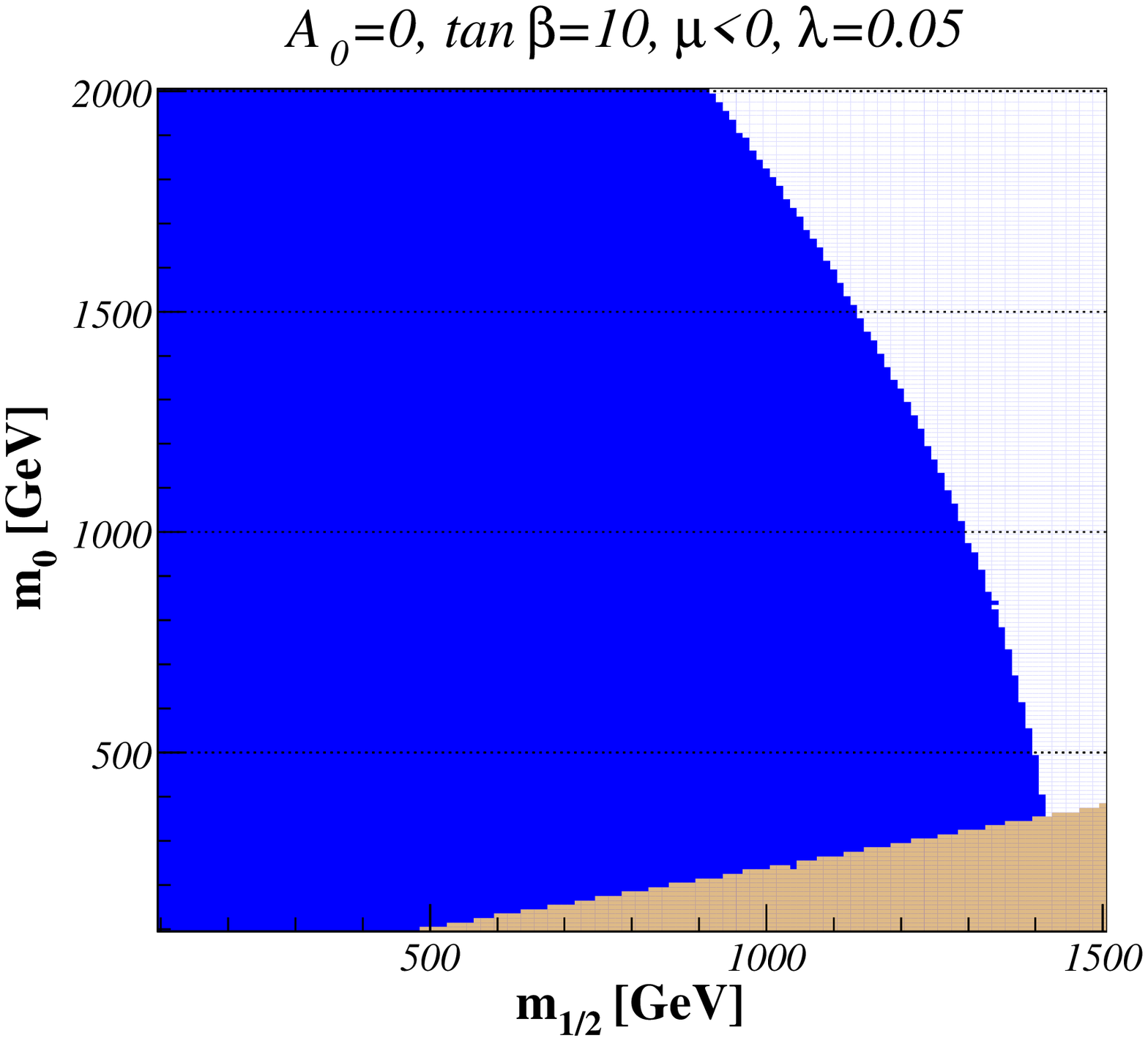}\hspace{1mm}
 \includegraphics[width=0.24\columnwidth]{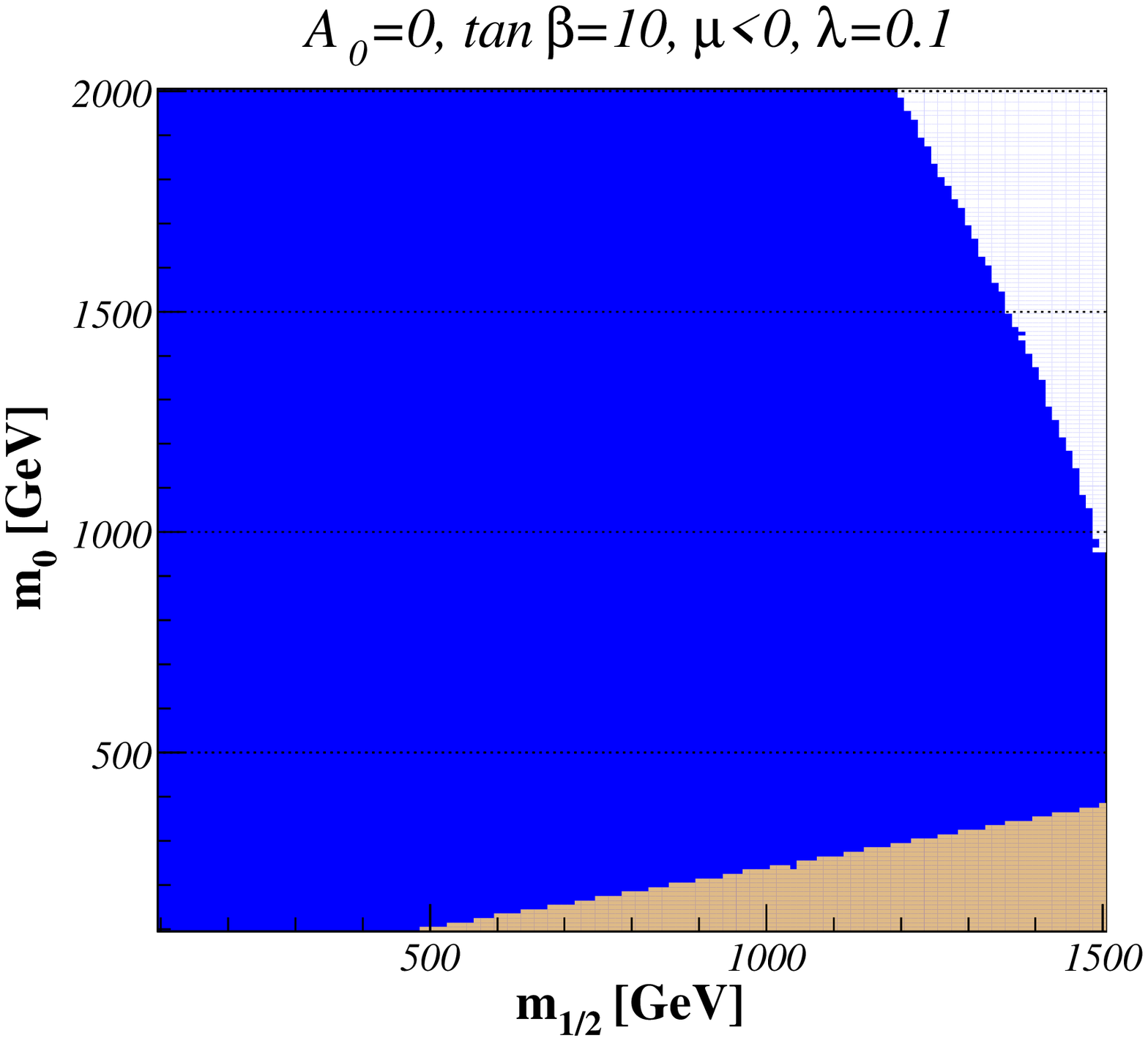}
 \caption{\label{fig:8}The $(m_0,m_{1/2})$-planes for $\tan\beta=10$,
          $A_0=0$ GeV, $\mu<0$, and $\lambda=0$, 0.03, 0.05 and 0.1. We show
          WMAP (black) favoured as well as $b\to s\gamma$ (blue) and charged
          LSP (beige) excluded regions of mSUGRA parameter space in minimal
          ($\lambda=0$) and non-minimal ($\lambda>0$) flavour violation.}
\end{figure}

\begin{figure}
 \centering
 \includegraphics[width=0.24\columnwidth]{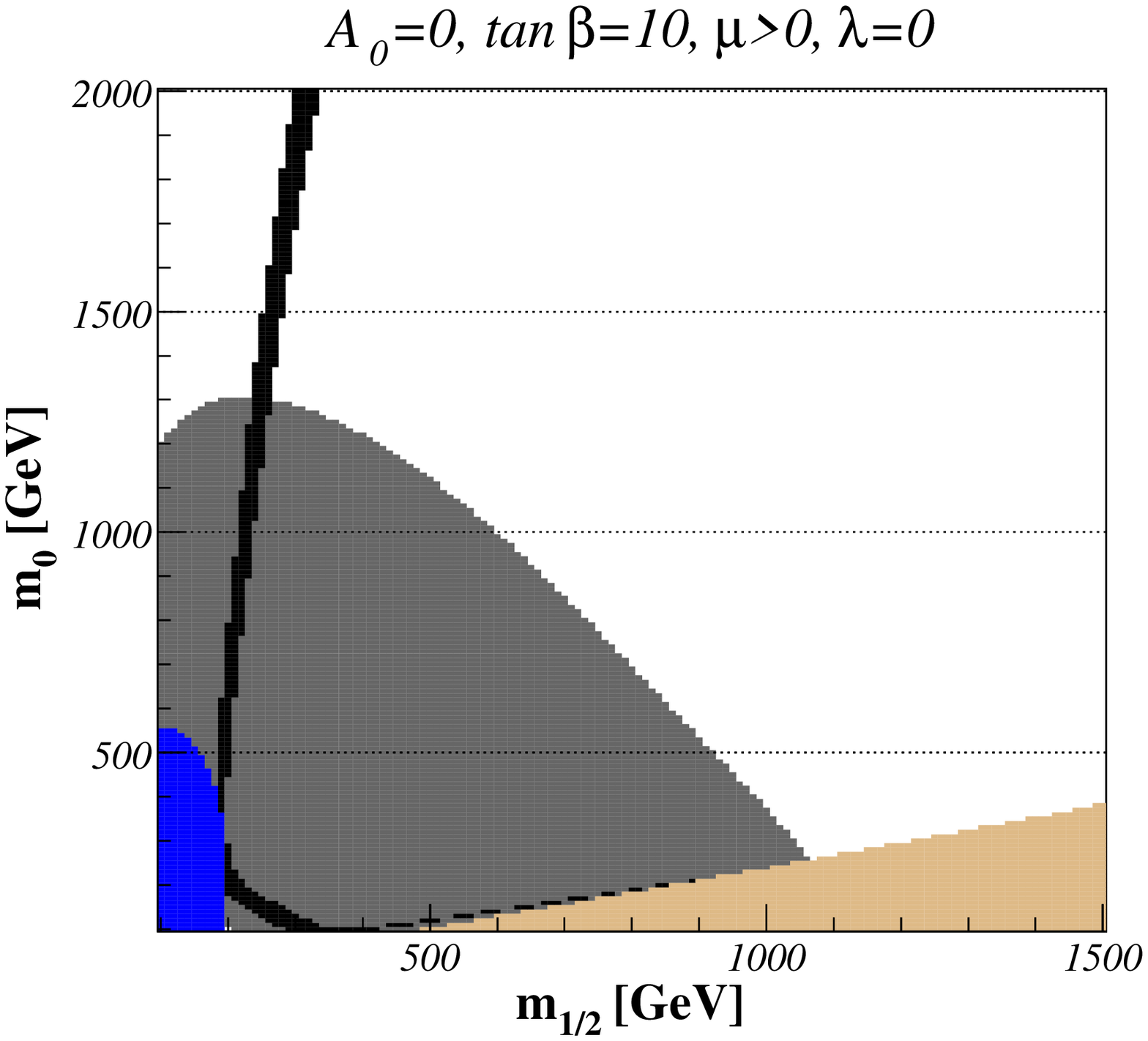}\hspace{1mm}
 \includegraphics[width=0.24\columnwidth]{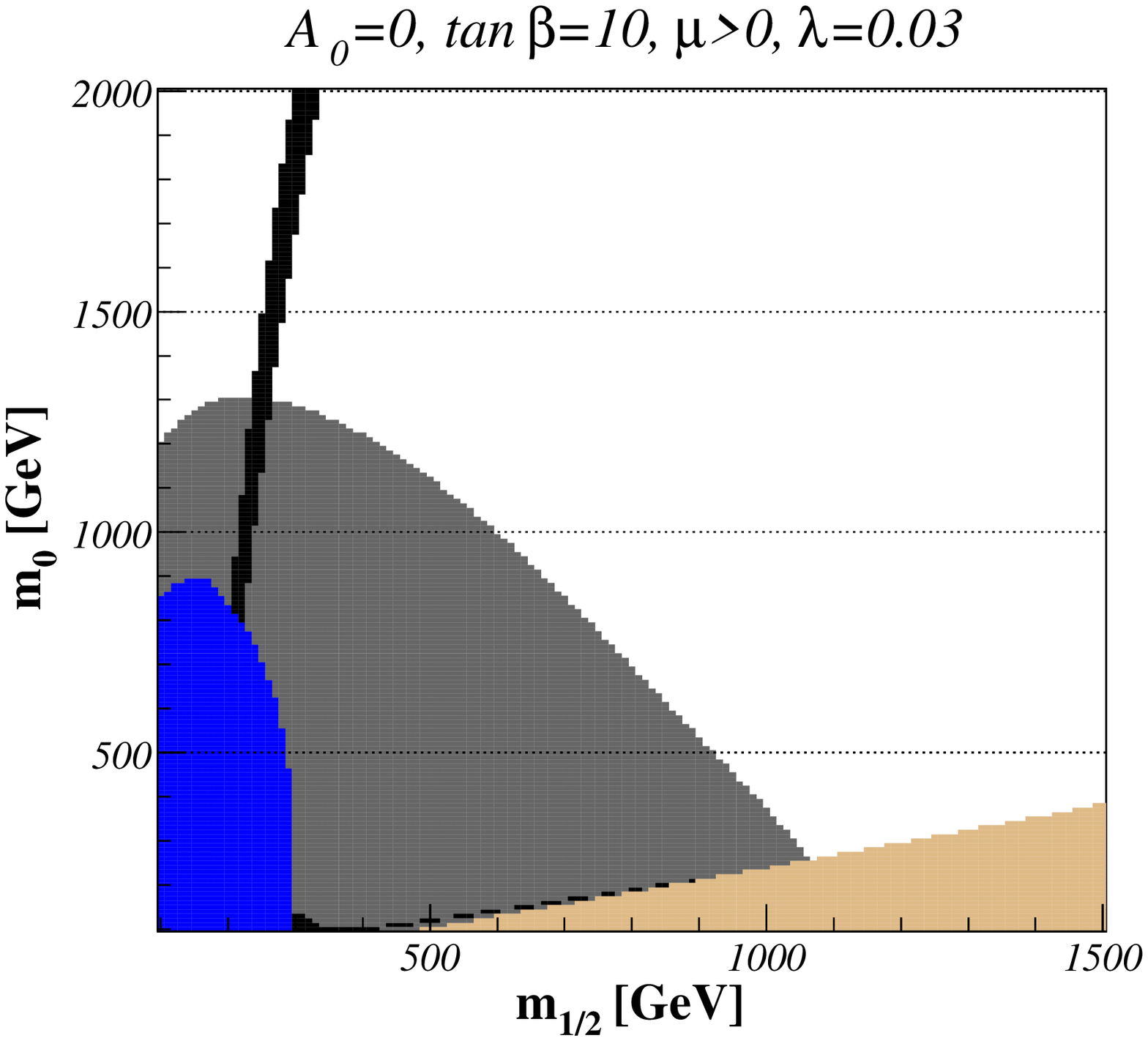}\hspace{1mm}
 \includegraphics[width=0.24\columnwidth]{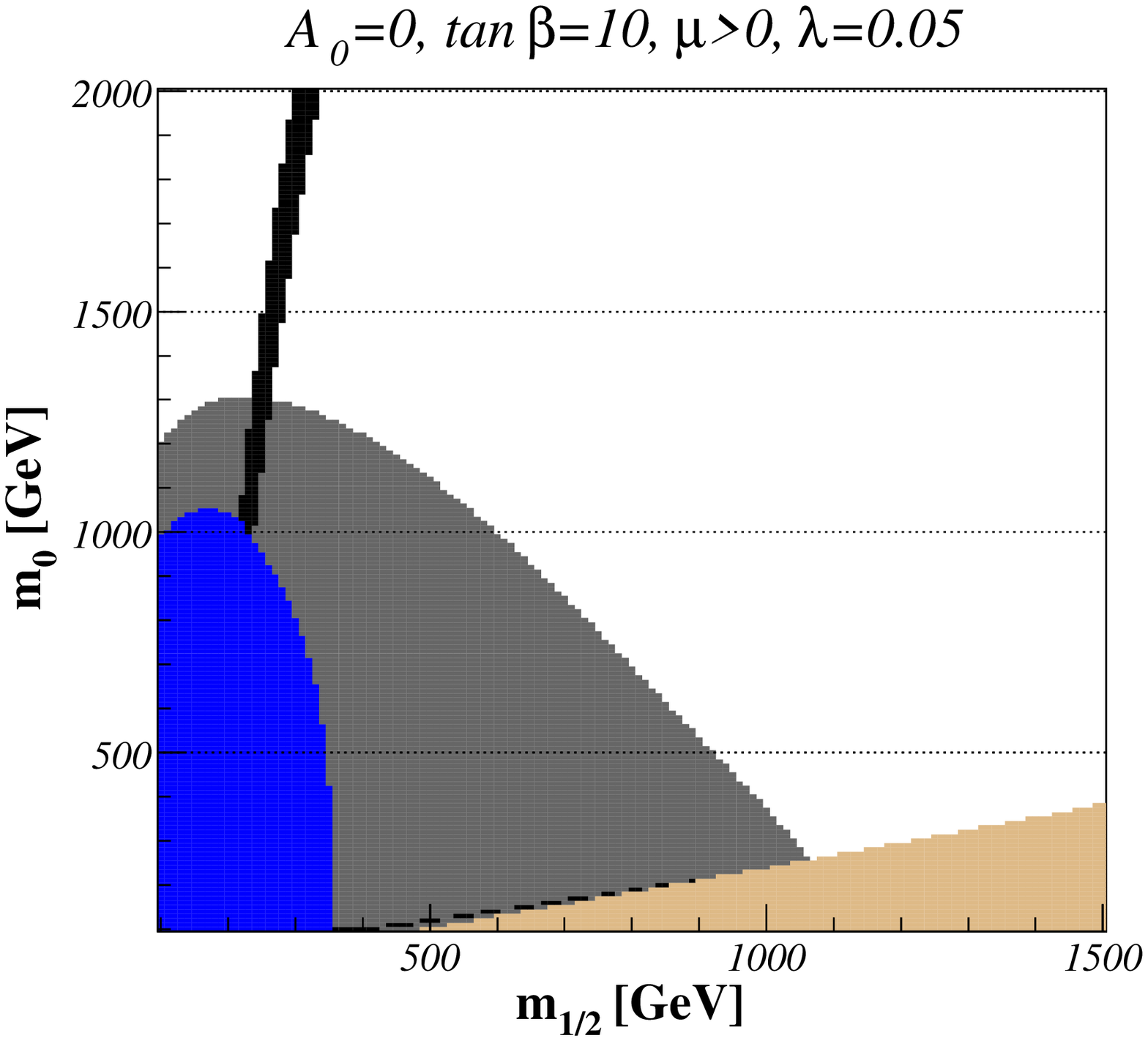}\hspace{1mm}
 \includegraphics[width=0.24\columnwidth]{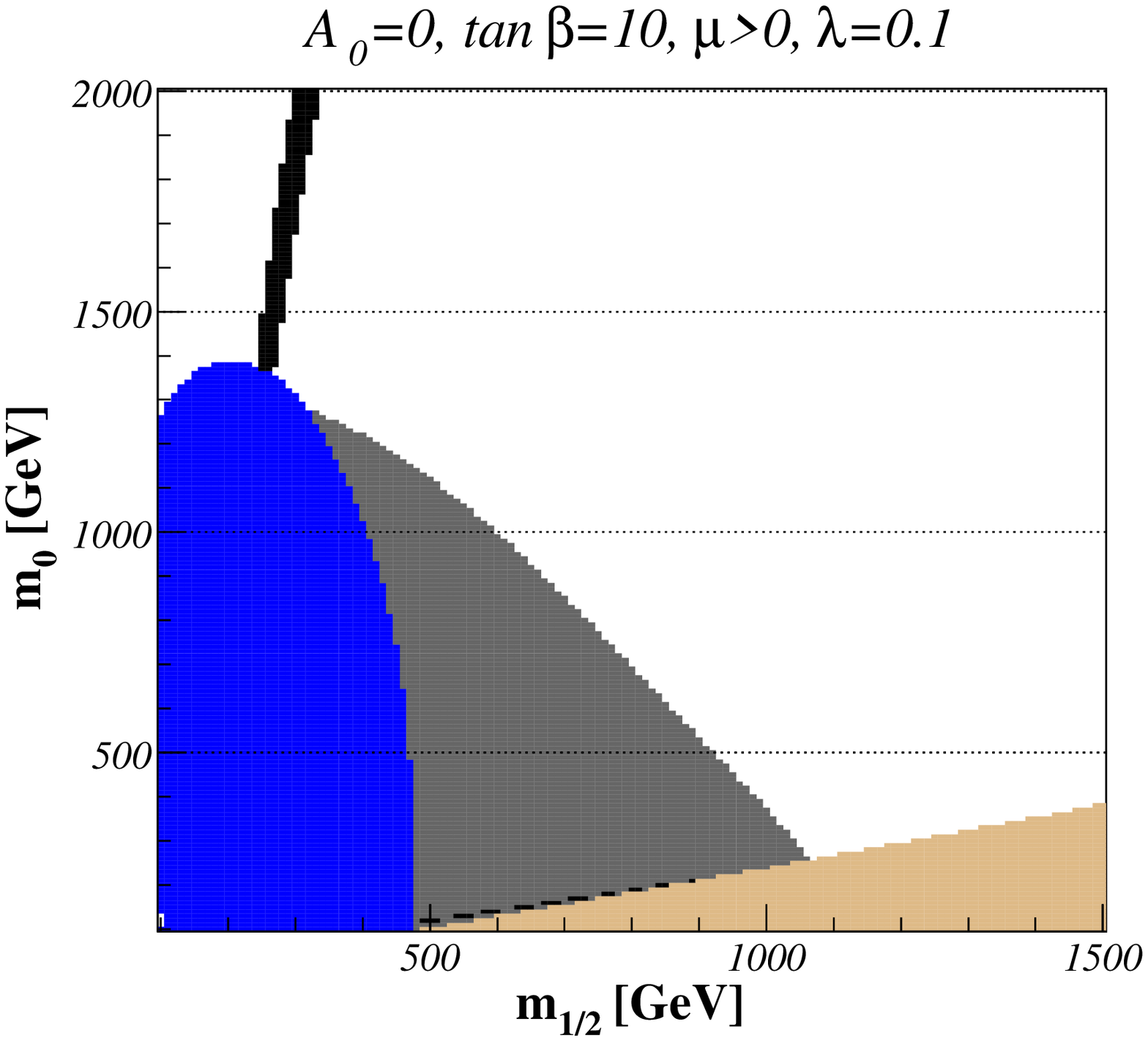}
 \caption{\label{fig:9}The $(m_0,m_{1/2})$-planes for $\tan\beta=10$,
          $A_0=0$ GeV, $\mu>0$, and $\lambda=0$, 0.03, 0.05 and 0.1. We show
          $a_\mu$ (grey) and WMAP (black) favoured as well as $b\to s\gamma$
          (blue) and charged LSP (beige) excluded regions of mSUGRA
          parameter space in minimal ($\lambda=0$) and non-minimal
          ($\lambda>0$) flavour violation.}
\end{figure}

Typical scans of the cMSSM parameter space in $m_0$ and $m_{1/2}$ with a
relatively small value of $\tan\beta = 10$ and $A_0 = 0$ are shown in Figs.\
\ref{fig:8} and \ref{fig:9} for $\mu < 0$ and $\mu > 0$, respectively.
%
%
%
%
All experimental limits described in Sec.\ \ref{sec:const} are imposed at
the $2\sigma$-level. The $b\to s \gamma$ excluded region depends strongly on
flavour mixing, while the regions favoured by $g_\mu - 2$ and the dark
matter relic density are quite insensitive to variations of the
$\lambda$-parameter. $\Delta\rho$ constrains the parameter space only for
heavy scalar masses $m_0>2000$ GeV and heavy gaugino masses $m_{1/2}>1500$
GeV, so that the corresponding excluded regions are not shown here.

The dominant SUSY effects in the calculation of the anomalous magnetic
moment of the muon come from induced quantum loops of a gaugino and a
slepton. Squarks contribute only at the two-loop level. This reduces
the dependence on flavour violation in the squark sector considerably.
Furthermore, the region $\mu<0$ is disfavoured in all SUSY models, since the
one-loop SUSY contributions are approximatively given by \cite{Moroi:1995yh}
\bea
 a_\mu^{{\rm SUSY,~1-loop}} \simeq 13 \times
 10^{-10}\, \lr \frac{100~{\rm GeV}}{M_{\rm SUSY}} \rr^2
 \tan\beta \ {\rm sgn}(\mu),
\eea
if all SUSY particles (the relevant ones are the smuon, sneutralino,
chargino, and neutralino) have a common mass $M_{\rm SUSY}$. Negative values
of $\mu$ would then increase, not decrease, the disagreement between the
experimental measurements and the theoretical SM value of $a_\mu$.
Furthermore, the measured $b \to s \gamma$ branching ratio excludes
virtually all of the region favoured by the dark matter relic density,
except for very high scalar SUSY masses. We therefore do not consider
negative values of $\mu$ in the rest of this work. As stated above, we have
also checked that the shape of the different regions depends extremely
weakly on the trilinear coupling $A_0$.

%
\begin{figure}
 \centering
 \includegraphics[width=0.24\columnwidth]{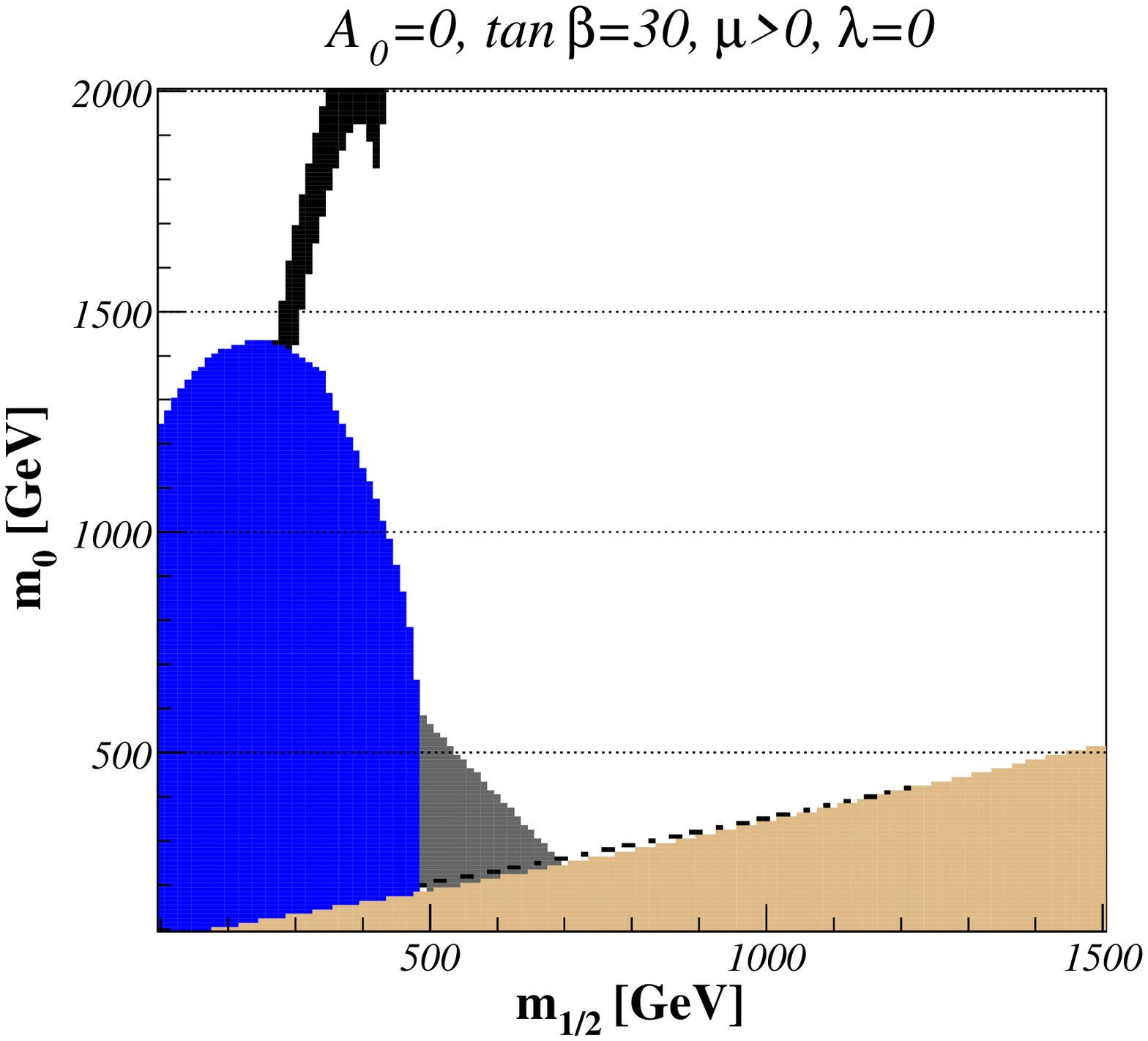}\hspace{1mm}
 \includegraphics[width=0.24\columnwidth]{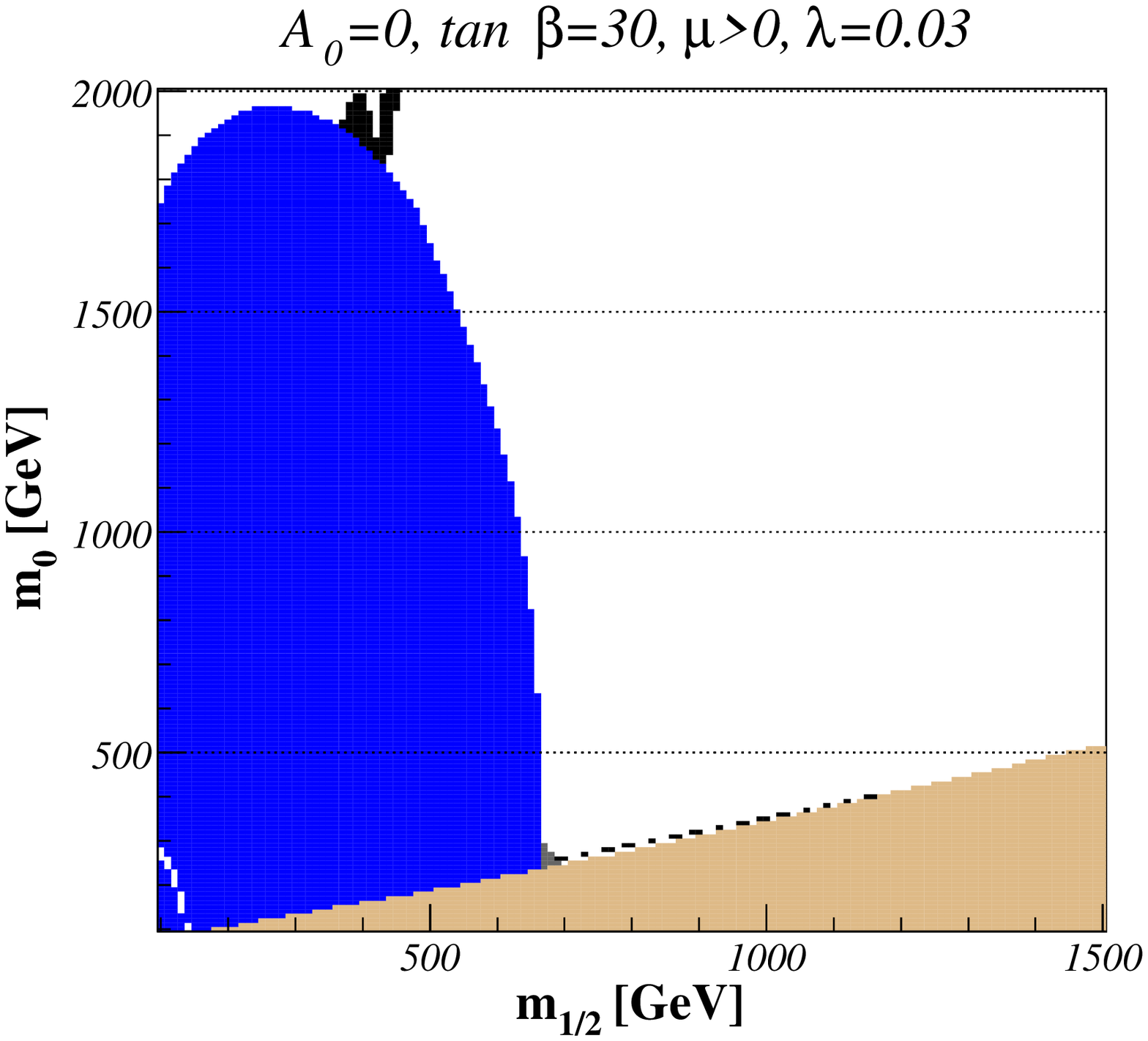}\hspace{1mm}
 \includegraphics[width=0.24\columnwidth]{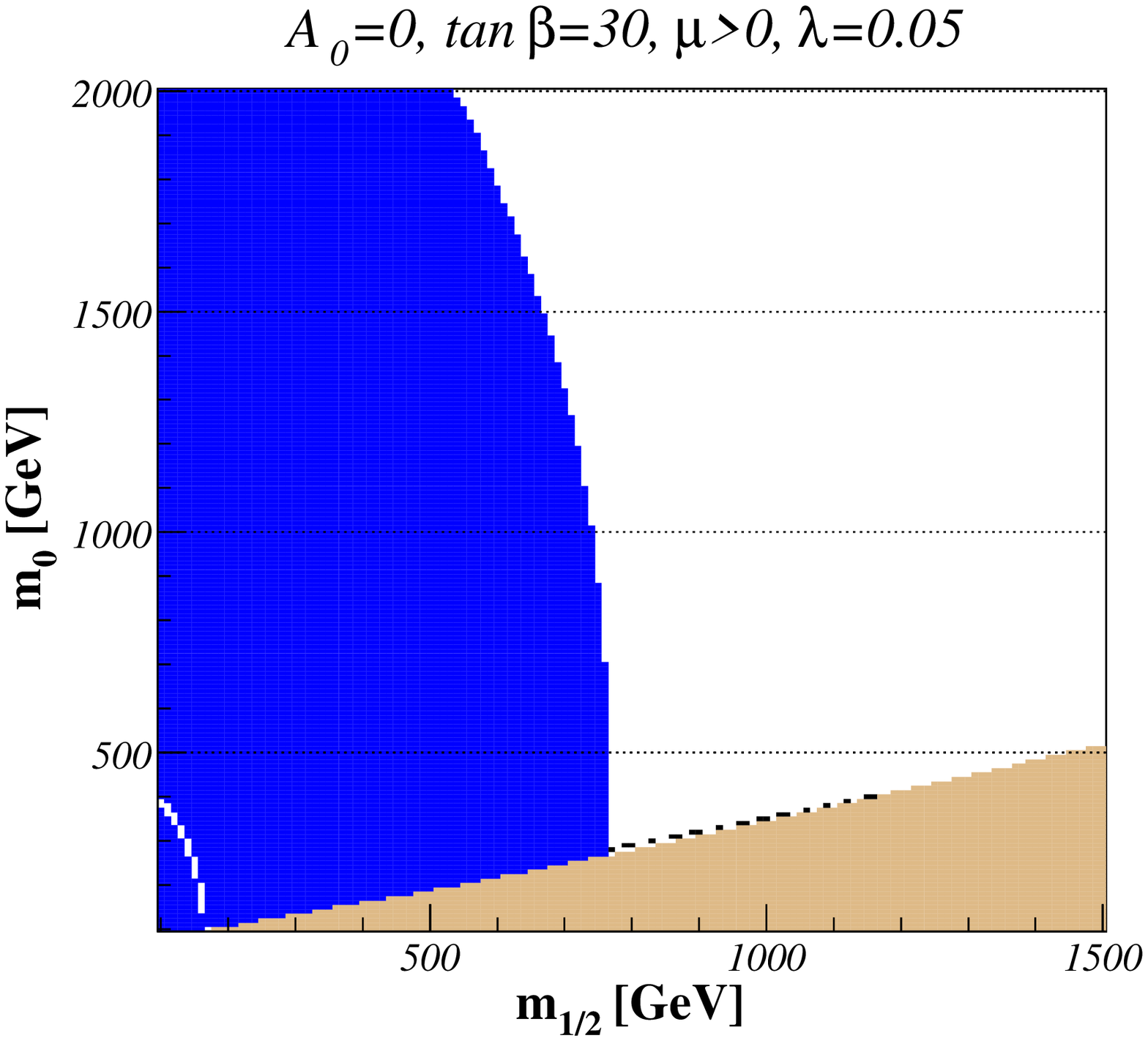}\hspace{1mm}
 \includegraphics[width=0.24\columnwidth]{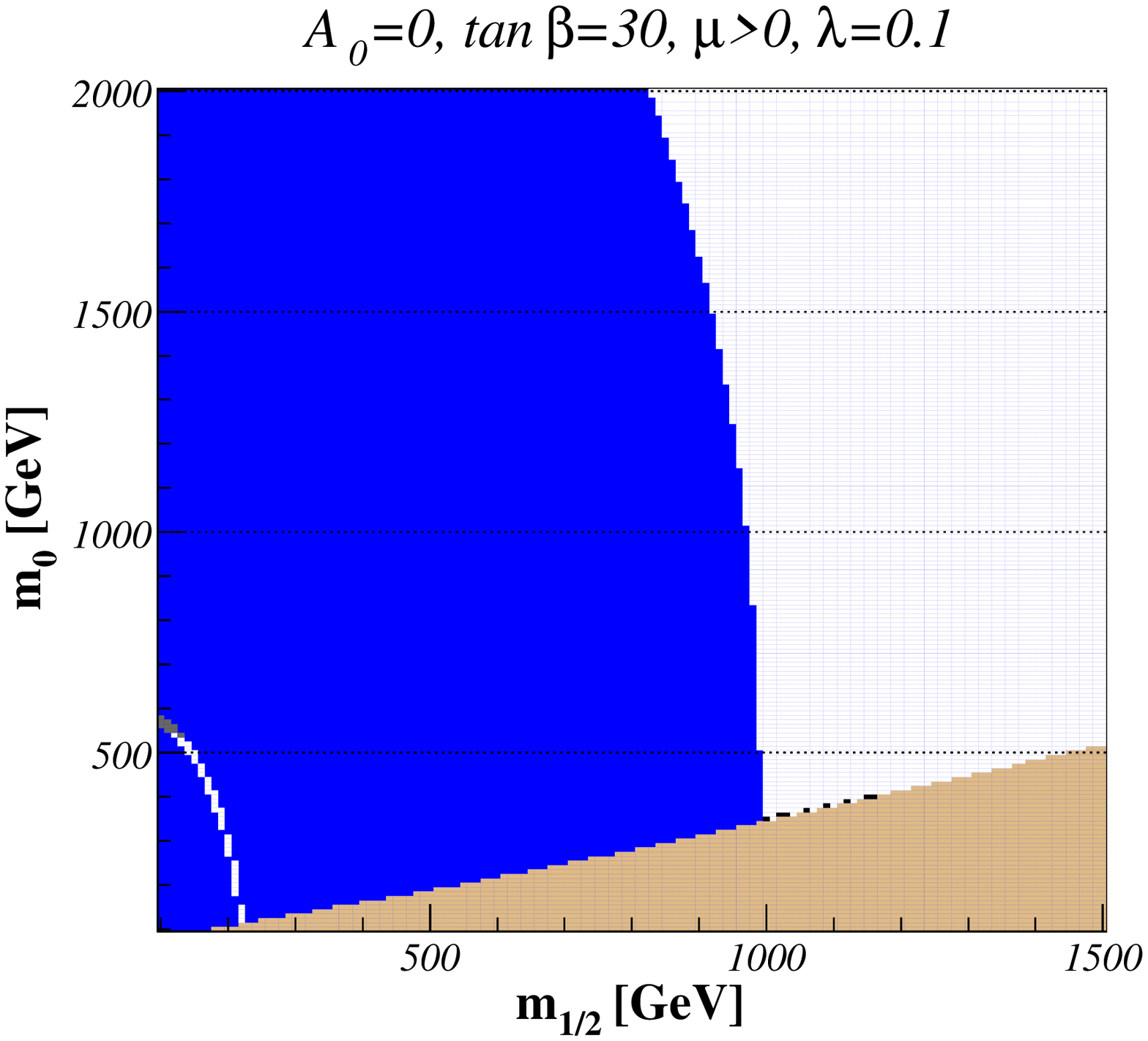}
 \caption{\label{fig:10}The $(m_0,m_{1/2})$ planes for $\tan\beta=30$,
          $A_0=0$ GeV, $\mu>0$, and $\lambda=0$, 0.03, 0.05 and 0.1. We show
          $a_\mu$ (grey) and WMAP (black) favoured as well as $b\to s\gamma$
          (blue) and charged LSP (beige) excluded regions of mSUGRA
          parameter space in minimal ($\lambda=0$) and non-minimal
          ($\lambda>0$) flavour violation.}
\end{figure}
%
%
\begin{figure}
 \centering
 \includegraphics[width=0.24\columnwidth]{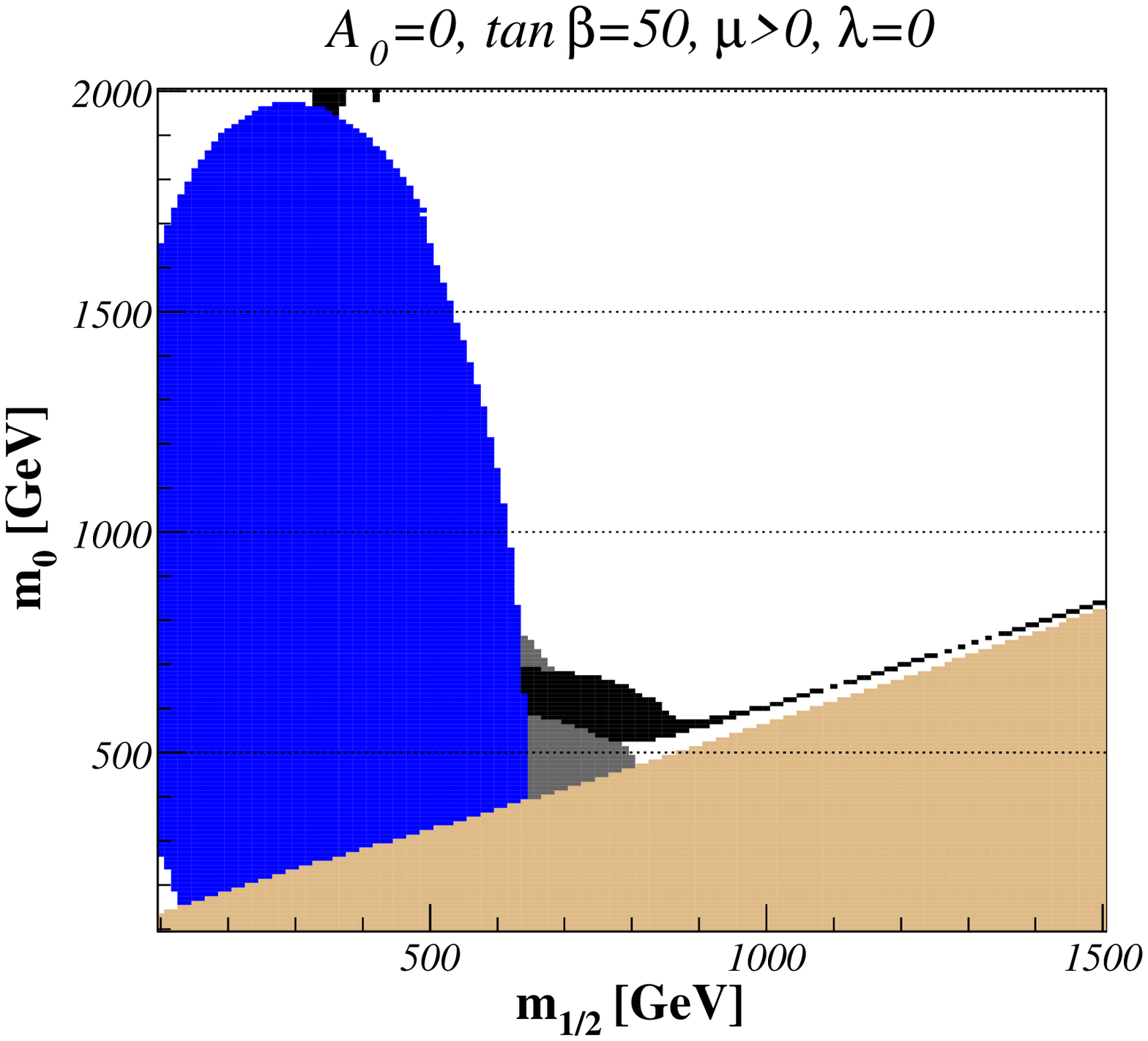}\hspace{1mm}
 \includegraphics[width=0.24\columnwidth]{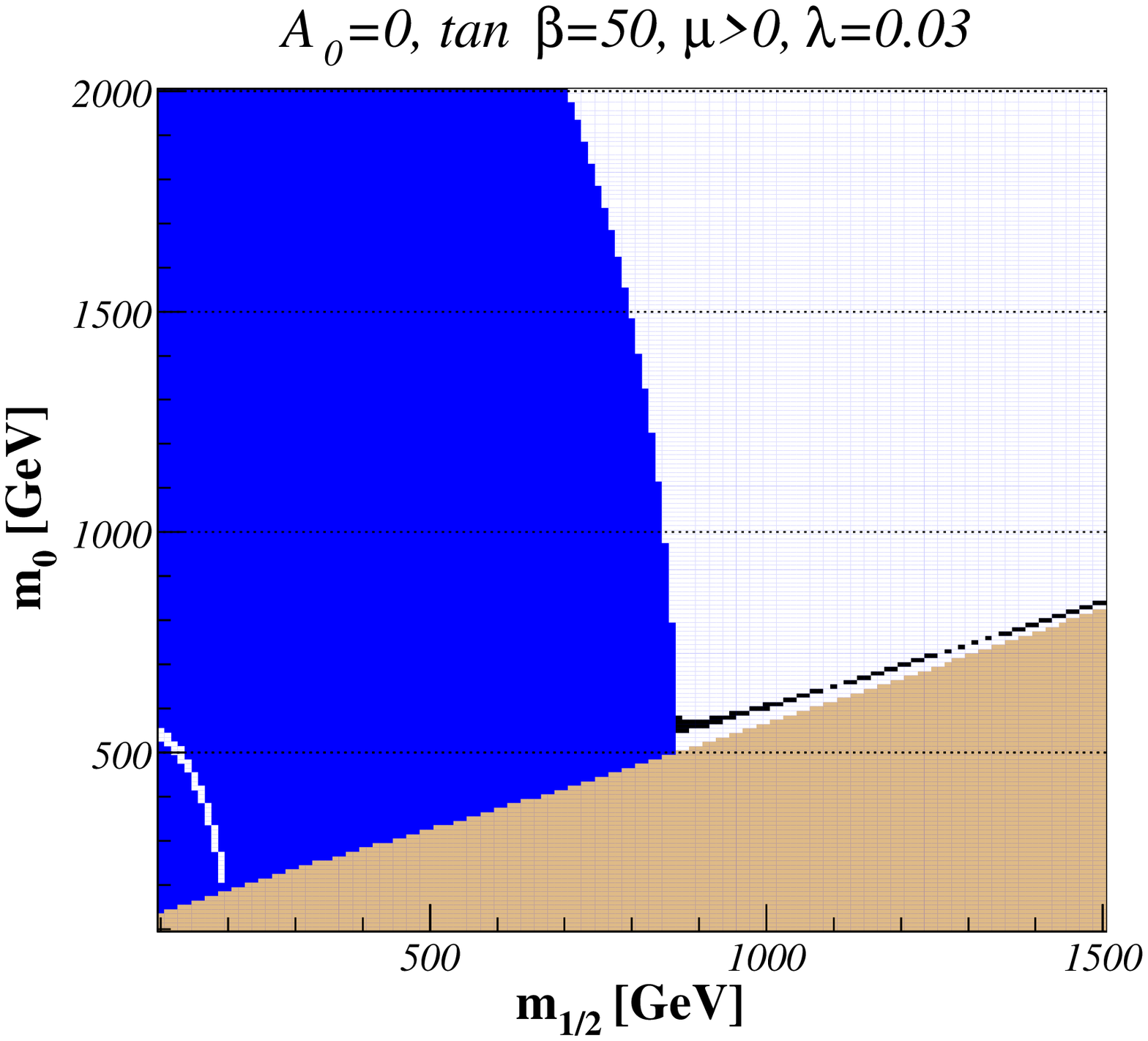}\hspace{1mm}
 \includegraphics[width=0.24\columnwidth]{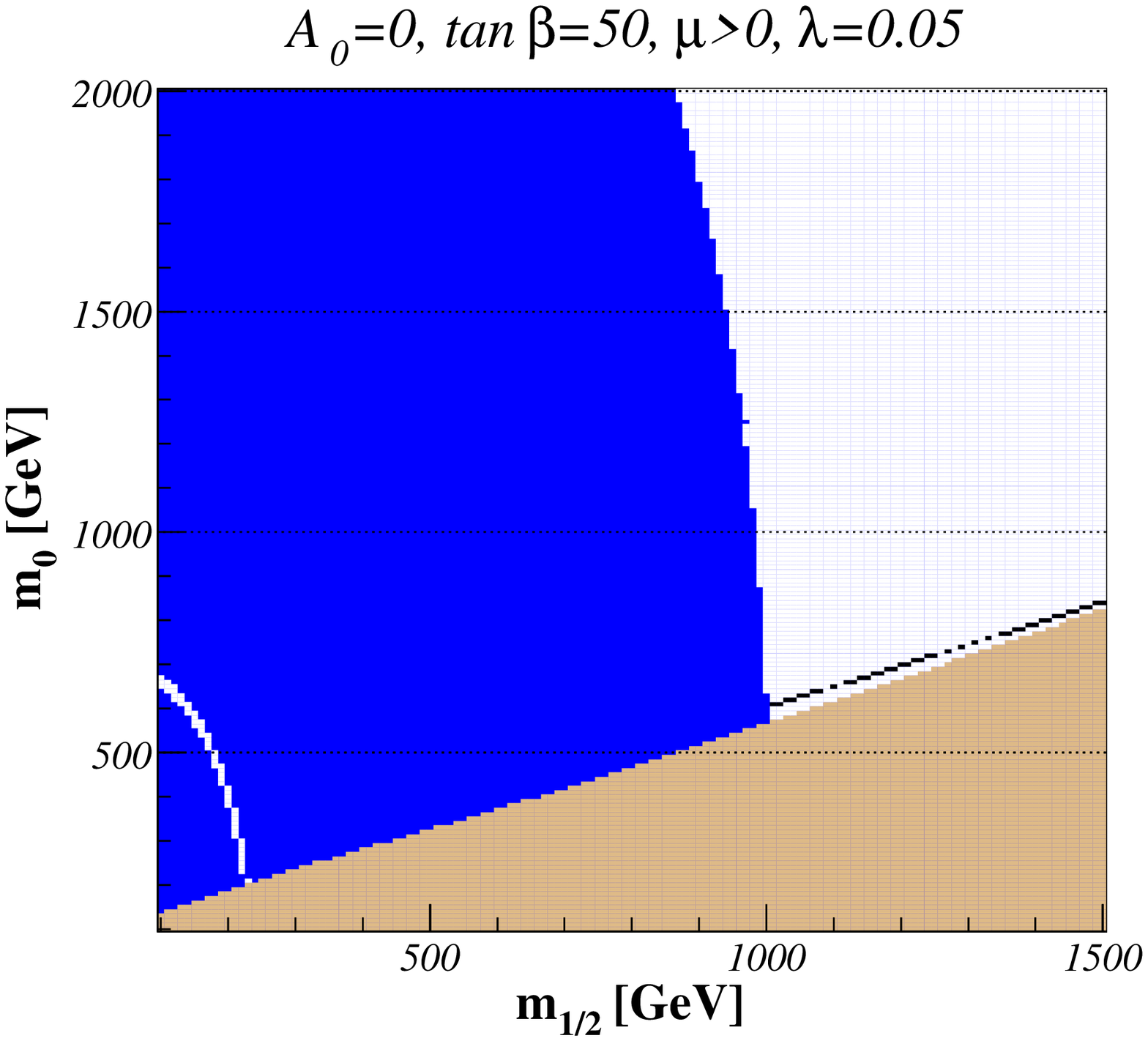}\hspace{1mm}
 \includegraphics[width=0.24\columnwidth]{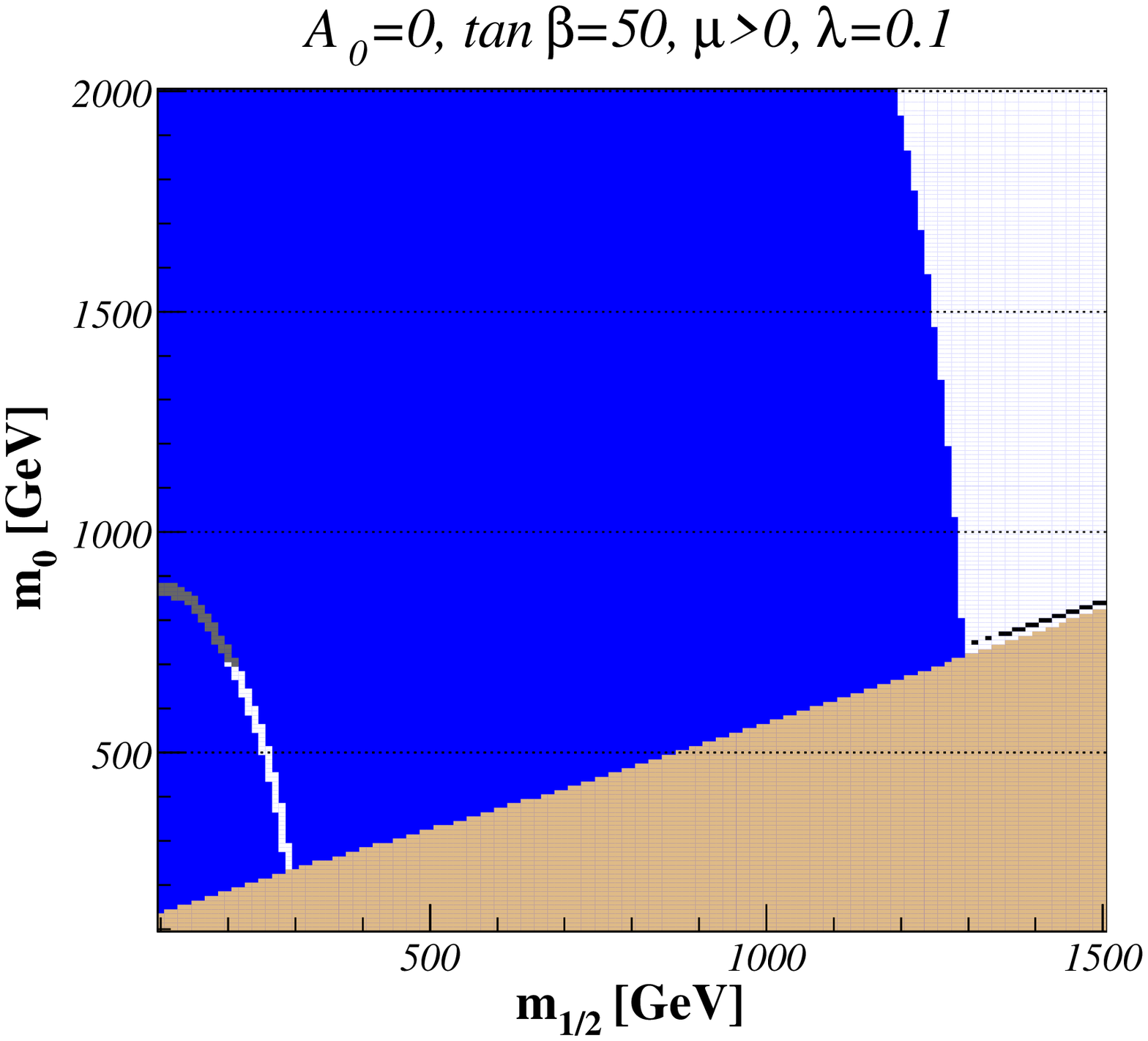}
 \caption{\label{fig:11}The $(m_0,m_{1/2})$ planes for $\tan\beta=50$,
          $A_0=0$ GeV, $\mu>0$, and $\lambda=0$, 0.03, 0.05 and 0.1. We show
          $a_\mu$ (grey) and WMAP (black) favoured as well as $b\to s\gamma$
          (blue) and charged LSP (beige) excluded regions of mSUGRA
          parameter space in minimal ($\lambda=0$) and non-minimal
          ($\lambda>0$) flavour violation.}
\end{figure}
%
In Figs.\ \ref{fig:10} and \ref{fig:11}, we show the ($m_0,m_{1/2}$)-planes
for larger $\tan\beta$, namely $\tan\beta=30$ and $\tan\beta=50$, and
for $\mu>0$. The regions which are favoured both by the anomalous magnetic
moment of the muon and by the cold dark matter relic density, and which are
not excluded by the $b\to s \gamma$ measurements, are stringently
constrained and do not allow for large flavour violation.

\subsection{(c)MFV and NMFV Benchmark Points and Slopes}
\label{sec:bench}

Restricting ourselves to non-negative values of $\mu$, we now inspect the
$(m_0,m_{1/2})$-planes in Figs.\ \ref{fig:9}-\ref{fig:11} for cMSSM
scenarios that
\begin{itemize}
 \item are allowed/favoured by low-energy, electroweak precision, and
       cosmological constraints,
 \item permit non-minimal flavour violation among left-chiral squarks of the
       second and third generation up to $\lambda\leq0.1$,
 \item and are at the same time collider-friendly, i.e.\ have relatively
  low values of $m_0$ and $m_{1/2}$.
\end{itemize}
Our choices are presented in Tab.\ \ref{tab:2}, together with the nearest
%
\begin{table}
 \caption{\label{tab:2}Benchmark points allowing for flavour violation
          among the second and third generations for $A_0=0$, $\mu>0$, and
          three different values of $\tan\beta$. For comparison we also
          show the nearest pre-WMAP SPS \cite{Allanach:2002nj,%
          Aguilar-Saavedra:2005pw} and post-WMAP BDEGOP
          \cite{Battaglia:2003ab} benchmark points and indicate the
          relevant cosmological regions.}
 \begin{tabular}{c|ccccc|ccc}
 & $m_0$ [GeV] & $m_{1/2}$ [GeV] & $A_0$ [GeV] & $\tan\beta$ & sgn($\mu$)&
 SPS & BDEGOP & Cosmol.\ Region \\
 \hline
 A & 700 & 200 & 0 & 10 & 1 & 2  & E' & Focus Point \\
 B & 100 & 400 & 0 & 10 & 1 & 3  & C' & Co-Annihilation \\
 C & 230 & 590 & 0 & 30 & 1 & 1b & I' & Co-Annihilation  \\
 D & 600 & 700 & 0 & 50 & 1 & 4  & L' & Bulk/Higgs-funnel \\
 \end{tabular}
\end{table}
%
pre-WMAP Snowmass Points (and Slopes, SPS) \cite{Allanach:2002nj,%
Aguilar-Saavedra:2005pw} and the nearest post-WMAP scenarios proposed
in Ref.\ \cite{Battaglia:2003ab}. We also indicate the relevant cosmological
region for each point and attach a model line (slope) to it, given by
\bea
{\rm A:} & 180~{\rm GeV}~\leq~m_{1/2}~\leq~250~{\rm GeV}~, & m_0~=~-1936~{\rm GeV}+~12.9\,m_{1/2},\nonumber\\
{\rm B:} & 400~{\rm GeV}~\leq~m_{1/2}~\leq~900~{\rm GeV}~, & m_0~=~~~~~4.93~{\rm GeV}+0.229\,m_{1/2},\nonumber\\
{\rm C:} & 500~{\rm GeV}~\leq~m_{1/2}~\leq~700~{\rm GeV}~, & m_0~=~~~~~~~\,54~{\rm GeV}+0.297\,m_{1/2},\nonumber\\
{\rm D:} & 575~{\rm GeV}~\leq~m_{1/2}~\leq~725~{\rm GeV}~, &
m_0~=~~~~~~600~{\rm GeV}. \eea These slopes trace the
allowed/favoured regions from lower to higher masses and can, of
course, also be used in cMFV scenarios with $\lambda=0$. We have
verified that in the case of MFV \cite{Altmannshofer:2007cs} the
hierarchy $\Delta_{\rm LL}^{qq'}\gg\Delta_{\rm LR,RL}^{qq'}\gg
\Delta_{\rm RR}^{qq'}$ and the equality of $\lambda^{sb}_{LL} =
\lambda^{ct}_{LL}$ are still reasonably well fulfilled numerically
with the values of $\lambda^{sb}_{LL} \approx \lambda^{ct}_{LL}$
ranging from zero to $5\times 10^{-3}\ldots 1\times 10^{-2}$ for
our four typical benchmark points.

Starting with Fig.\ \ref{fig:9} and $\tan\beta=10$, the bulk region of
equally low scalar and fermion masses is all but excluded by the $b\to s
\gamma$ branching ratio. This leaves as a favoured region first the
so-called focus point region of low fermion masses $m_{1/2}$, where the
lightest neutralinos are relatively heavy, have a significant Higgsino
component, and annihilate dominantly into pairs of electroweak gauge bosons.
Our benchmark point A lies in this region, albeit at smaller masses than
 SPS 2     ($m_0=1450$ GeV, $m_{1/2}=300$ GeV) and
 BDEGOP E' ($m_0=1530$ GeV, $m_{1/2}=300$ GeV),
which lie outside the region favoured by $a_\mu$ (grey-shaded) and lead to
collider-unfriendly squark and gaugino masses.

The second favoured region for small $\tan\beta$ is the
co-annihilation branch of low scalar masses $m_0$, where the
lighter tau-slepton mass eigenstate is not much heavier than the
lightest neutralino and the two have a considerable
co-annihilation cross section. This is where we have chosen our
benchmark point B, which differs from the points
 SPS 3     ($m_0=90$ GeV, $m_{1/2}=400$ GeV) and
 BDEGOP C' ($m_0=85$ GeV, $m_{1/2}=400$ GeV)
only very little in the scalar mass. This minor difference may be traced to
the fact that we use DarkSUSY 4.1 \cite{Gondolo:2004sc} instead of the
private dark matter program SSARD of Ref.\ \cite{Battaglia:2003ab}.

At the larger value of $\tan\beta=30$ in Fig.\ \ref{fig:10}, only the
co-annihilation region survives the constraints coming from $b\to s\gamma$
decays. Here we choose our point C, which has slightly higher masses than
both
 SPS 1b    ($m_0=200$ GeV, $m_{1/2}=400$ GeV) and
 BDEGOP I' ($m_0=175$ GeV, $m_{1/2}=350$ GeV),
due to the ever more stringent constraints from the above-mentioned rare
$B$-decay.

For the very large value of $\tan\beta=50$ in Fig.\ \ref{fig:11}, the bulk
region reappears at relatively heavy scalar and fermion masses. Here, the
couplings of the heavier scalar and pseudo-scalar Higgses $H^0$ and $A^0$ to
bottom quarks and tau-leptons and the charged-Higgs coupling to top-bottom
pairs are significantly enhanced, resulting e.g.\ in increased dark matter
annihilation cross sections through $s$-channel Higgs-exchange into
bottom-quark final states. So as $\tan \beta$ increases further, the
so-called Higgs-funnel region eventually makes its appearance on the
diagonal of large scalar and fermion masses. We choose our point D in the
concentrated (bulky) region favoured by cosmology and $a_\mu$ at masses,
that are slightly higher than those of
 SPS 4     ($m_0=400$ GeV, $m_{1/2}=300$ GeV) and
 BDEGOP L' ($m_0=300$ GeV, $m_{1/2}=450$ GeV).
We do so in order to escape again from the constraints of the $b\to s\gamma$
decay, which are stronger today than they were a few years ago. In this
scenario, squarks and gluinos are very heavy with masses above 1 TeV.

\subsection{Dependence of Precision Observables and Squark-Mass Eigenvalues
 on Flavour Violation}

Let us now turn to the dependence of the precision variables discussed in
Sec.\ \ref{sec:const} on the flavour violating parameter $\lambda$ in the
four benchmark scenarios defined in Sec.\ \ref{sec:bench}. As already
mentioned, we expect the leptonic observable $a_\mu$ to depend weakly (at
two loops only) on the squark sector, and this is confirmed by our numerical
analysis. We find constant values of 6, 14, 16, and 13$\times10^{-10}$ for
the benchmarks A, B, C, and D, all of which lie well within $2\sigma$ (the
latter three even within $1\sigma$) of the experimentally favoured range
$(22\pm10)\times10^{-10}$.

\begin{figure}
 \centering
 \includegraphics[width=0.32\columnwidth]{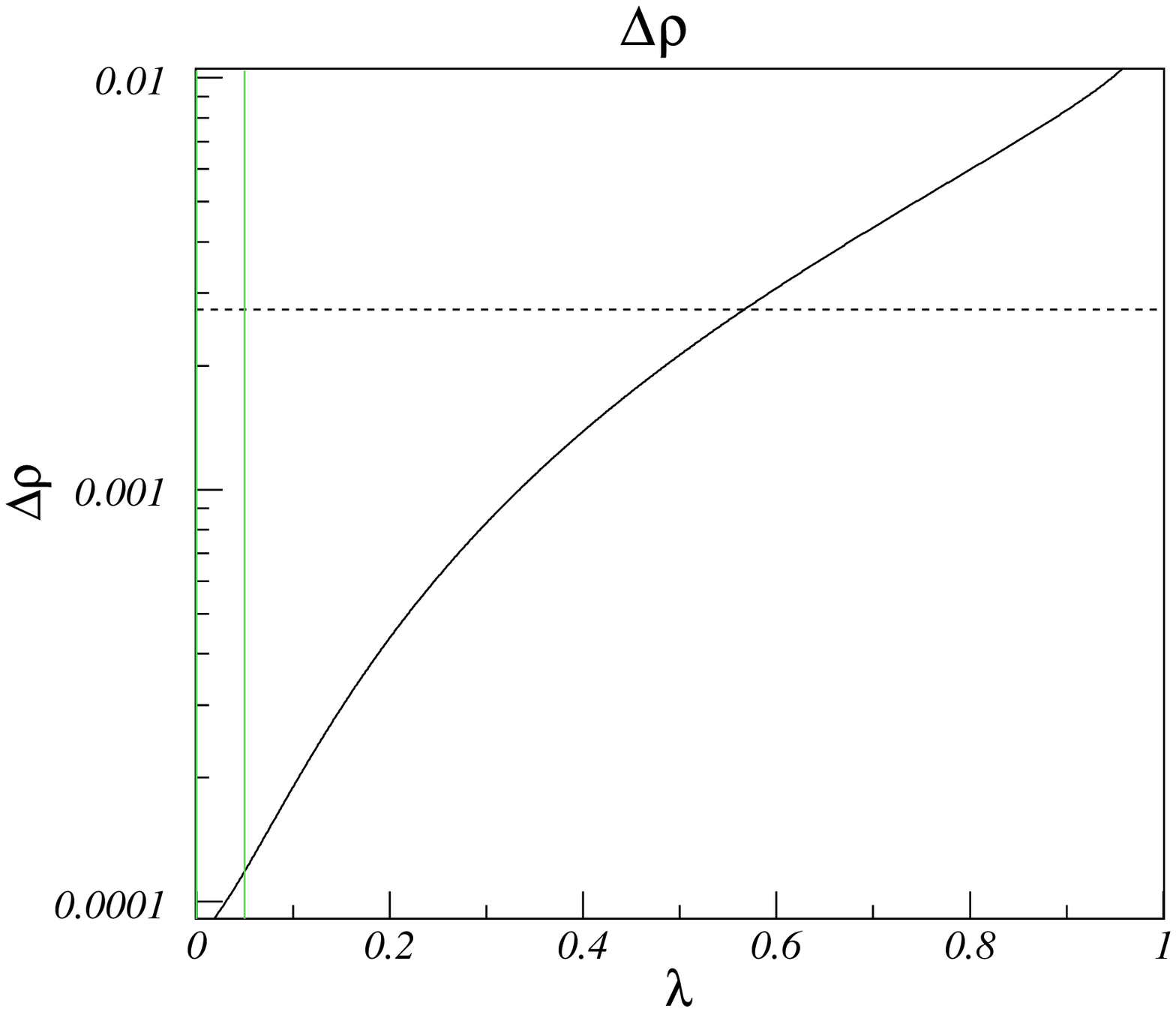}
 \includegraphics[width=0.32\columnwidth]{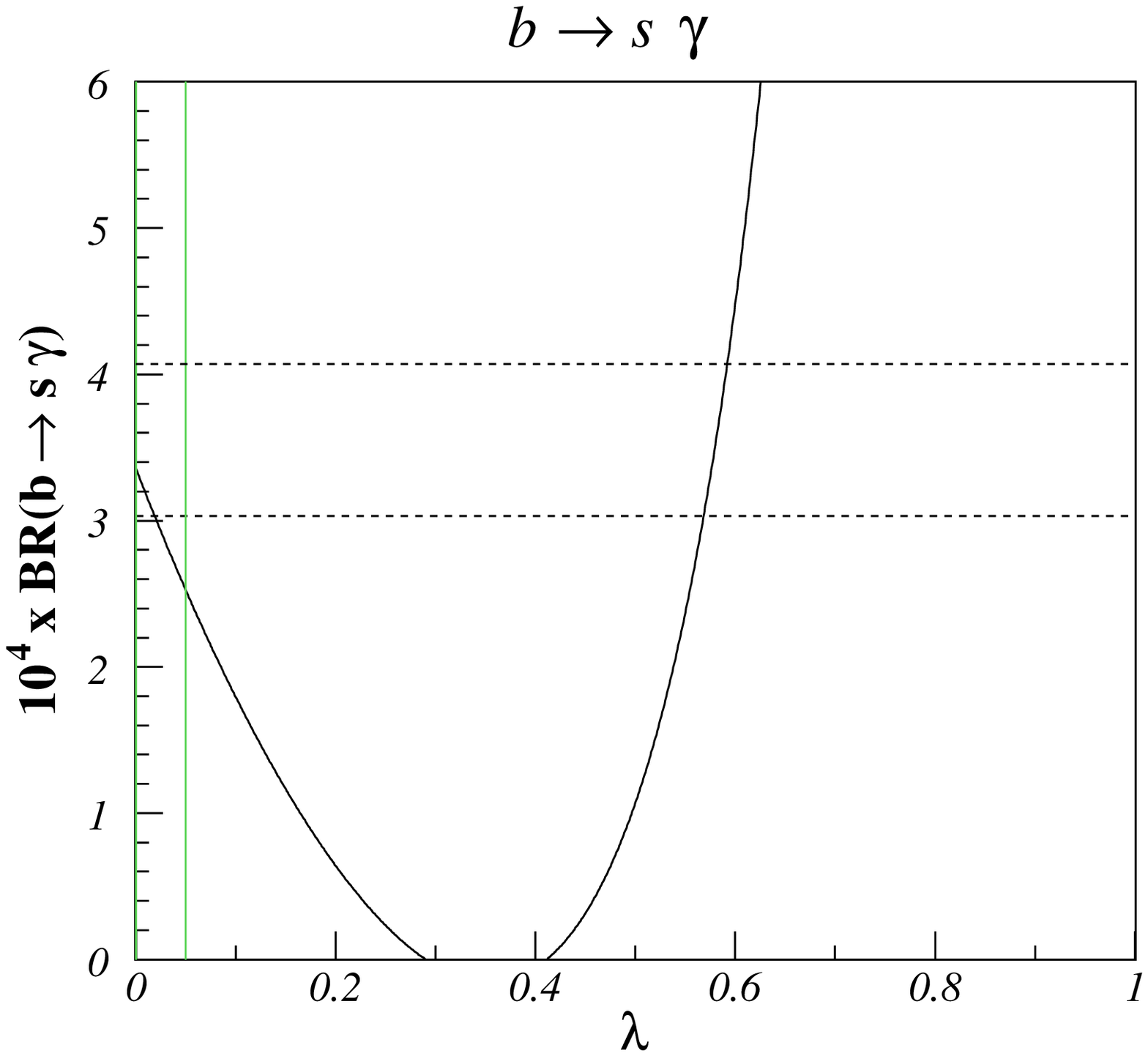}
 \includegraphics[width=0.32\columnwidth]{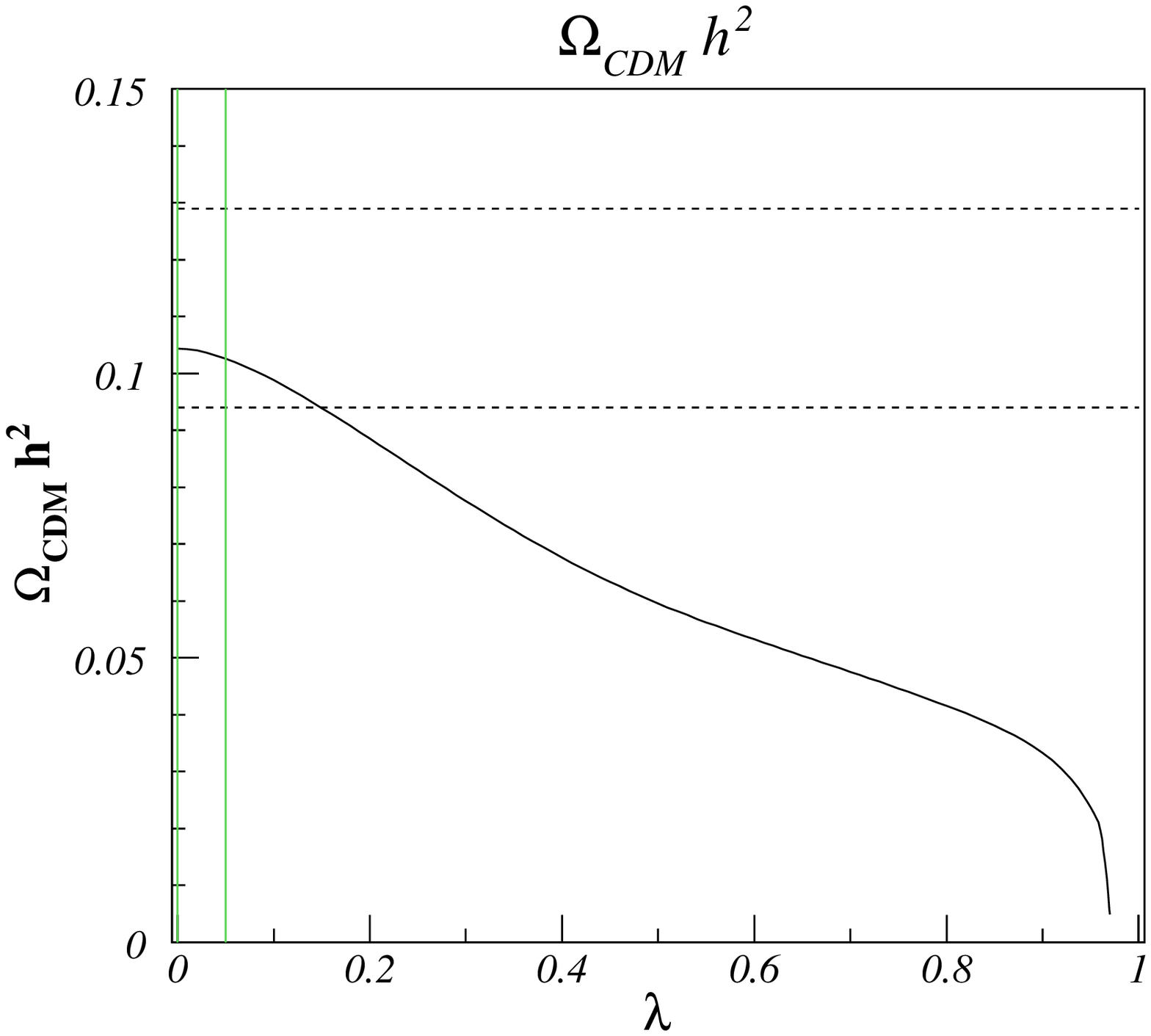}\vspace*{2mm}
 \includegraphics[width=0.32\columnwidth]{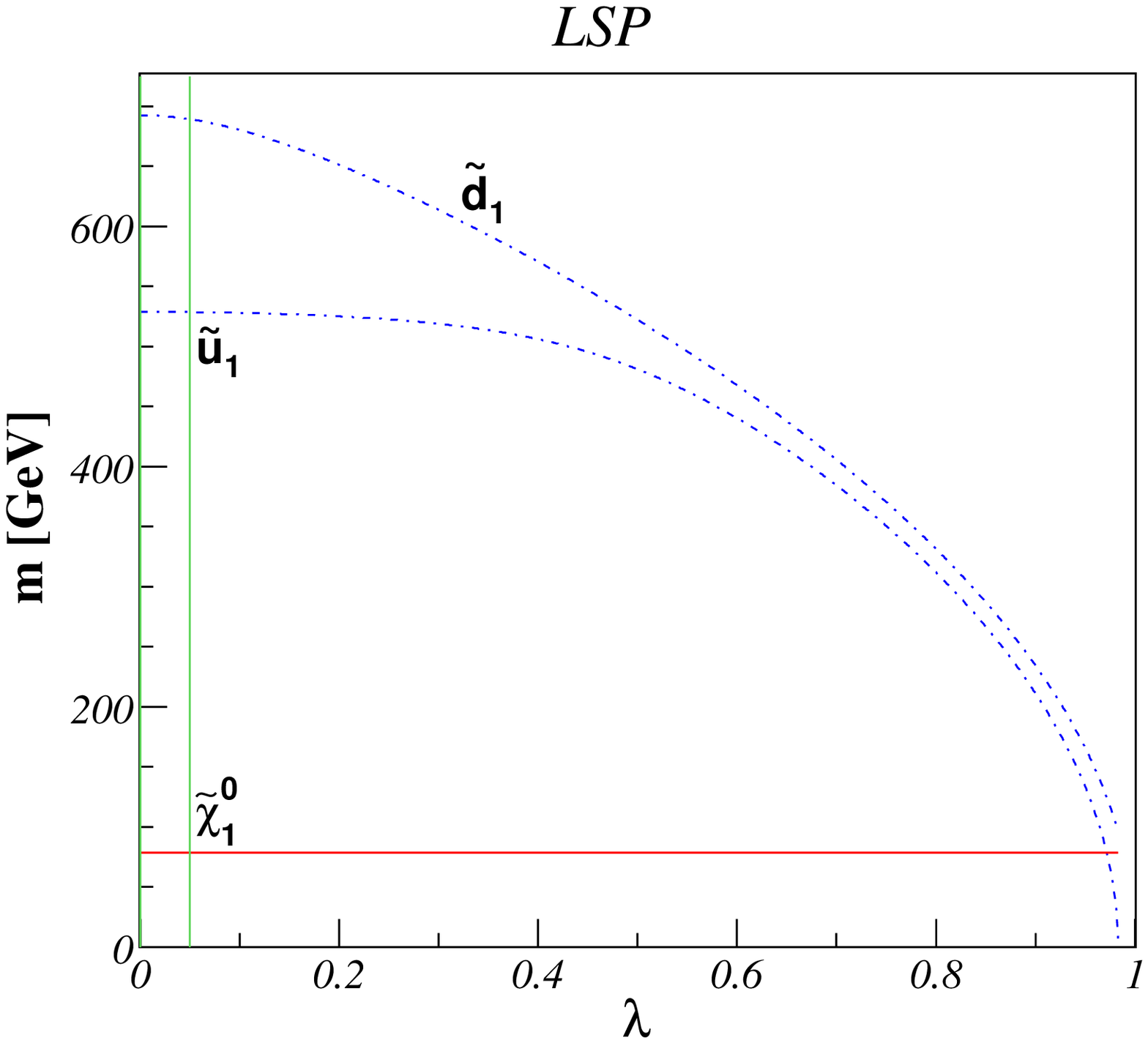}
 \includegraphics[width=0.32\columnwidth]{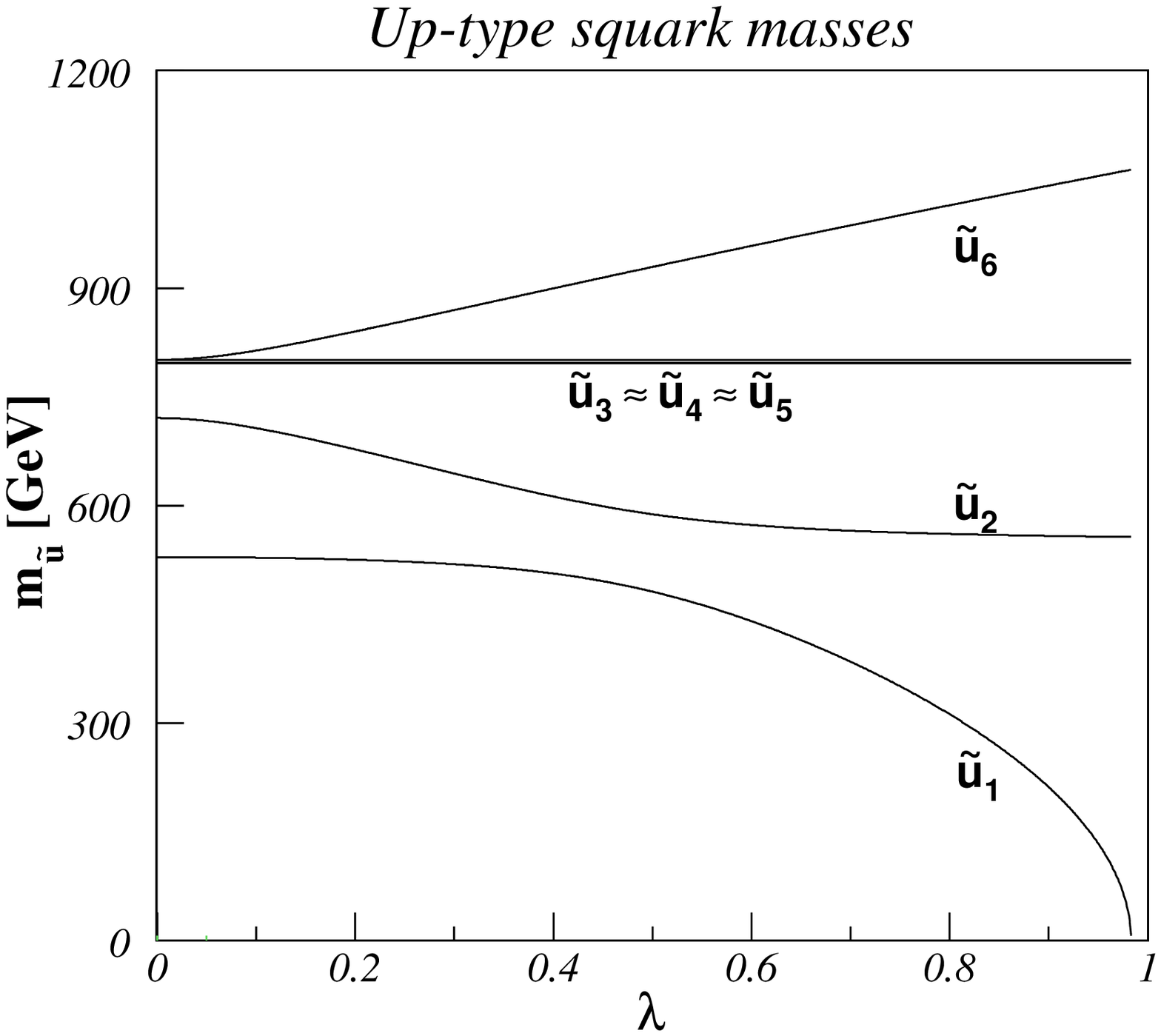}
 \includegraphics[width=0.32\columnwidth]{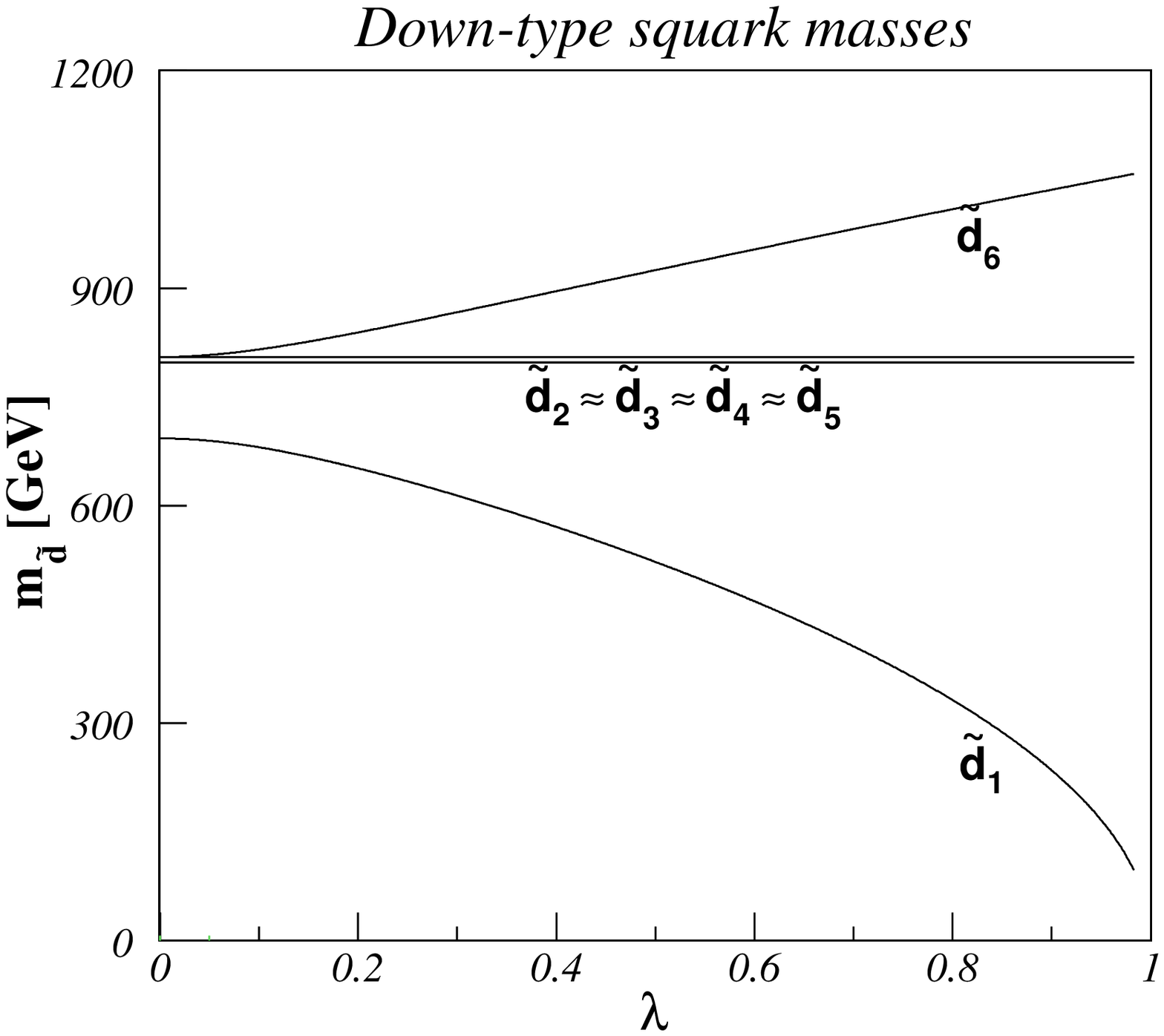}
 \caption{\label{fig:12}Dependence of the precision variables BR$(b\to
          s\gamma)$, $\Delta\rho$, and the cold dark matter relic density
          $\Omega_{CDM}h^2$ (top) as well as of the lightest SUSY particle,
          up- and down-type squark masses (bottom) on the NMFV parameter
          $\lambda$ in our benchmark scenario A. The experimentally allowed
          ranges (within $2\sigma$) are indicated by horizontal dashed
          lines.}
\end{figure}

\begin{figure}
 \centering
 \includegraphics[width=0.32\columnwidth]{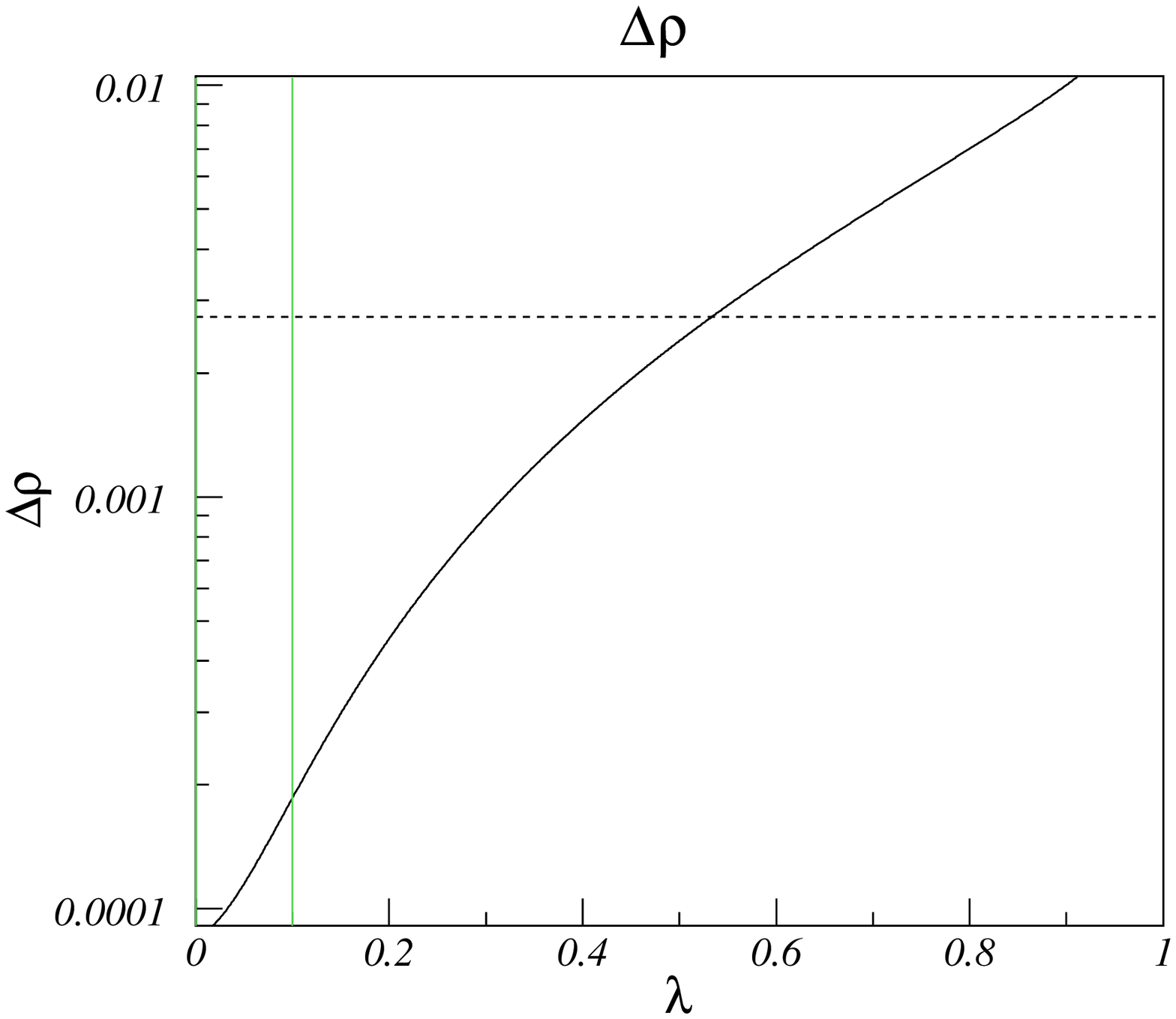}
 \includegraphics[width=0.32\columnwidth]{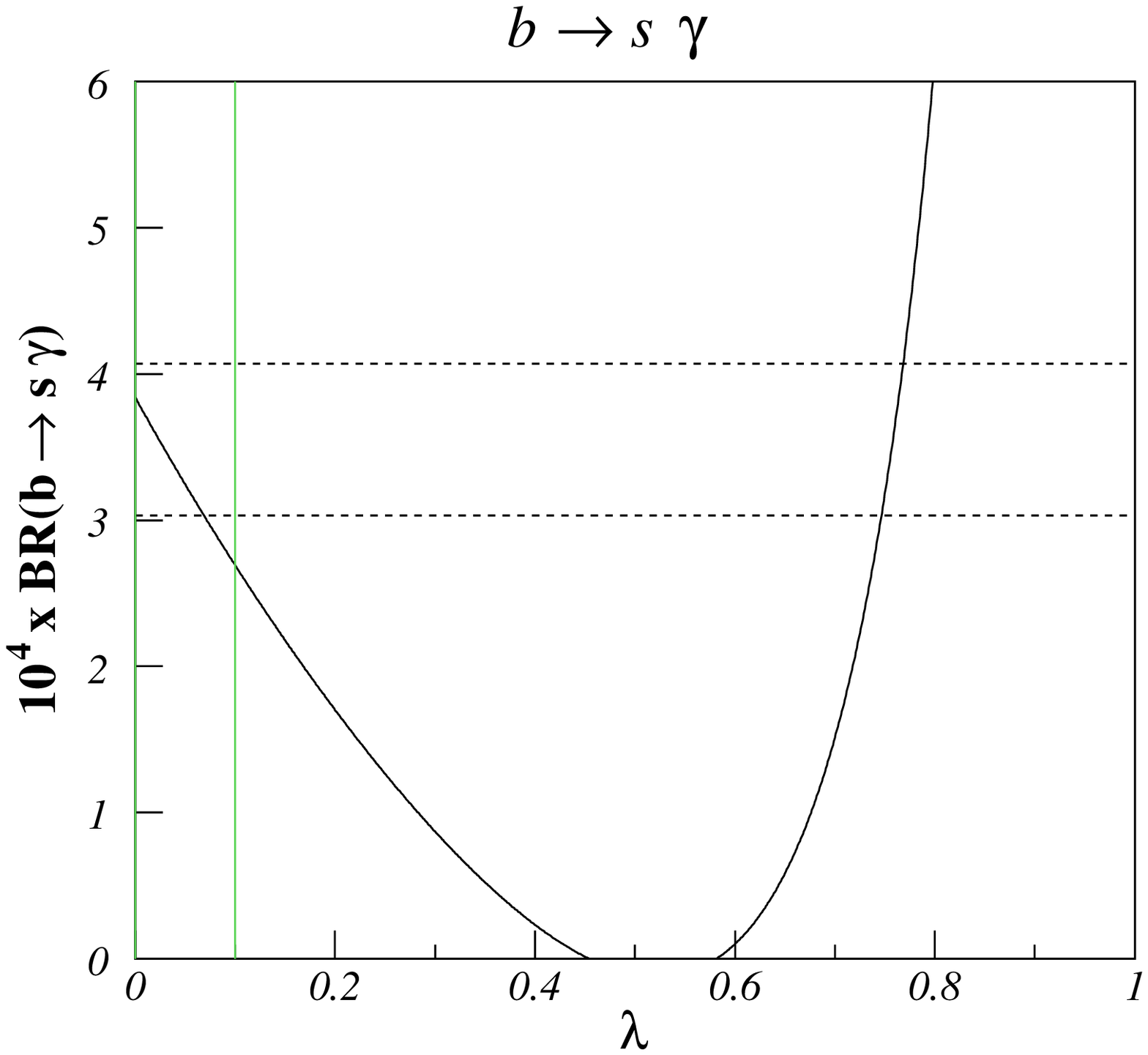}
 \includegraphics[width=0.32\columnwidth]{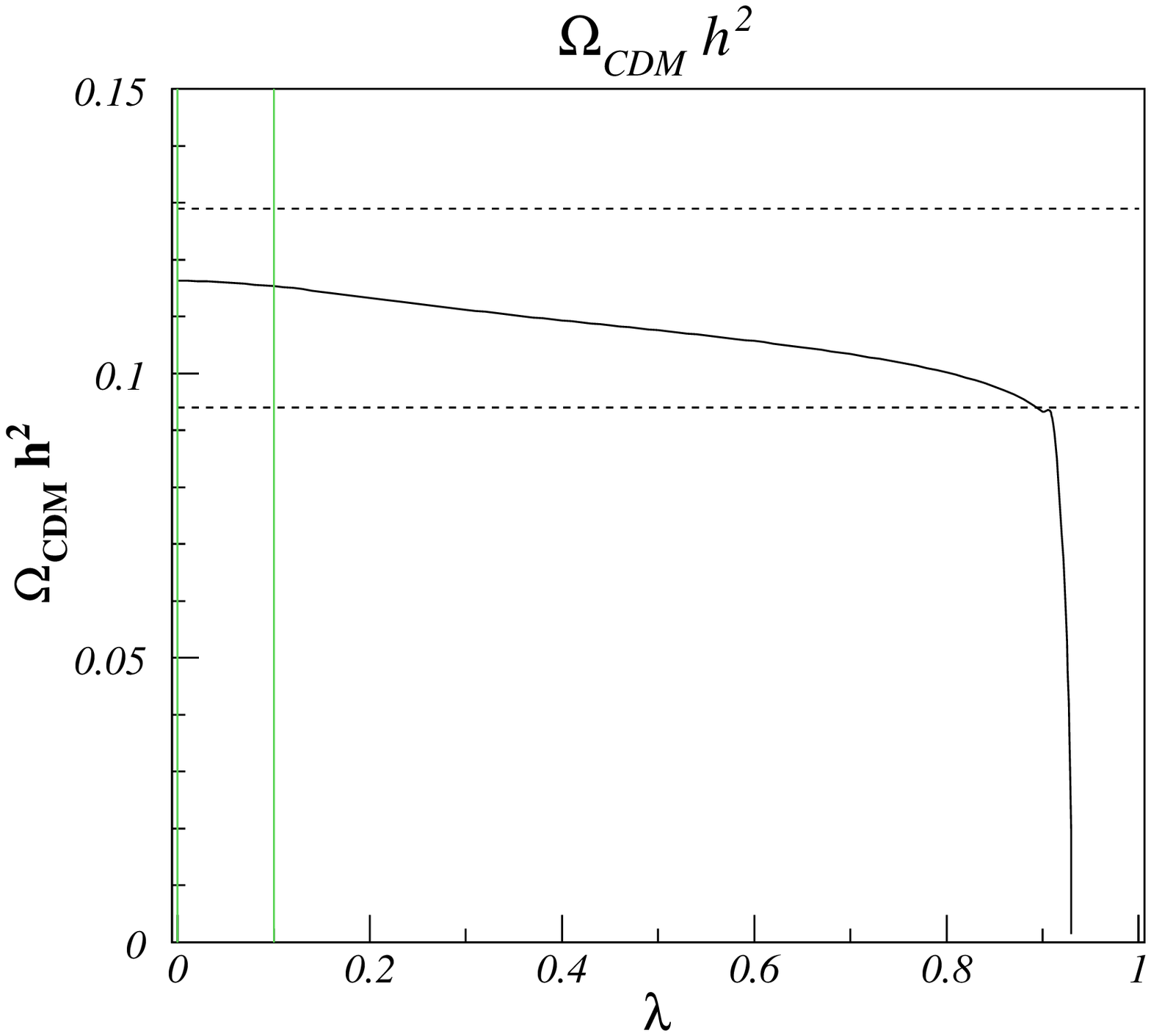}\vspace*{2mm}
 \includegraphics[width=0.32\columnwidth]{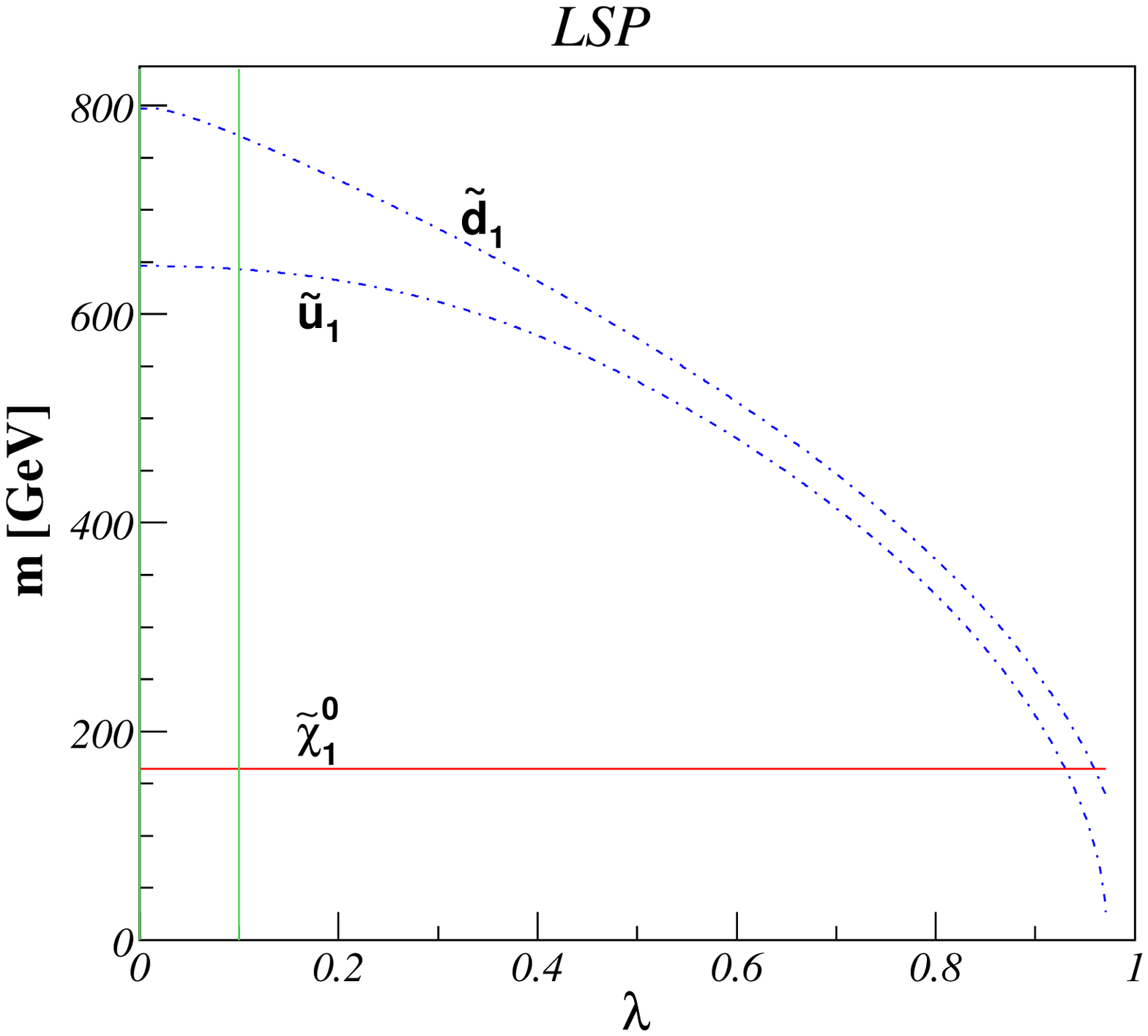}
 \includegraphics[width=0.32\columnwidth]{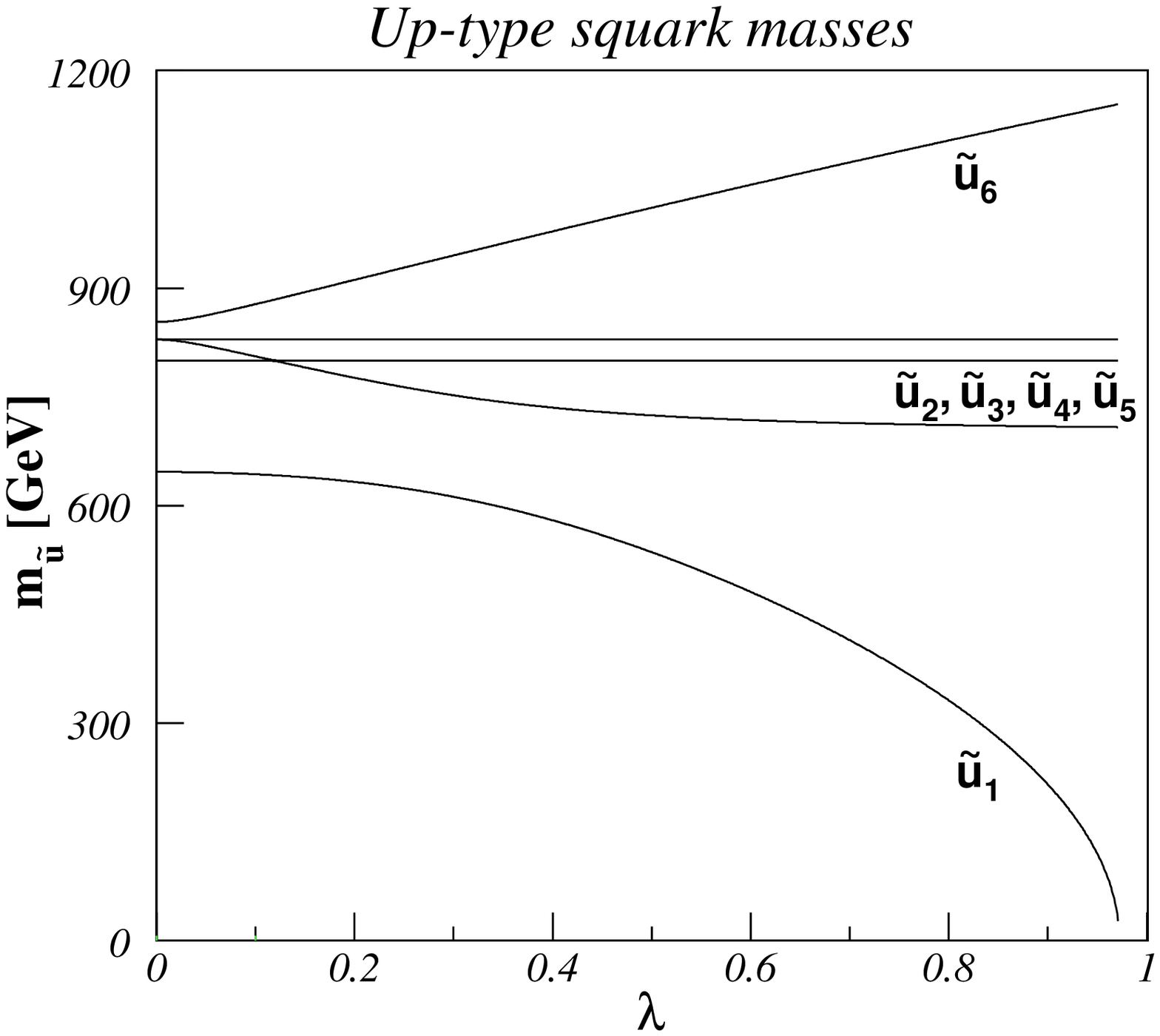}
 \includegraphics[width=0.32\columnwidth]{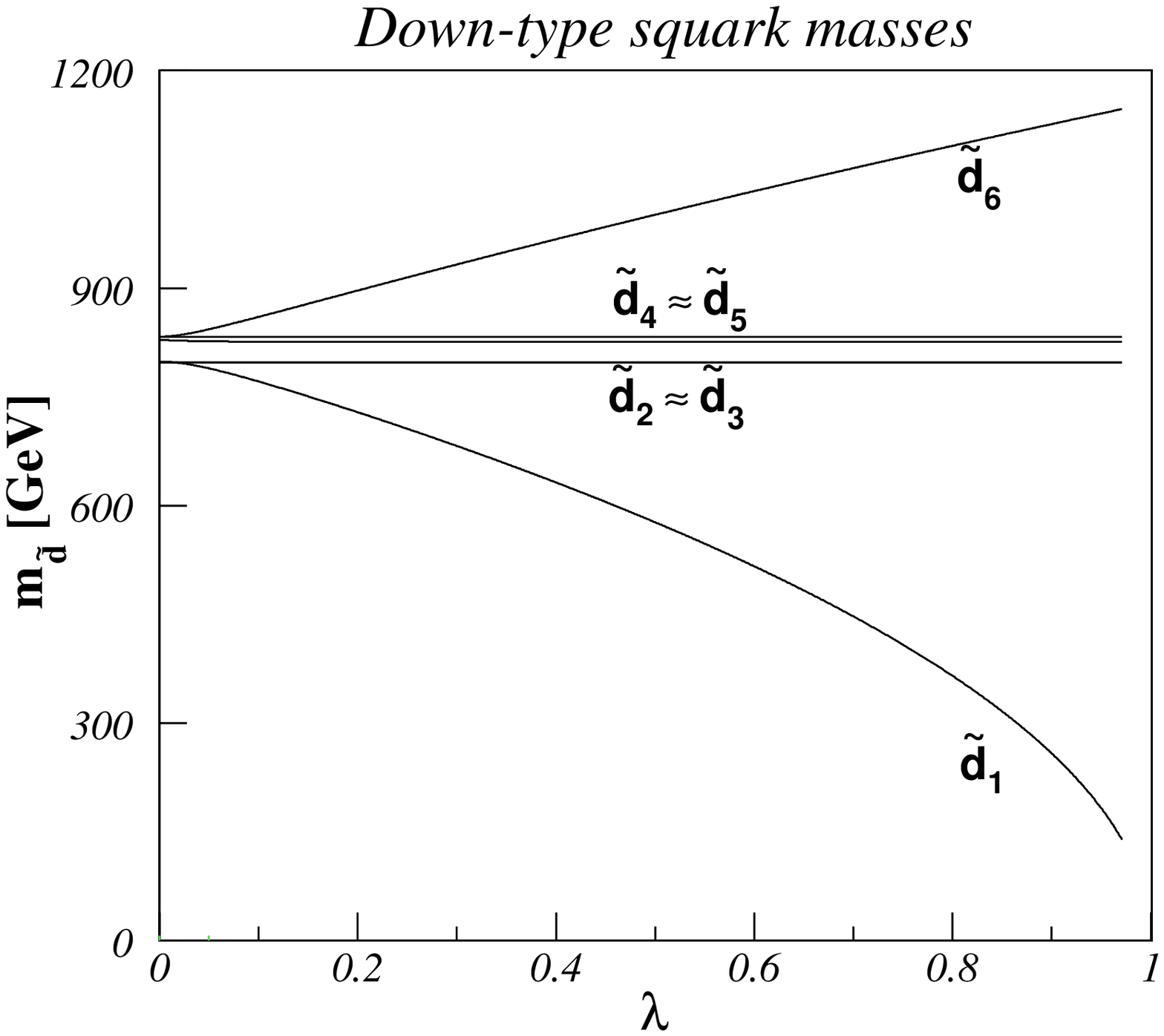}
 \caption{\label{fig:13}Same as Fig.\ \ref{fig:12} for our benchmark
          scenario B.}
\end{figure}

\begin{figure}
 \centering
 \includegraphics[width=0.32\columnwidth]{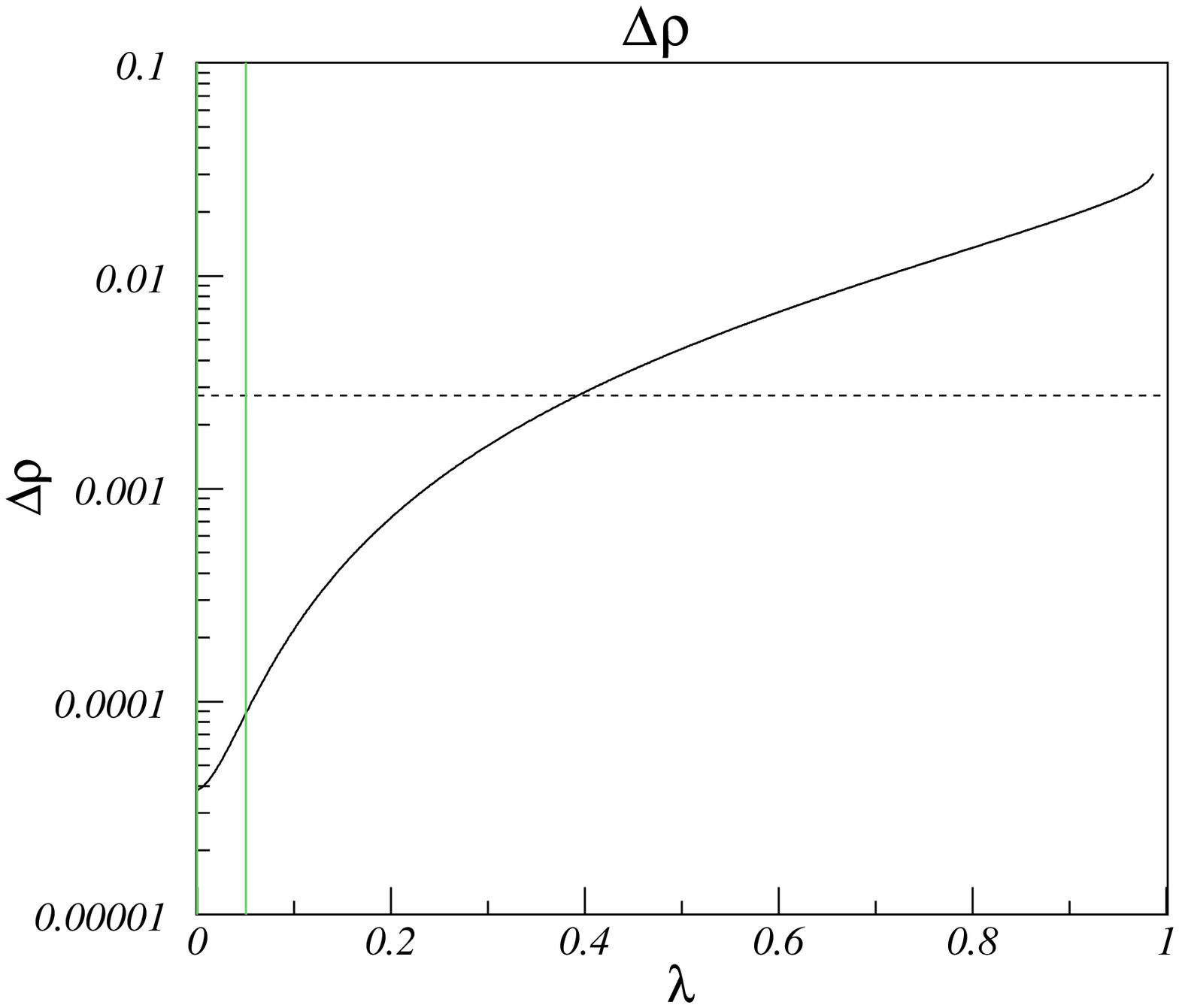}
 \includegraphics[width=0.32\columnwidth]{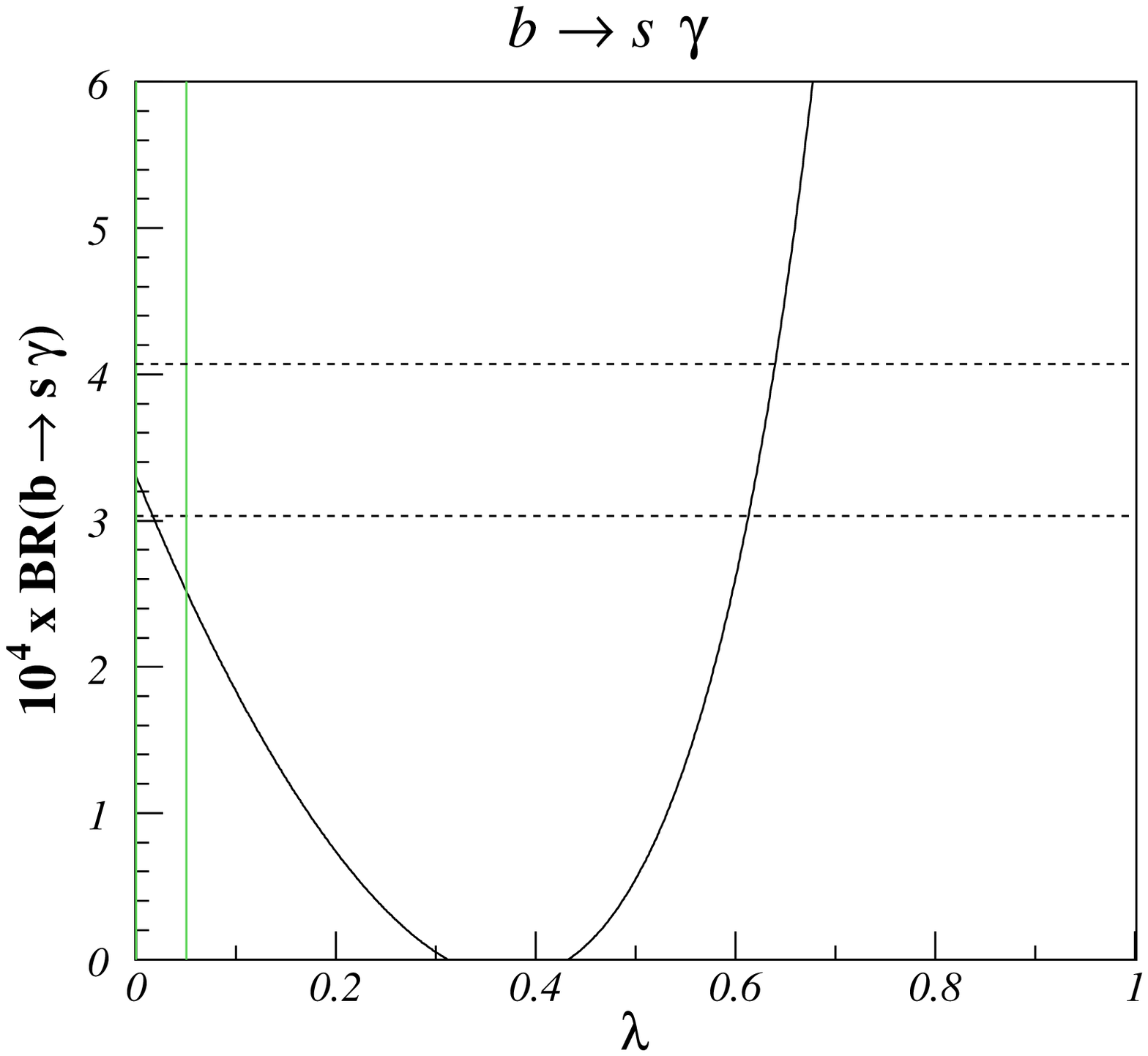}
 \includegraphics[width=0.32\columnwidth]{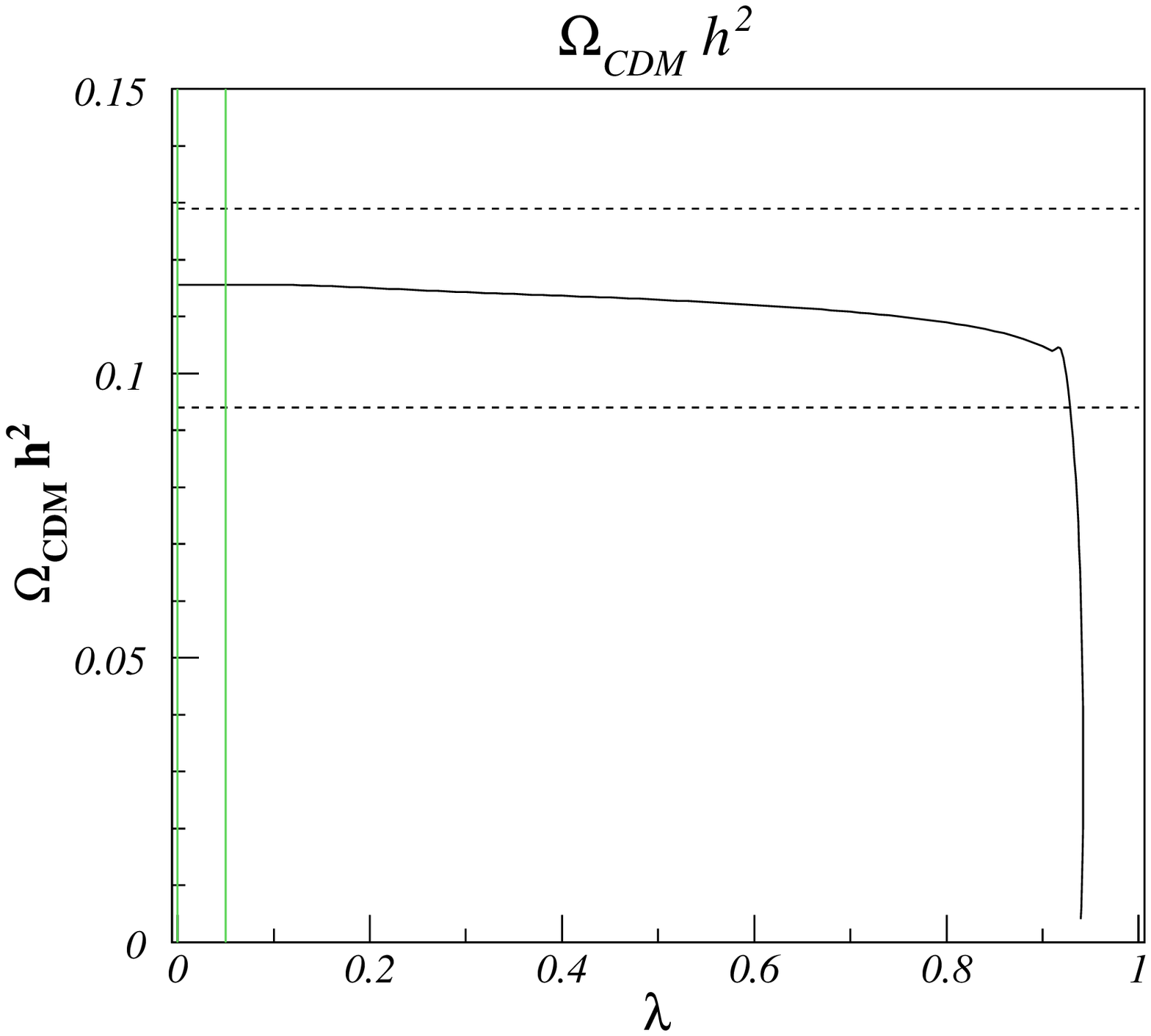}\vspace*{2mm}
 \includegraphics[width=0.32\columnwidth]{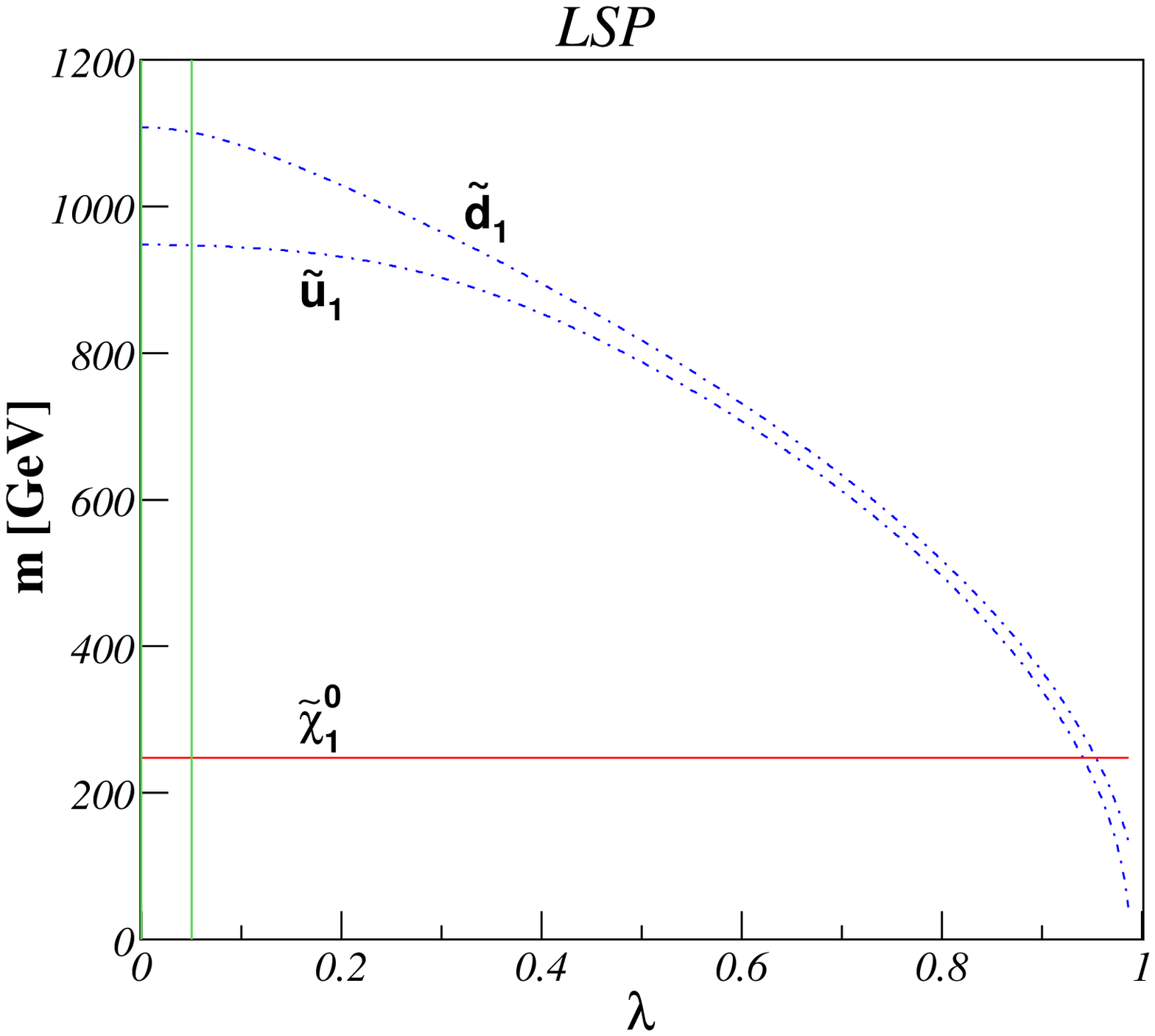}
 \includegraphics[width=0.32\columnwidth]{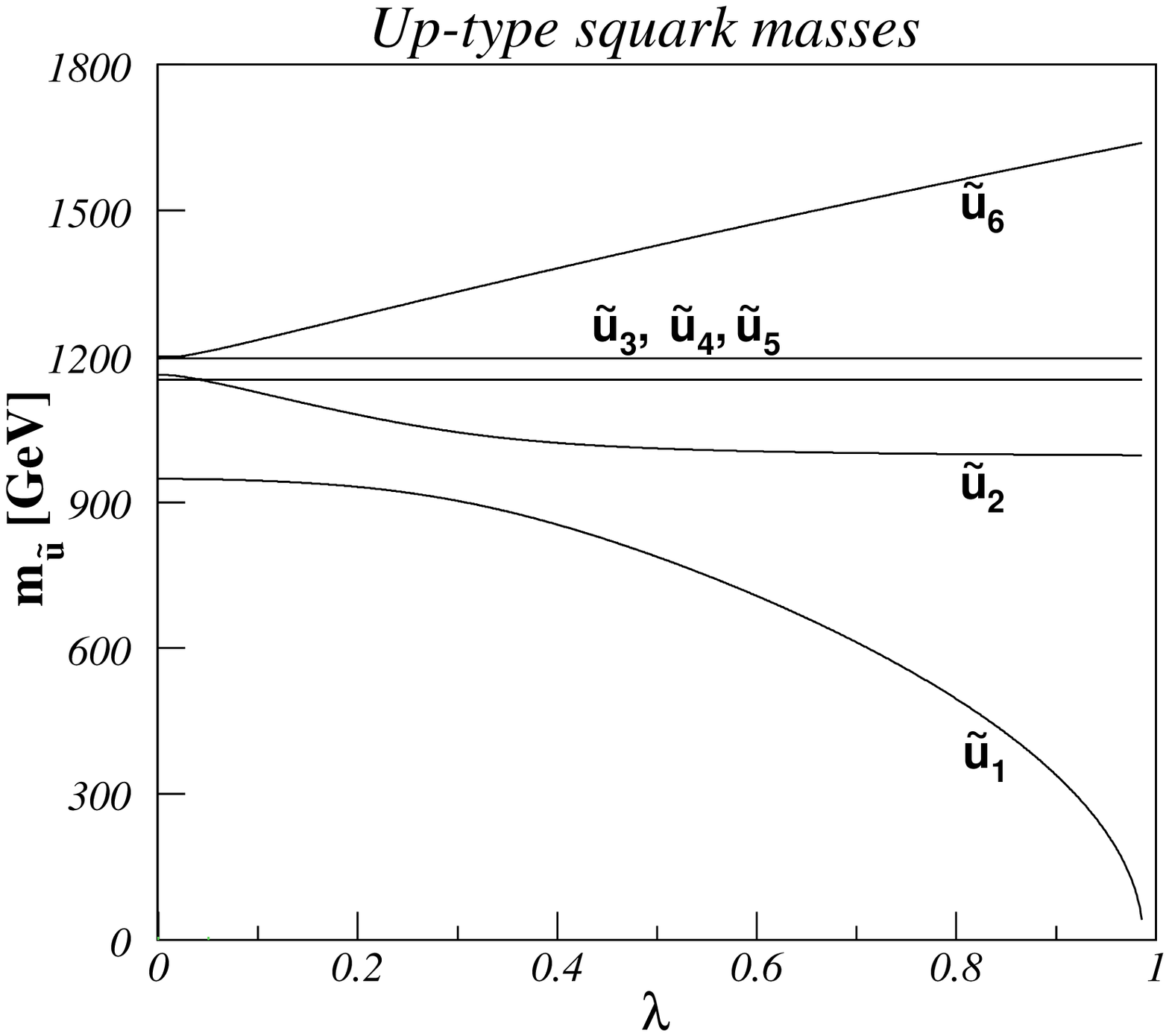}
 \includegraphics[width=0.32\columnwidth]{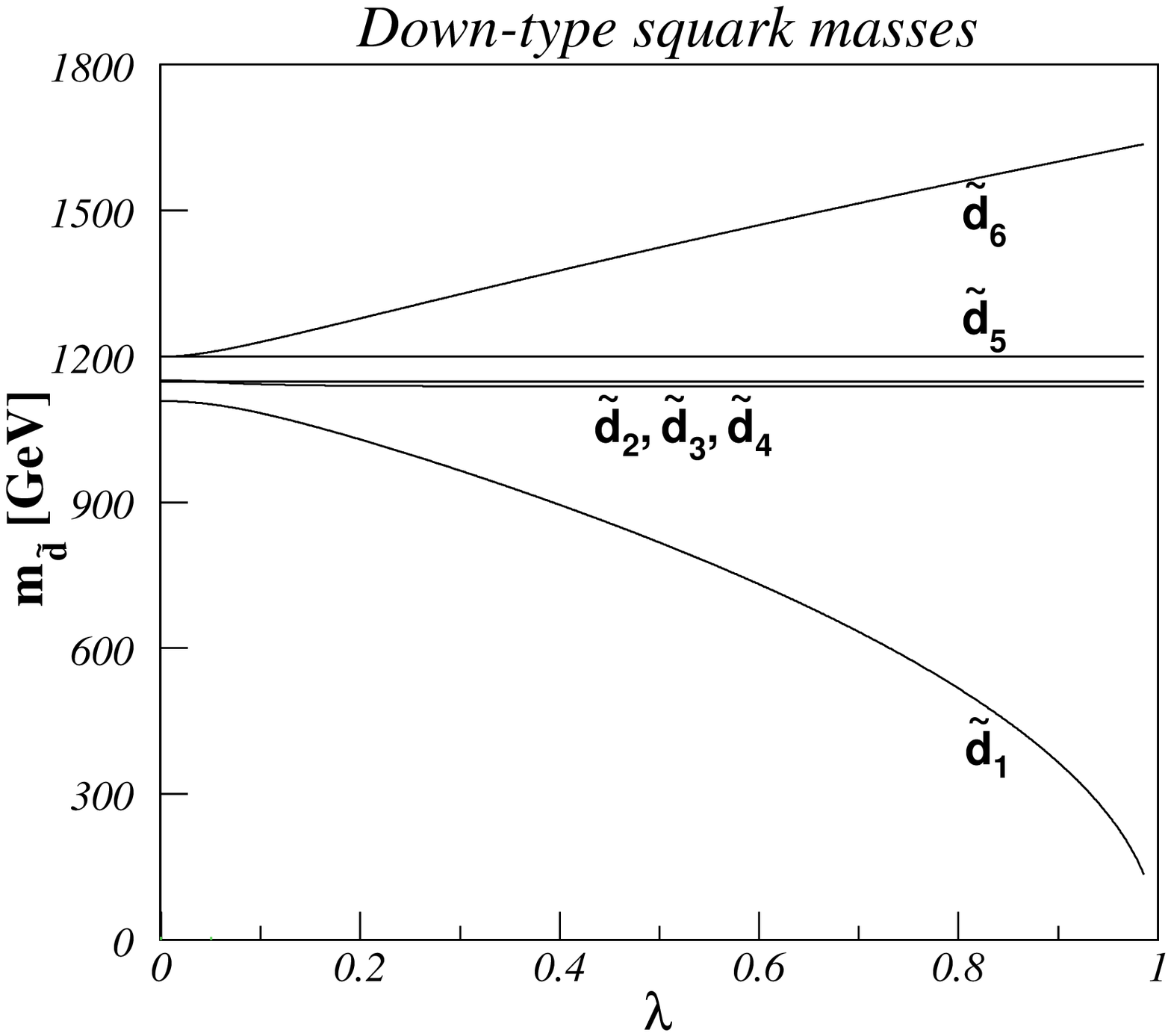}
 \caption{\label{fig:14}Same as Fig.\ \ref{fig:12} for our benchmark
          scenario C.}
\end{figure}

\begin{figure}
 \centering
 \includegraphics[width=0.32\columnwidth]{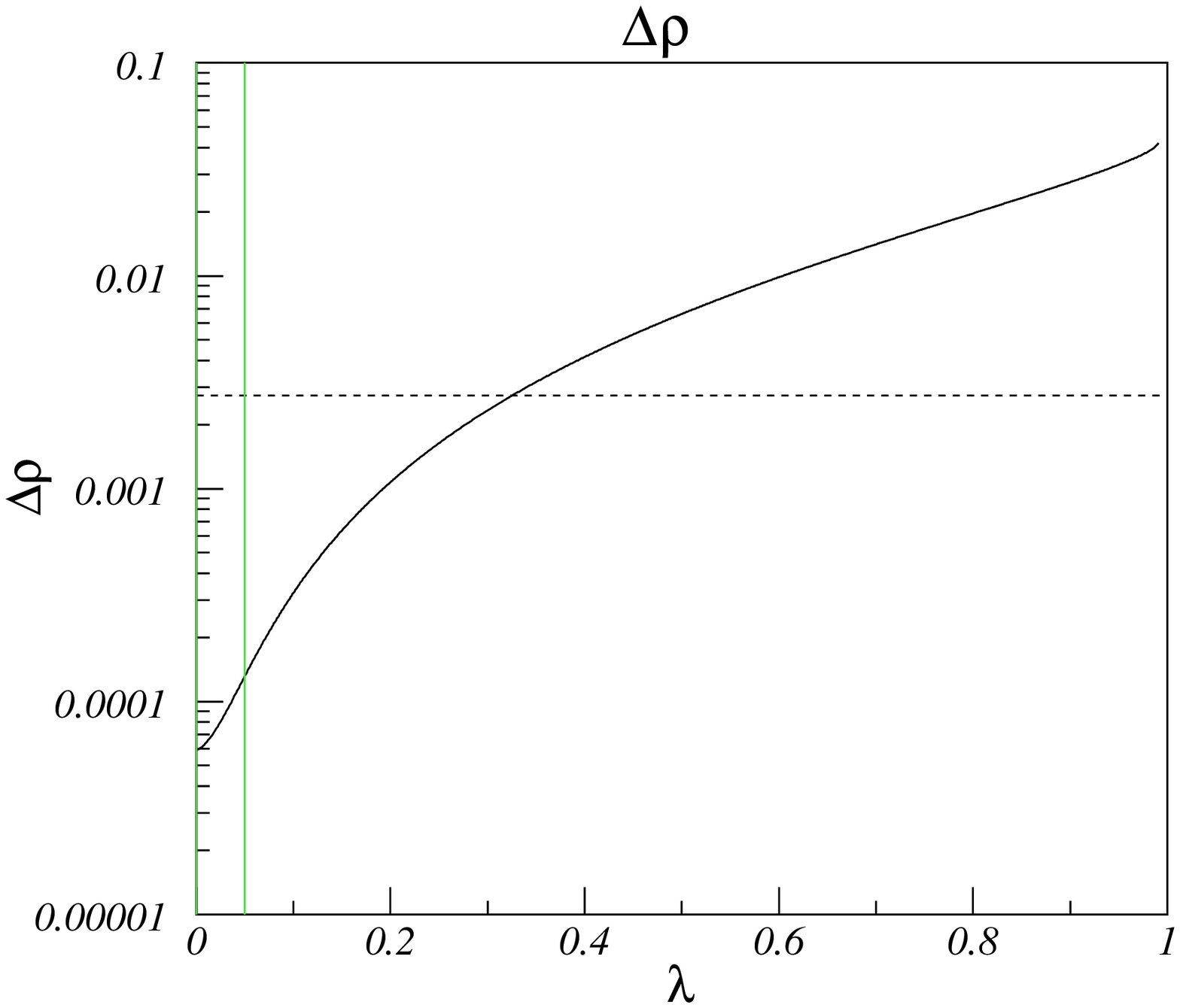}
 \includegraphics[width=0.32\columnwidth]{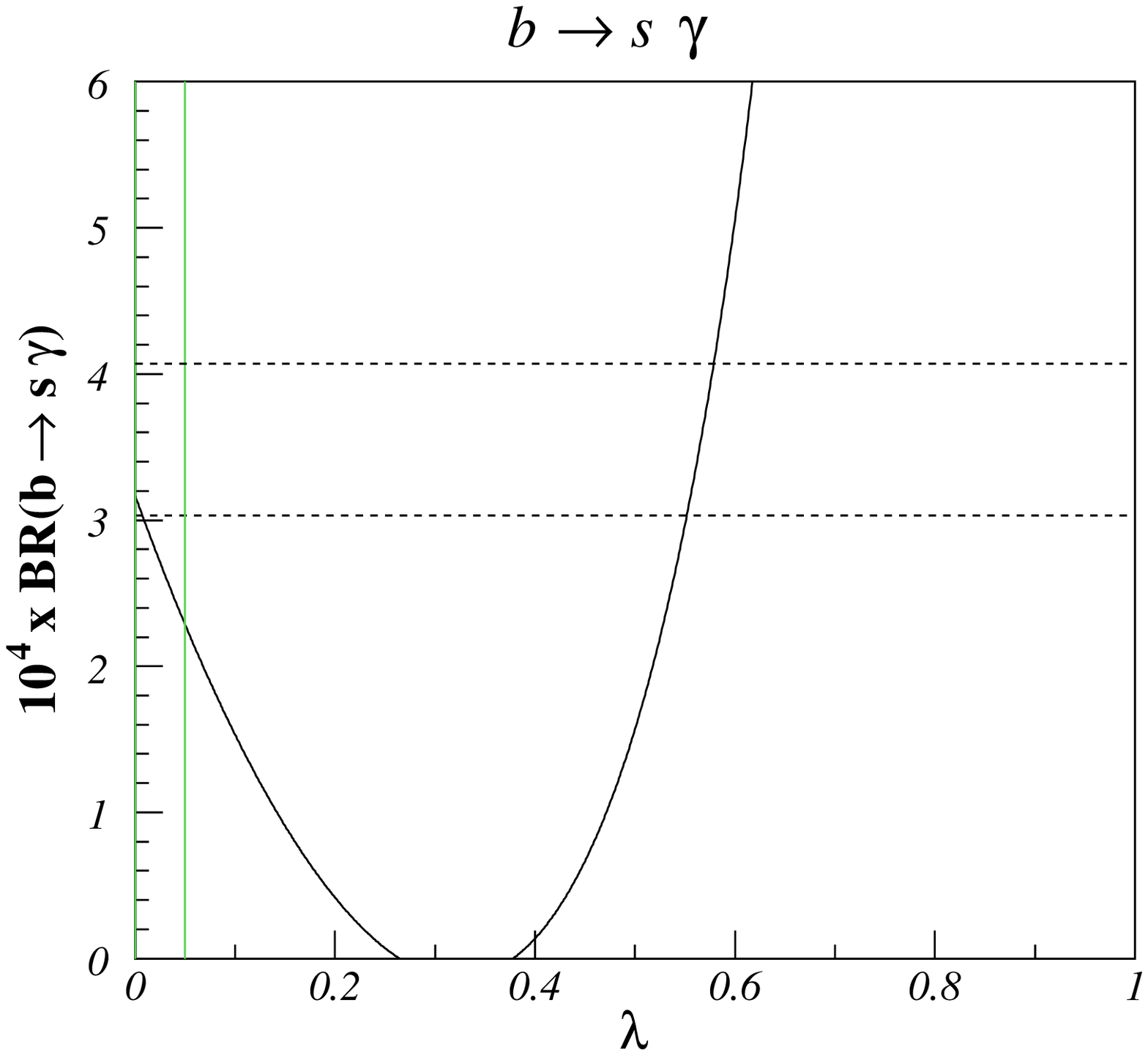}
 \includegraphics[width=0.32\columnwidth]{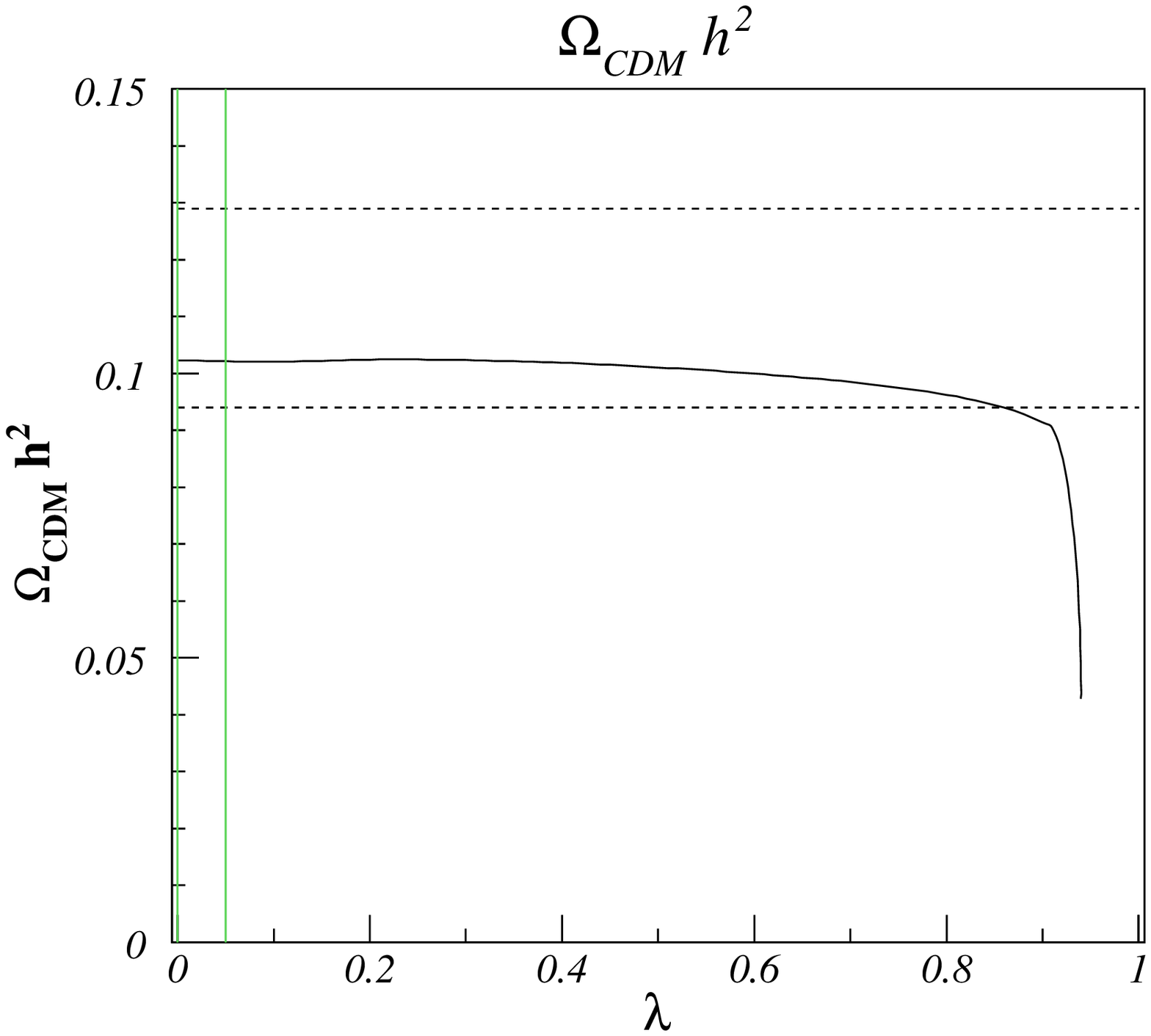}\vspace*{2mm}
 \includegraphics[width=0.32\columnwidth]{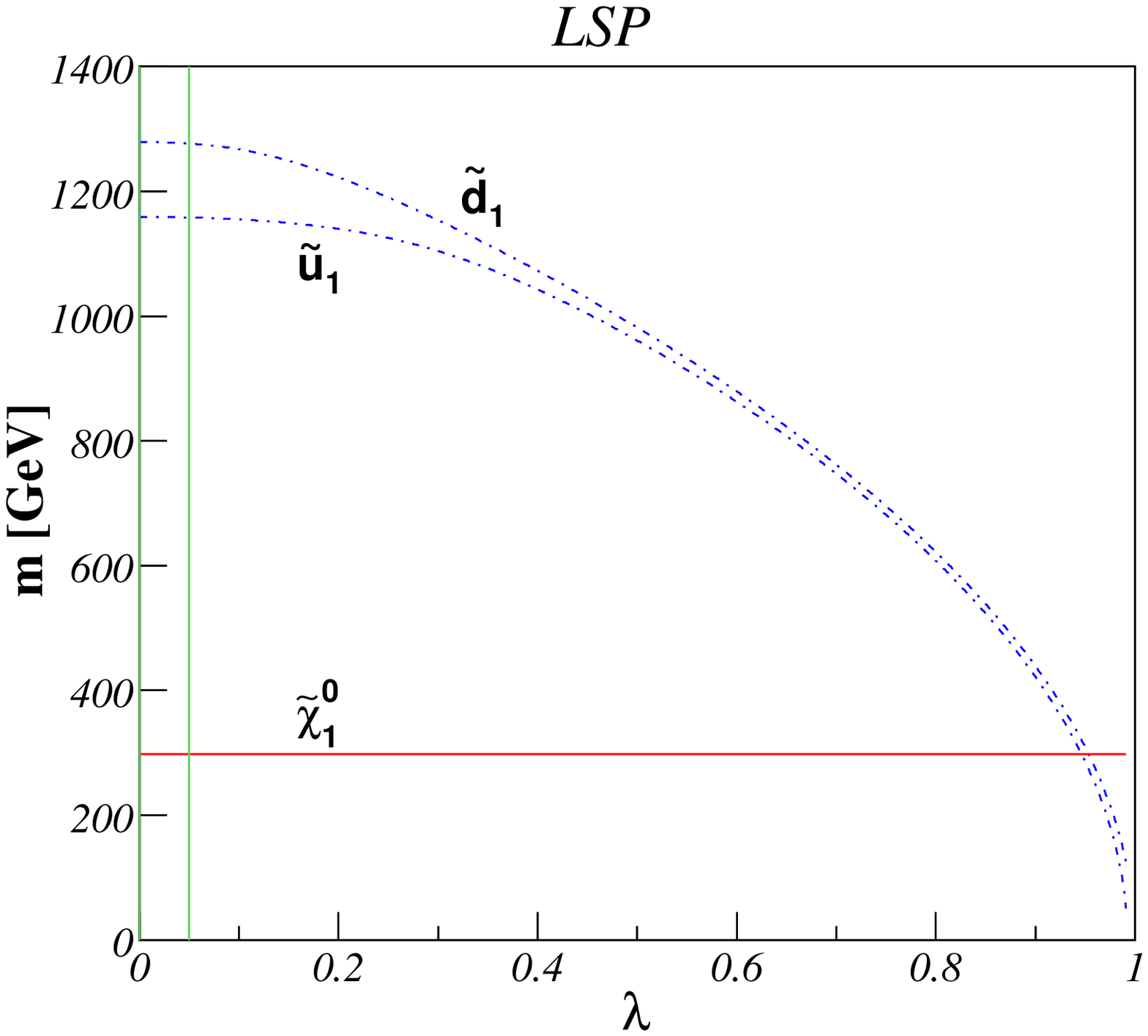}
 \includegraphics[width=0.32\columnwidth]{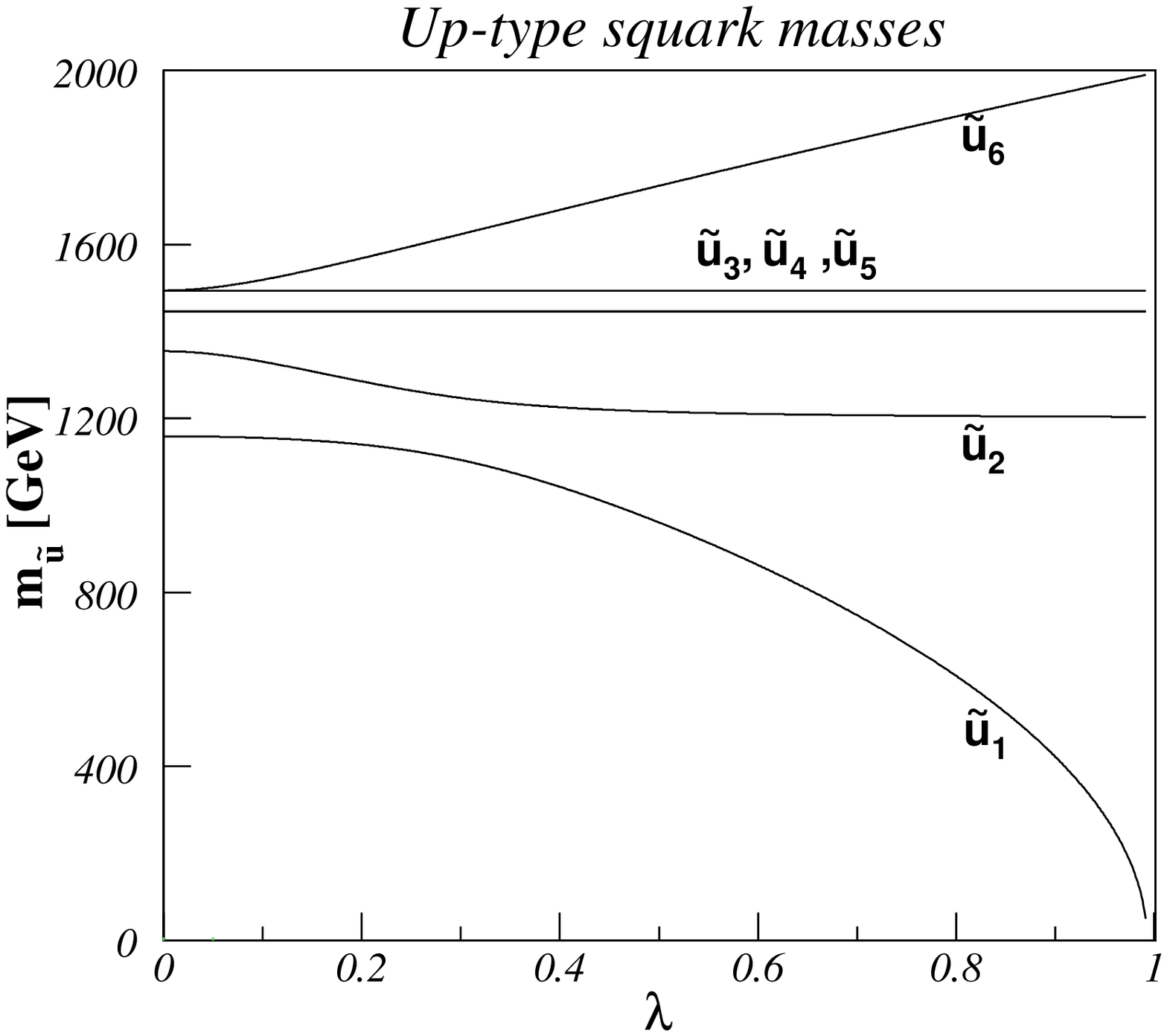}
 \includegraphics[width=0.32\columnwidth]{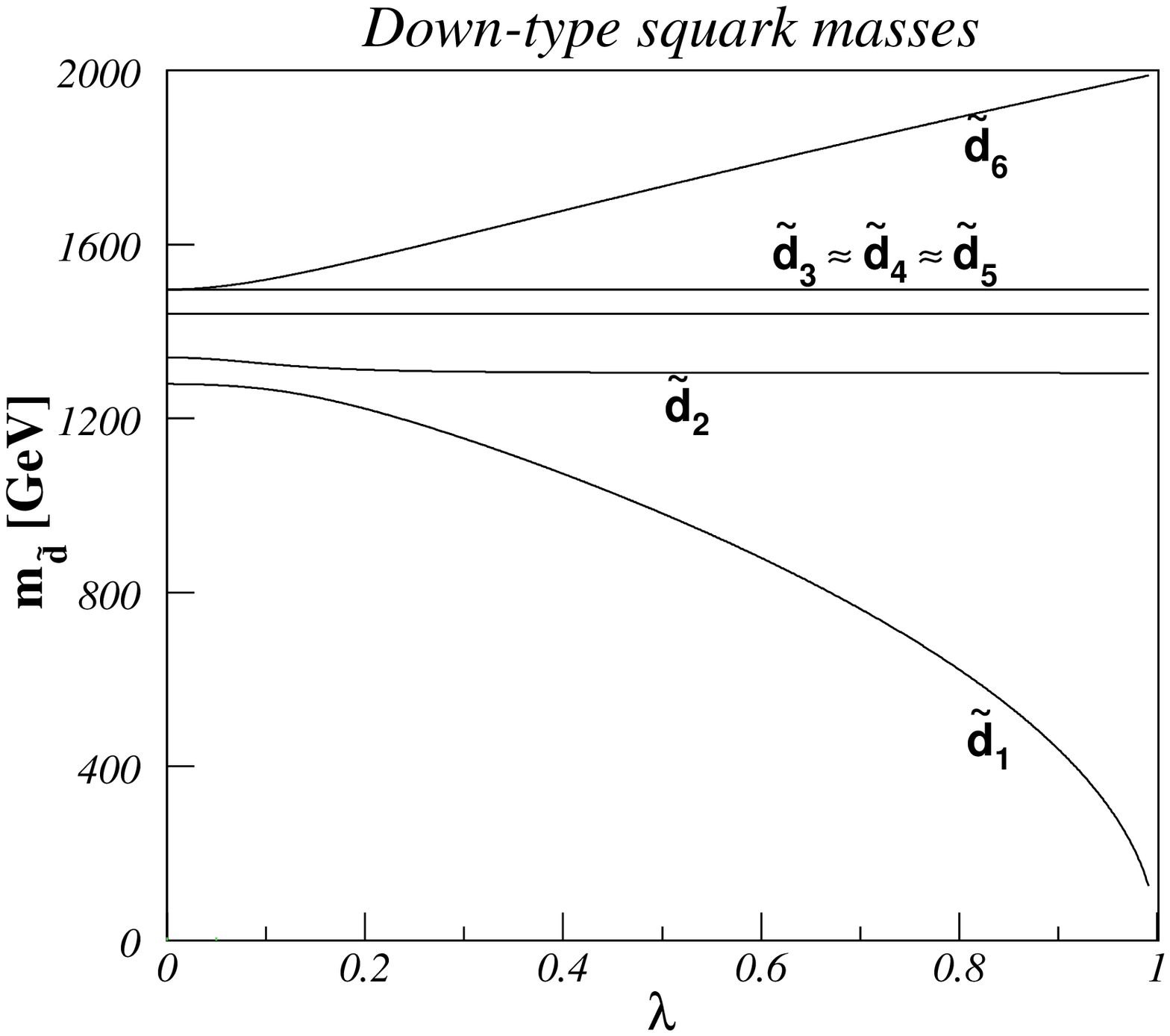}
 \caption{\label{fig:15}Same as Fig.\ \ref{fig:12} for our benchmark
          scenario D.}
\end{figure}

The electroweak precision observable $\Delta\rho$ is shown first in
Figs.\ \ref{fig:12}-\ref{fig:15} for the four benchmark scenarios A, B, C,
and D.
%
%
%
%
%
%
%
%
On our logarithmic scale, only the experimental upper bound of the
$2\sigma$-range is visible as a dashed line. While the self-energy
diagrams of the electroweak gauge bosons depend obviously strongly
on the helicities, flavours, and mass eigenvalues of the squarks
in the loop, the SUSY masses in our scenarios are sufficiently
small and the experimental error is still sufficiently large to
allow for relatively large values of $\lambda\leq 0.57$, 0.52,
0.38, and 0.32 for the benchmark points A, B, C, and D,
respectively. As mentioned above, $\Delta\rho$ conversely
constrains SUSY models in cMFV only for masses above 2000 GeV for
$m_0$ and 1500 GeV for $m_{1/2}$.

The next diagram in Figs.\ \ref{fig:12}-\ref{fig:15} shows the
dependence of the most stringent low-energy constraint, coming
from the good agreement between the measured $b\to s\gamma$
branching ratio and the two-loop SM prediction, on the NMFV
parameter $\lambda$. The dashed lines of the $2\sigma$-bands
exhibit two allowed regions, one close to $\lambda=0$ (vertical
green line) and a second one around $\lambda\simeq0.57$, 0.75,
0.62, and 0.57, respectively. As is well-known, the latter are,
however, disfavoured by $b\to s\mu^+\mu^-$ data constraining the
sign of the $b\to s \gamma$ amplitude to be the same as in the SM
\cite{Gambino:2004mv}. We will therefore limit ourselves later to
the regions $\lambda\leq0.05$ (points A, C, and D) and
$\lambda\leq0.1$ (point B) in the vicinity of (c)MFV (see also
Tab.\ \ref{tab:1}).

The 95\% confidence-level (or $2\sigma$) region for the cold dark matter
density was given in absolute values in Ref.\ \cite{Hamann:2006pf} and is
shown as a dashed band in the upper right part of Figs.\
\ref{fig:12}-\ref{fig:15}. However, only the lower bound (0.094) is of
relevance, as the relic density falls with increasing $\lambda$. This is
not so pronounced in our model B as in our model A, where squark masses are
light and the lightest neutralino has a sizable Higgsino-component, so
that squark exchanges contribute significantly to the annihilation cross
sections. For models C and D there is little sensitivity of $\Omega_{CDM}
h^2$ (except at very large $\lambda\leq1$), as the squark masses are
generally larger.

The rapid fall-off of the relic density for very large $\lambda\leq1$
can be understood by looking at the resulting lightest up- and down-type
squark mass eigenvalues in the lower left part of Figs.\
\ref{fig:12}-\ref{fig:15}. For maximal flavour violation, the off-diagonal
squark mass matrix elements are of similar size as the diagonal ones,
leading to one squark mass eigenvalue that approaches and finally falls
below the lightest neutralino (dark matter) mass. Light squark propagators
and co-annihilation processes thus lead to a rapidly falling dark matter
relic density and finally to cosmologically excluded NMFV SUSY models,
since the LSP must be electrically neutral and a colour singlet.

An interesting phenomenon of level reordering between neighbouring states
can be observed in the lower central diagrams of Figs.\
\ref{fig:12}-\ref{fig:15} for the two lowest mass
eigenvalues of up-type squarks. The squark mass eigenstates are, by
definition, labeled in ascending order with the mass eigenvalues, so that
$\tilde{u}_1$ represents the lightest, $\tilde{u}_2$ the second-lightest,
and $\tilde{u}_6$ the heaviest up-type squark. As $\lambda$ and the
off-diagonal entries in the mass matrix increase, the splitting between the
lightest and highest mass eigenvalues naturally increases, whereas the
intermediate squark masses (of $\tilde{u}_{3,4,5}$) are practically
degenerate and insensitive to $\lambda$. These remarks also hold for the\
down-type squark masses shown in the lower right diagrams of Figs.\
\ref{fig:12}-\ref{fig:15}. However, for up-type squarks it is first the
second-lowest mass that decreases up to intermediate values of $\lambda
=0.2...0.5$, whereas the lowest mass is constant, and only at this point the
second-lowest mass becomes constant and takes approximately the value of the
until here lowest squark mass, whereas the lowest squark mass starts to
decrease further with $\lambda$. These ``avoided crossings'' are a common
phenomenon for Hermitian matrices and reminiscent of meta-stable systems in
quantum mechanics. At the point where the two levels should cross, the
corresponding squark eigenstates mix and change character, as will be
explained in the next subsection. For scenario C (Fig.\ \ref{fig:14}), the
phenomenon occurs even a second time with an additional avoided crossing
between the states $\tilde{u}_2$ and $\tilde{u}_3$ at $\lambda\simeq0.05$.
For scenario B (Fig.\ \ref{fig:13}), this takes place at $\lambda\simeq0.1$,
and there is even another crossing at $\lambda\simeq0.02$. For down-type
squarks, the level-reordering phenomenon is not so pronounced.

\subsection{Chirality and Flavour Decomposition of Squark Mass Eigenstates}

%
\begin{figure}
 \centering
 \includegraphics[width=0.21\columnwidth]{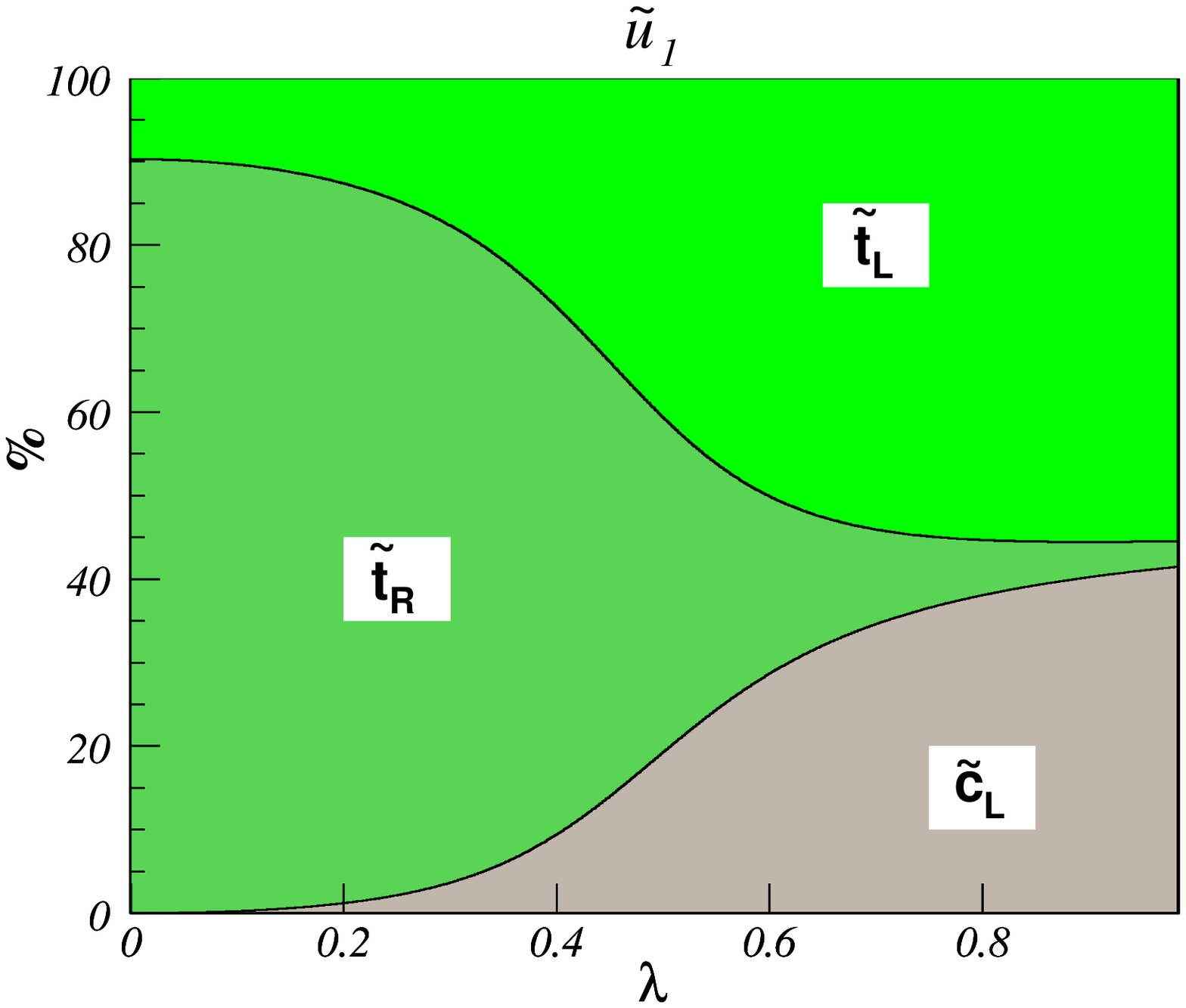}\hspace{2mm}
 \includegraphics[width=0.21\columnwidth]{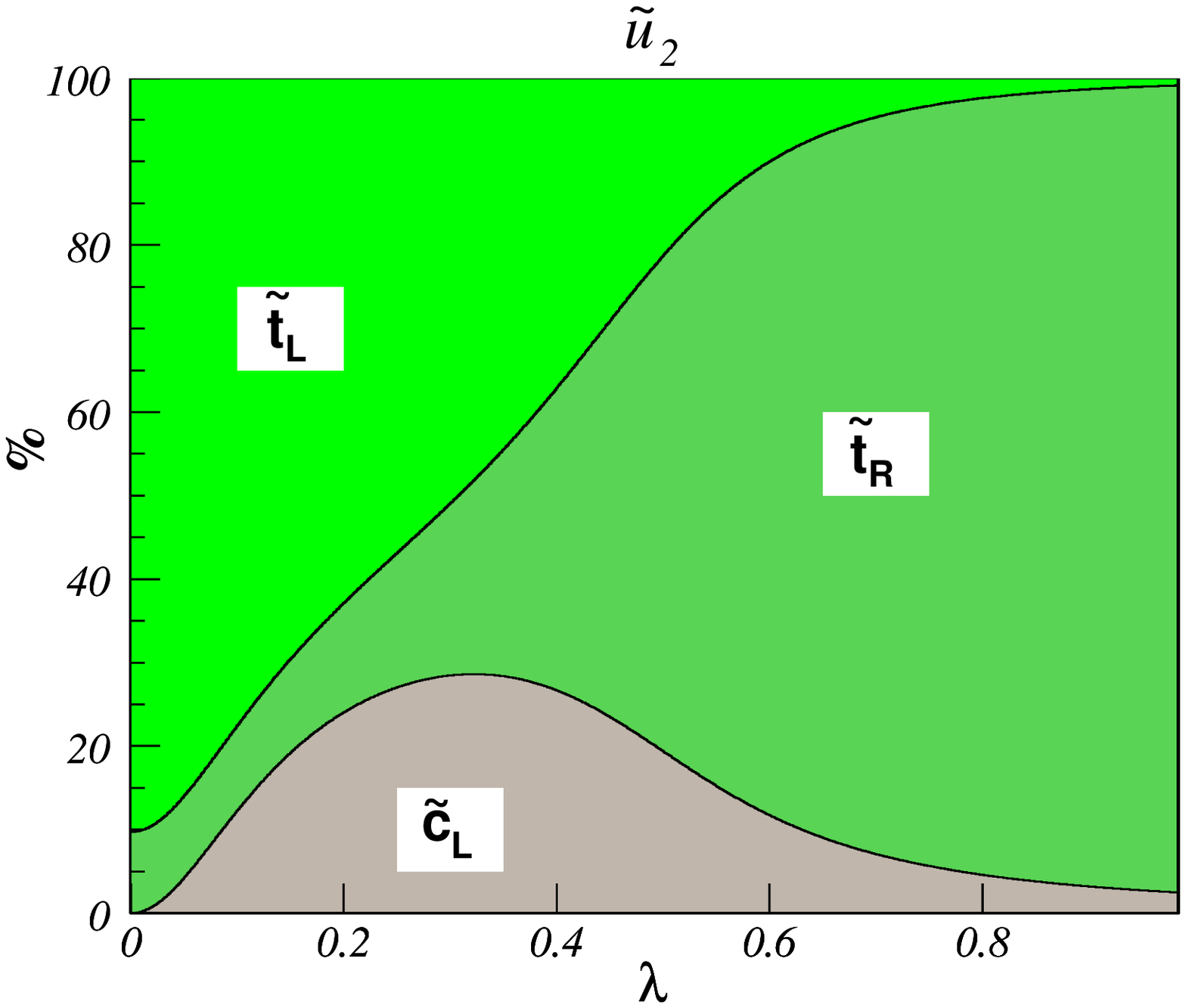}\hspace{2mm}
 \includegraphics[width=0.21\columnwidth]{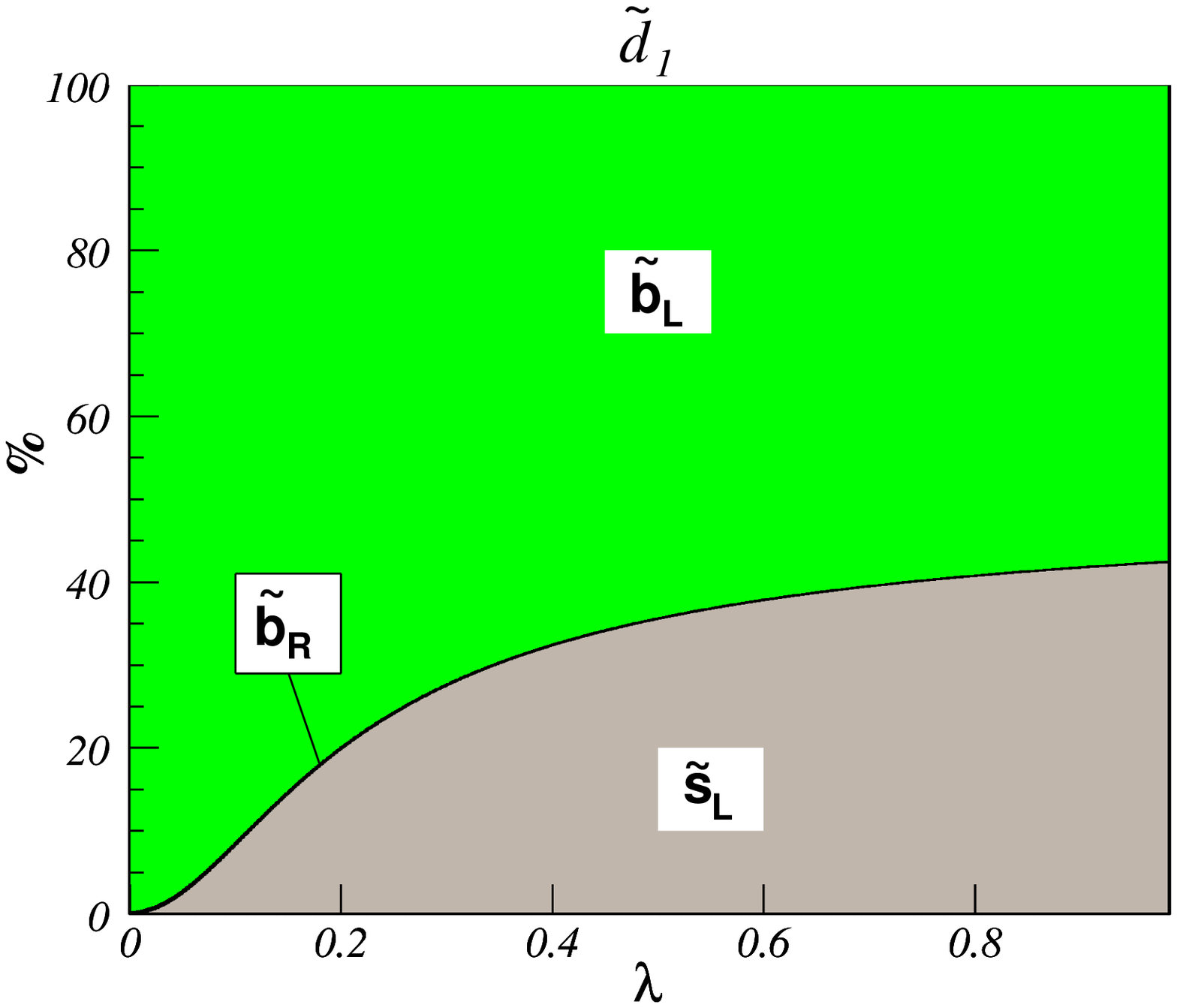}\hspace{2mm}
 \includegraphics[width=0.21\columnwidth]{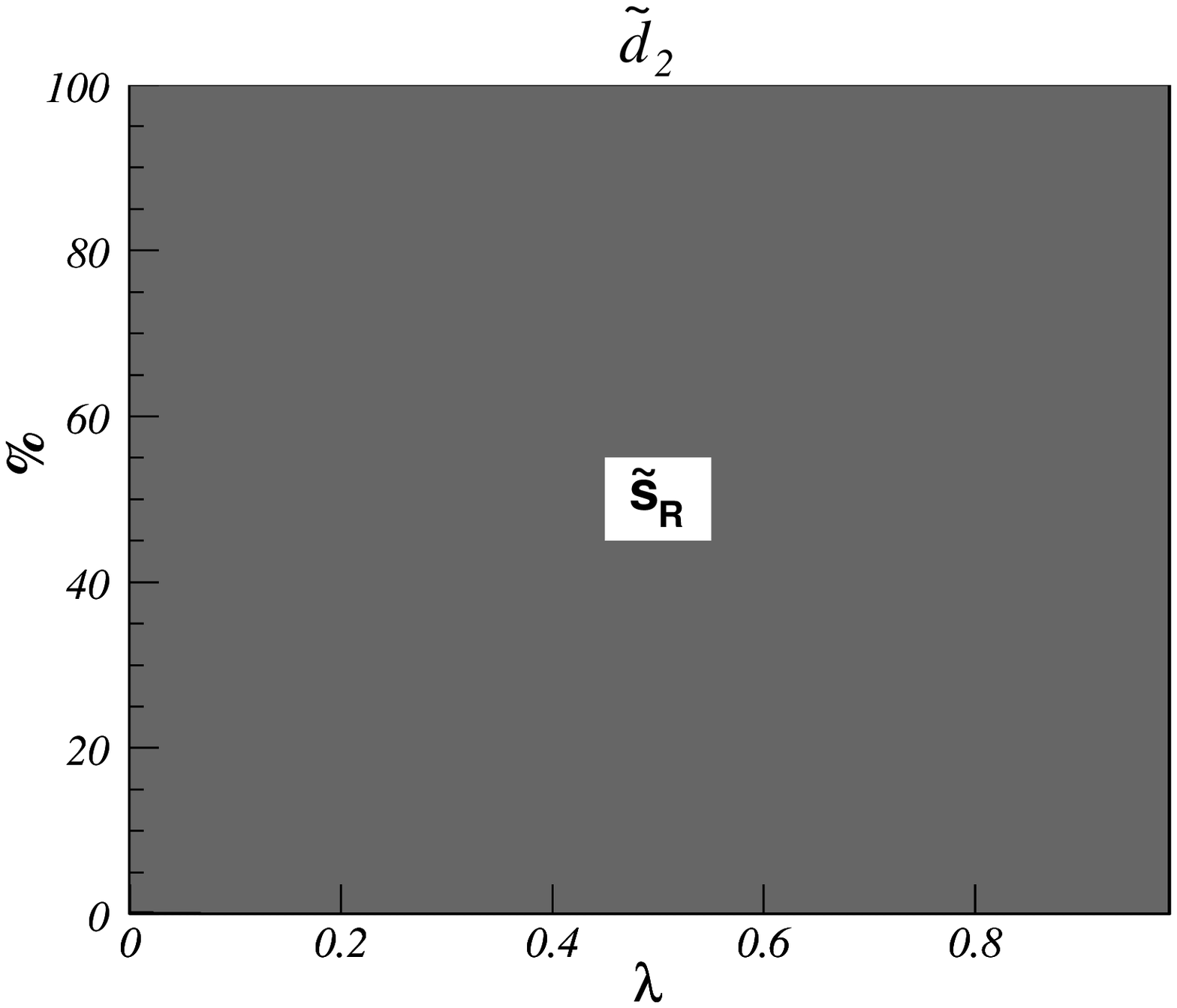}\vspace*{4mm}
 \includegraphics[width=0.21\columnwidth]{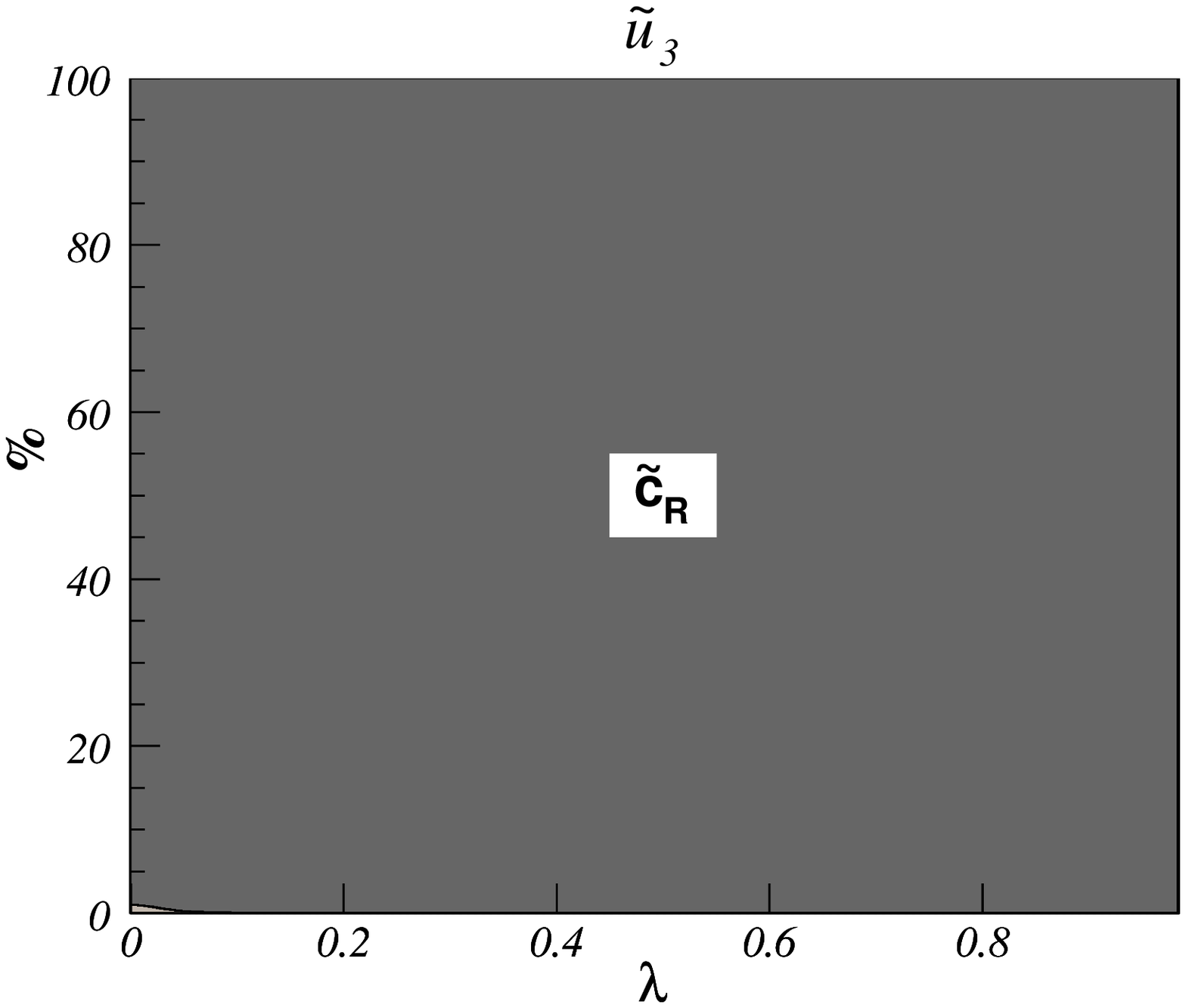}\hspace{2mm}
 \includegraphics[width=0.21\columnwidth]{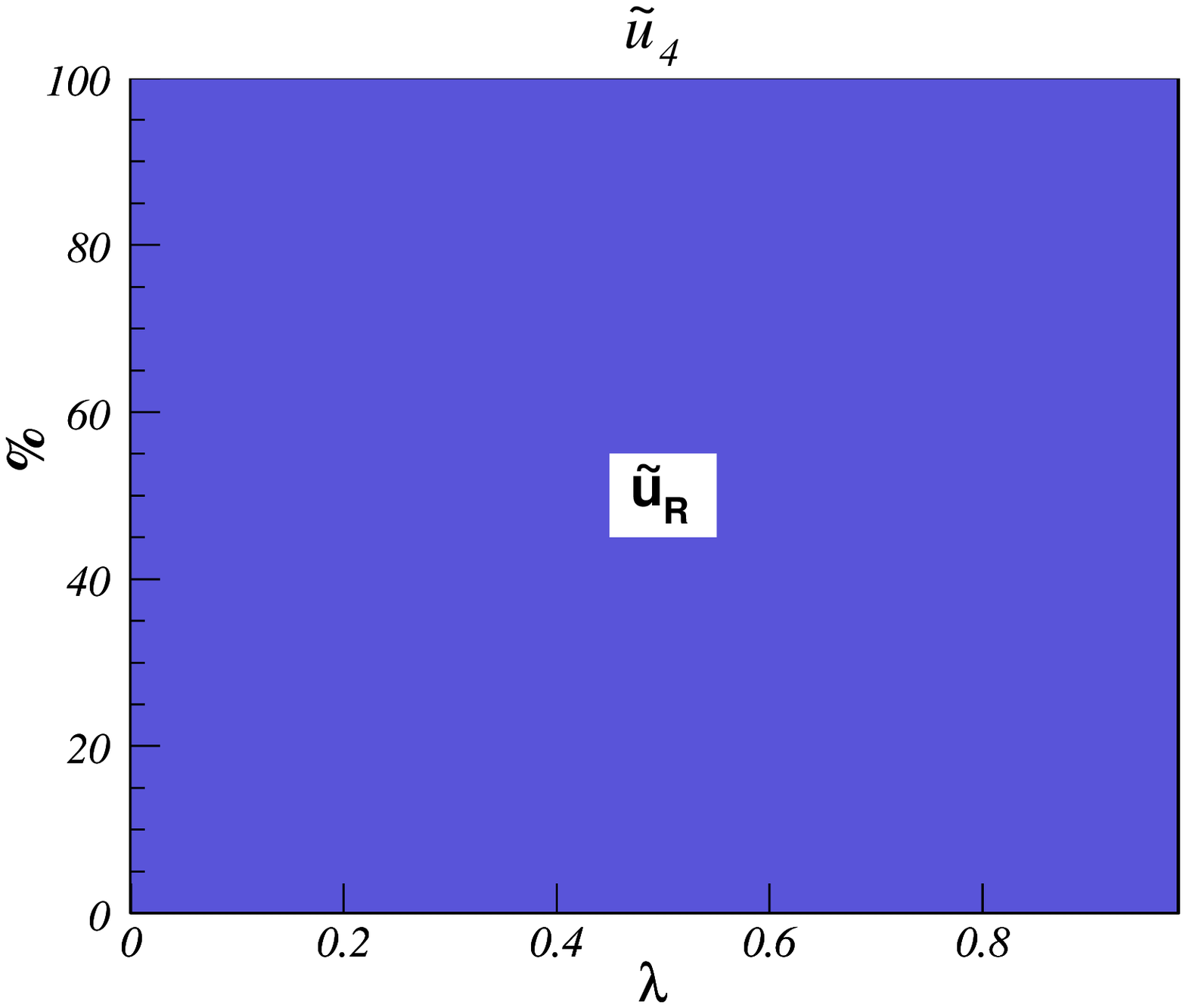}\hspace{2mm}
 \includegraphics[width=0.21\columnwidth]{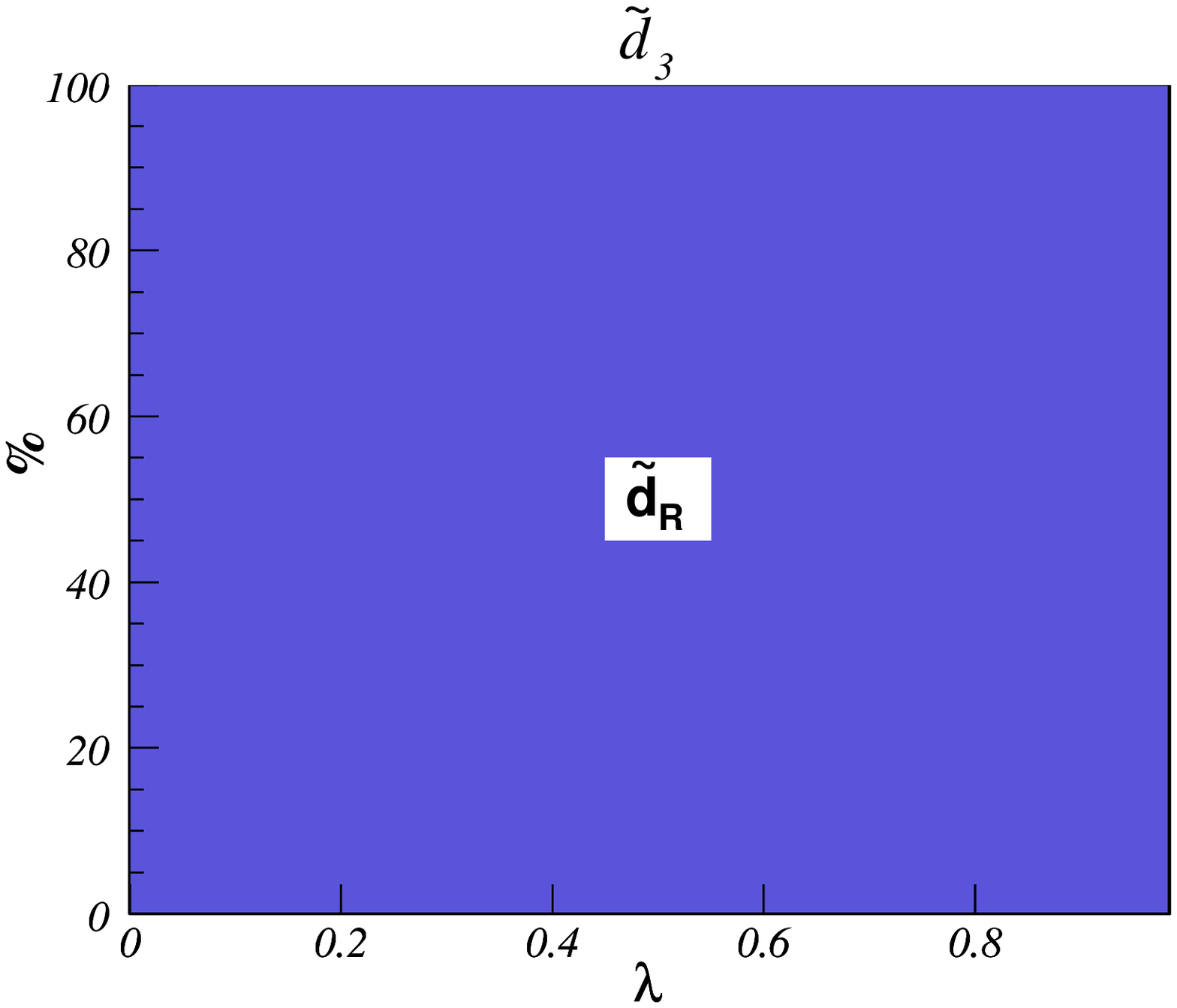}\hspace{2mm}
 \includegraphics[width=0.21\columnwidth]{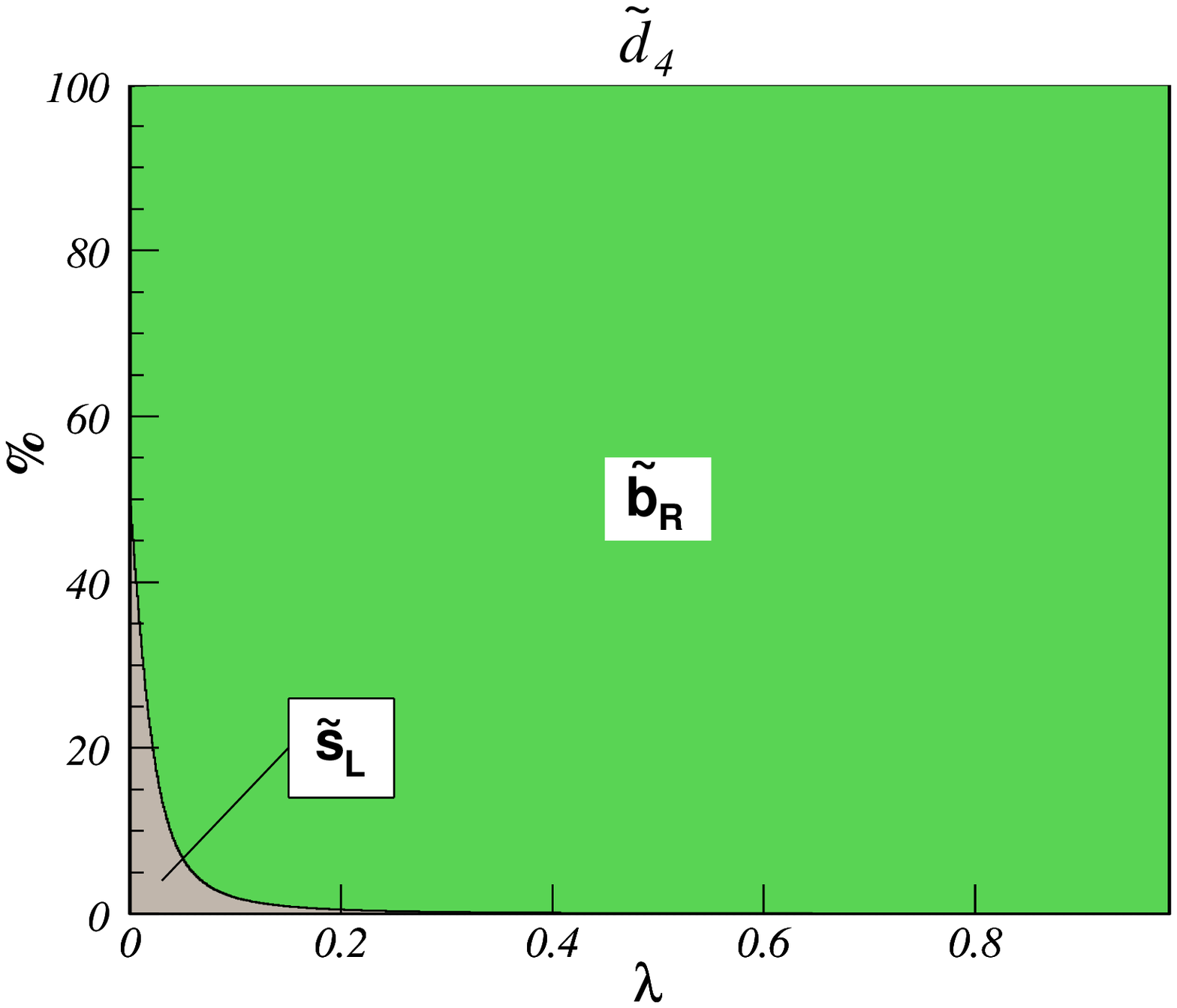}\vspace*{4mm}
 \includegraphics[width=0.21\columnwidth]{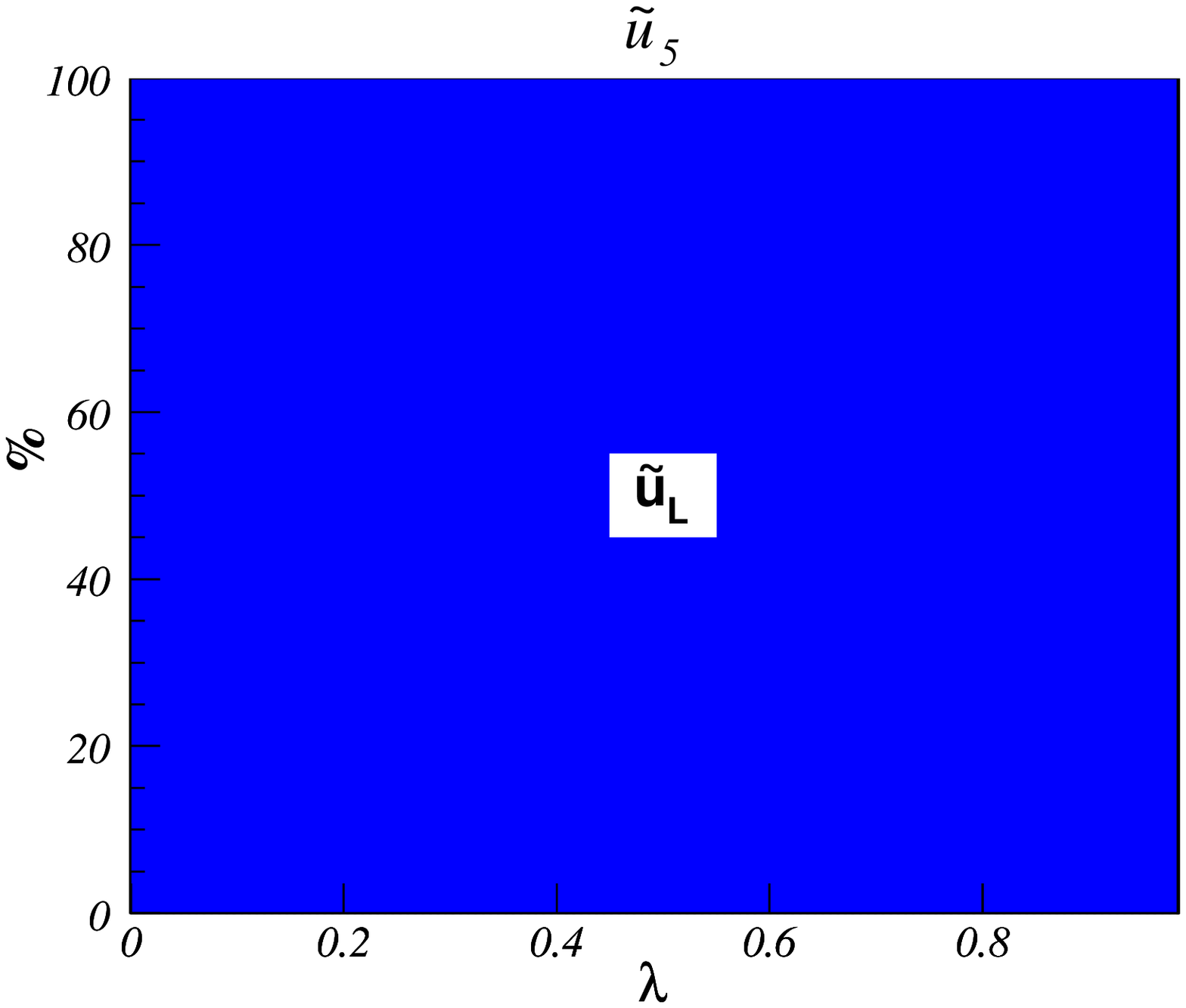}\hspace{2mm}
 \includegraphics[width=0.21\columnwidth]{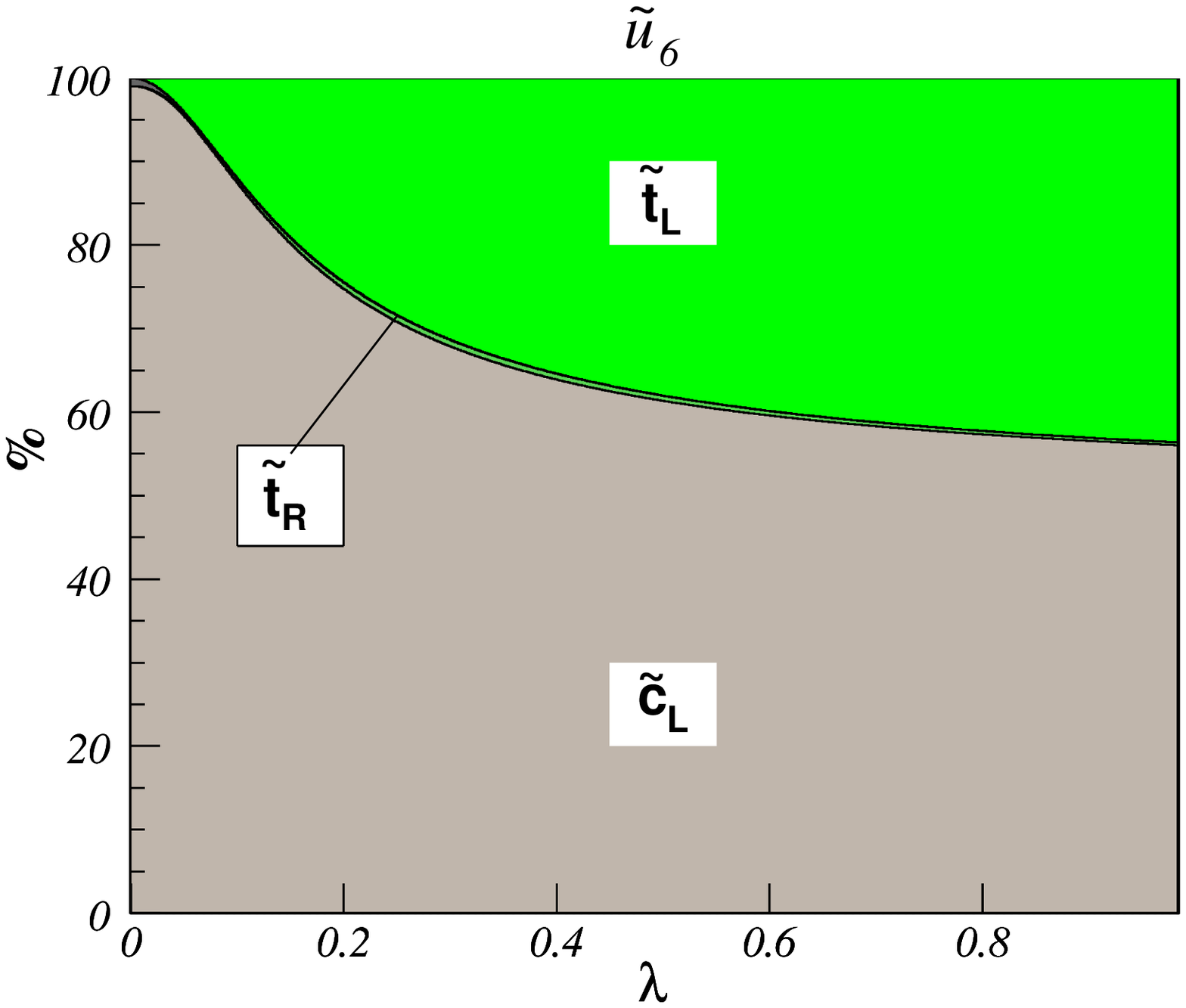}\hspace{2mm}
 \includegraphics[width=0.21\columnwidth]{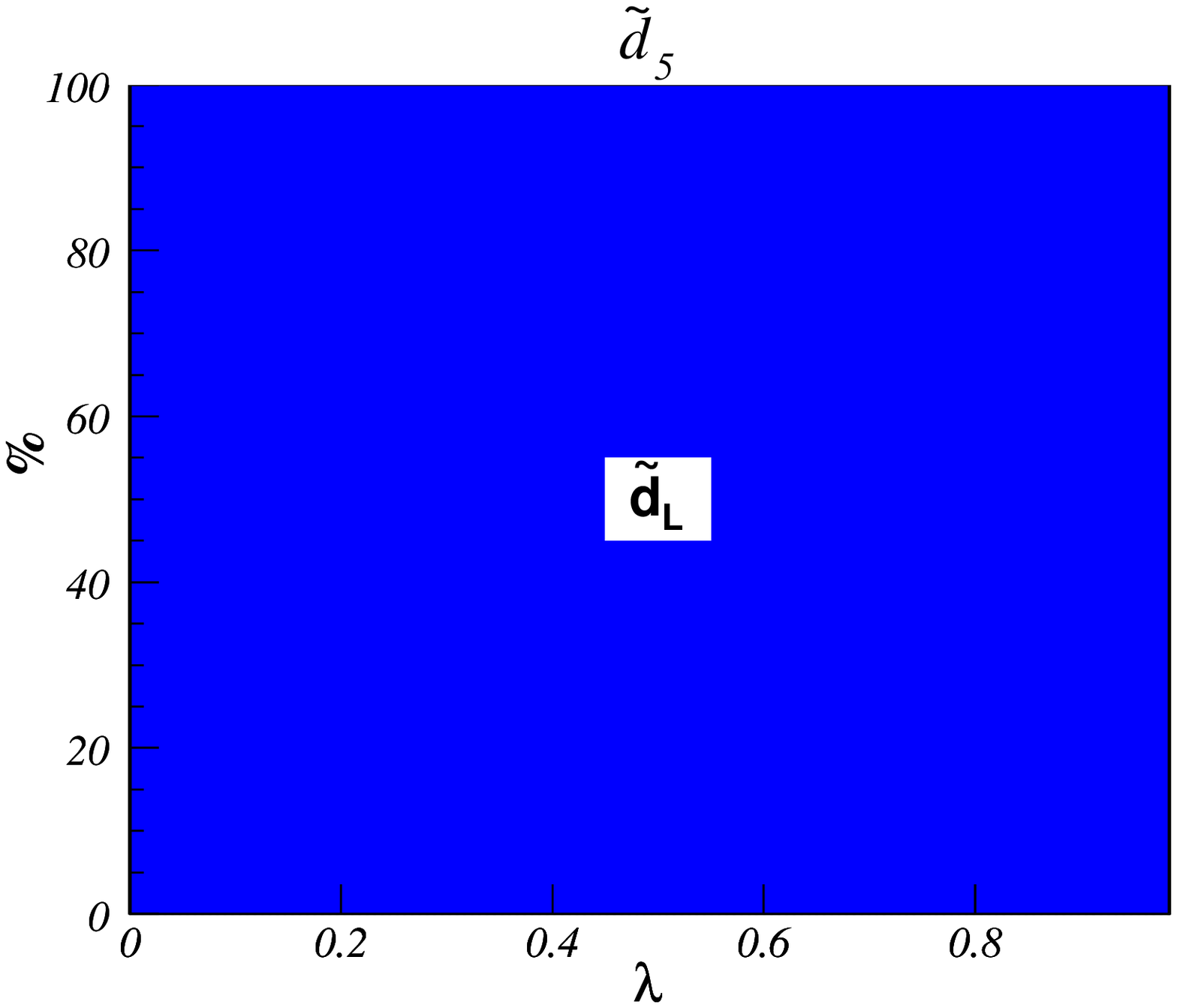}\hspace{2mm}
 \includegraphics[width=0.21\columnwidth]{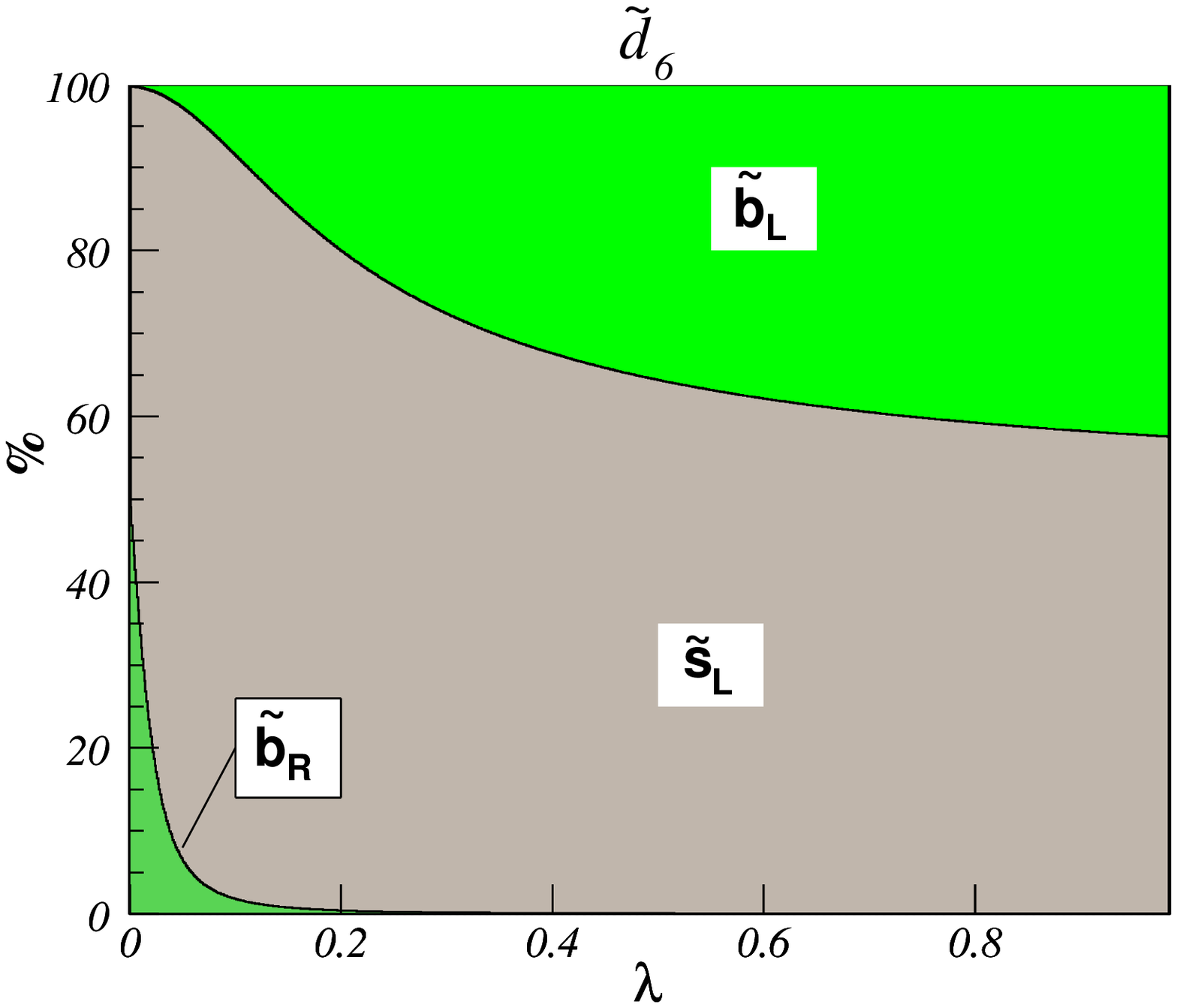}
 \caption{\label{fig:16}Dependence of the chirality (L, R) and flavour
          ($u$, $c$, $t$; $d$, $s$, and $b$) content of up- ($\tilde{u}_i$)
          and down-type ($\tilde{d}_i$) squark mass eigenstates on the NMFV
          parameter $\lambda\in[0;1]$ for benchmark point A.}
\end{figure}
%
%
\begin{figure}
 \centering
 \includegraphics[width=0.21\columnwidth]{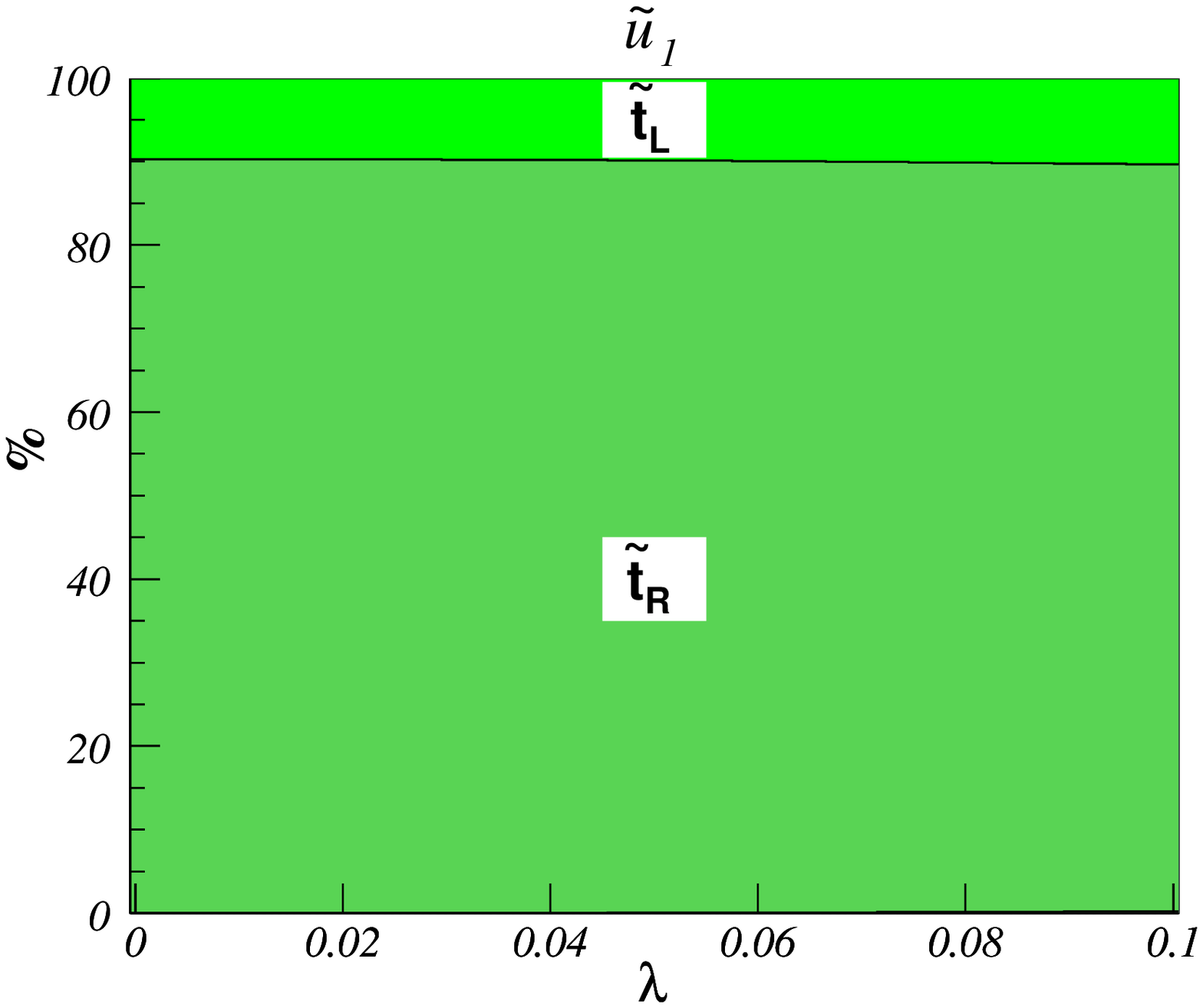}\hspace{2mm}
 \includegraphics[width=0.21\columnwidth]{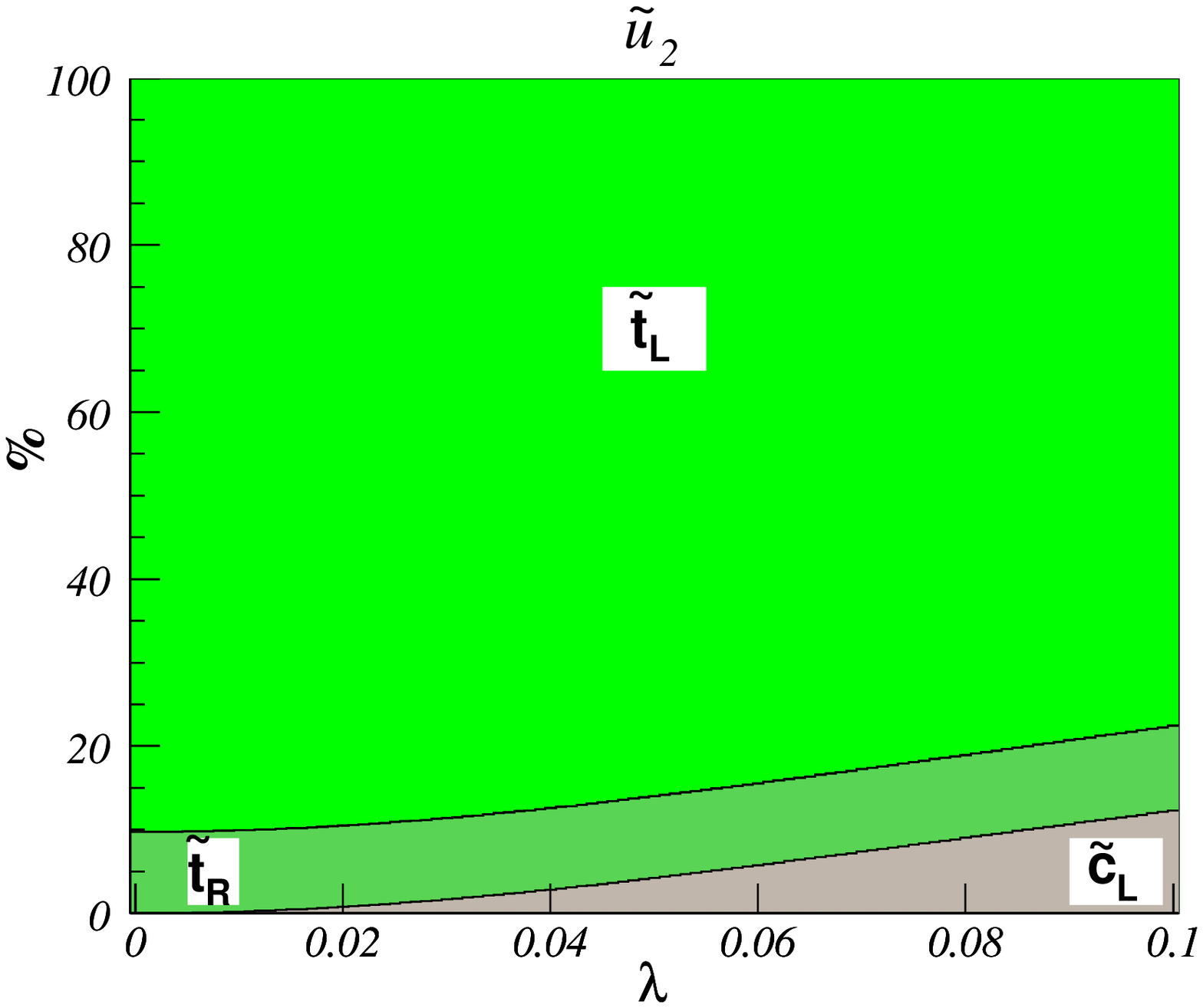}\hspace{2mm}
 \includegraphics[width=0.21\columnwidth]{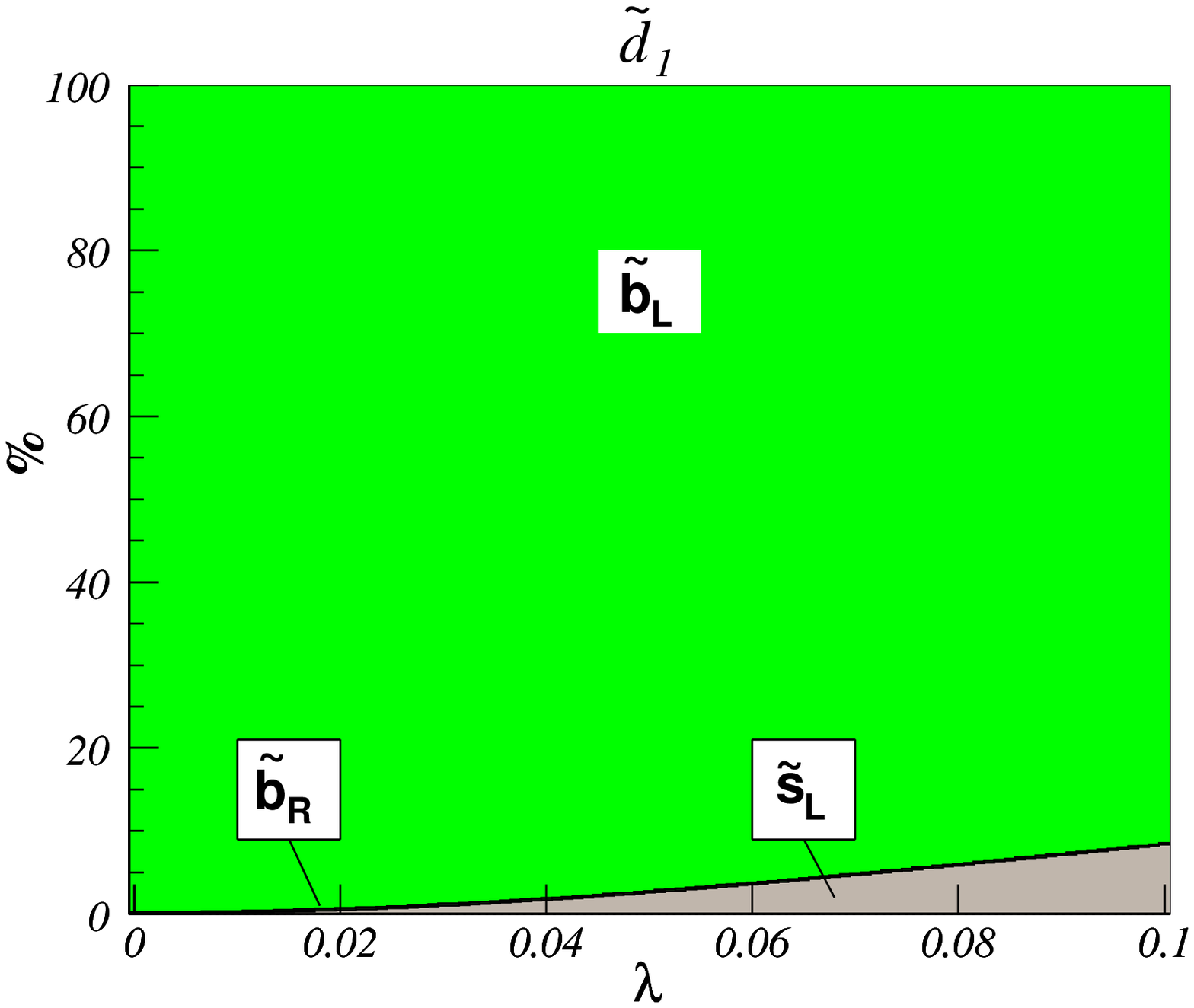}\hspace{2mm}
 \includegraphics[width=0.21\columnwidth]{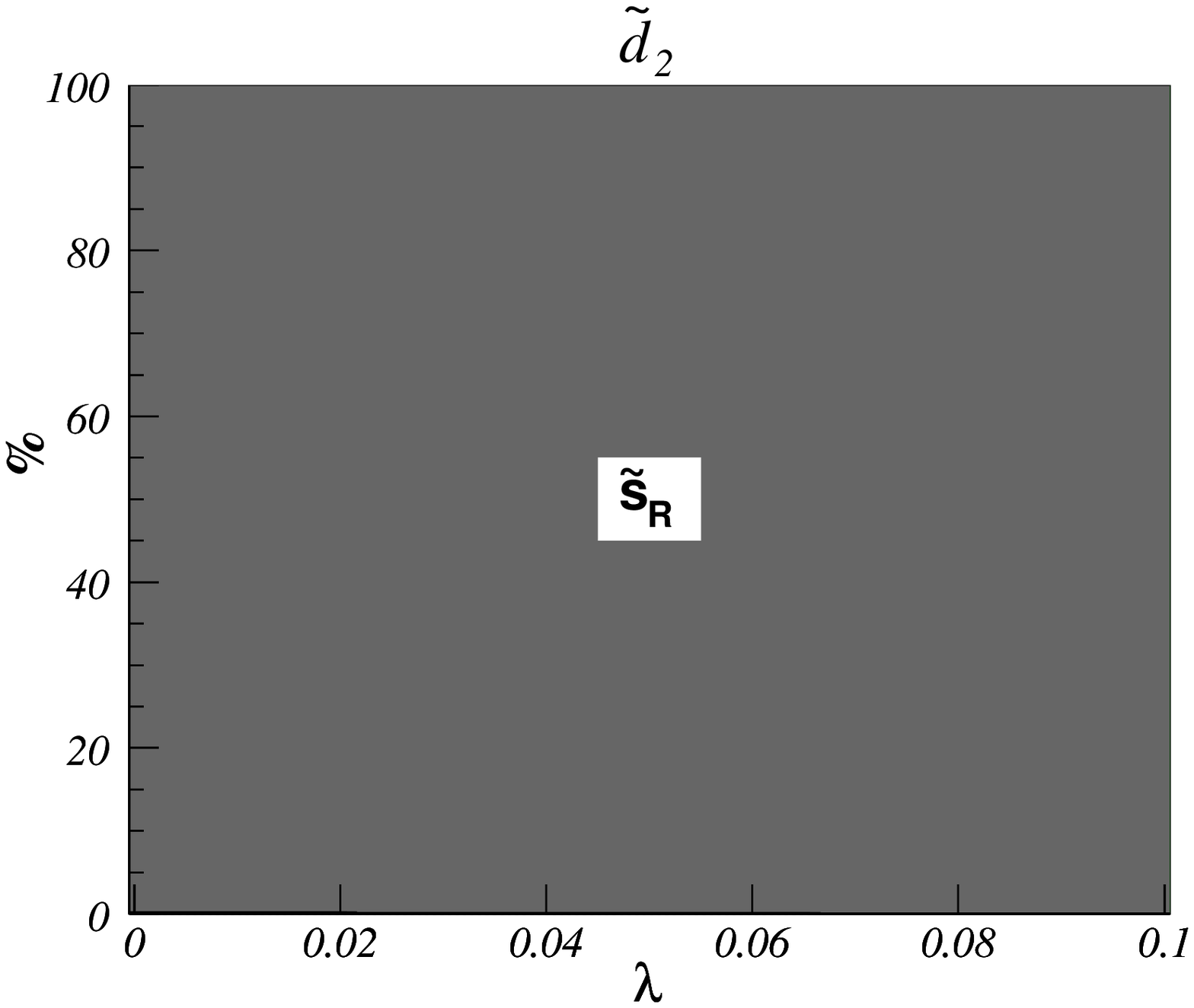}\vspace*{4mm}
 \includegraphics[width=0.21\columnwidth]{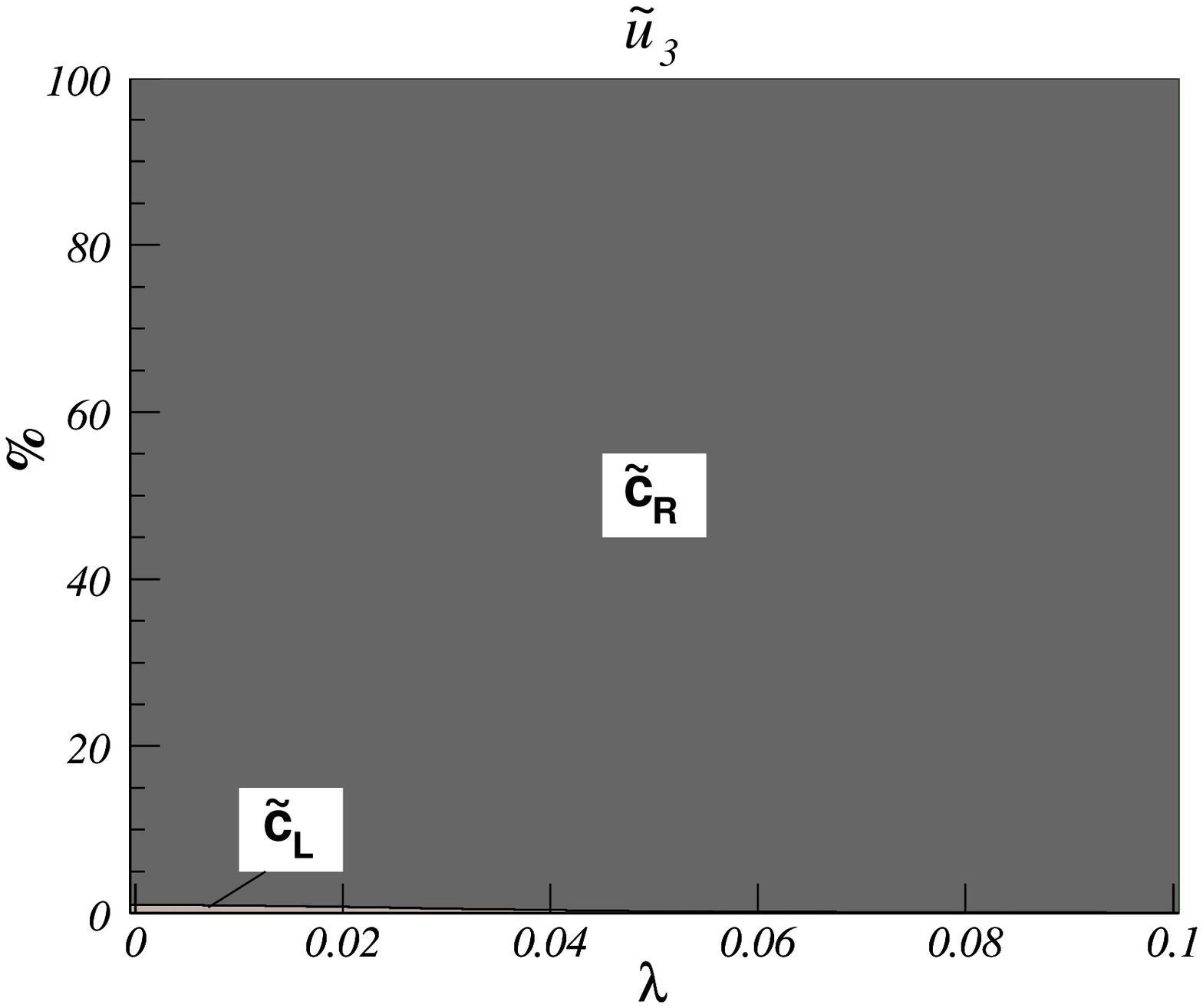}\hspace{2mm}
 \includegraphics[width=0.21\columnwidth]{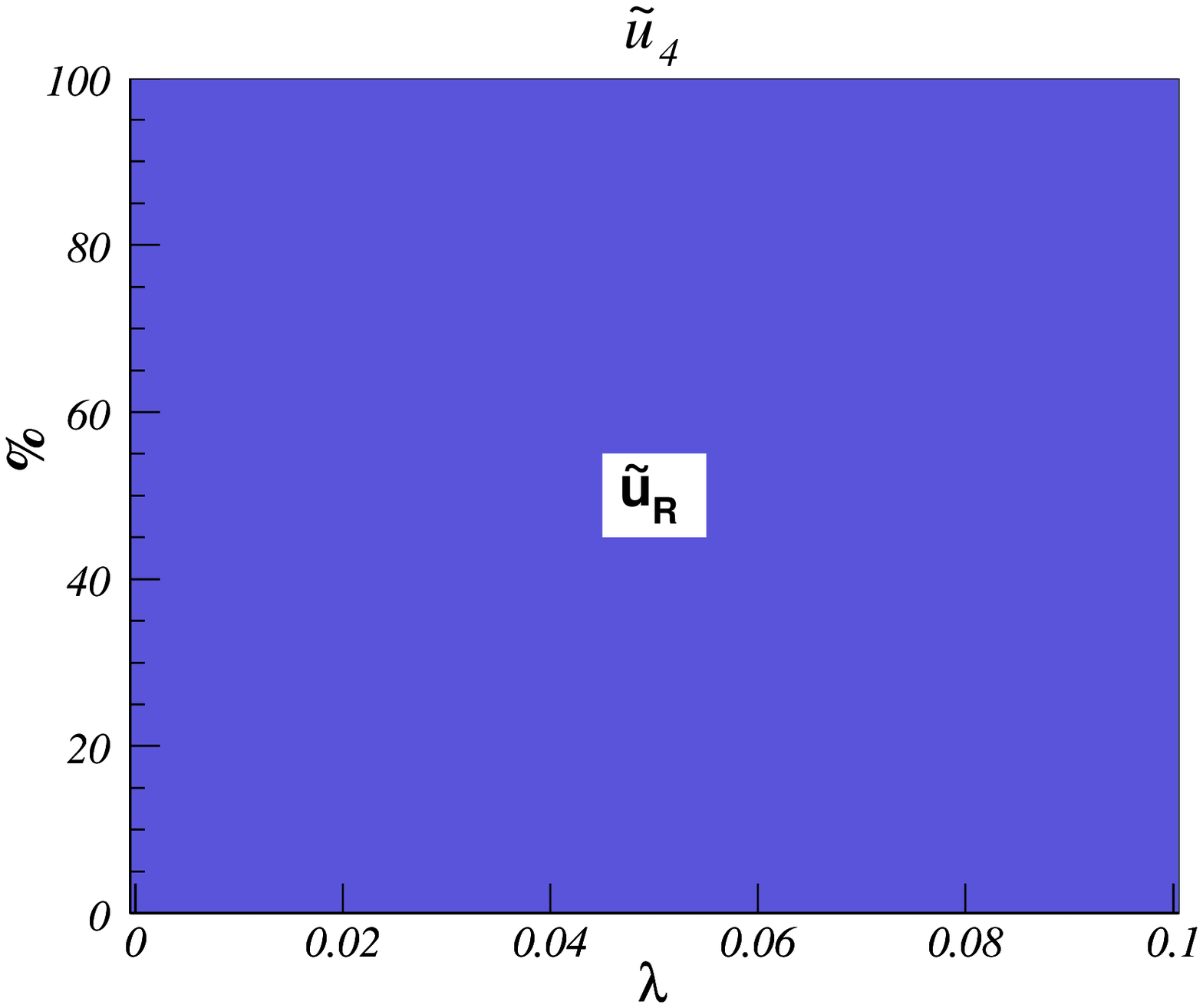}\hspace{2mm}
 \includegraphics[width=0.21\columnwidth]{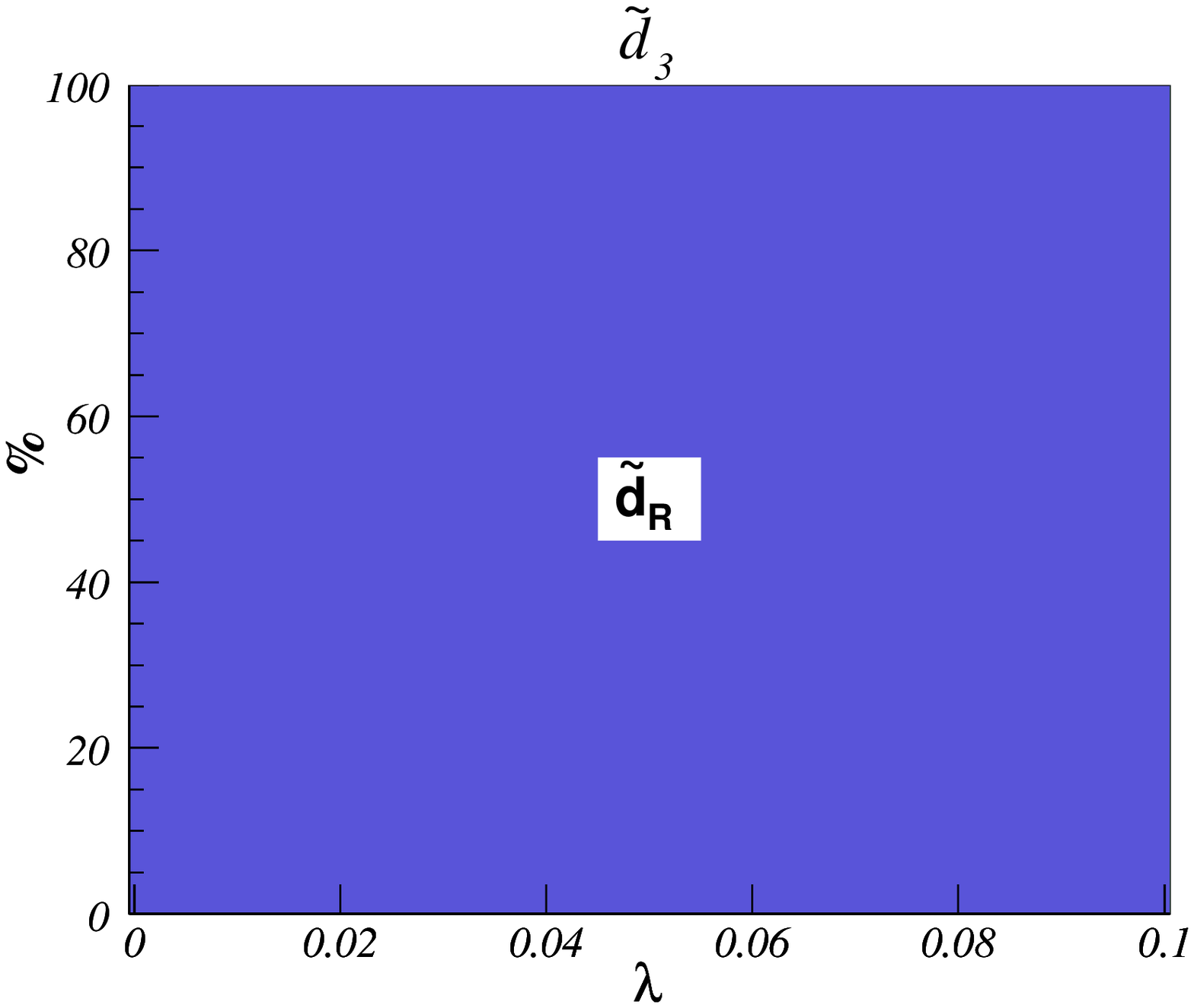}\hspace{2mm}
 \includegraphics[width=0.21\columnwidth]{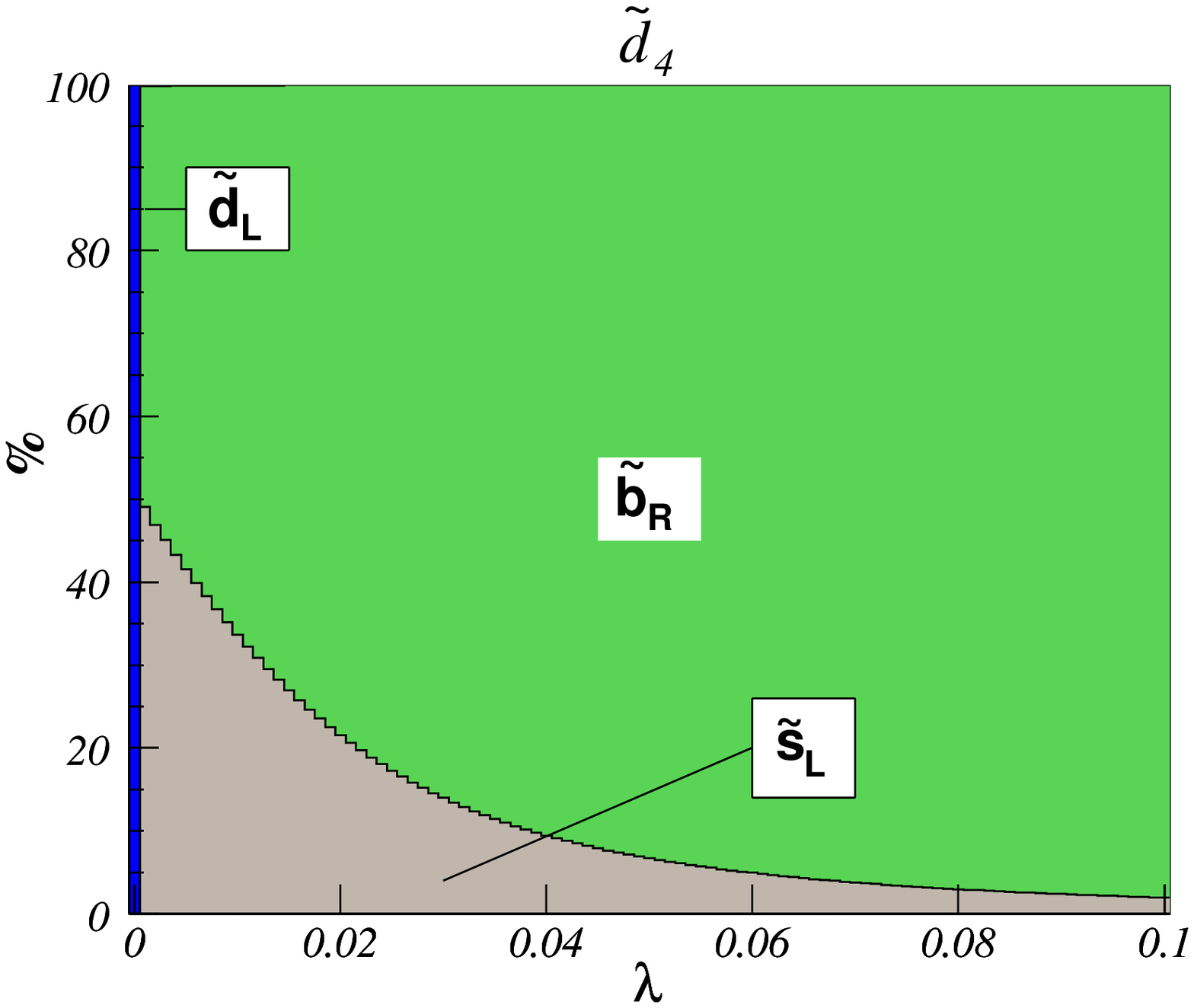}\vspace*{4mm}
 \includegraphics[width=0.21\columnwidth]{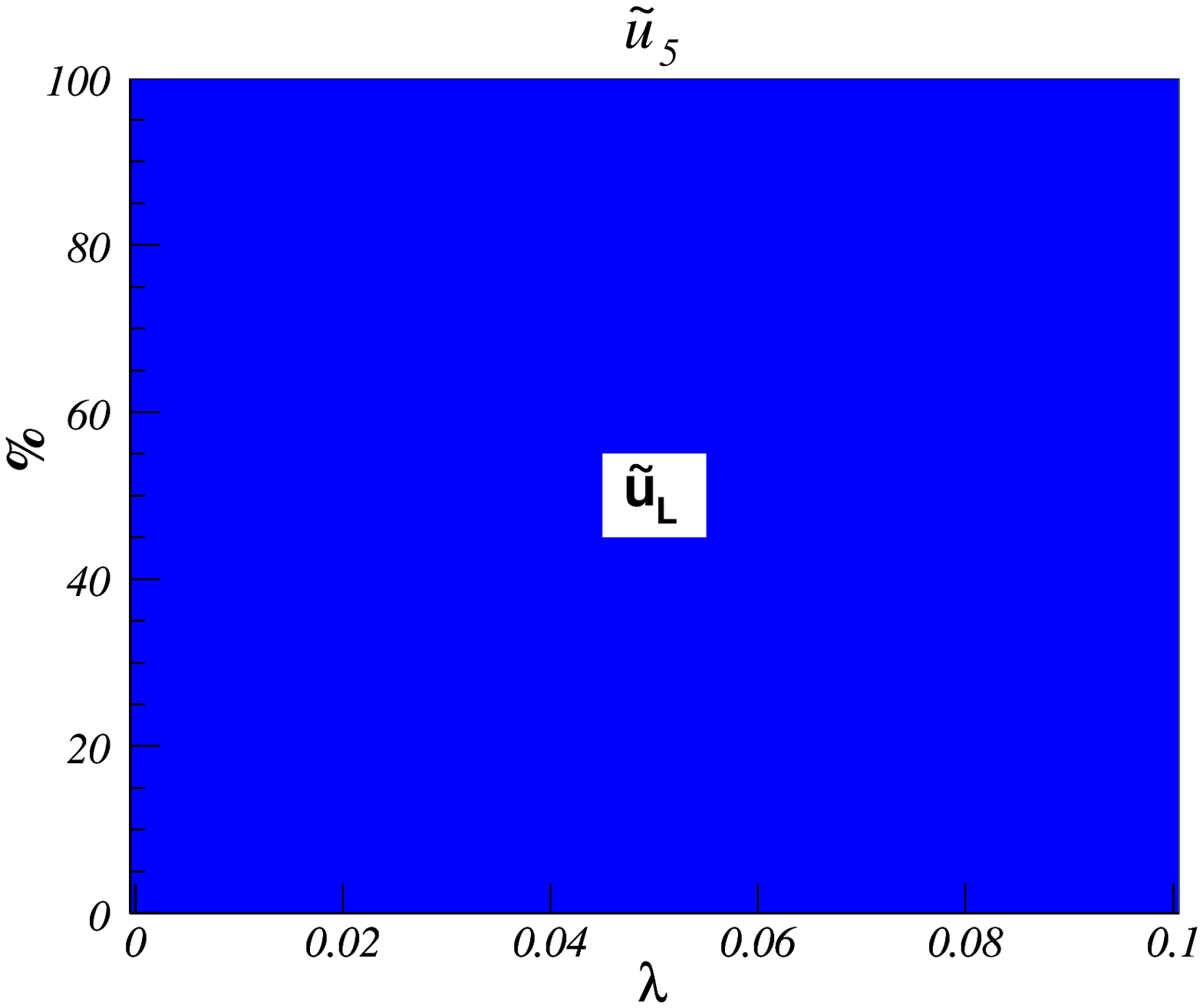}\hspace{2mm}
 \includegraphics[width=0.21\columnwidth]{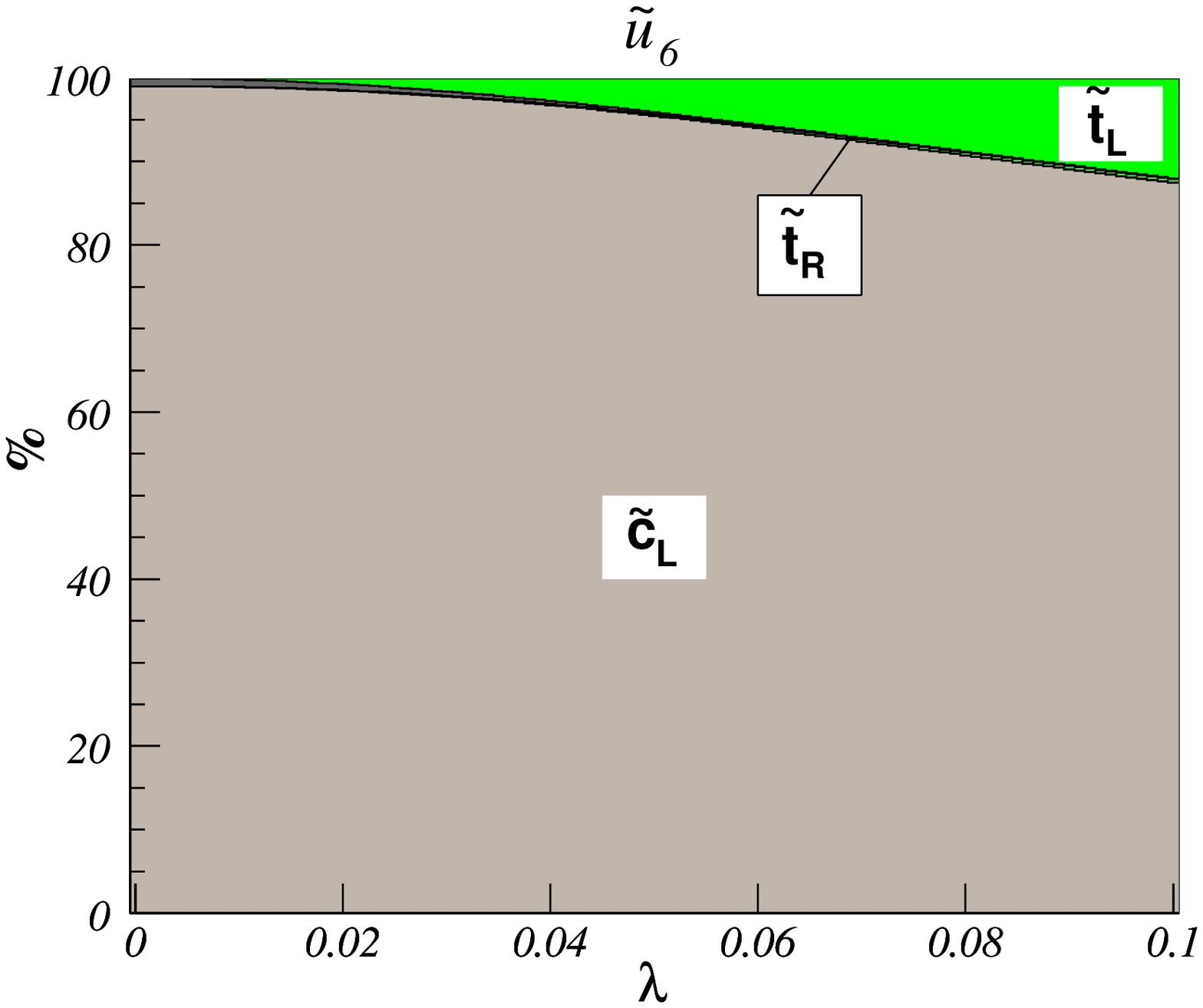}\hspace{2mm}
 \includegraphics[width=0.21\columnwidth]{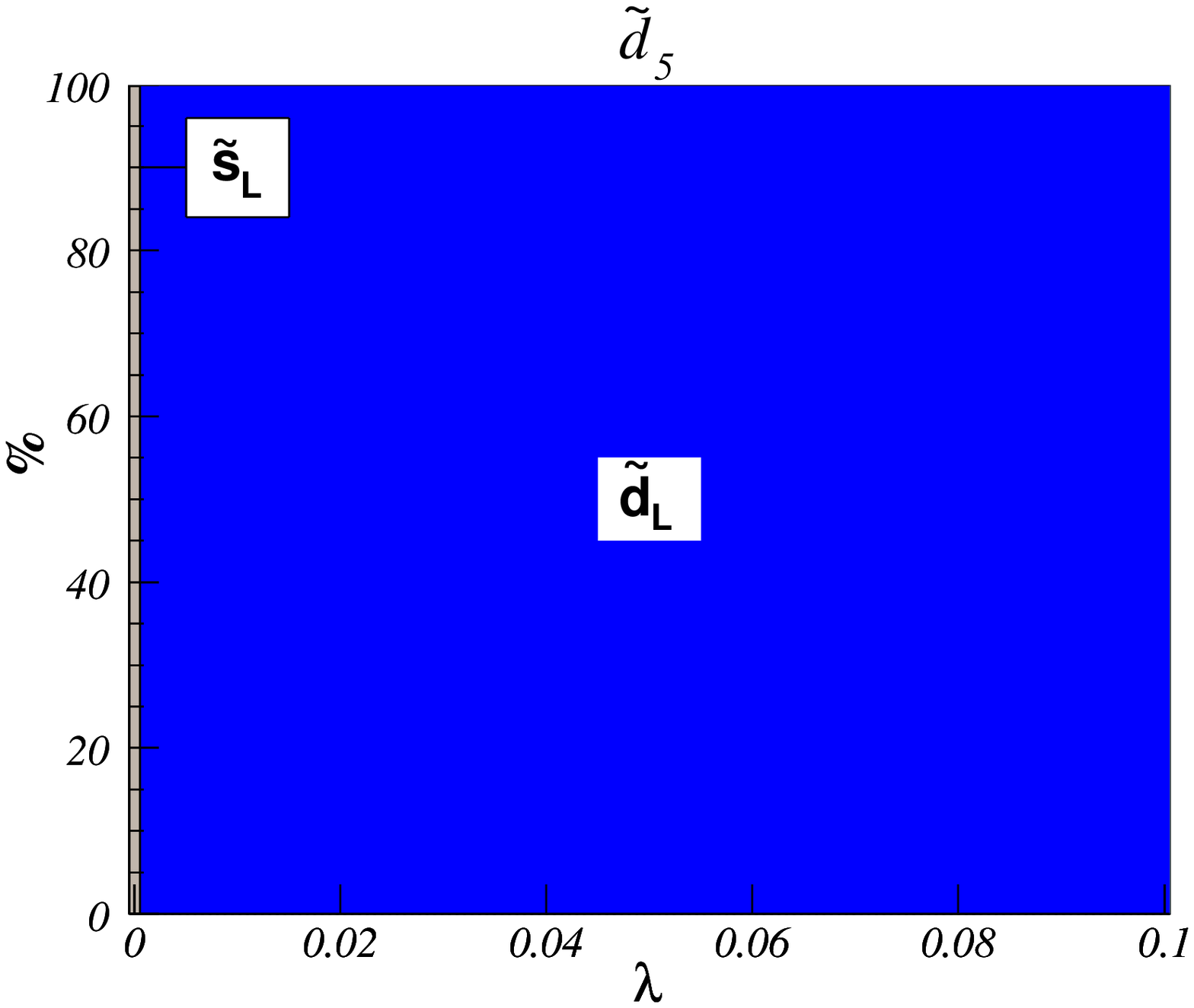}\hspace{2mm}
 \includegraphics[width=0.21\columnwidth]{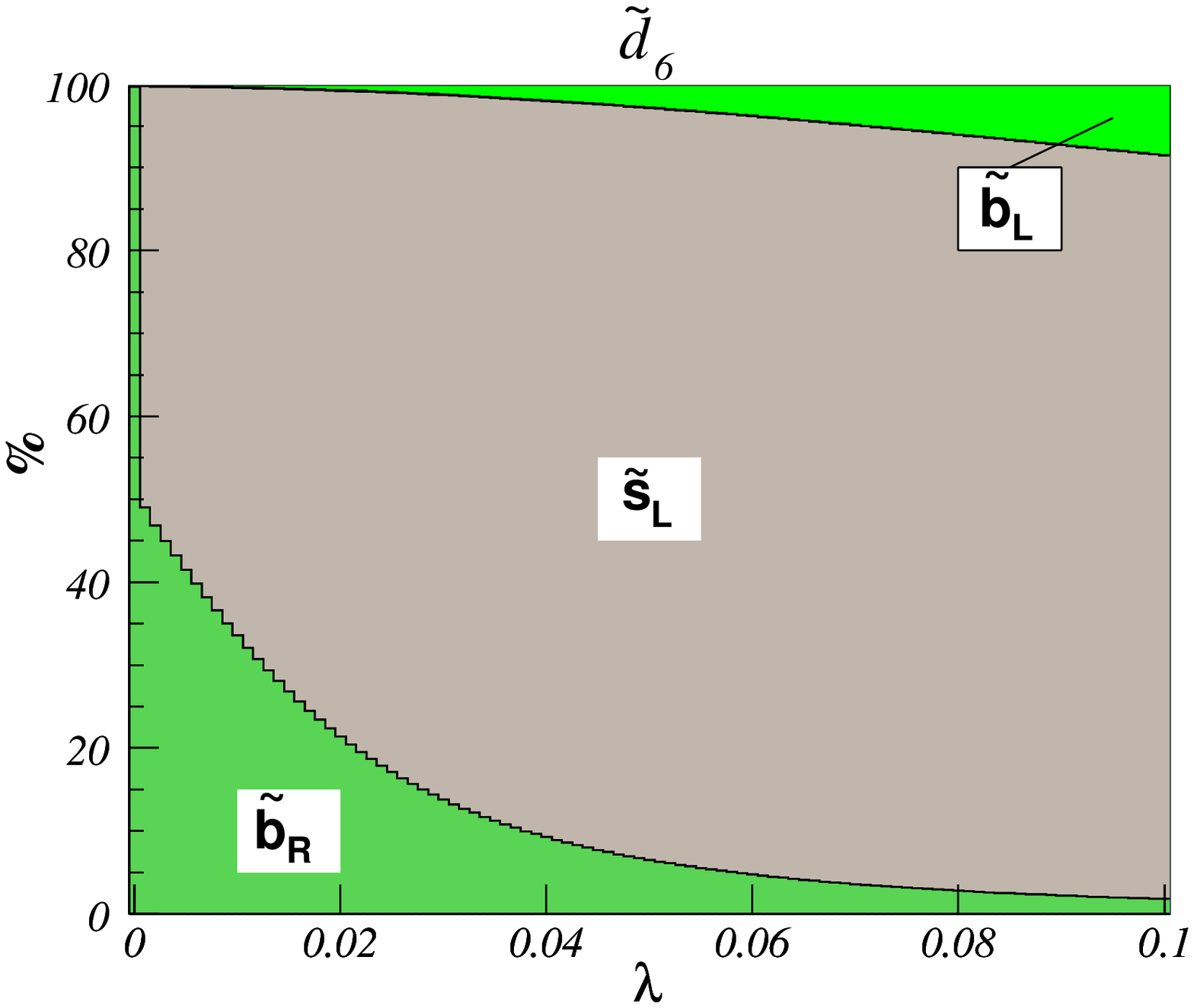}
 \caption{\label{fig:16p}Same as Fig.\ \ref{fig:16} for $\lambda\in
          [0;0.1]$.}
\end{figure}
%
%
\begin{figure}
 \centering
 \includegraphics[width=0.21\columnwidth]{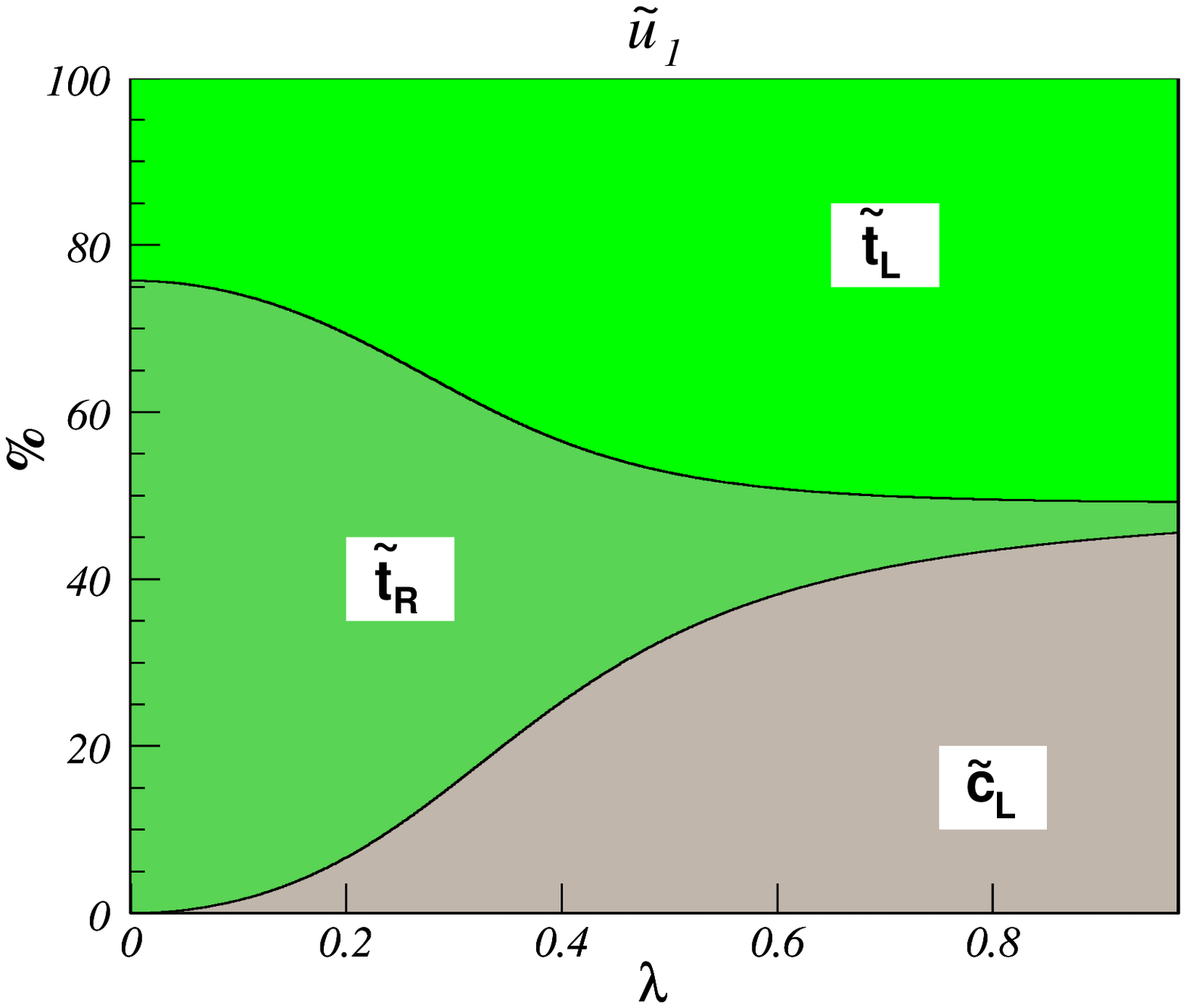}\hspace{2mm}
 \includegraphics[width=0.21\columnwidth]{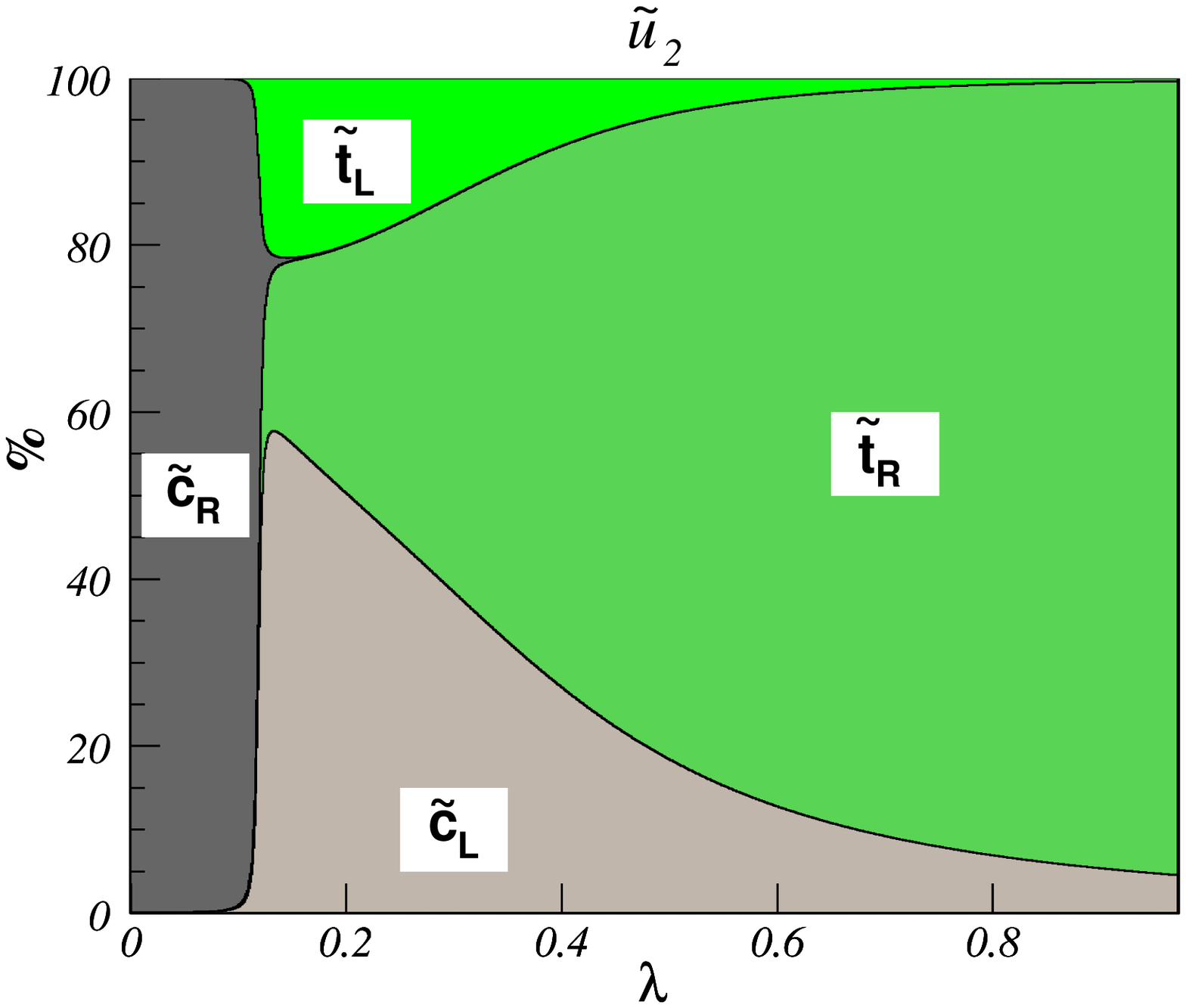}\hspace{2mm}
 \includegraphics[width=0.21\columnwidth]{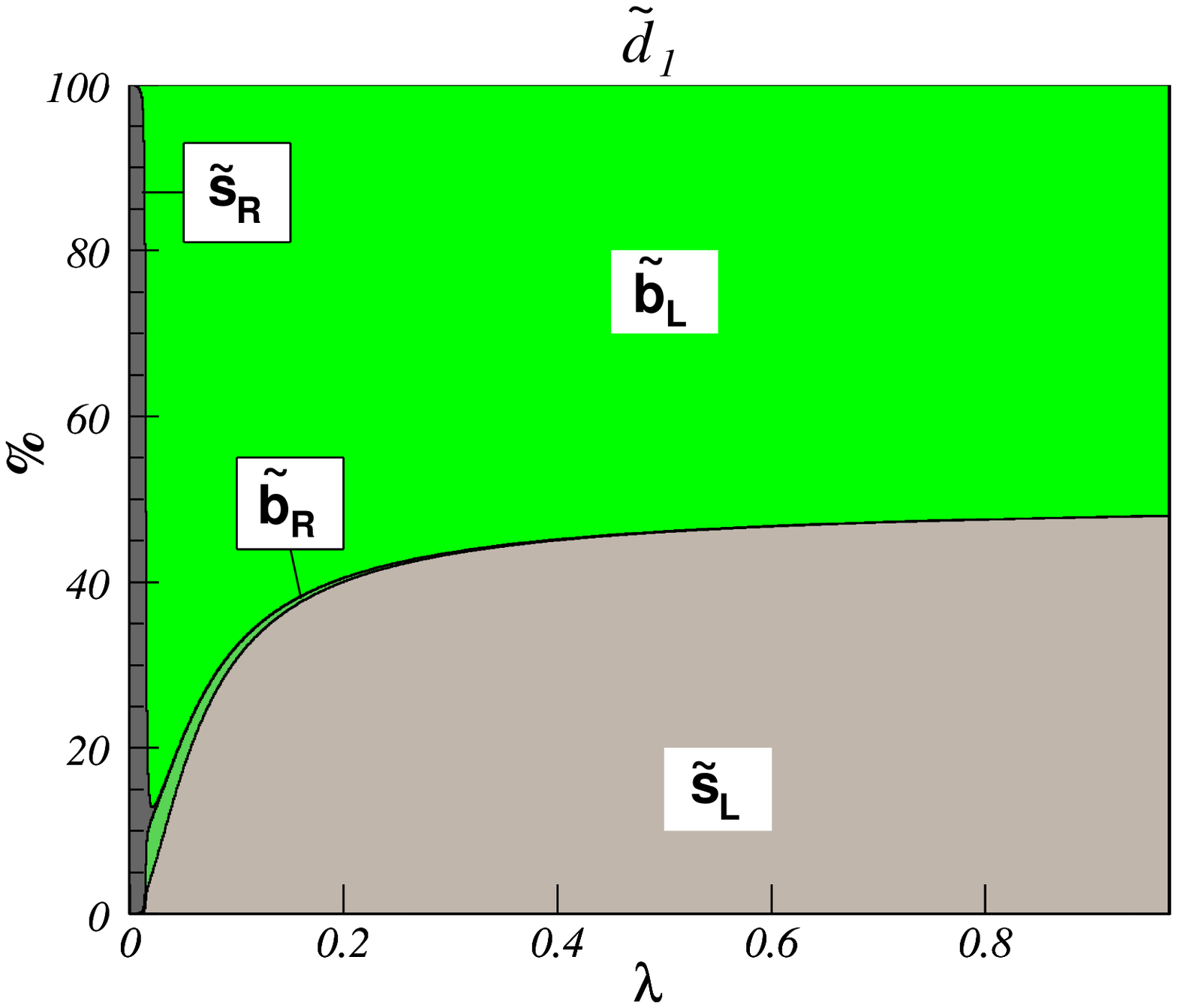}\hspace{2mm}
 \includegraphics[width=0.21\columnwidth]{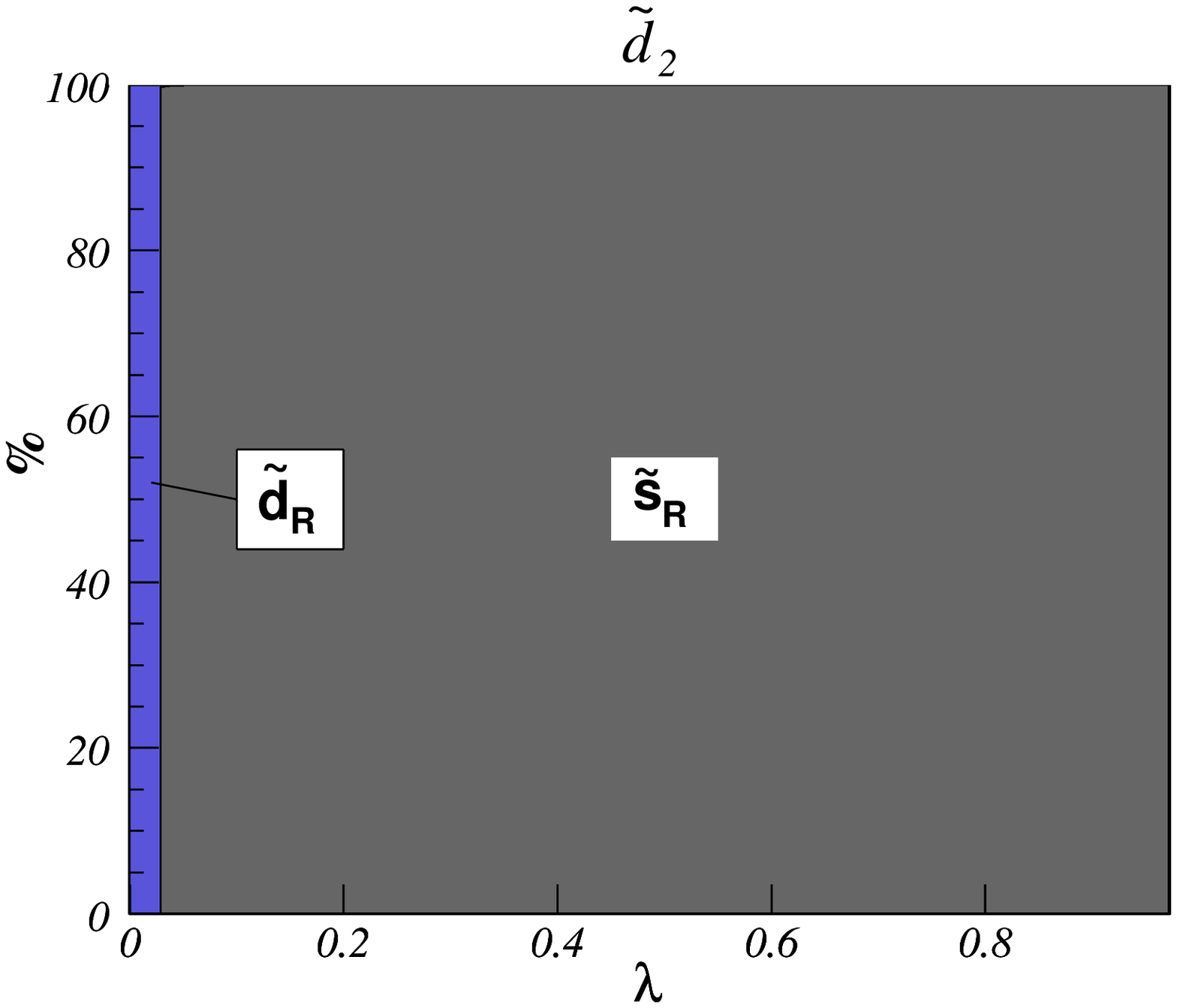}\vspace*{4mm}
 \includegraphics[width=0.21\columnwidth]{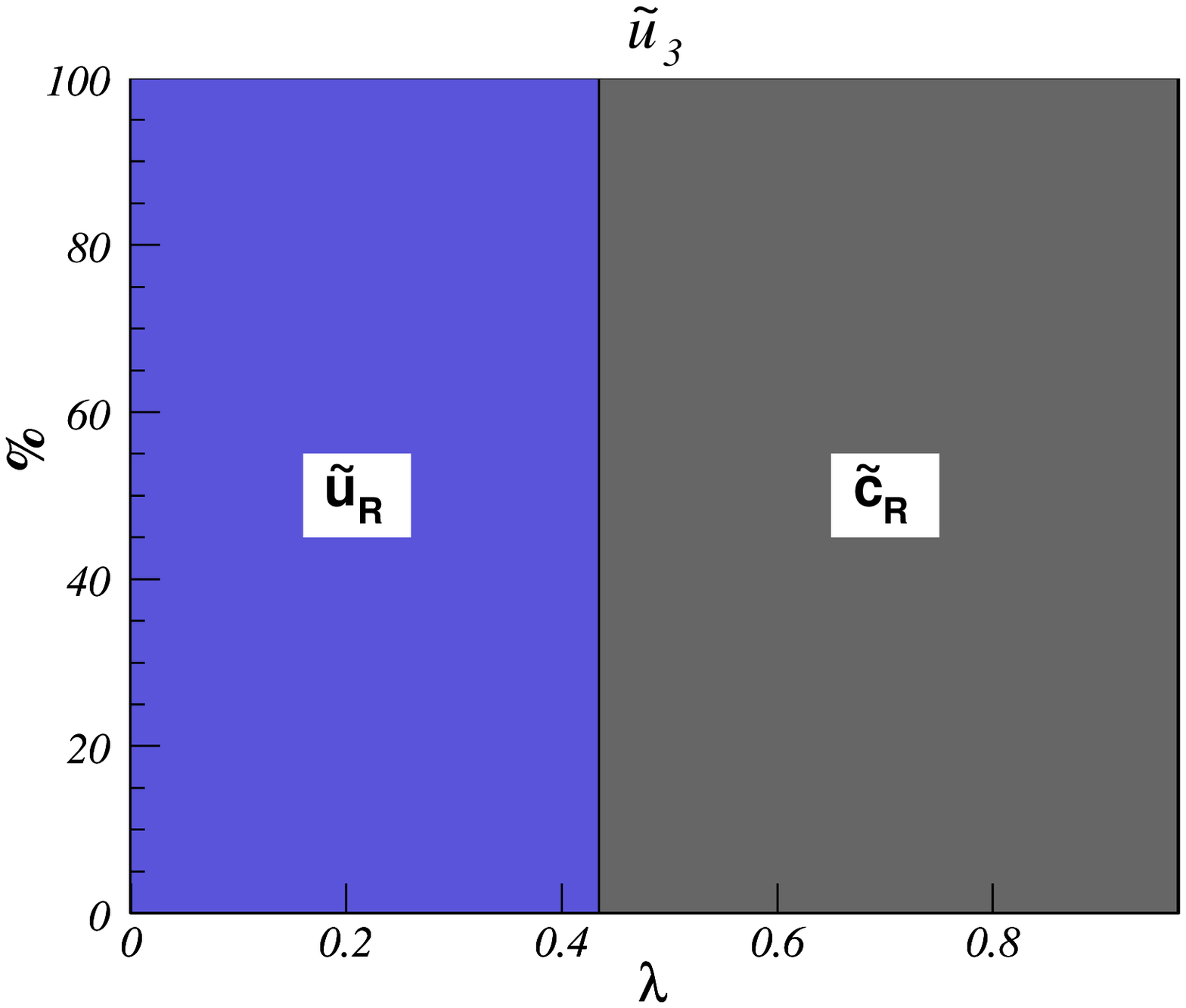}\hspace{2mm}
 \includegraphics[width=0.21\columnwidth]{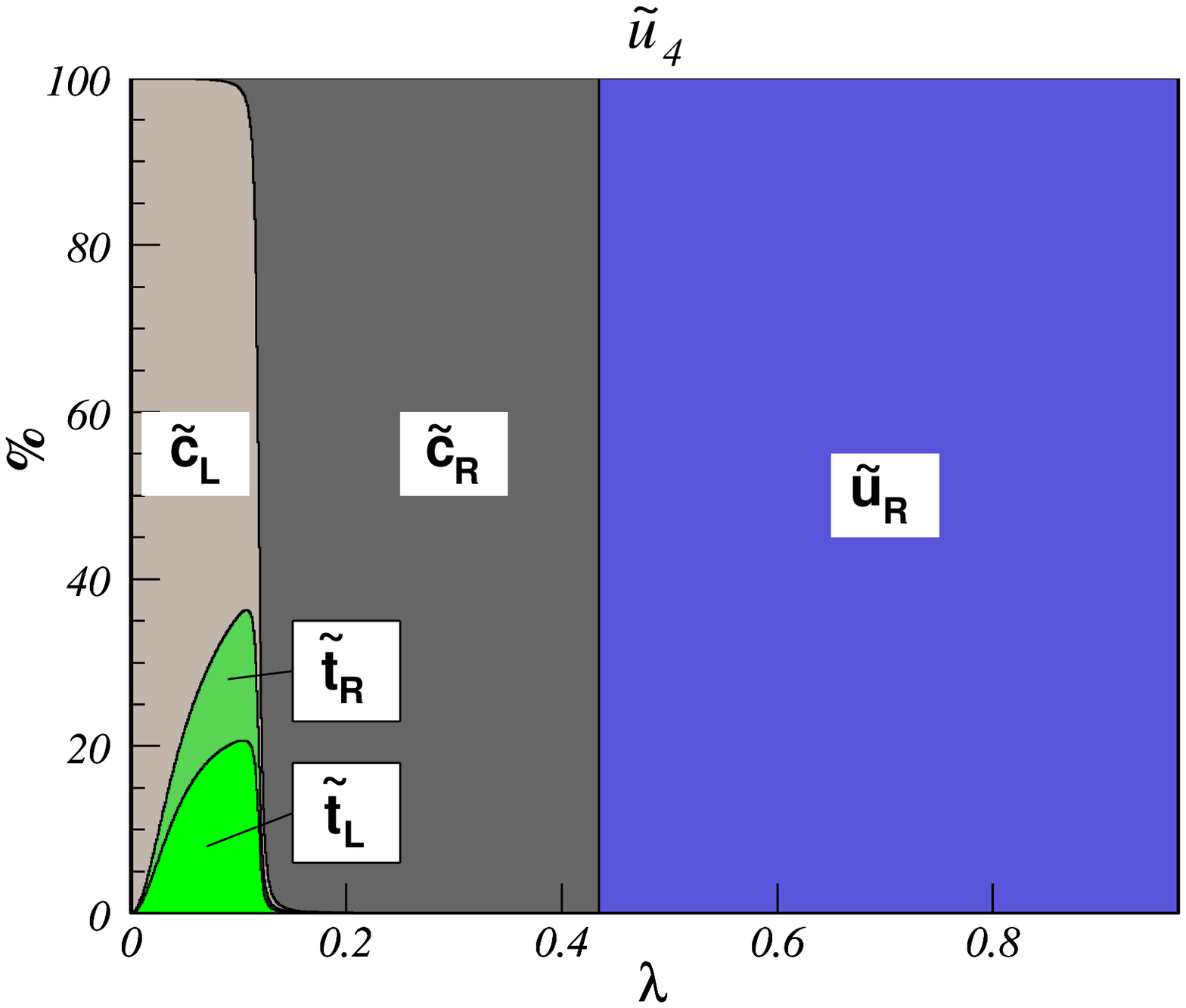}\hspace{2mm}
 \includegraphics[width=0.21\columnwidth]{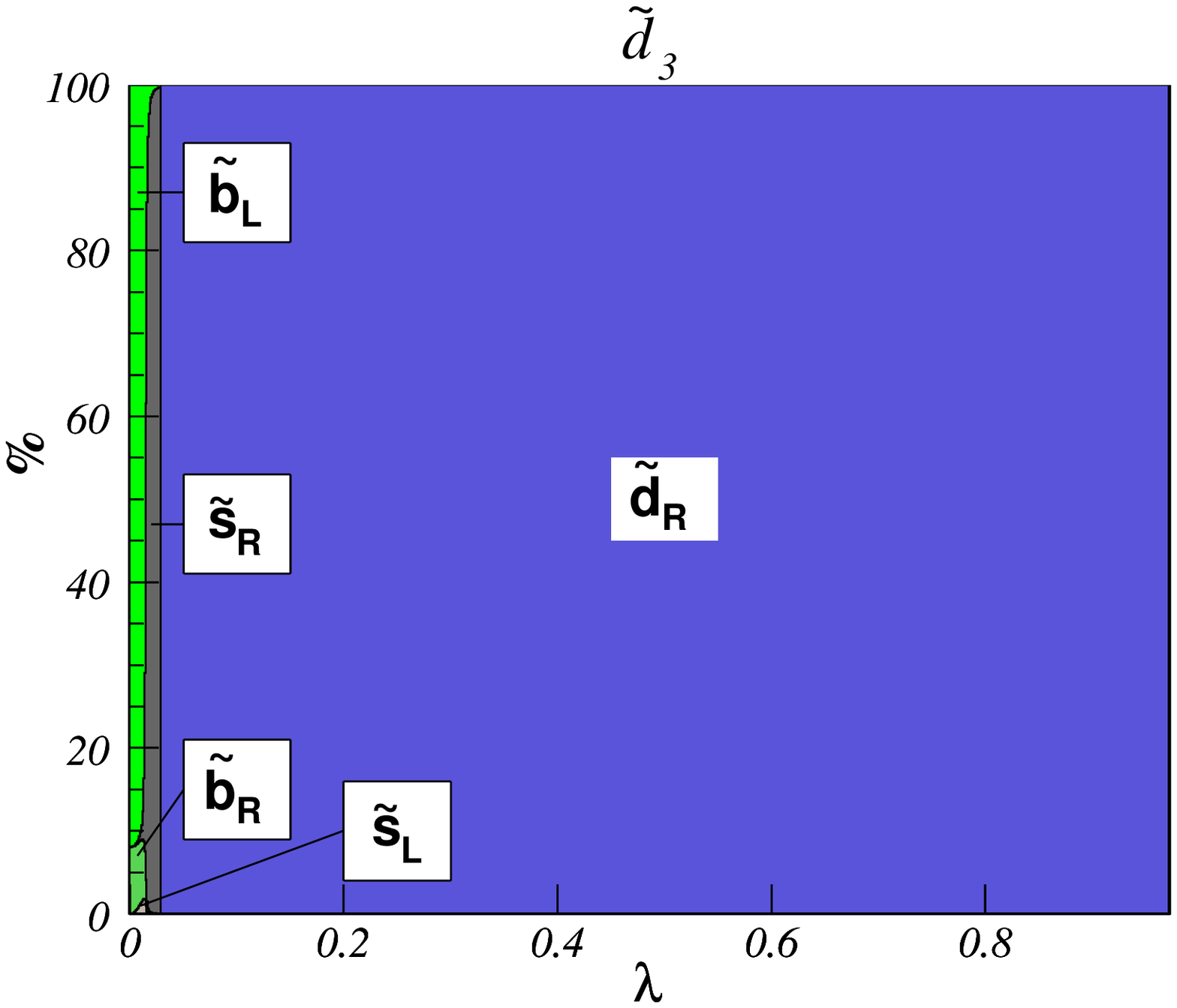}\hspace{2mm}
 \includegraphics[width=0.21\columnwidth]{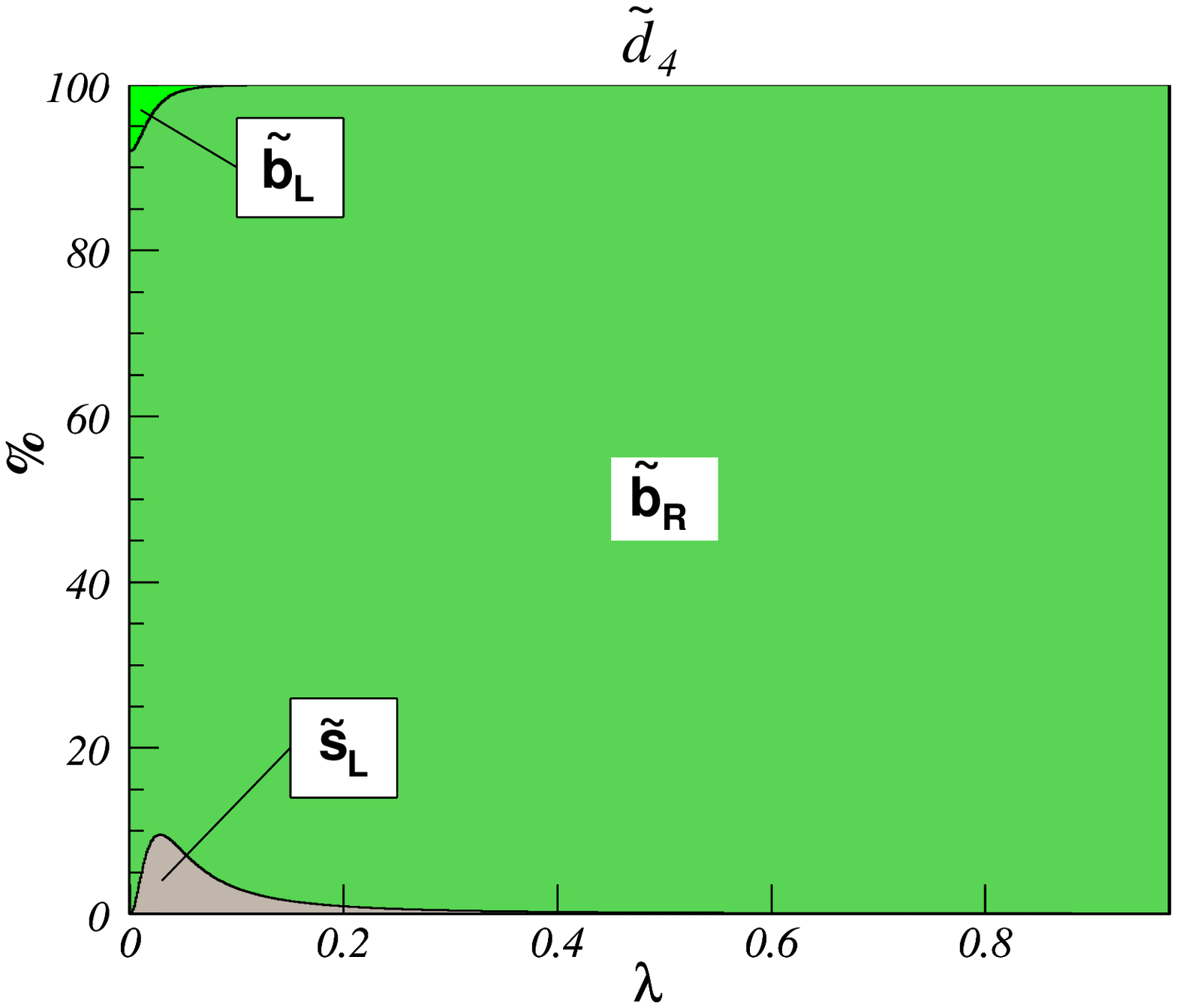}\vspace*{4mm}
 \includegraphics[width=0.21\columnwidth]{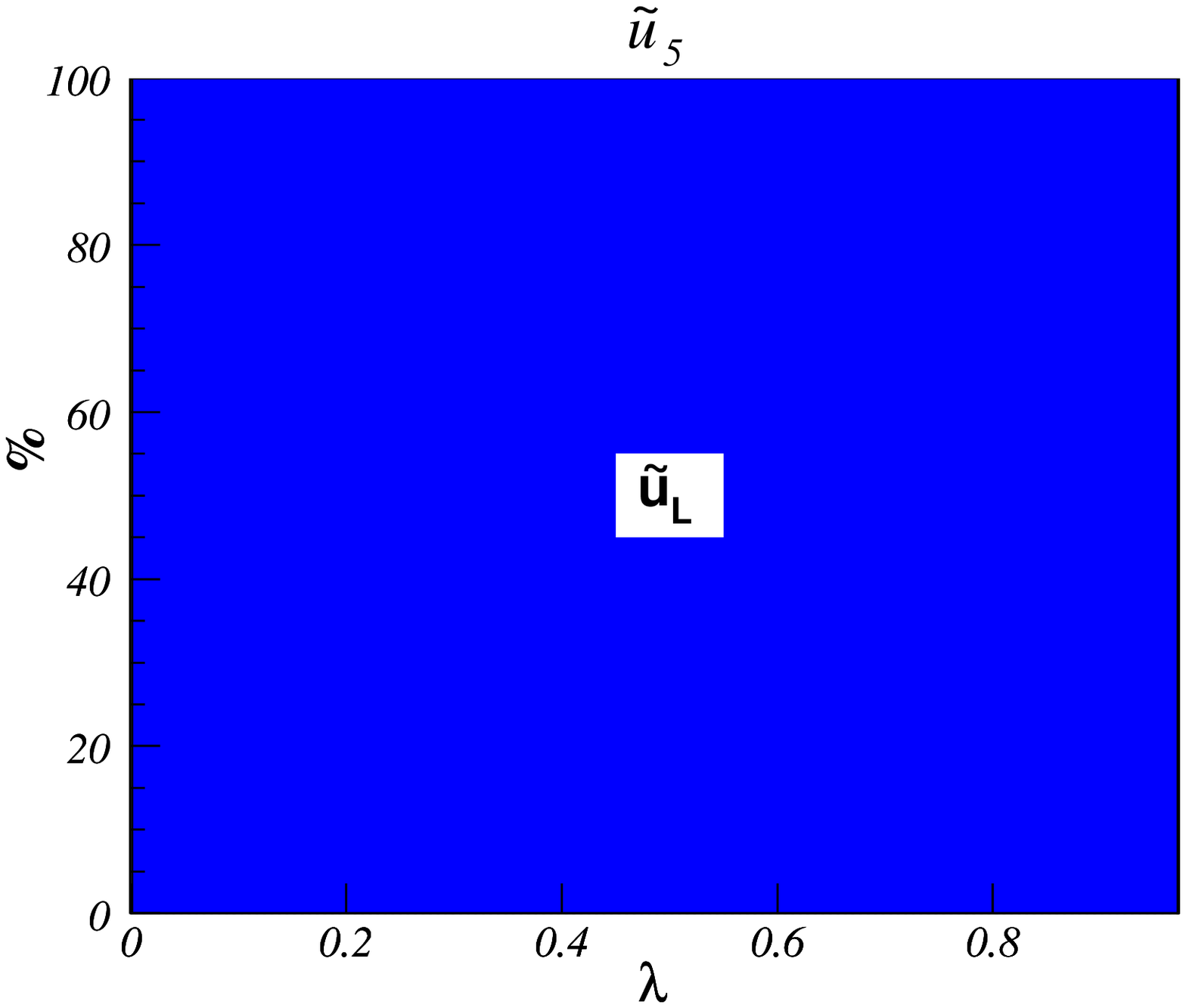}\hspace{2mm}
 \includegraphics[width=0.21\columnwidth]{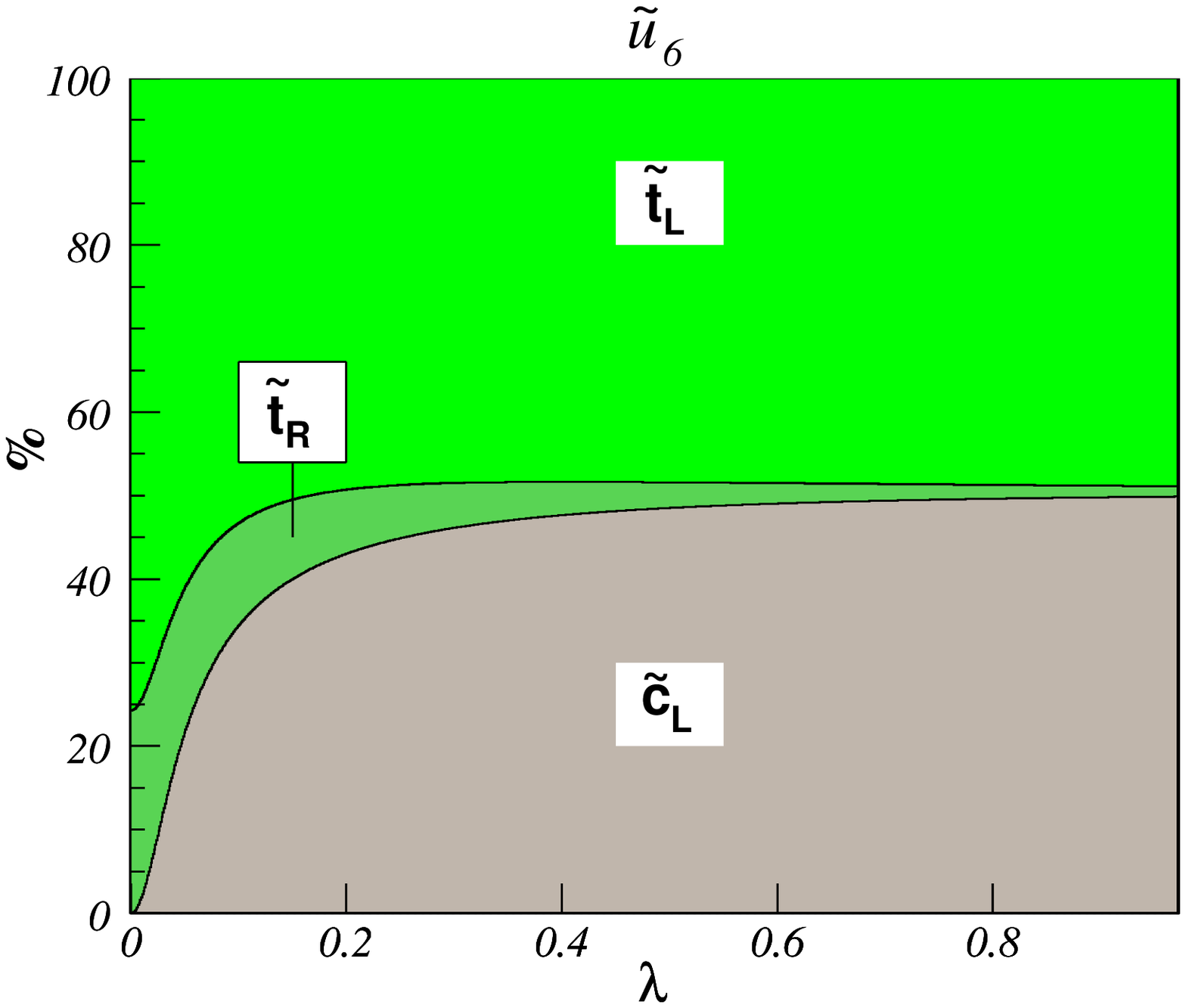}\hspace{2mm}
 \includegraphics[width=0.21\columnwidth]{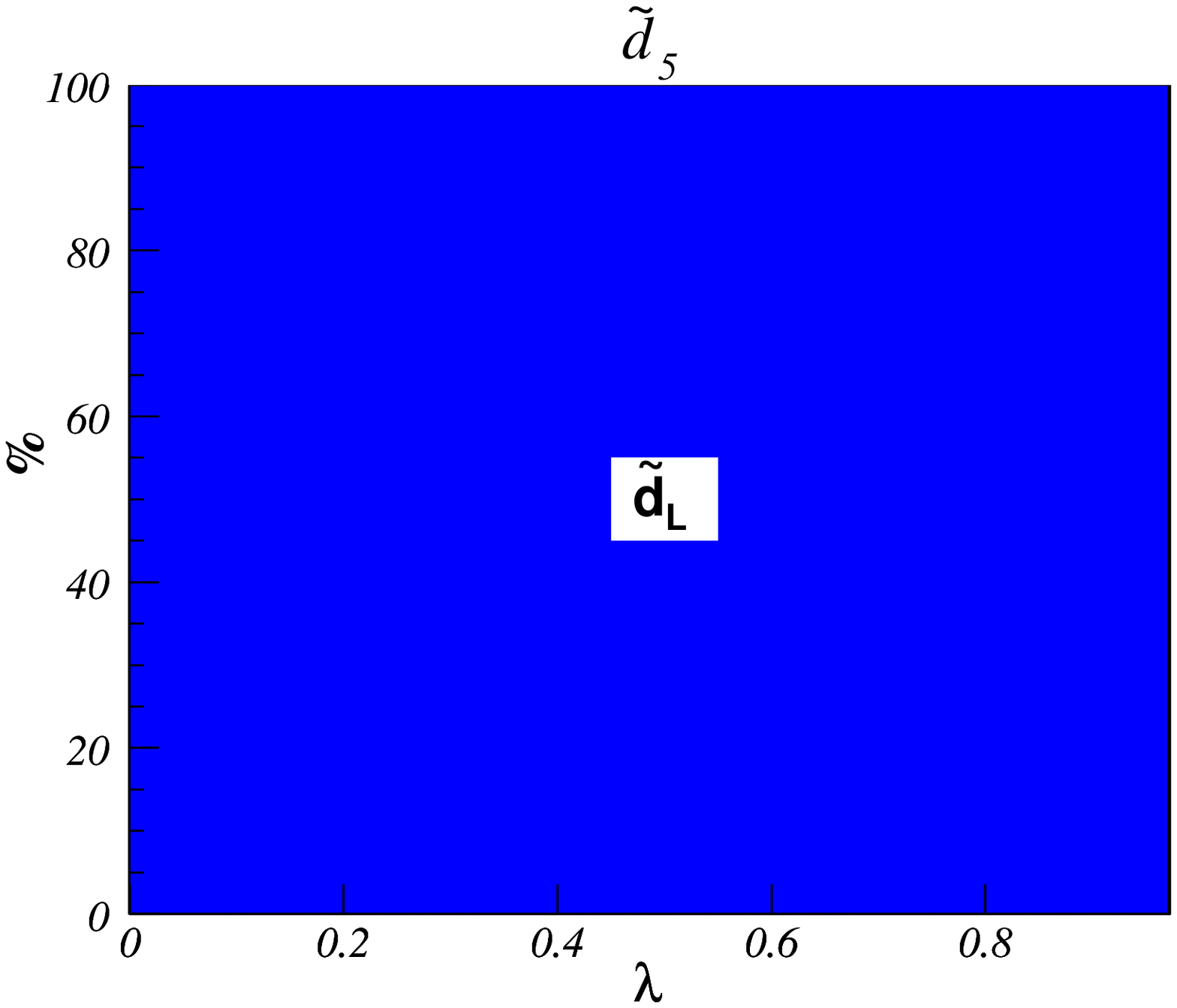}\hspace{2mm}
 \includegraphics[width=0.21\columnwidth]{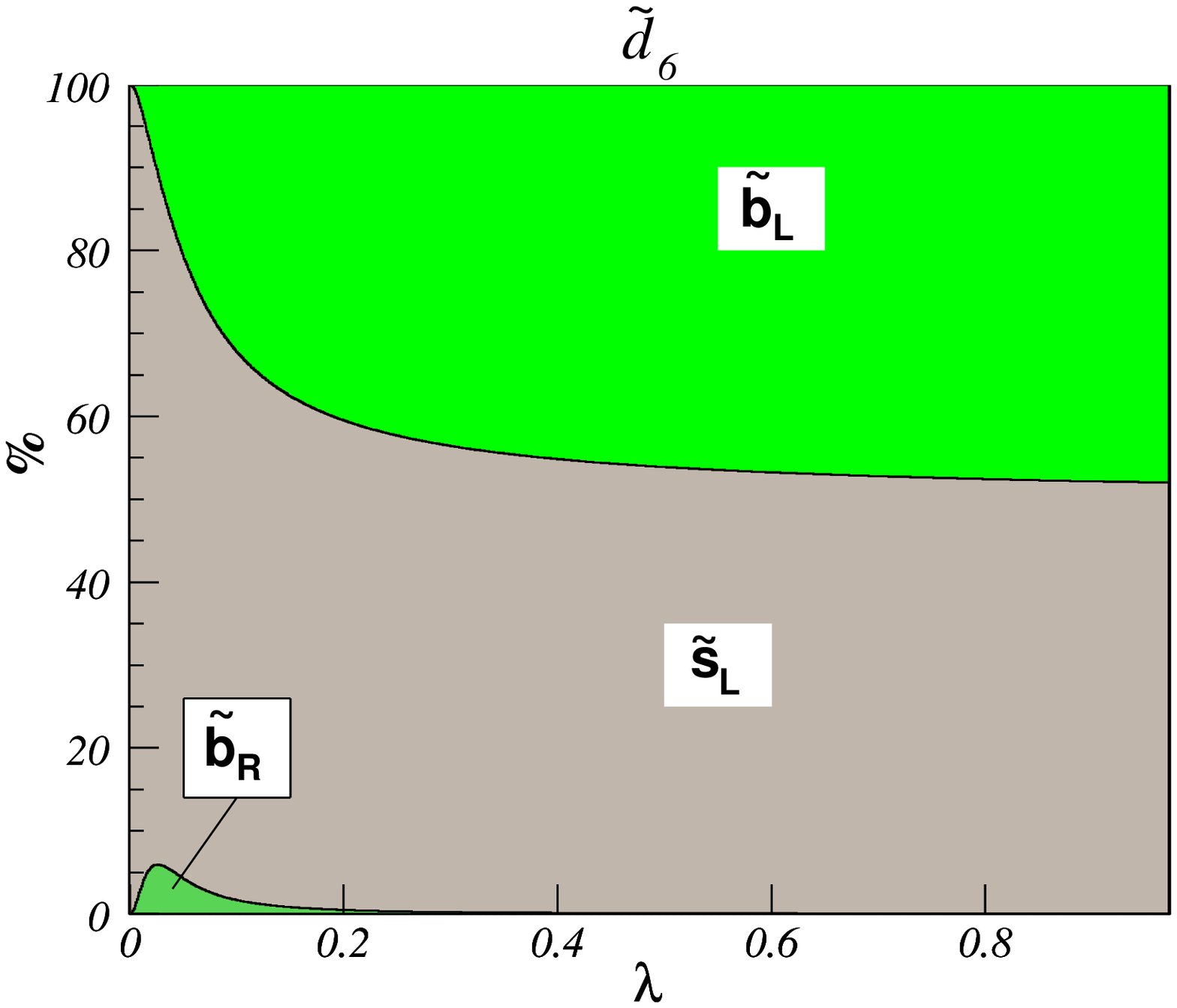}
 \caption{\label{fig:17}Same as Fig.\ \ref{fig:16} for benchmark point B.}
\end{figure}
%
%
\begin{figure}
 \centering
 \includegraphics[width=0.21\columnwidth]{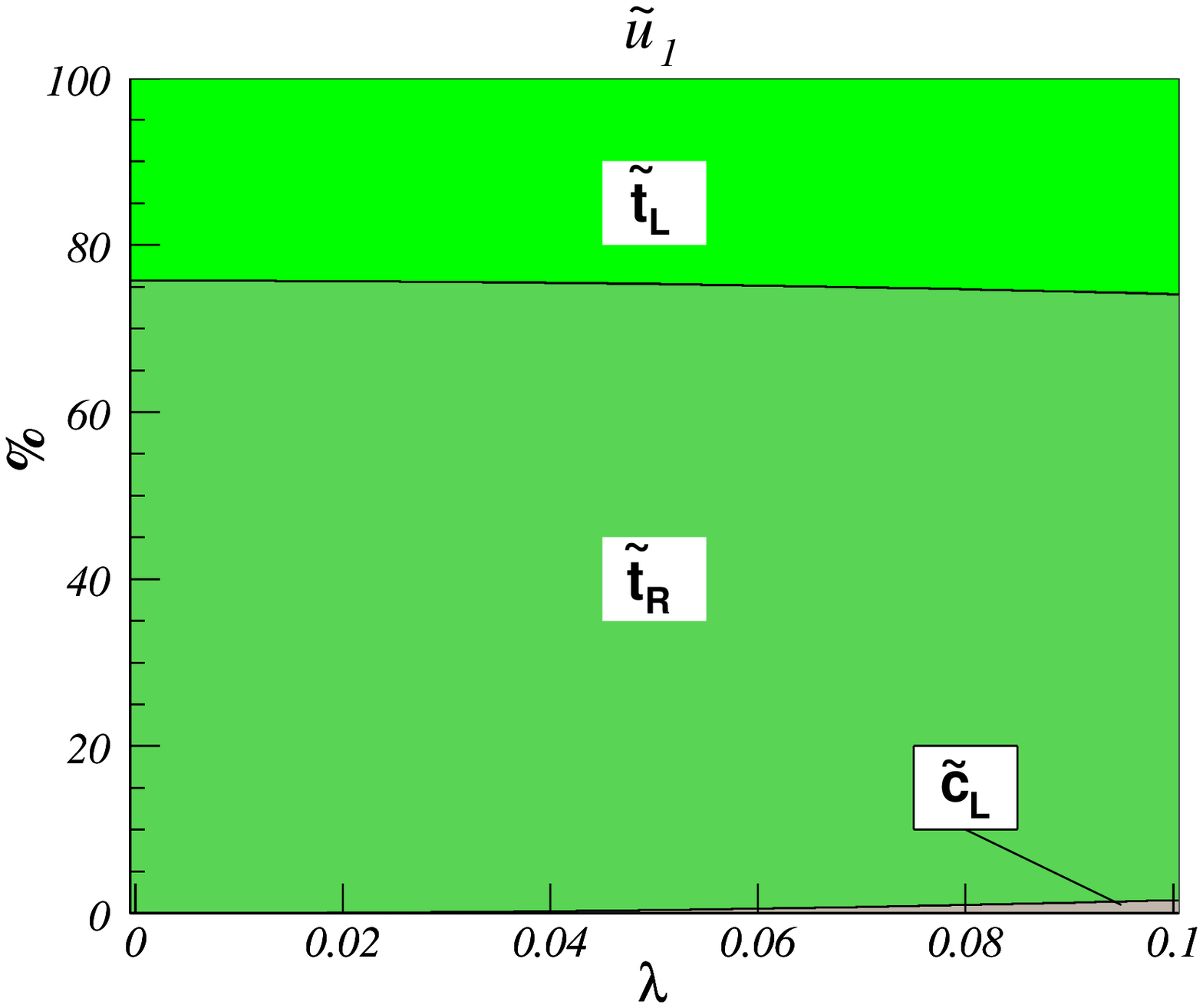}\hspace{2mm}
 \includegraphics[width=0.21\columnwidth]{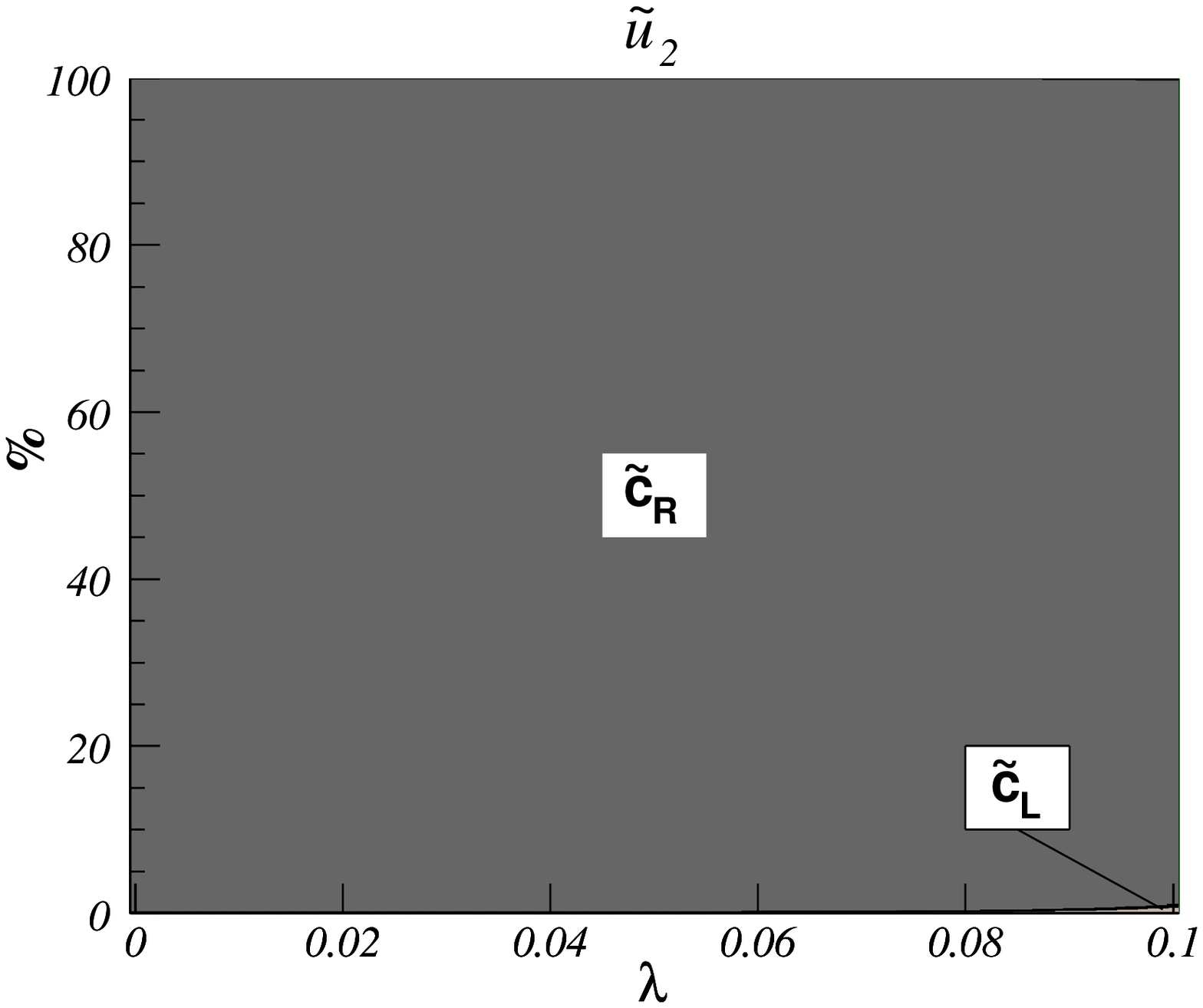}\hspace{2mm}
 \includegraphics[width=0.21\columnwidth]{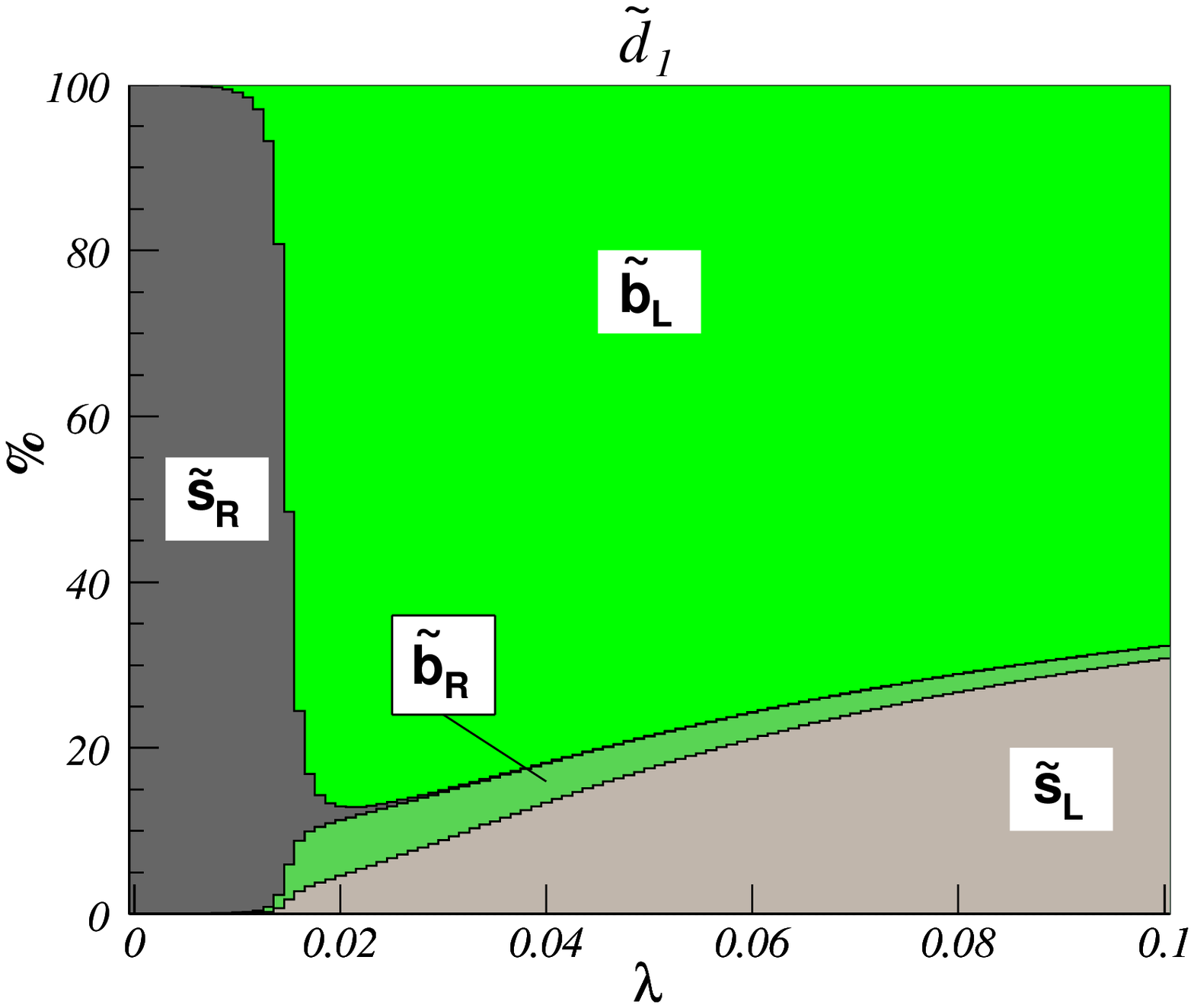}\hspace{2mm}
 \includegraphics[width=0.21\columnwidth]{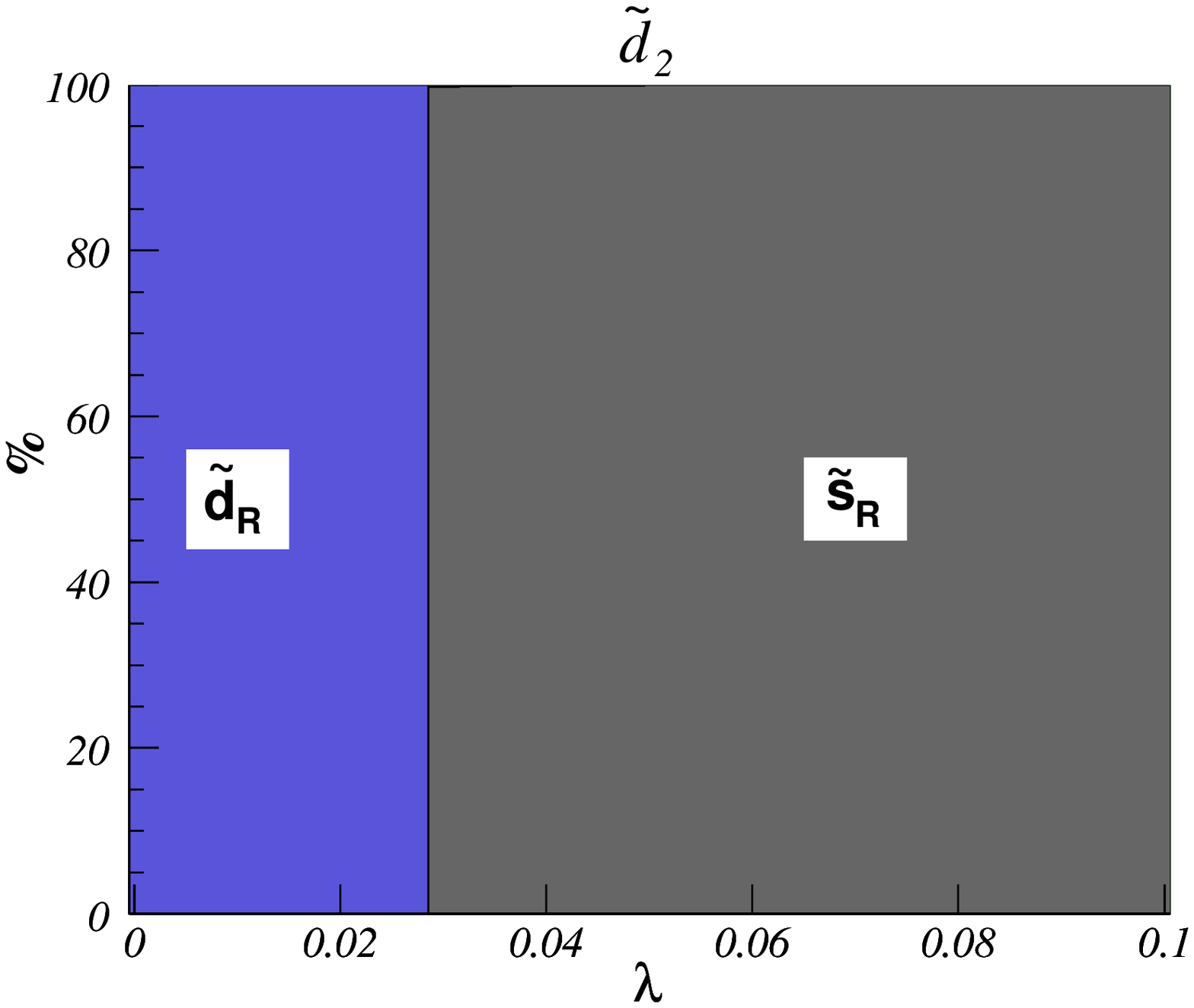}\vspace*{4mm}
 \includegraphics[width=0.21\columnwidth]{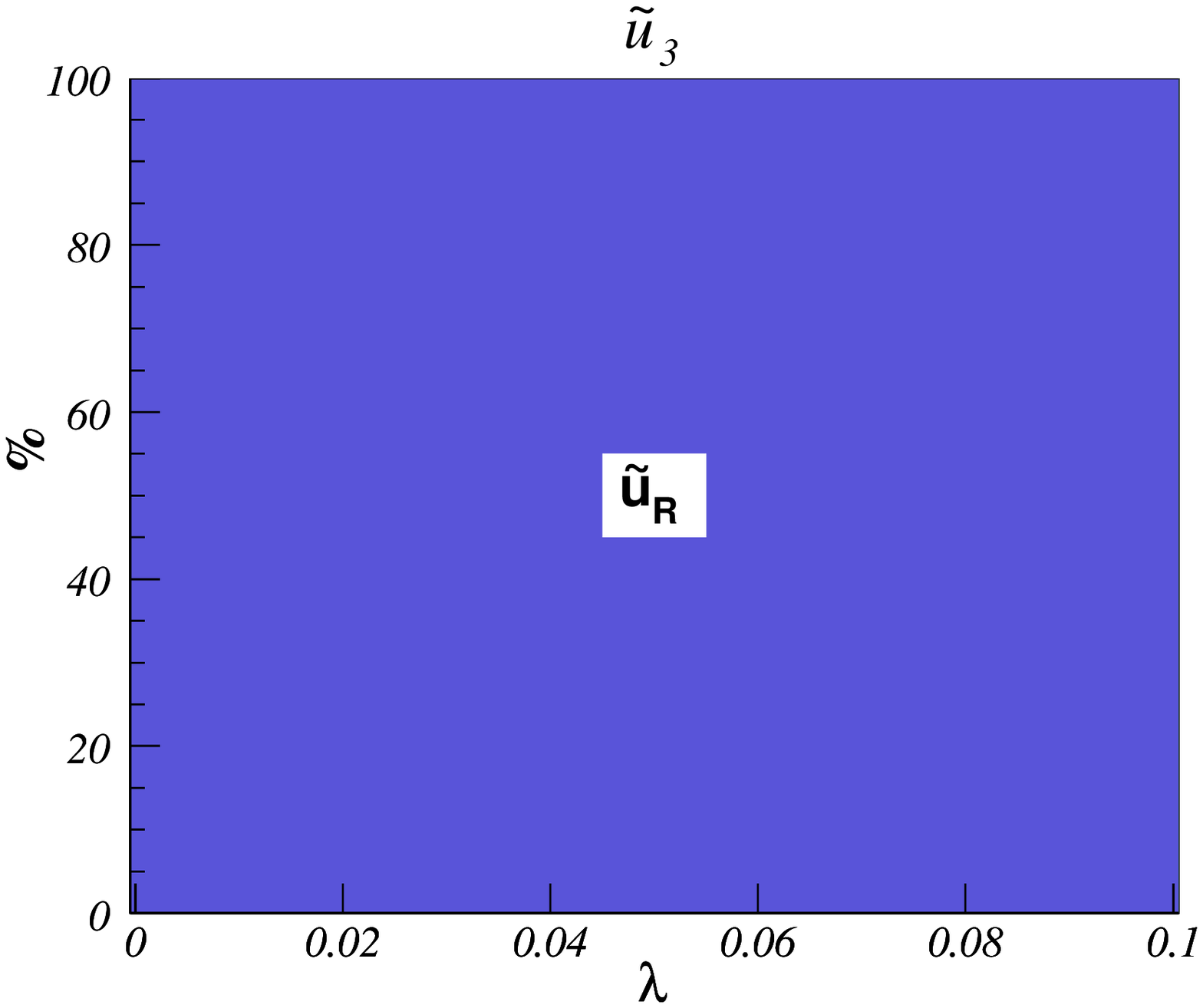}\hspace{2mm}
 \includegraphics[width=0.21\columnwidth]{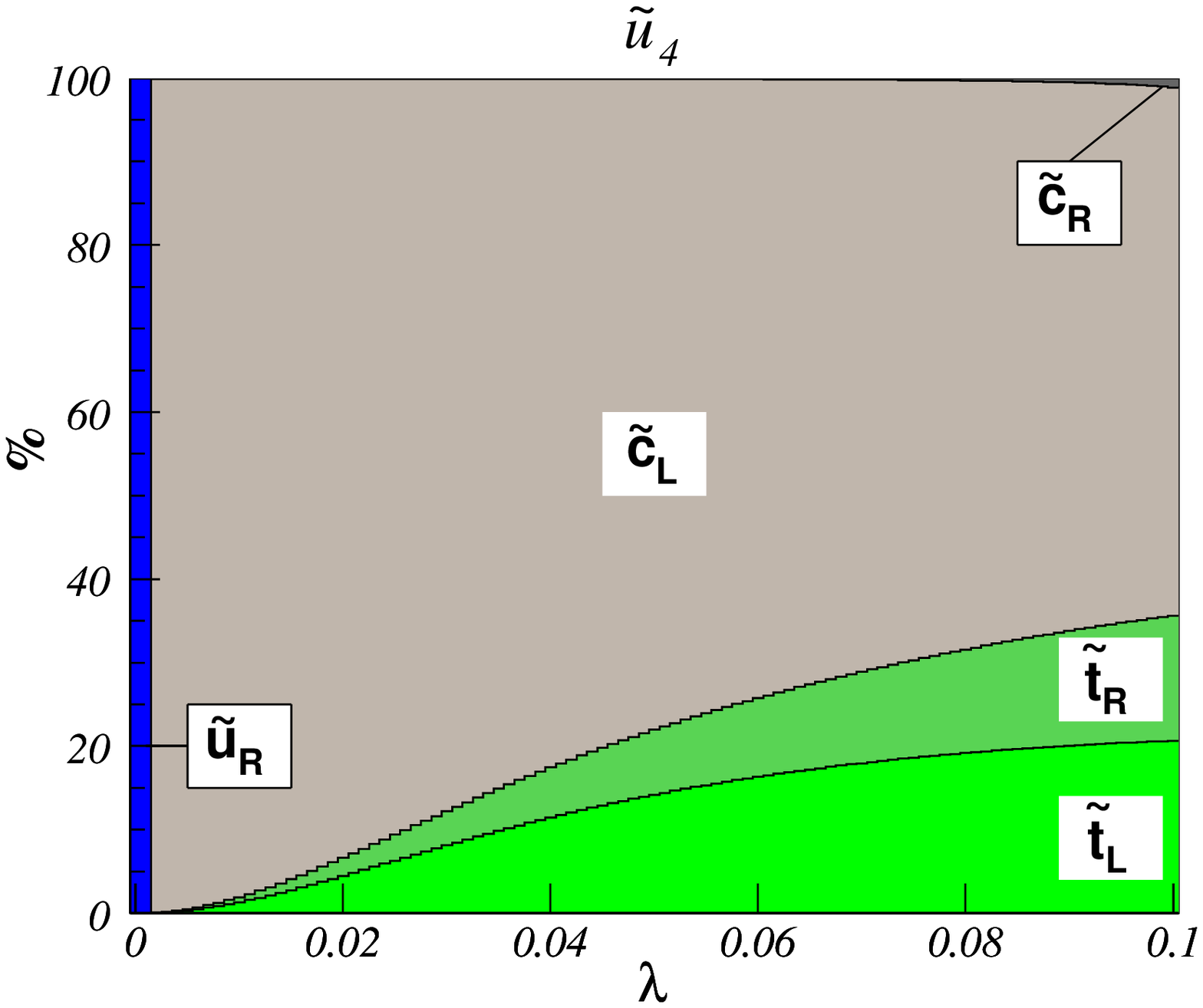}\hspace{2mm}
 \includegraphics[width=0.21\columnwidth]{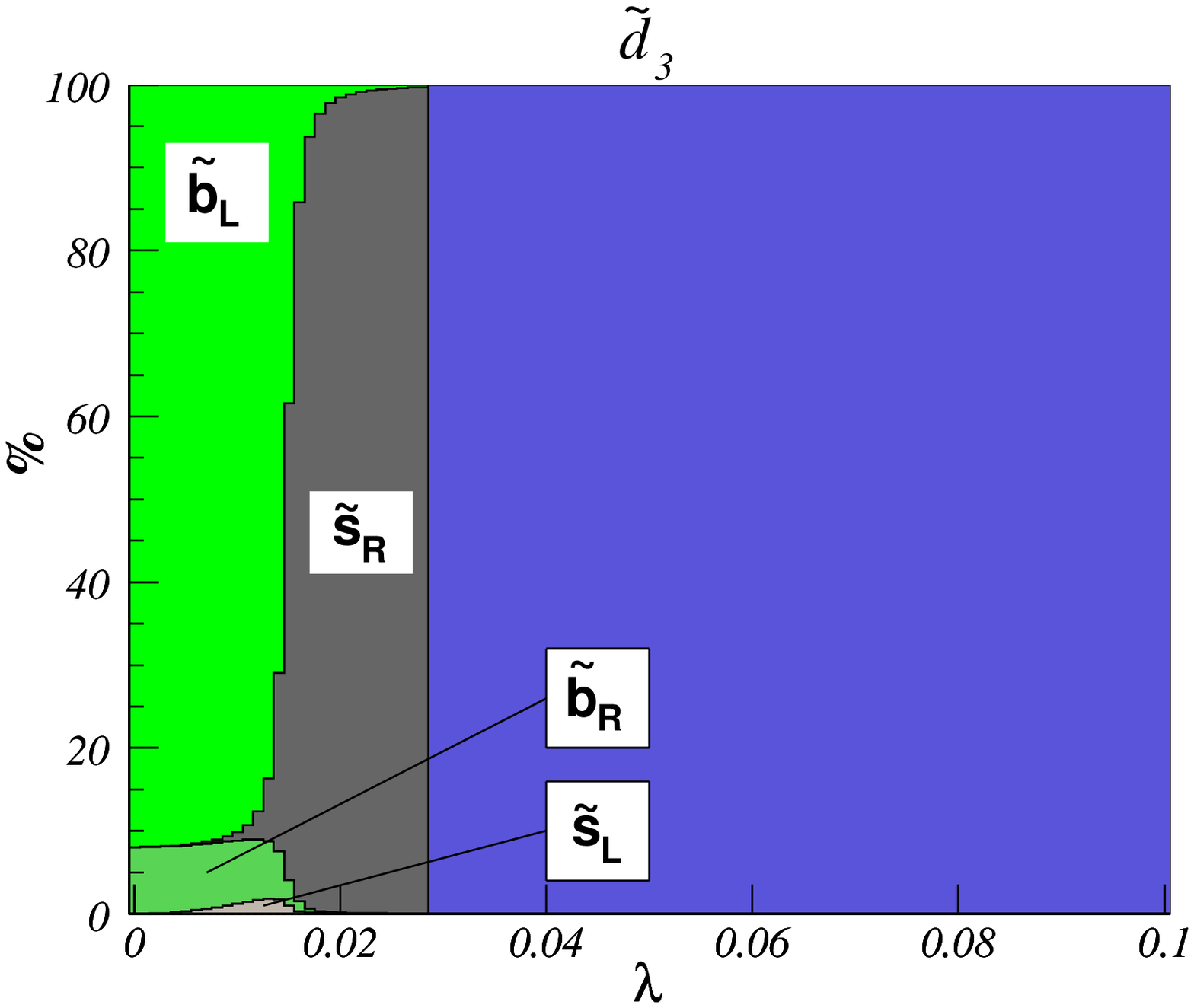}\hspace{2mm}
 \includegraphics[width=0.21\columnwidth]{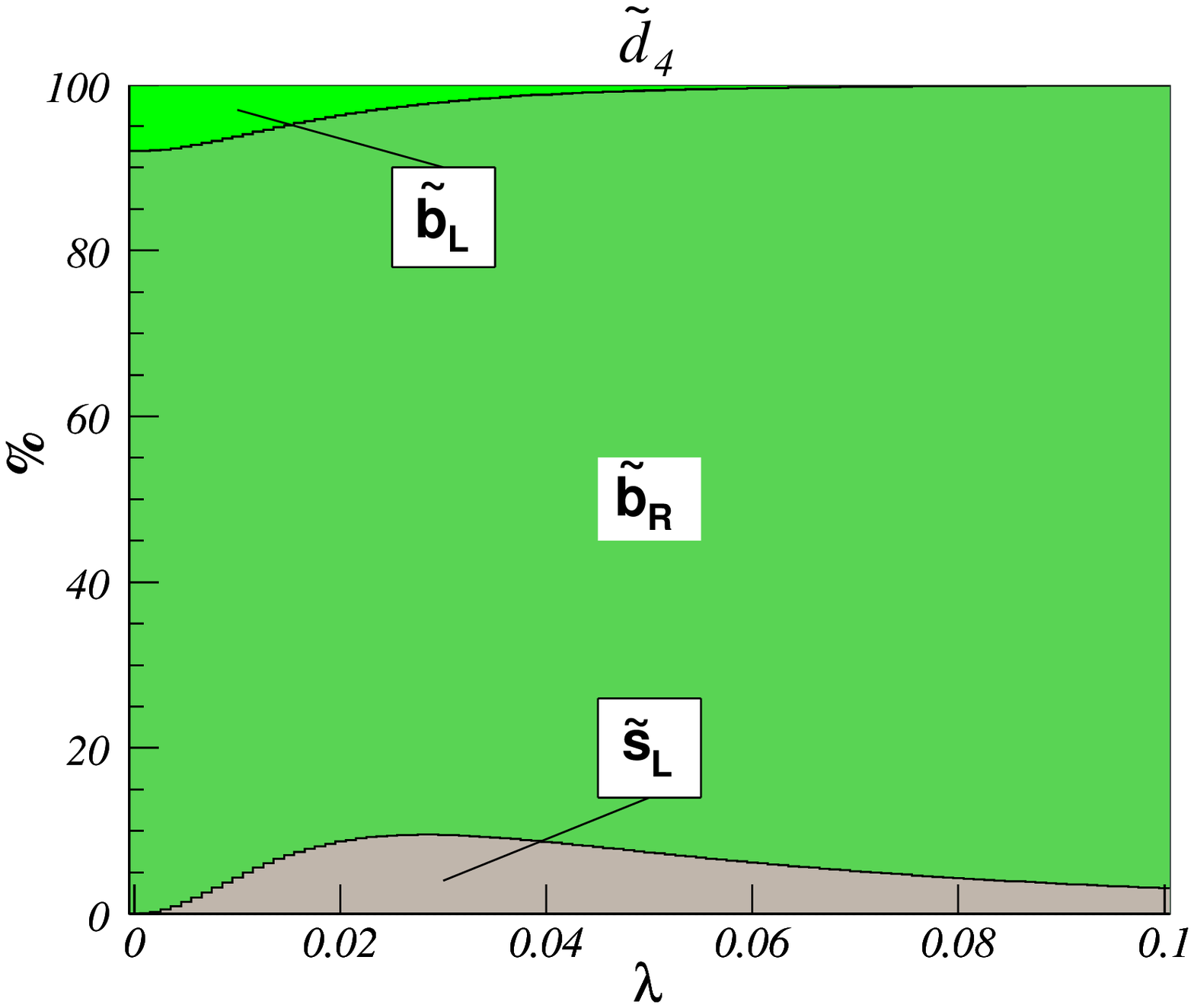}\vspace*{4mm}
 \includegraphics[width=0.21\columnwidth]{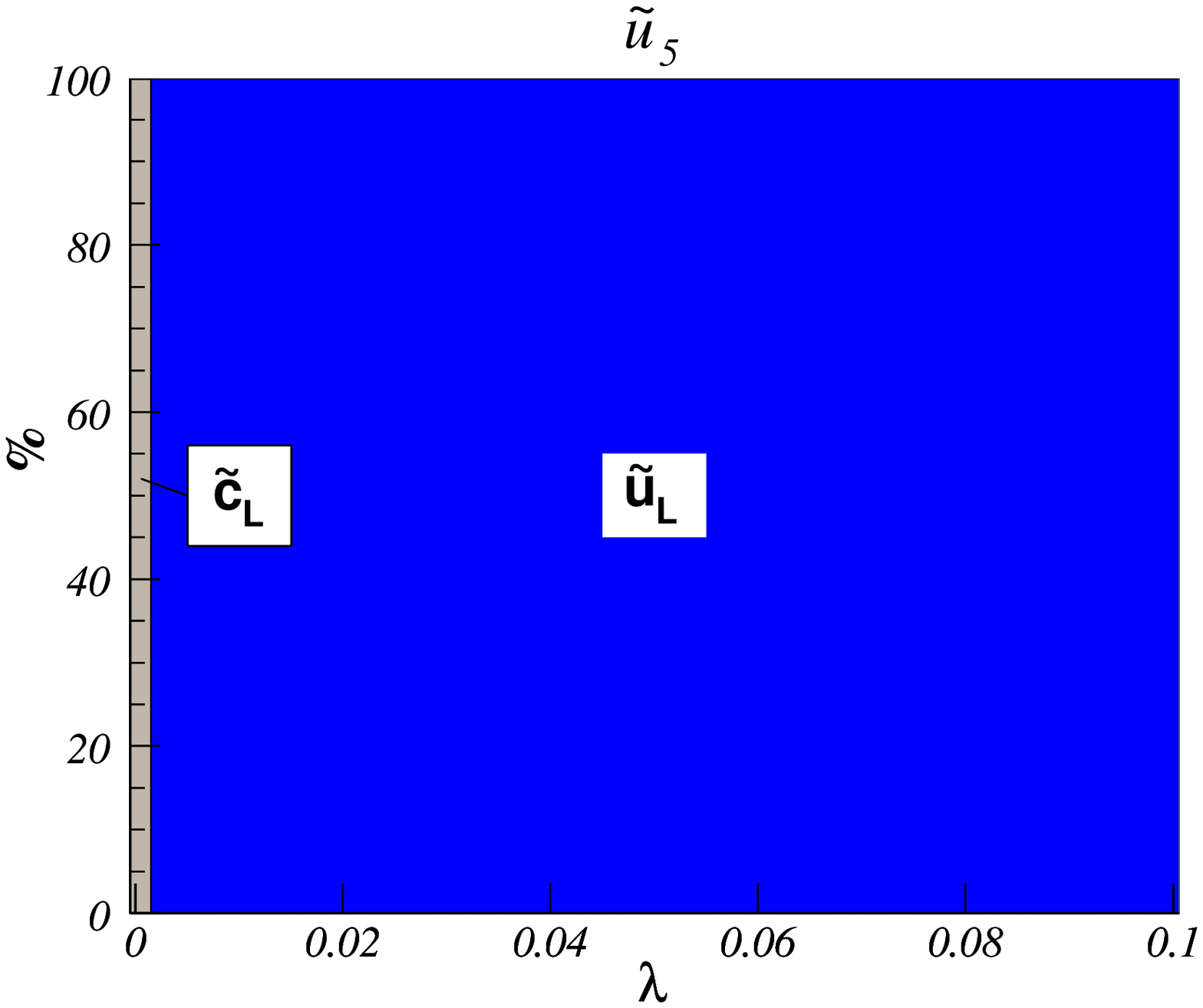}\hspace{2mm}
 \includegraphics[width=0.21\columnwidth]{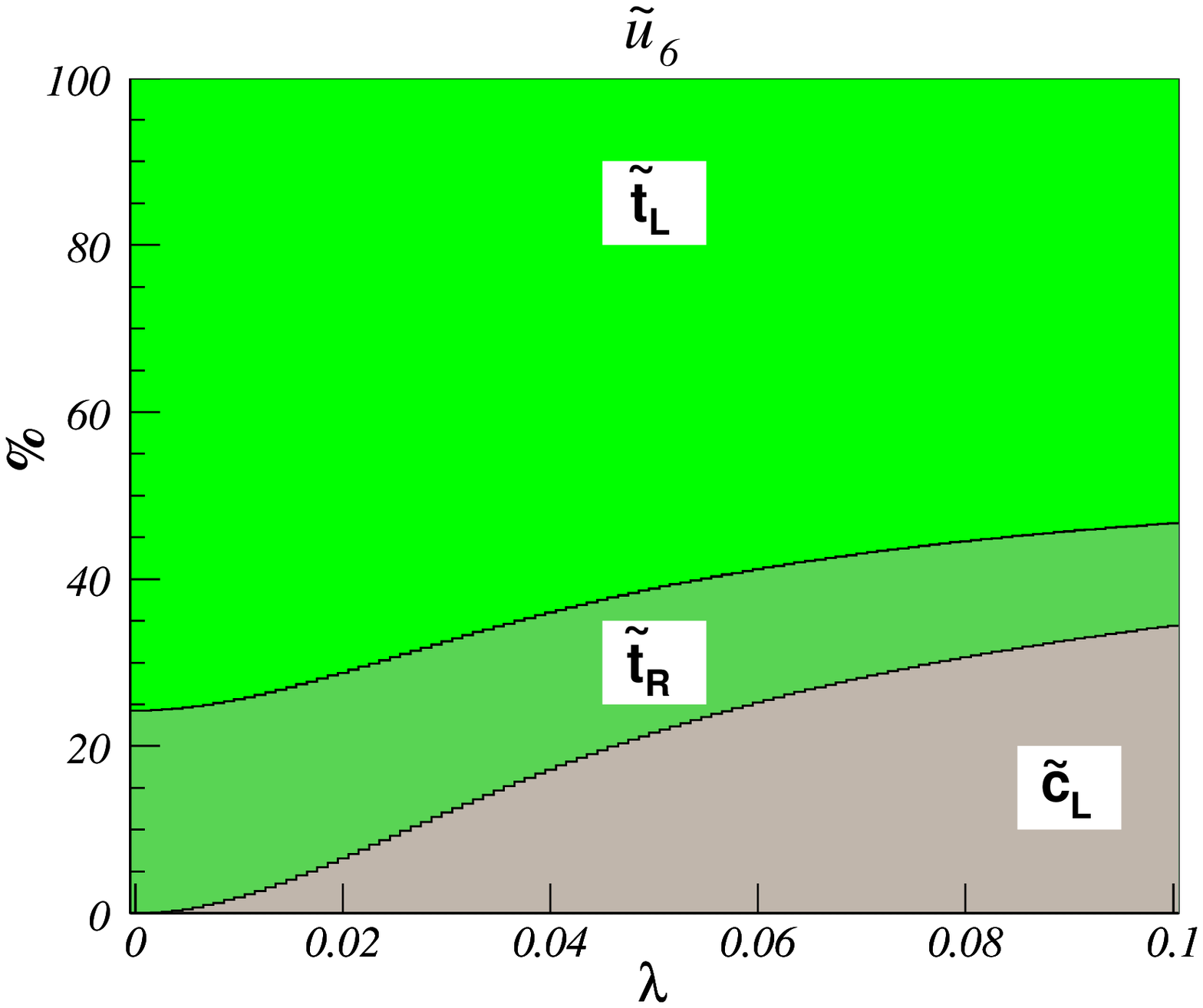}\hspace{2mm}
 \includegraphics[width=0.21\columnwidth]{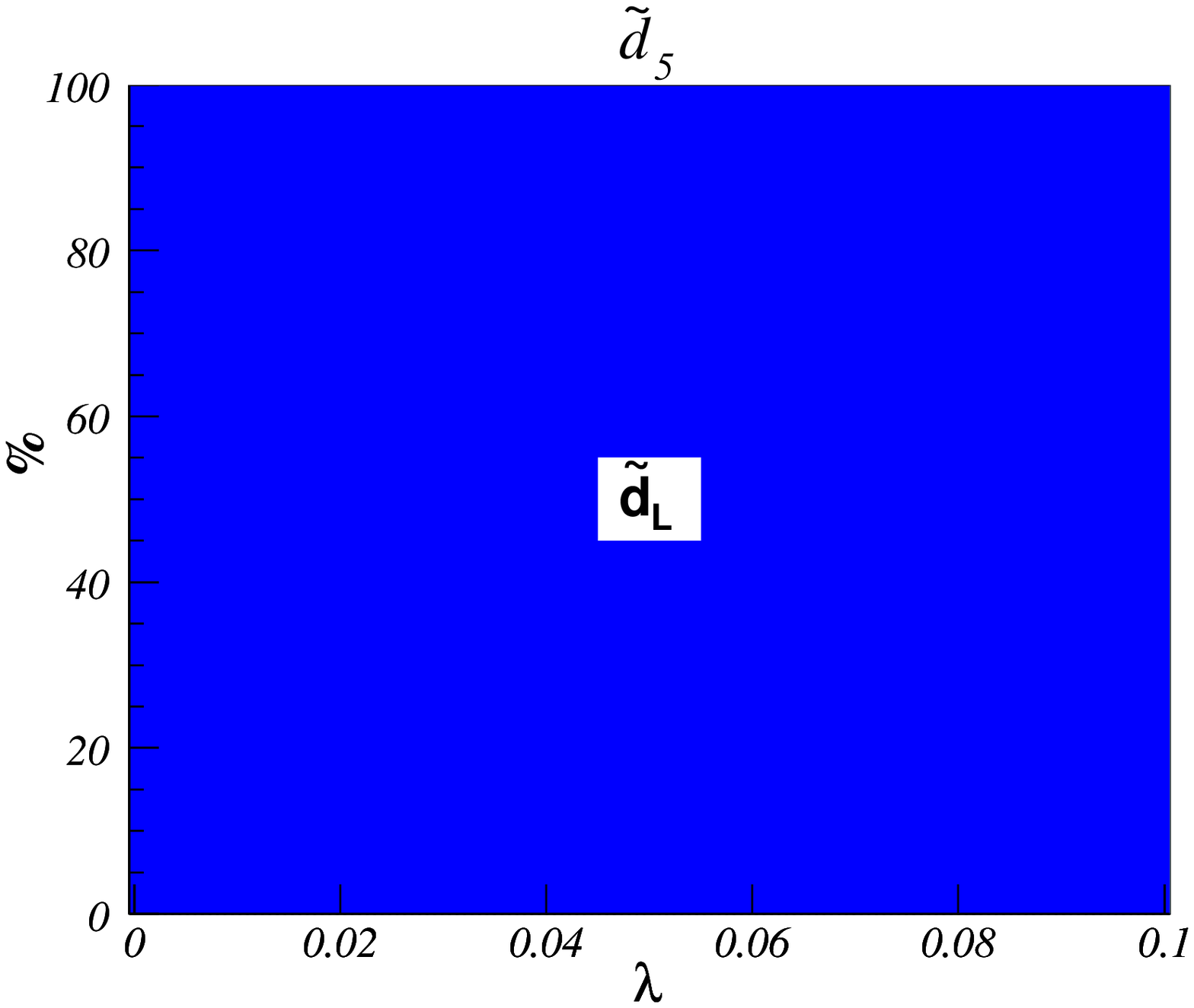}\hspace{2mm}
 \includegraphics[width=0.21\columnwidth]{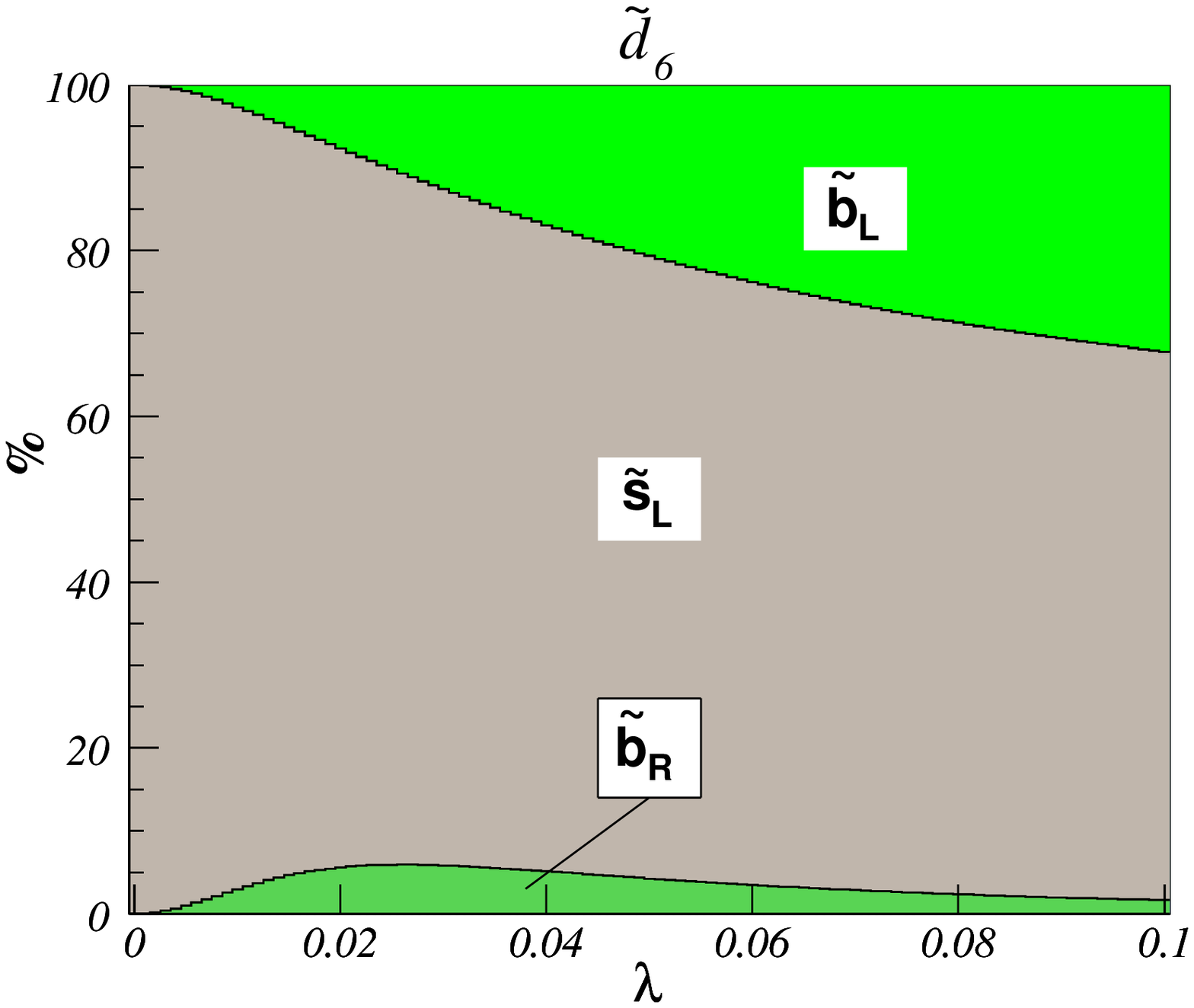}
 \caption{\label{fig:17p}Same as Fig.\ \ref{fig:17} for $\lambda\in
          [0;0.1]$.}
\end{figure}
%
%
\begin{figure}
 \centering
 \includegraphics[width=0.21\columnwidth]{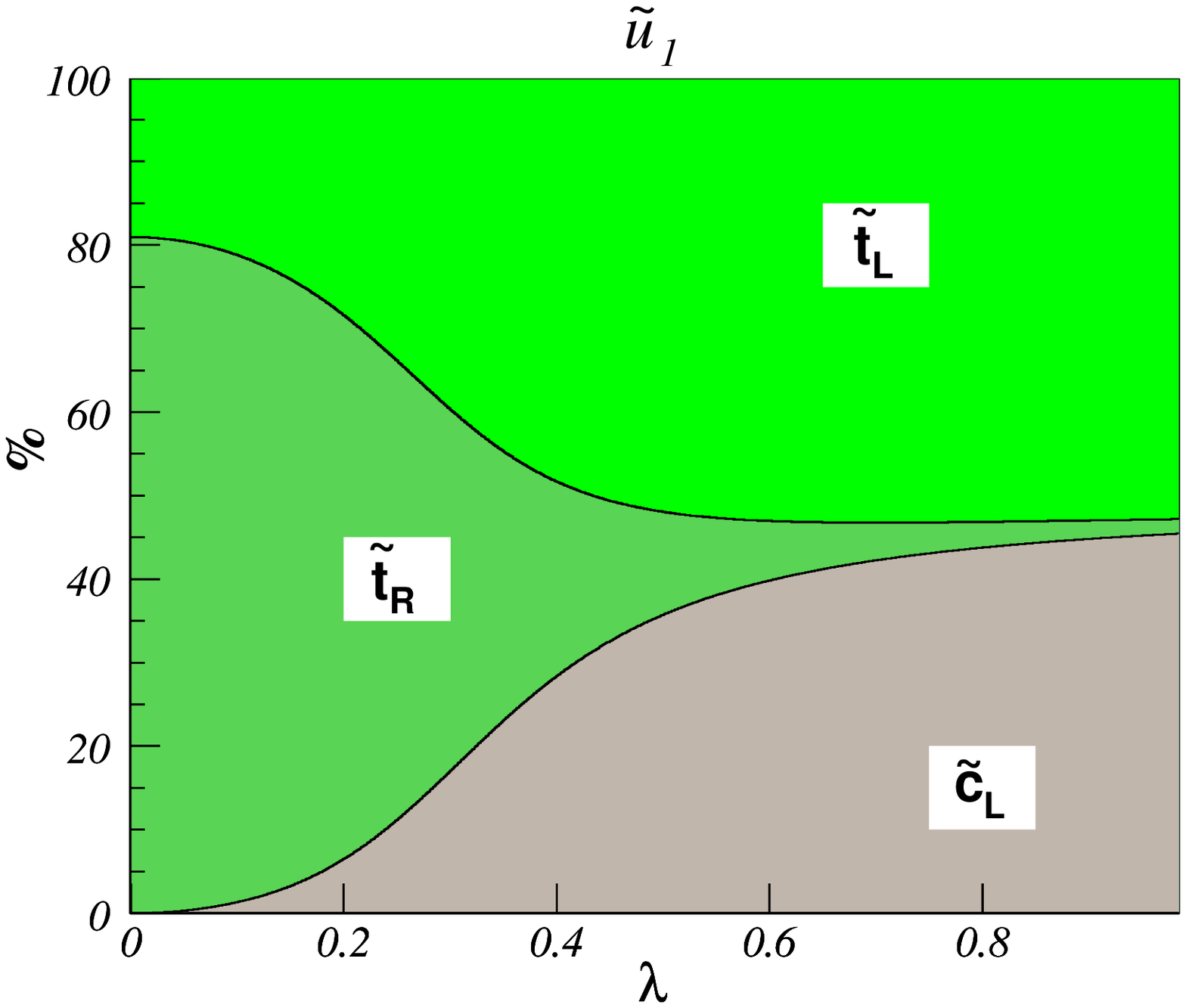}\hspace{2mm}
 \includegraphics[width=0.21\columnwidth]{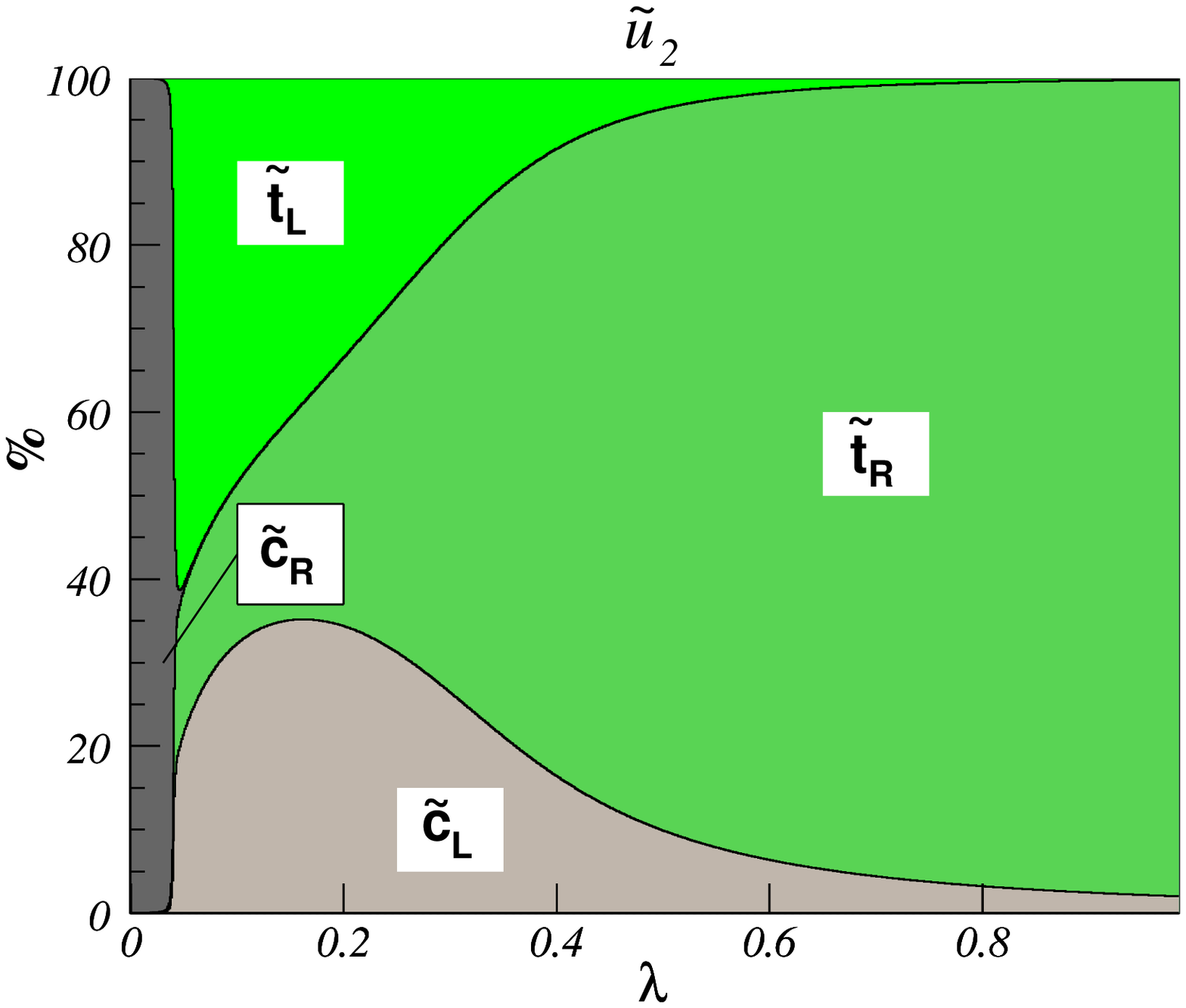}\hspace{2mm}
 \includegraphics[width=0.21\columnwidth]{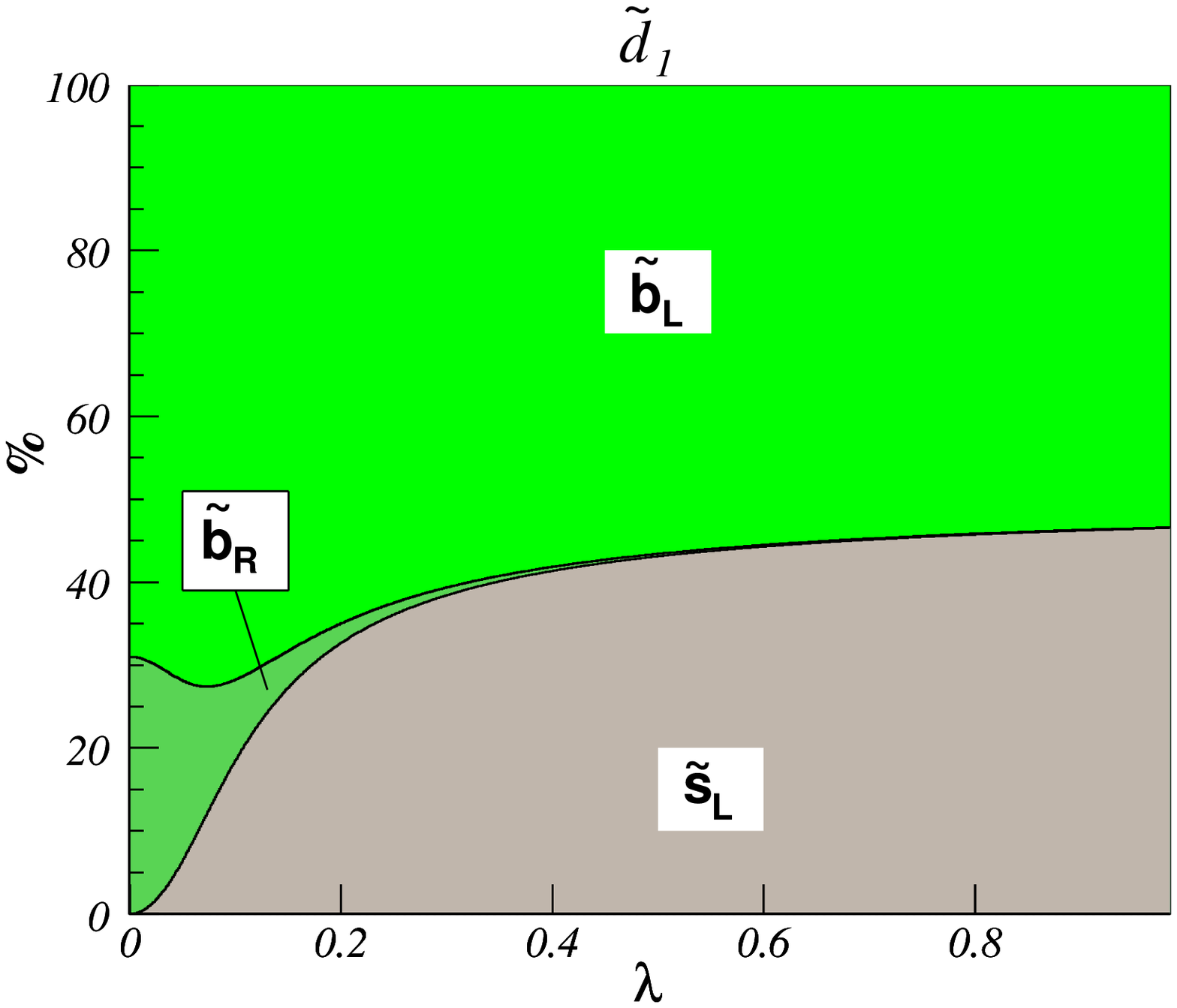}\hspace{2mm}
 \includegraphics[width=0.21\columnwidth]{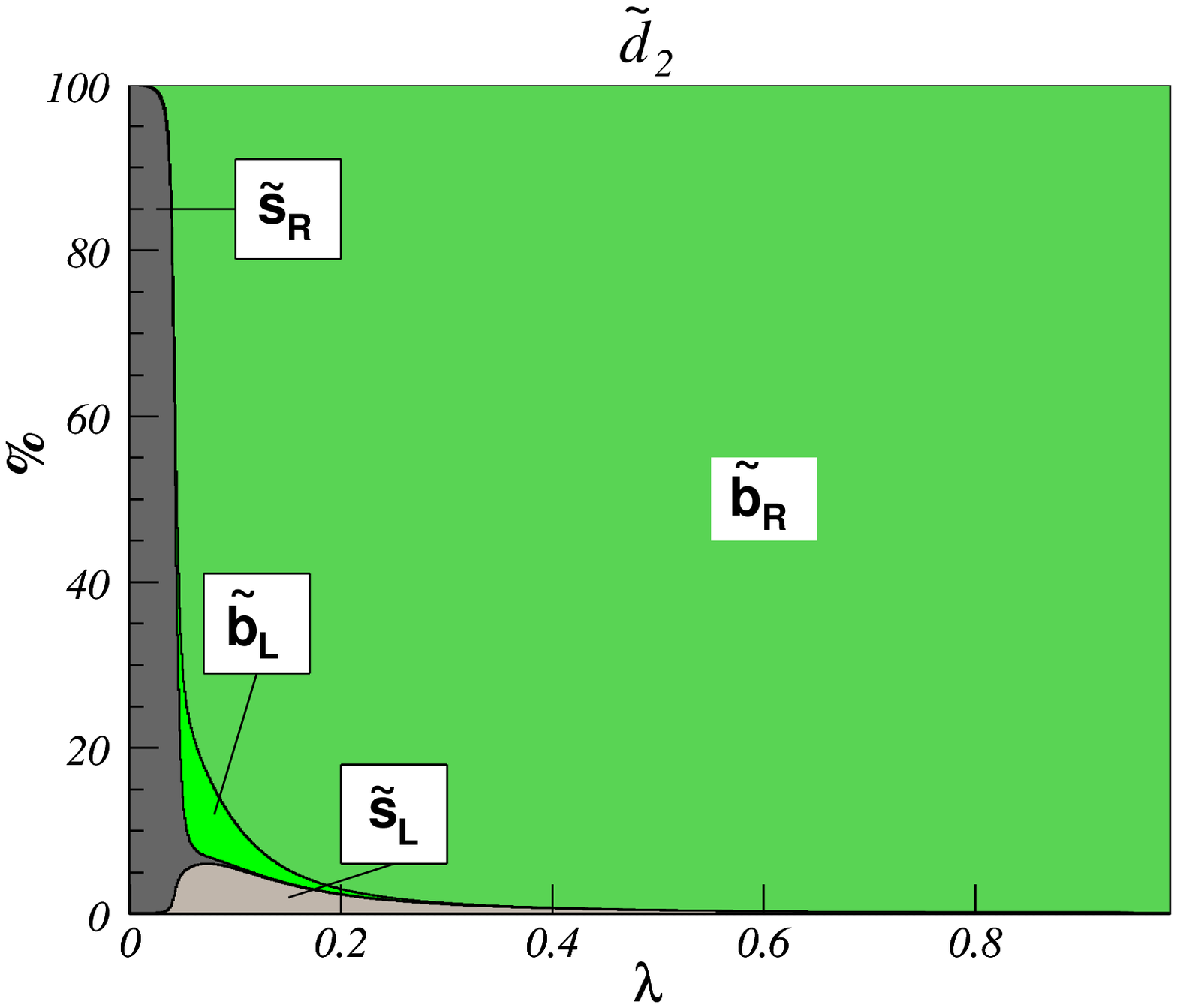}\vspace*{4mm}
 \includegraphics[width=0.21\columnwidth]{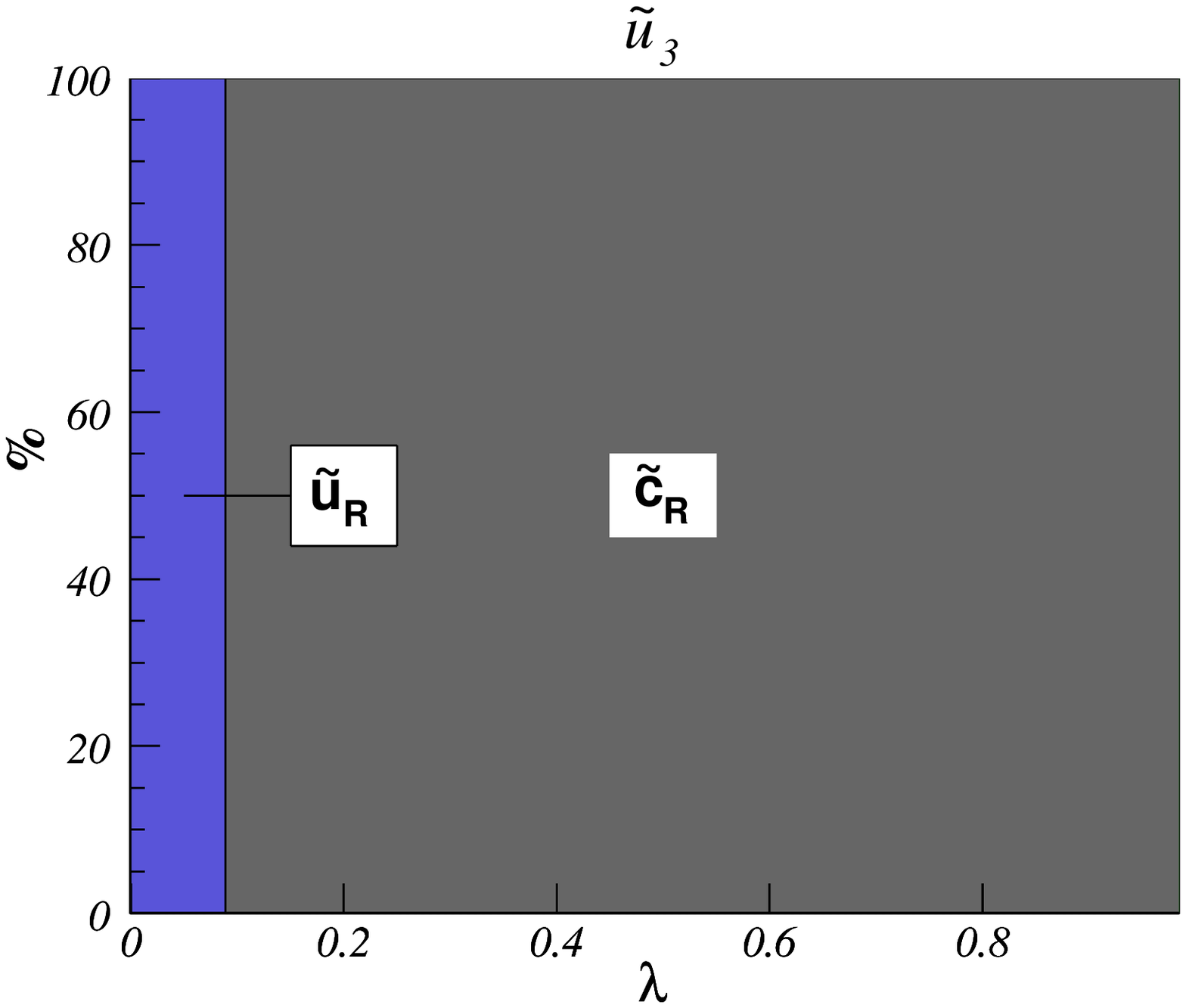}\hspace{2mm}
 \includegraphics[width=0.21\columnwidth]{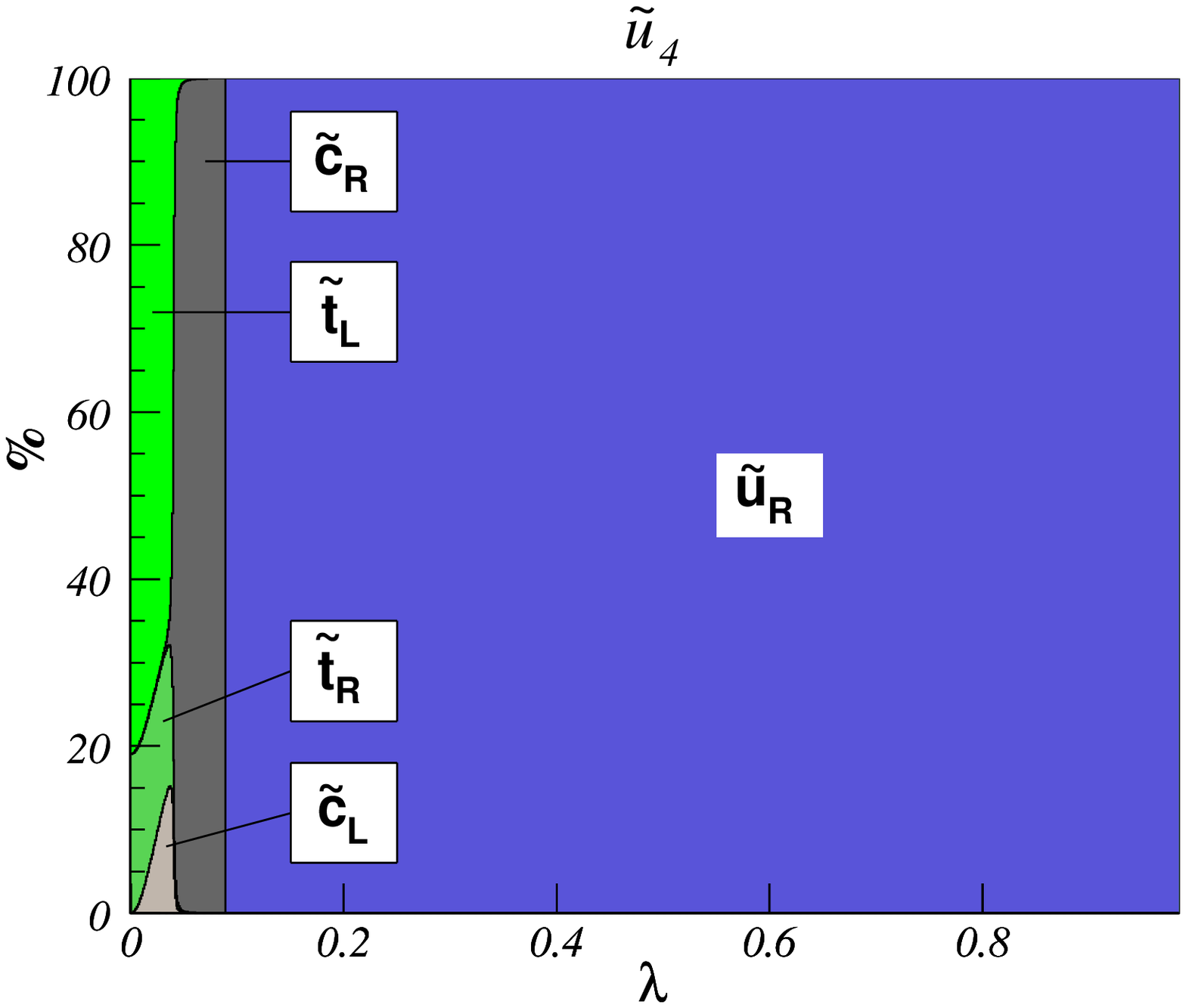}\hspace{2mm}
 \includegraphics[width=0.21\columnwidth]{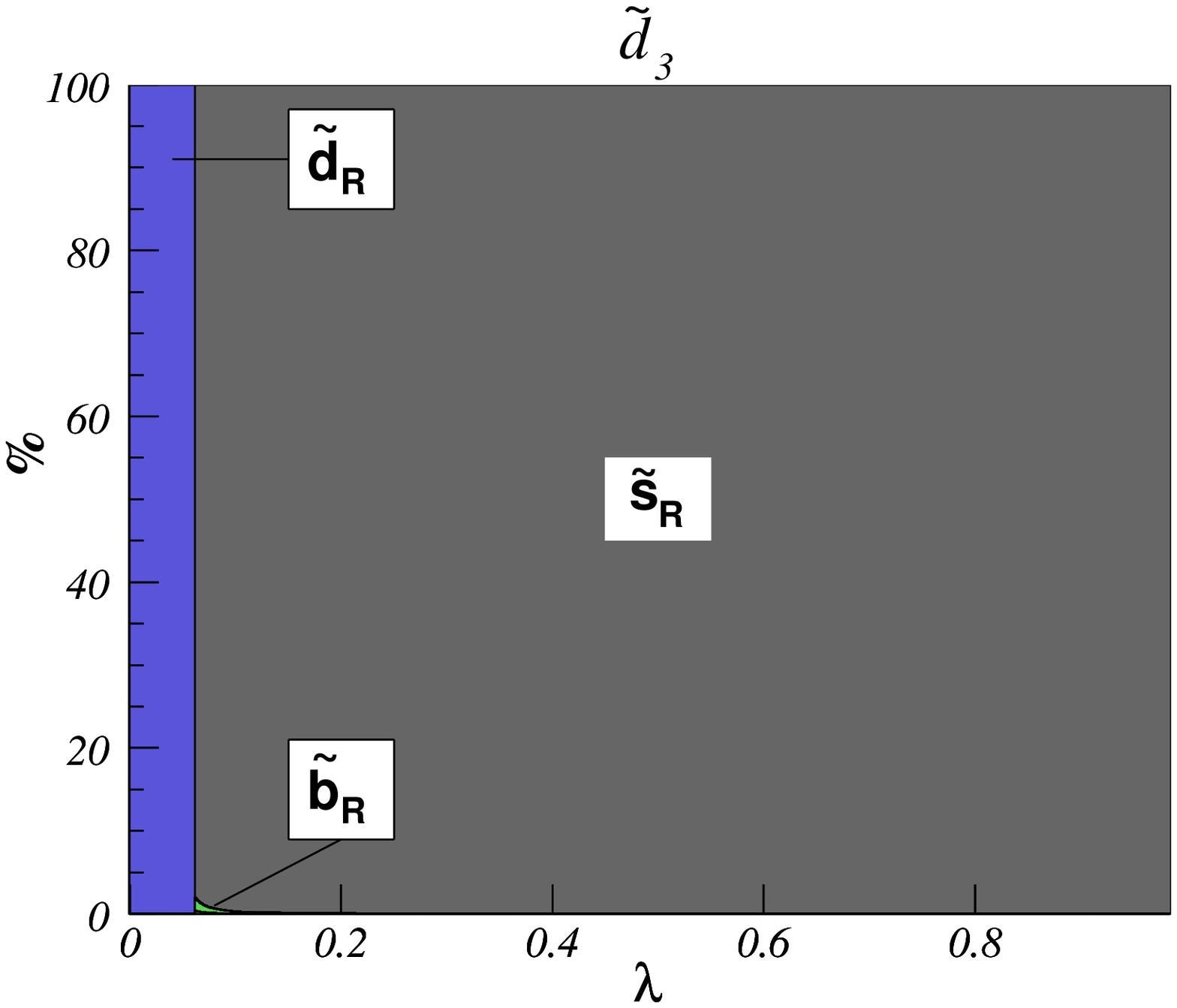}\hspace{2mm}
 \includegraphics[width=0.21\columnwidth]{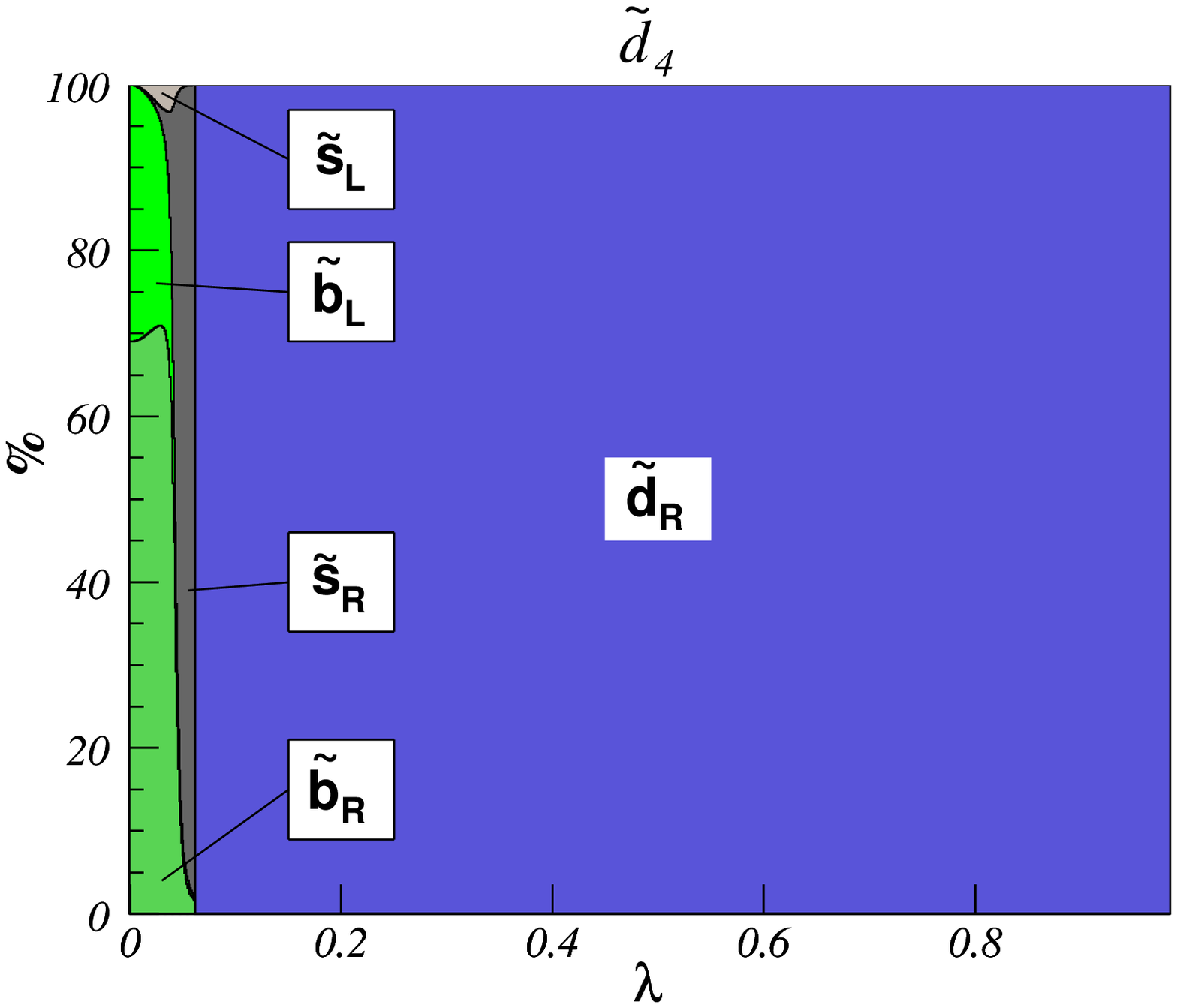}\vspace*{4mm}
 \includegraphics[width=0.21\columnwidth]{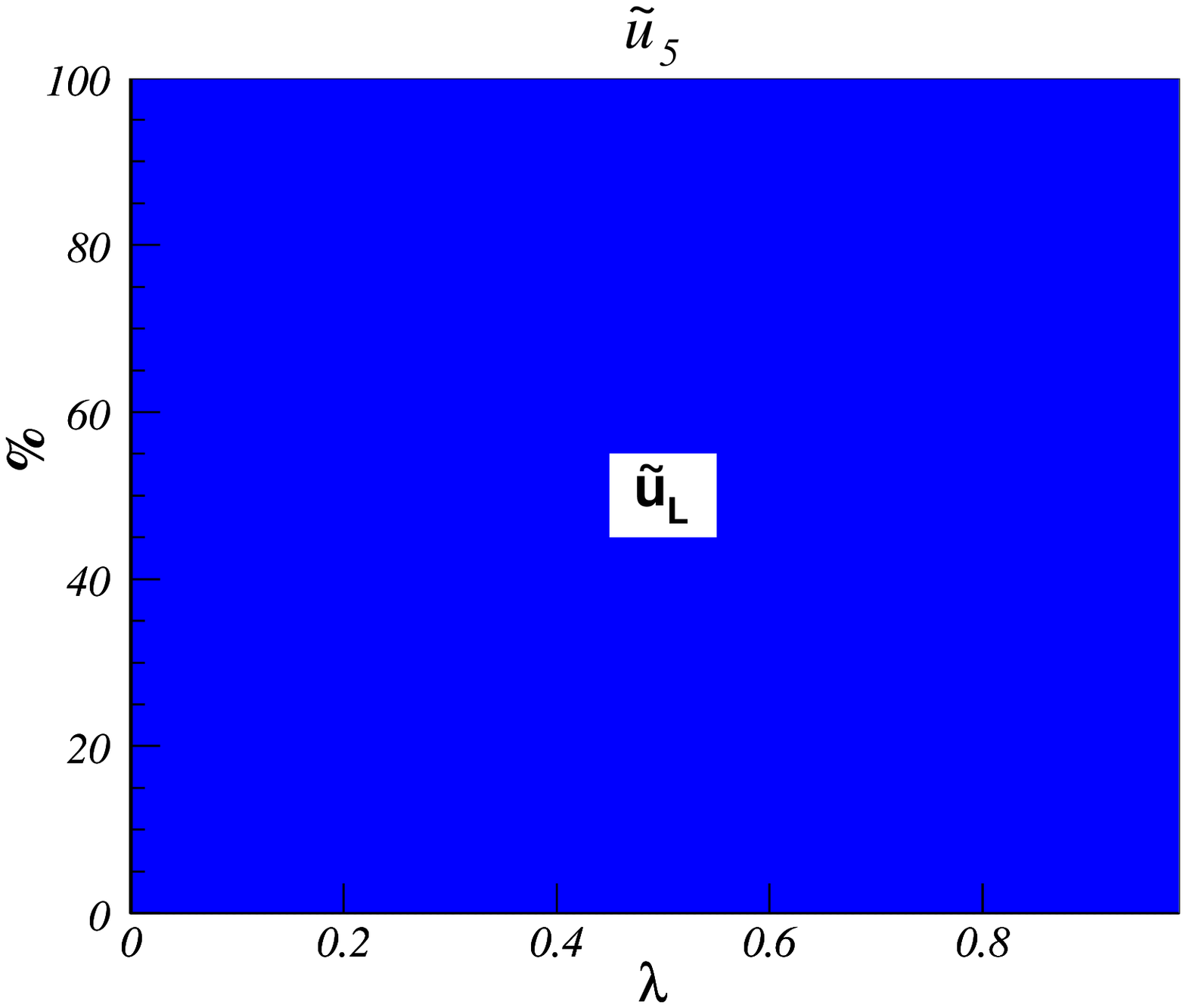}\hspace{2mm}
 \includegraphics[width=0.21\columnwidth]{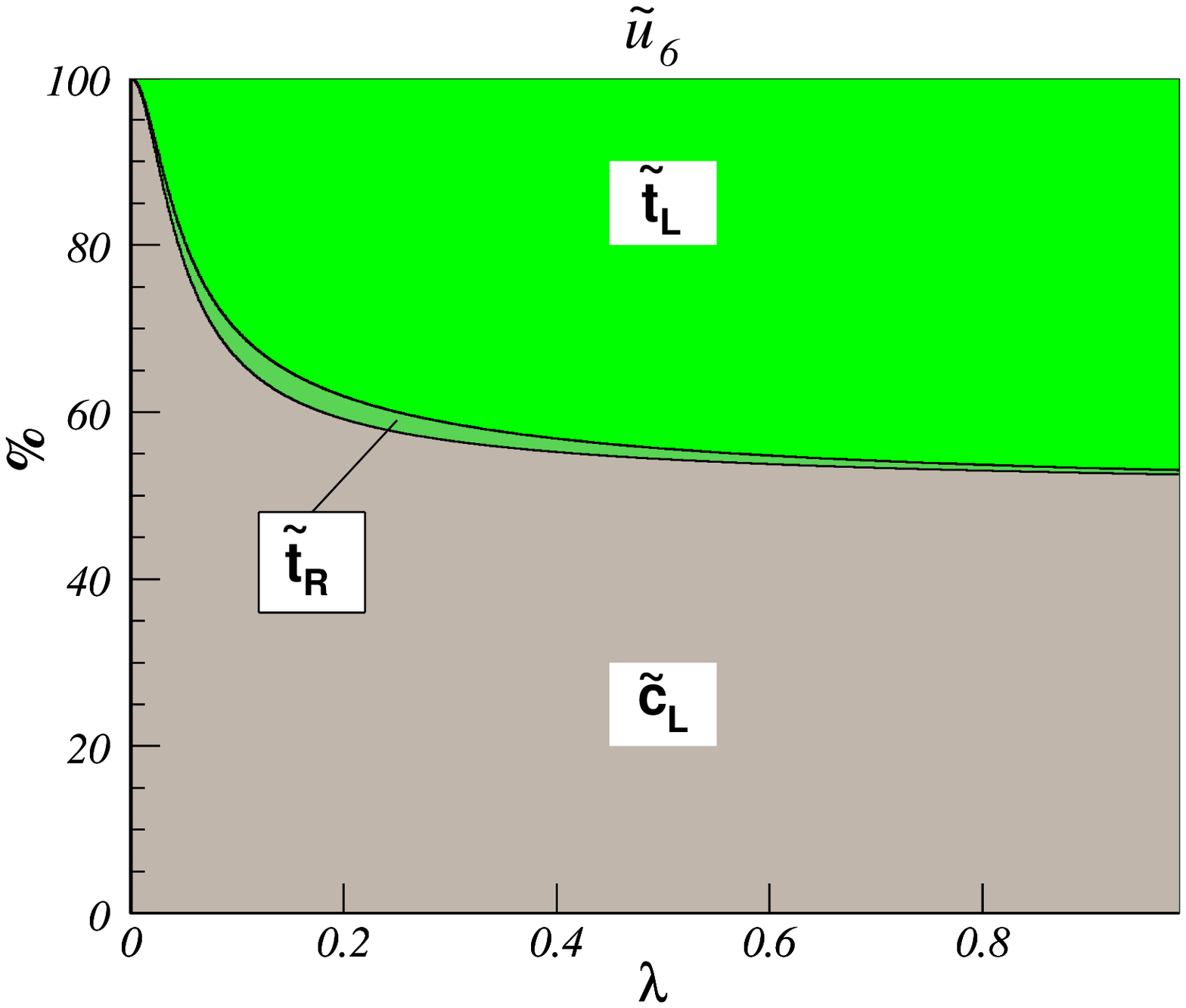}\hspace{2mm}
 \includegraphics[width=0.21\columnwidth]{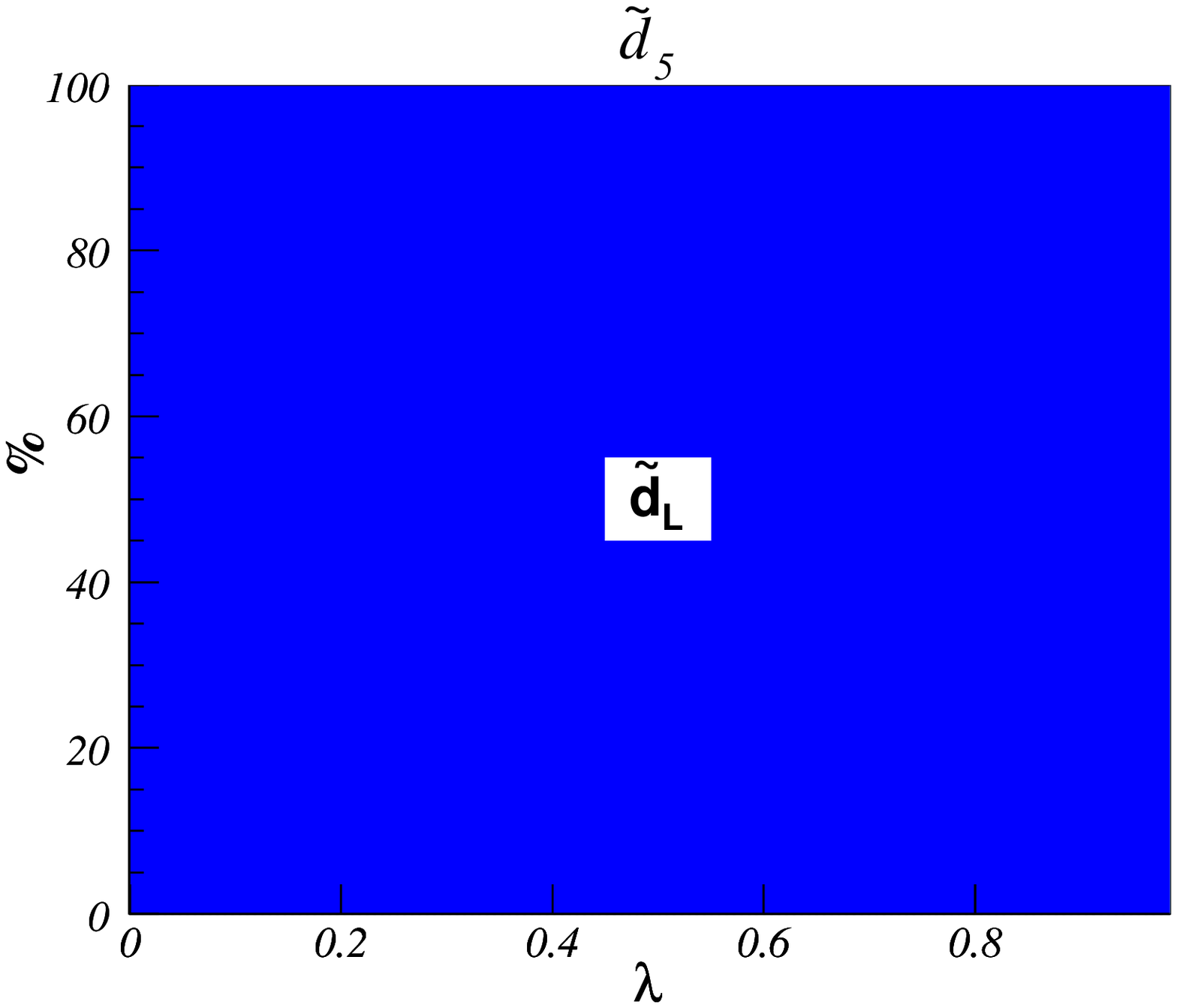}\hspace{2mm}
 \includegraphics[width=0.21\columnwidth]{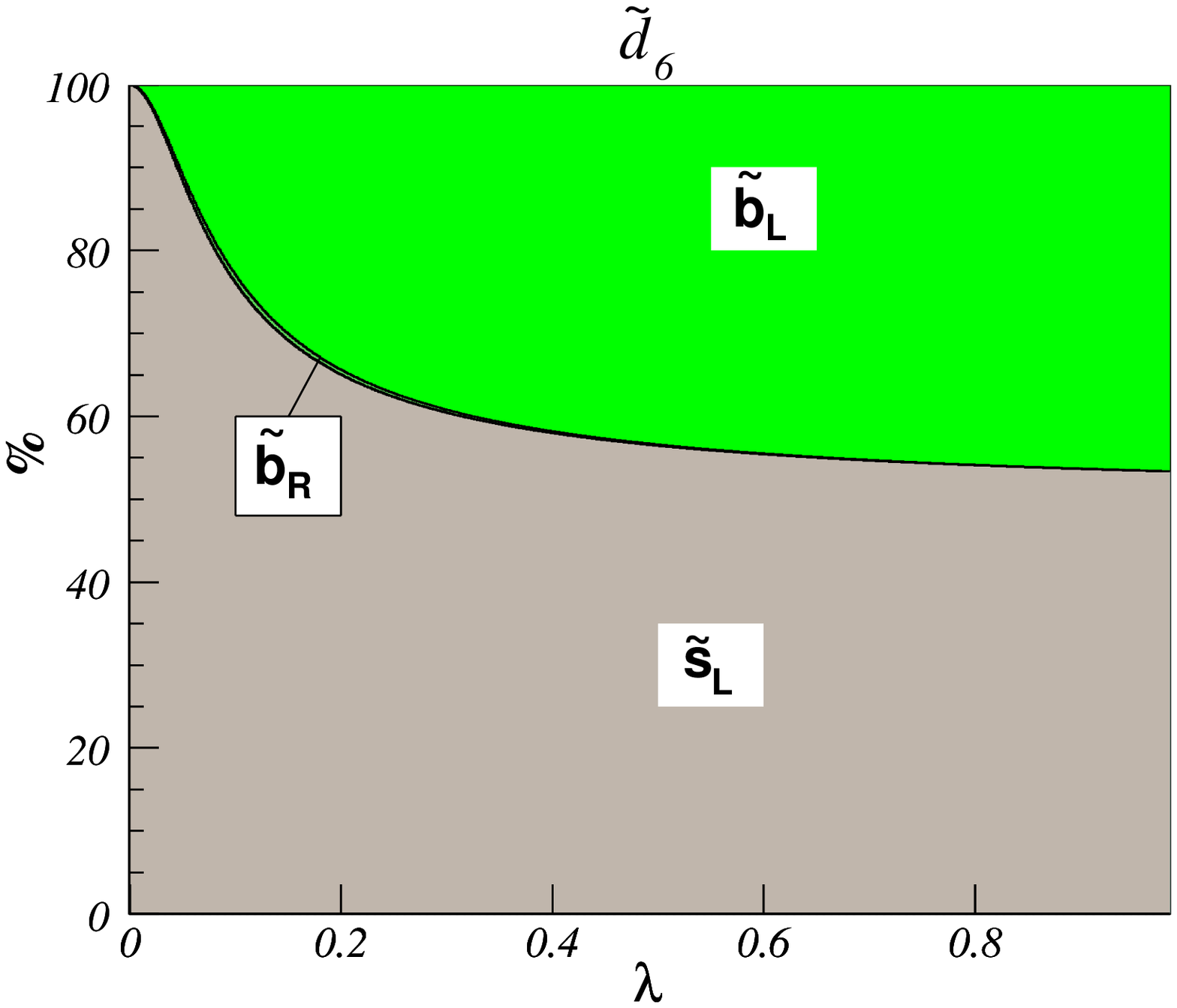}
 \caption{\label{fig:18}Same as Fig.\ \ref{fig:16} for benchmark point C.}
\end{figure}
%
%
\begin{figure}
 \centering
 \includegraphics[width=0.21\columnwidth]{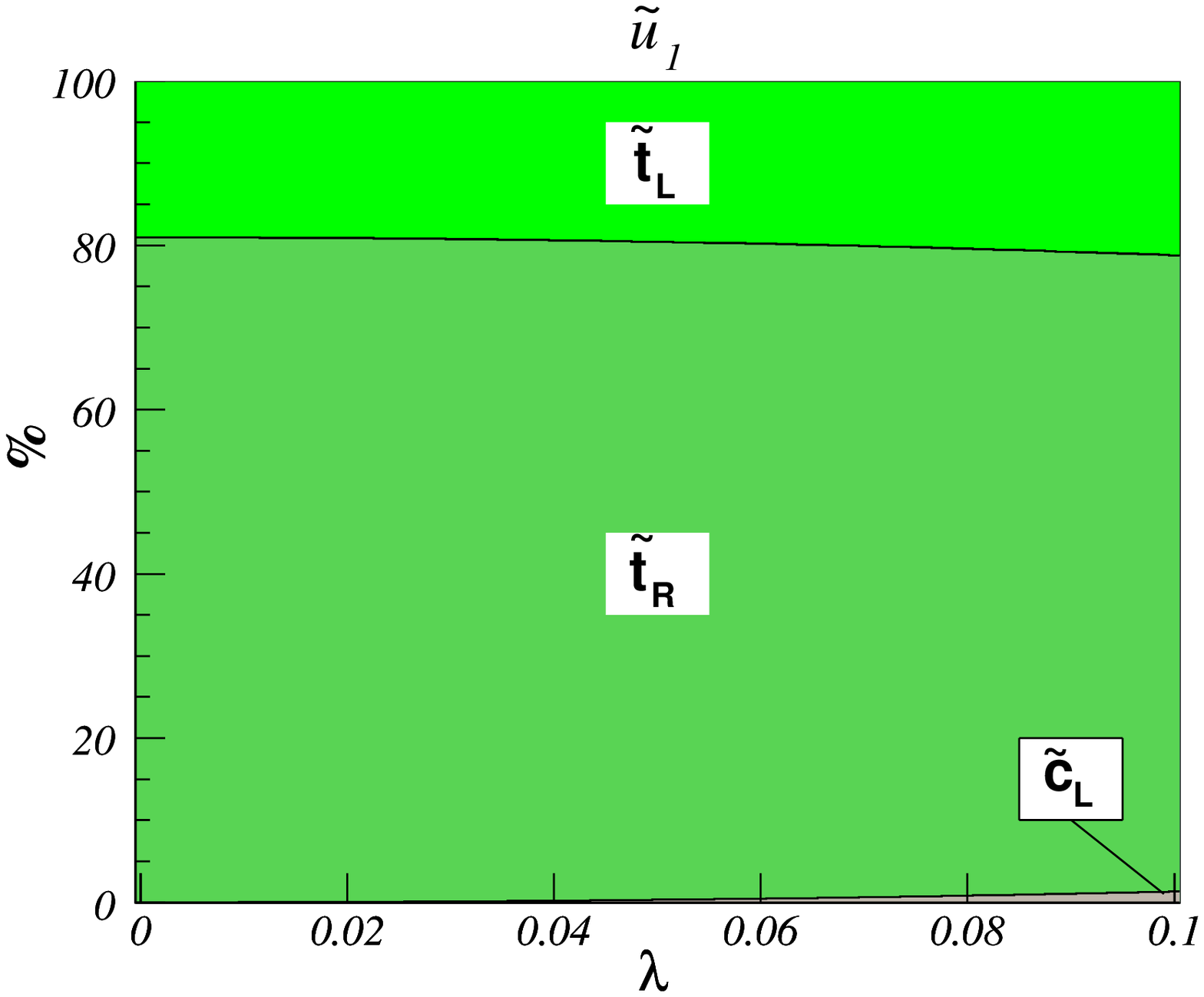}\hspace{2mm}
 \includegraphics[width=0.21\columnwidth]{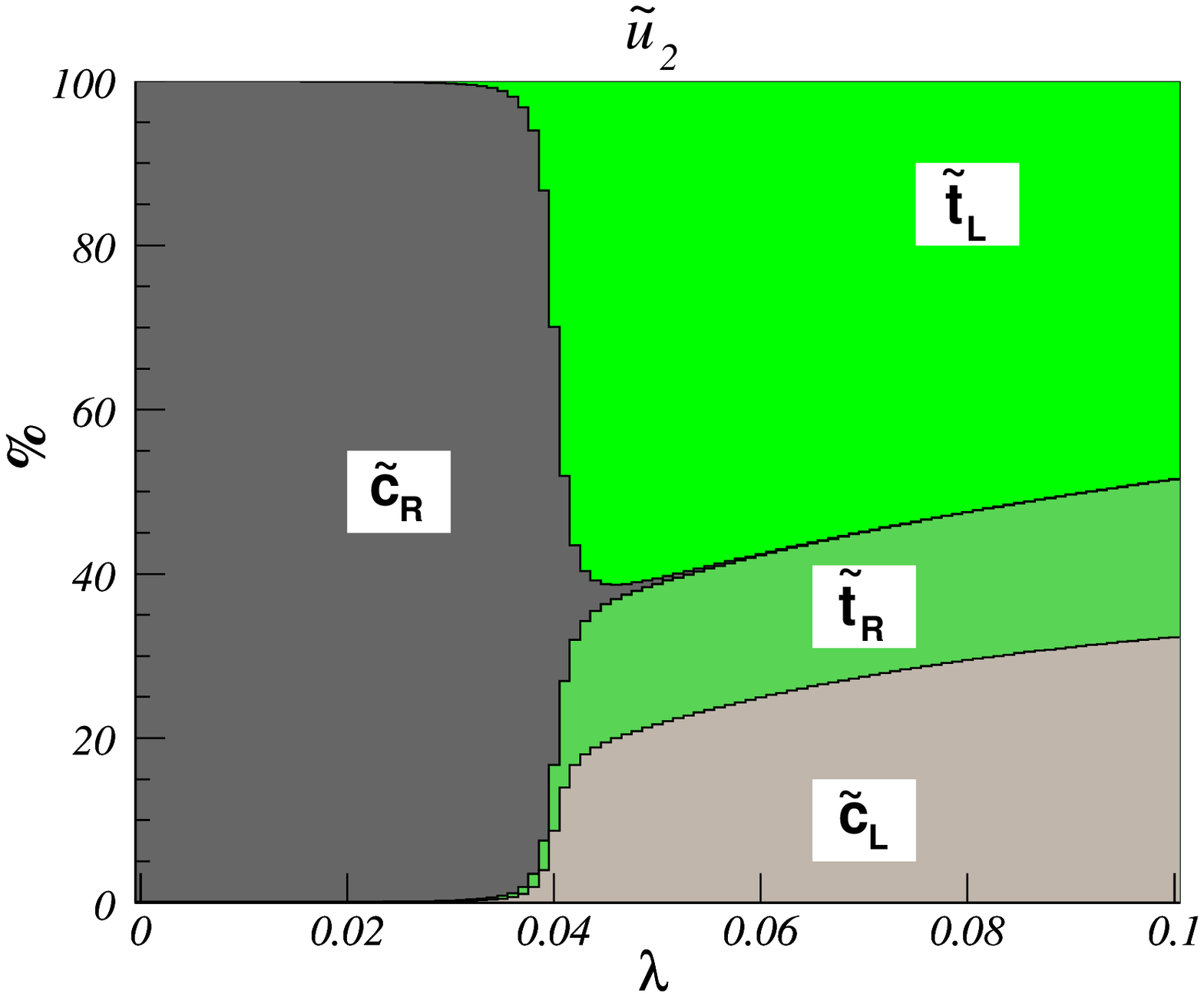}\hspace{2mm}
 \includegraphics[width=0.21\columnwidth]{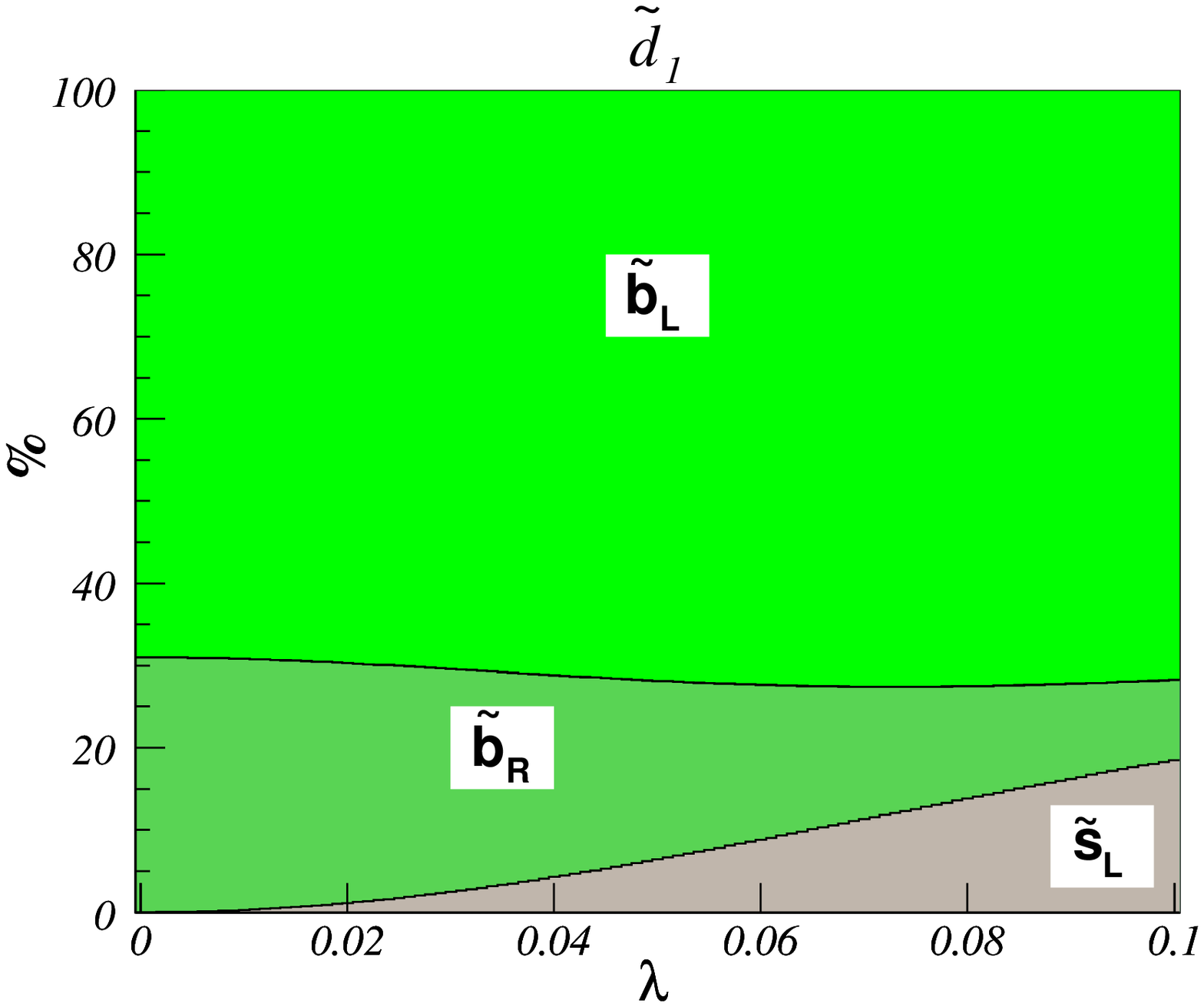}\hspace{2mm}
 \includegraphics[width=0.21\columnwidth]{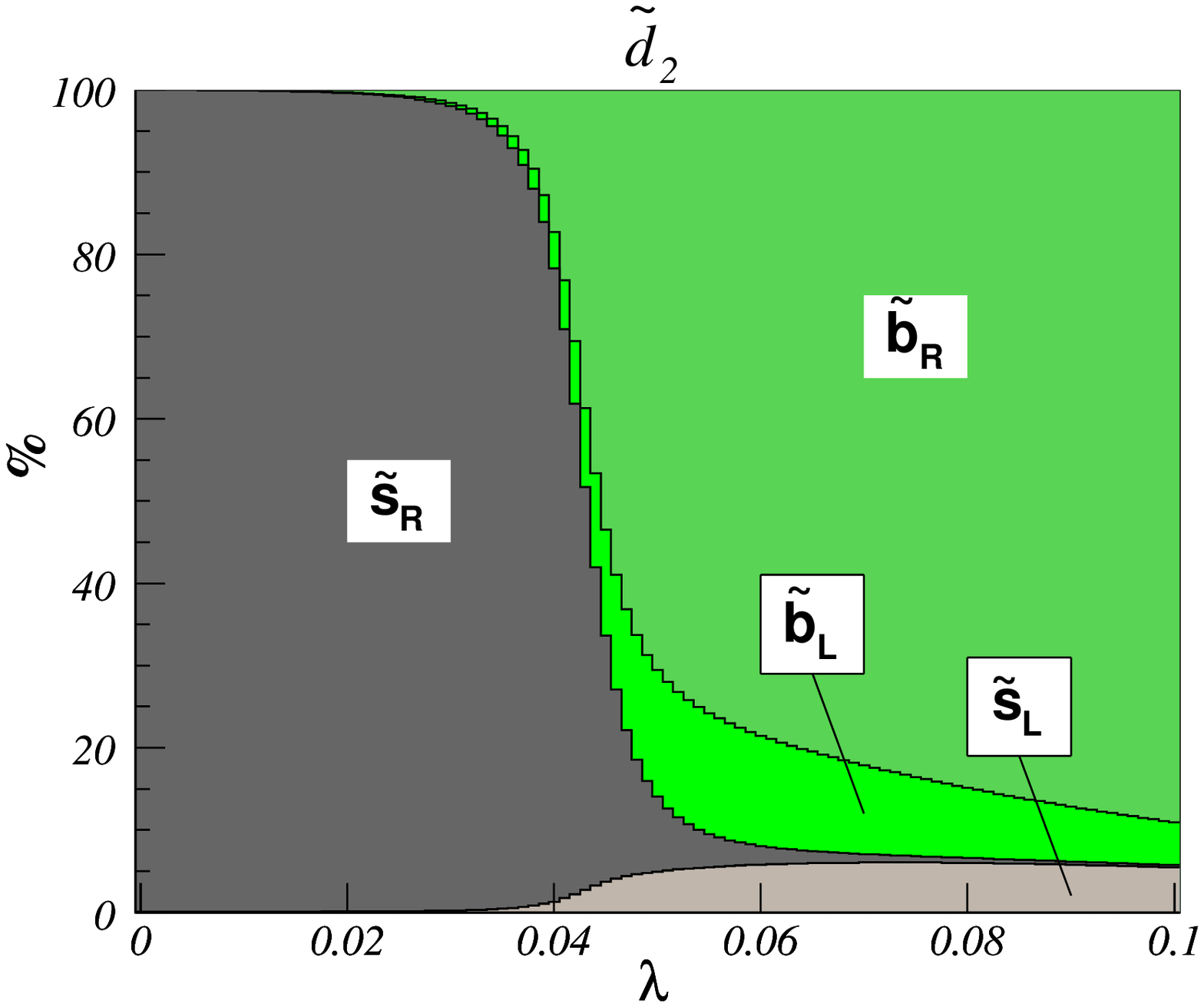}\vspace*{4mm}
 \includegraphics[width=0.21\columnwidth]{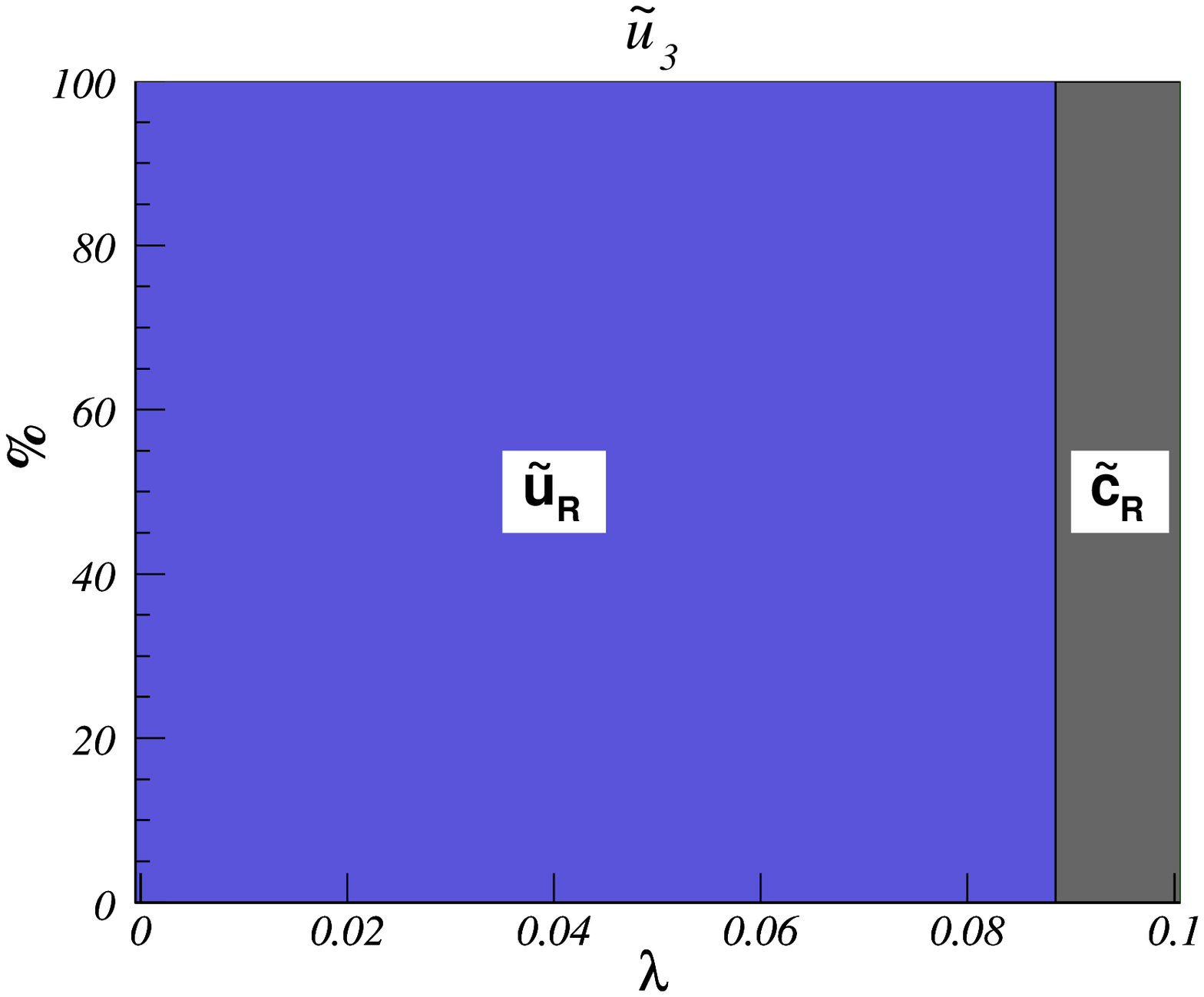}\hspace{2mm}
 \includegraphics[width=0.21\columnwidth]{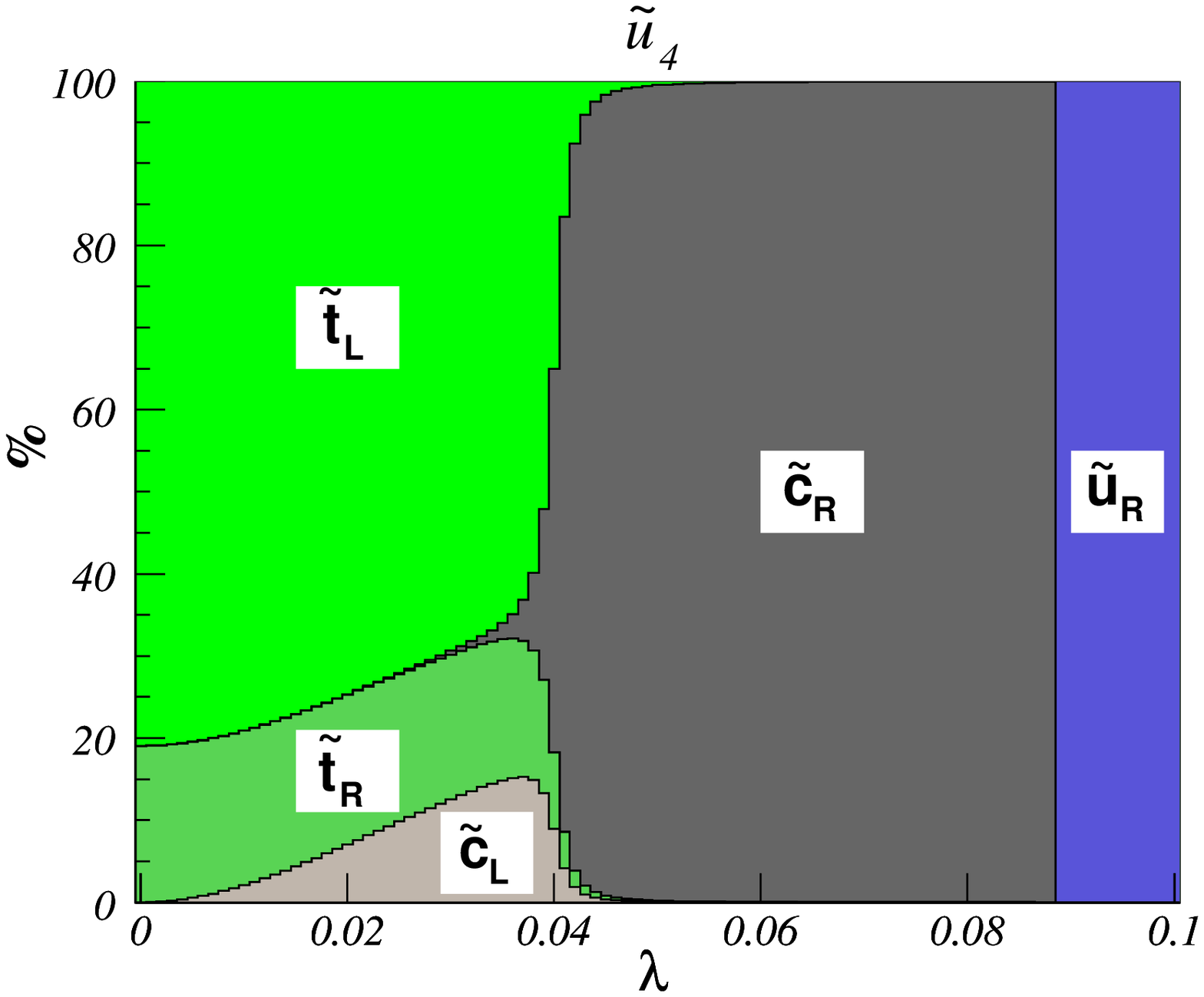}\hspace{2mm}
 \includegraphics[width=0.21\columnwidth]{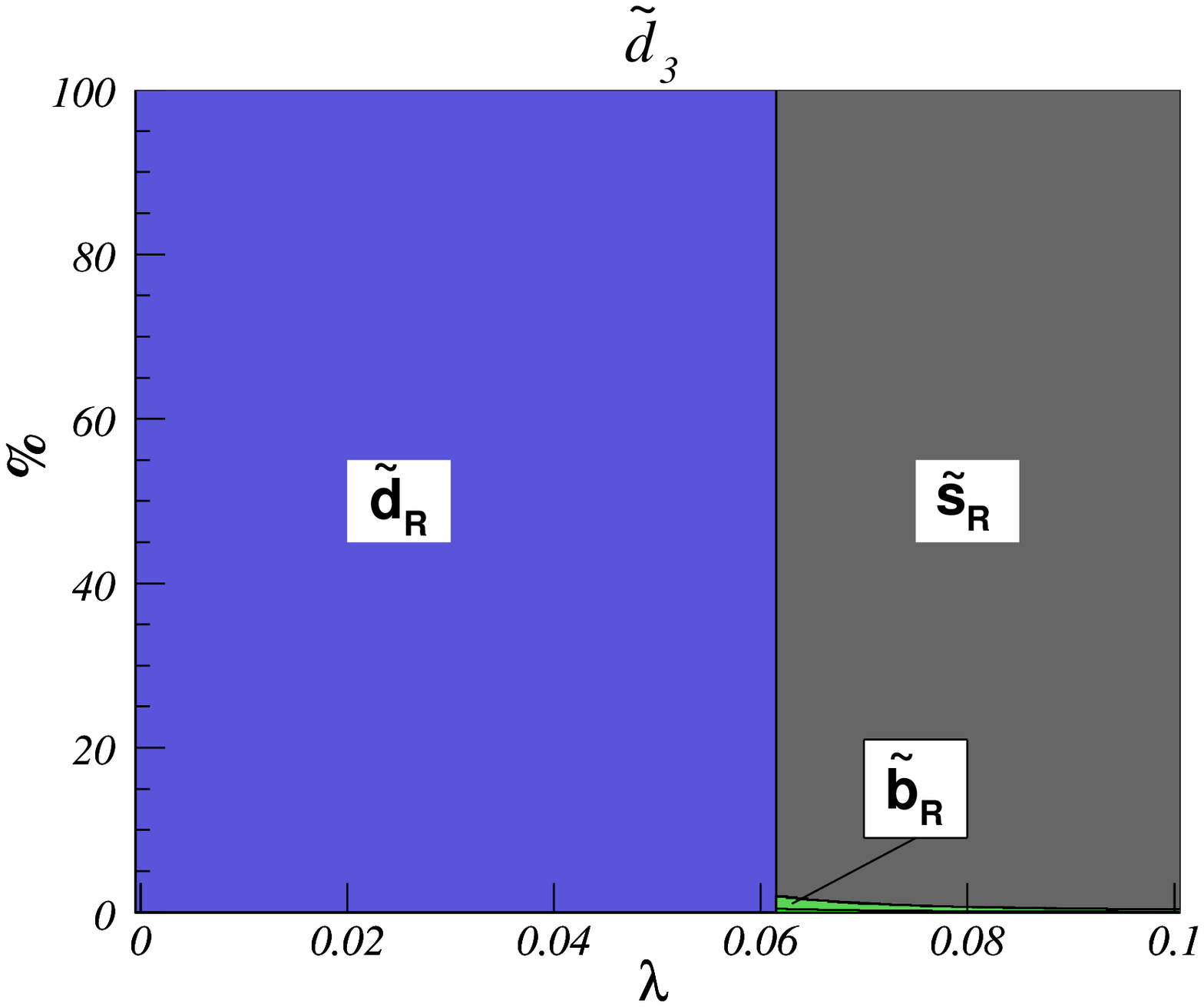}\hspace{2mm}
 \includegraphics[width=0.21\columnwidth]{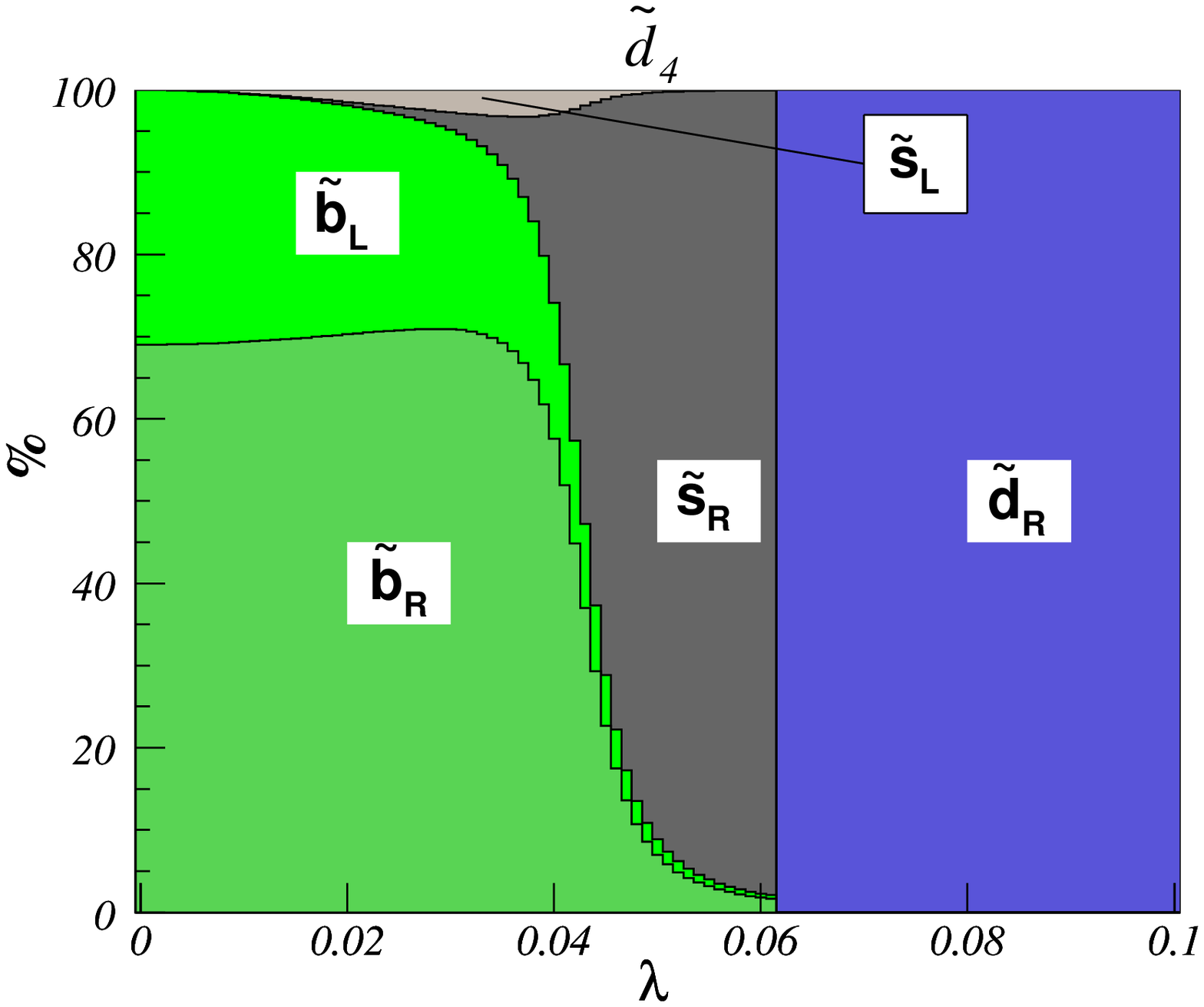}\vspace*{4mm}
 \includegraphics[width=0.21\columnwidth]{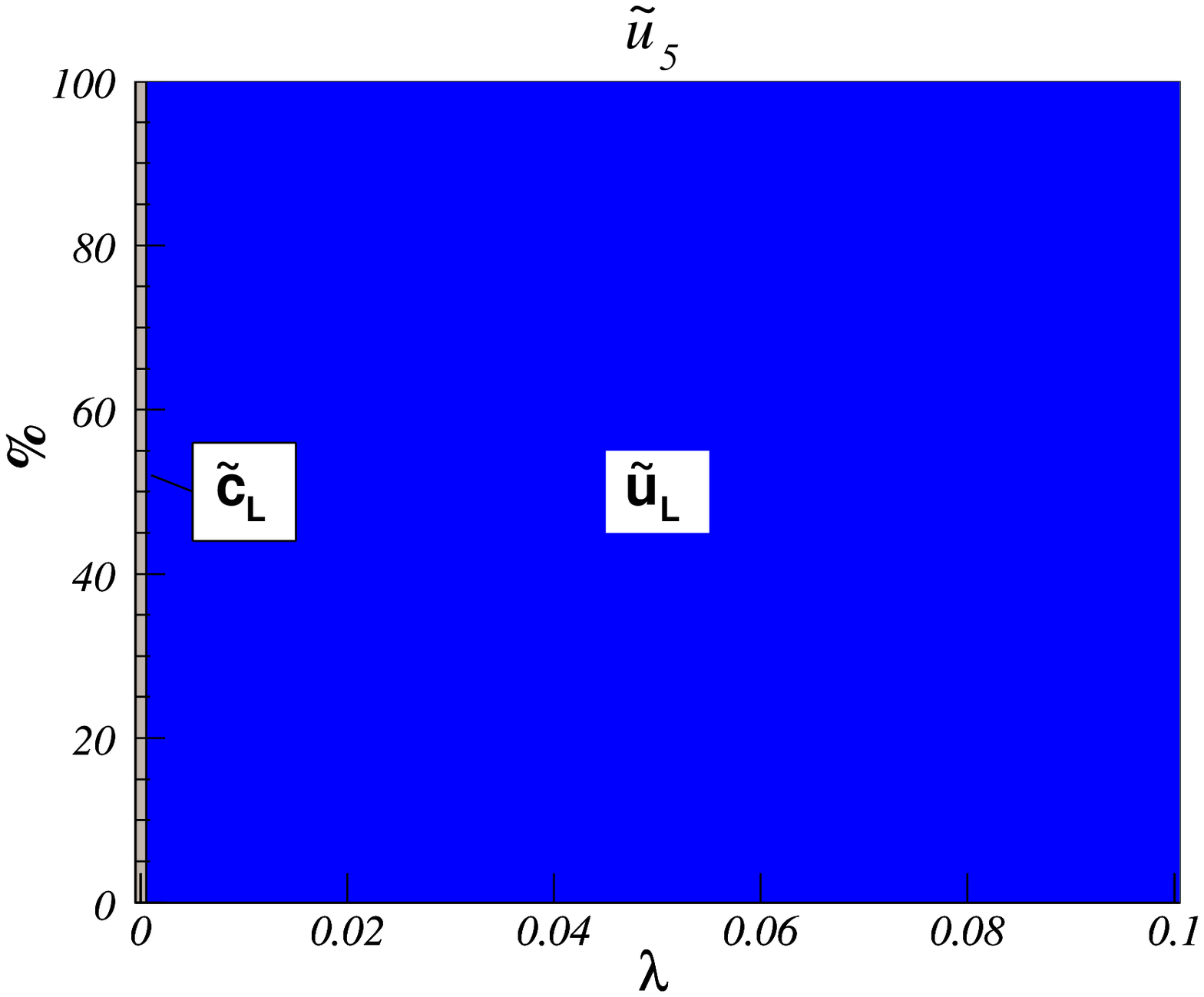}\hspace{2mm}
 \includegraphics[width=0.21\columnwidth]{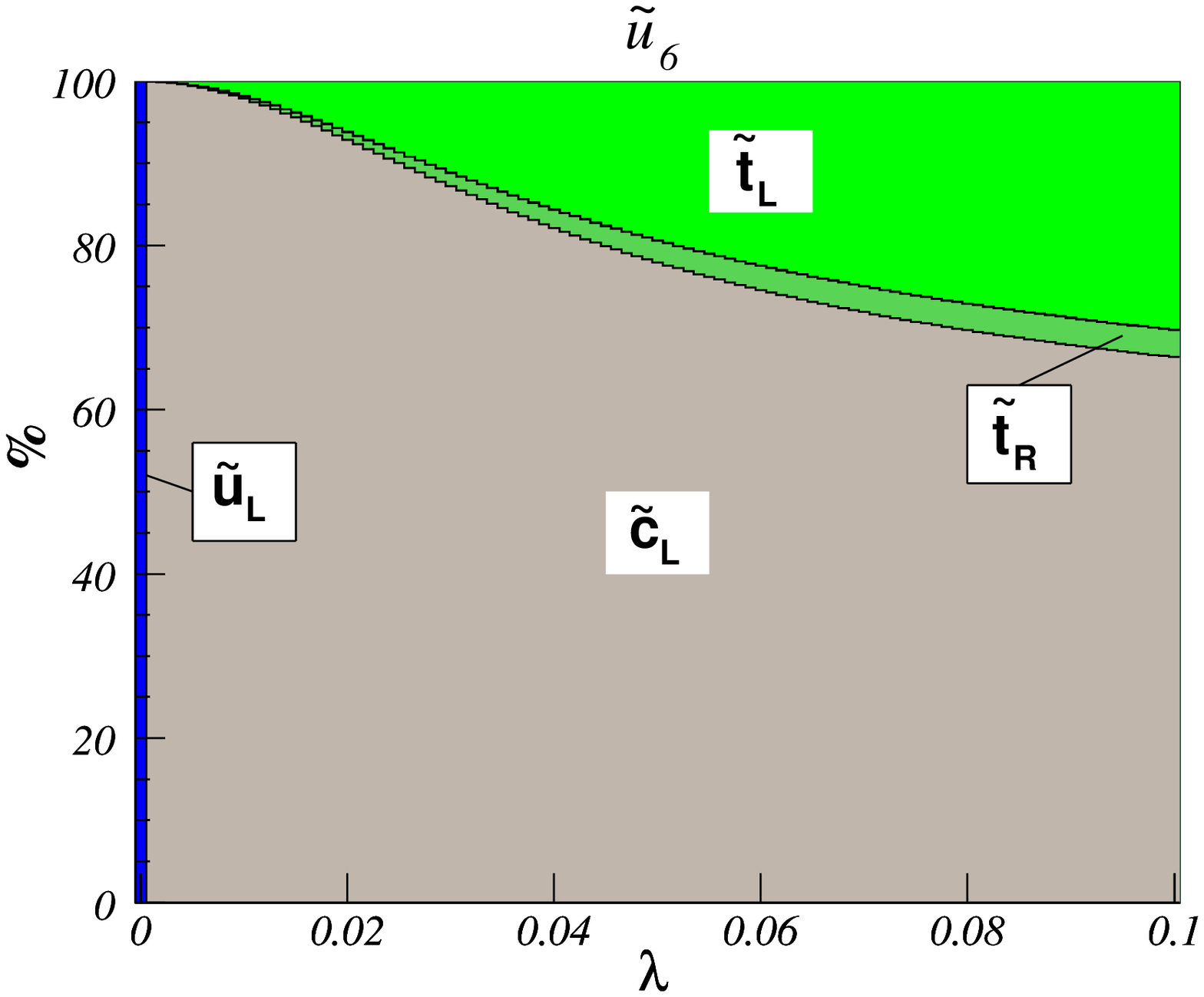}\hspace{2mm}
 \includegraphics[width=0.21\columnwidth]{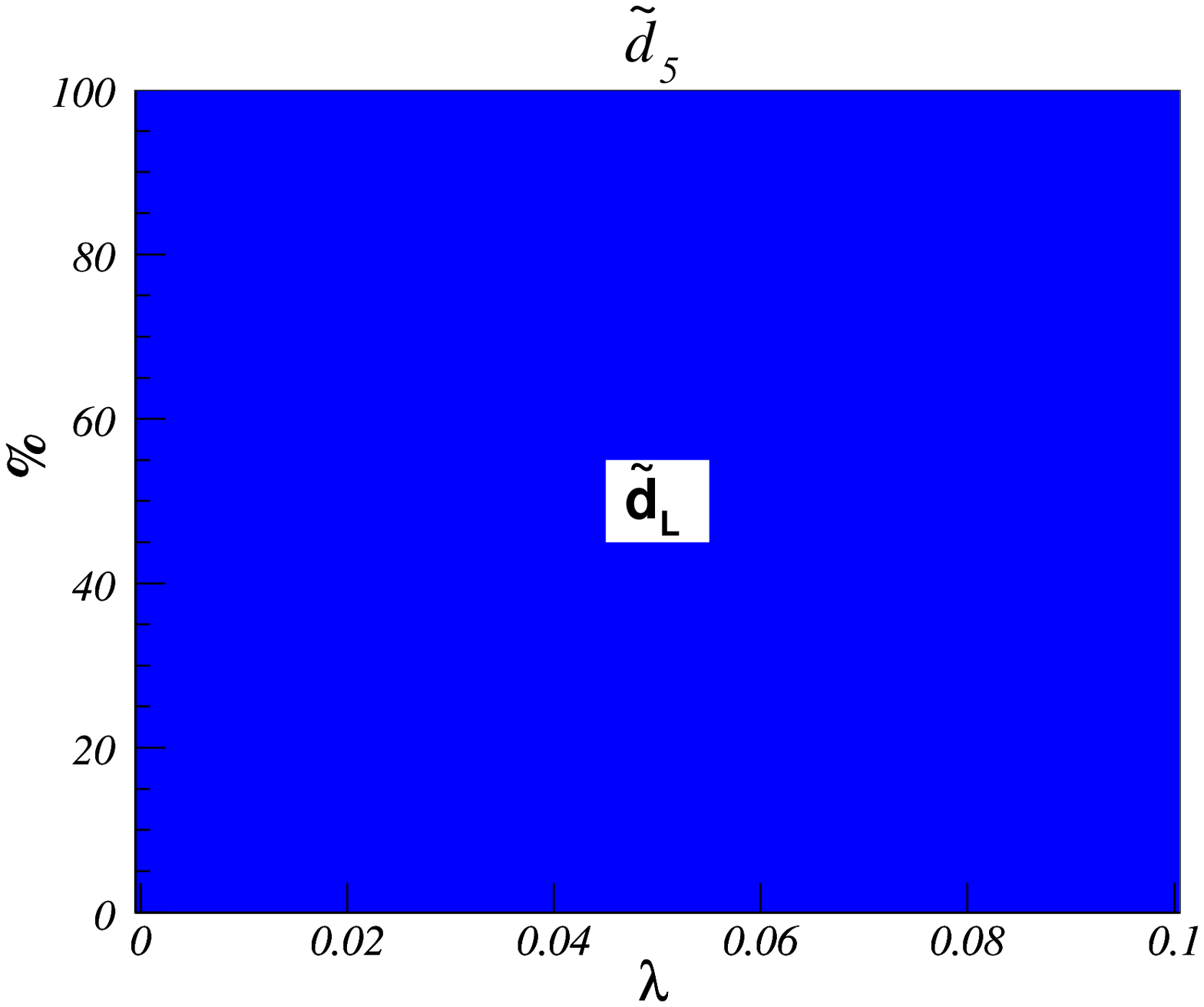}\hspace{2mm}
 \includegraphics[width=0.21\columnwidth]{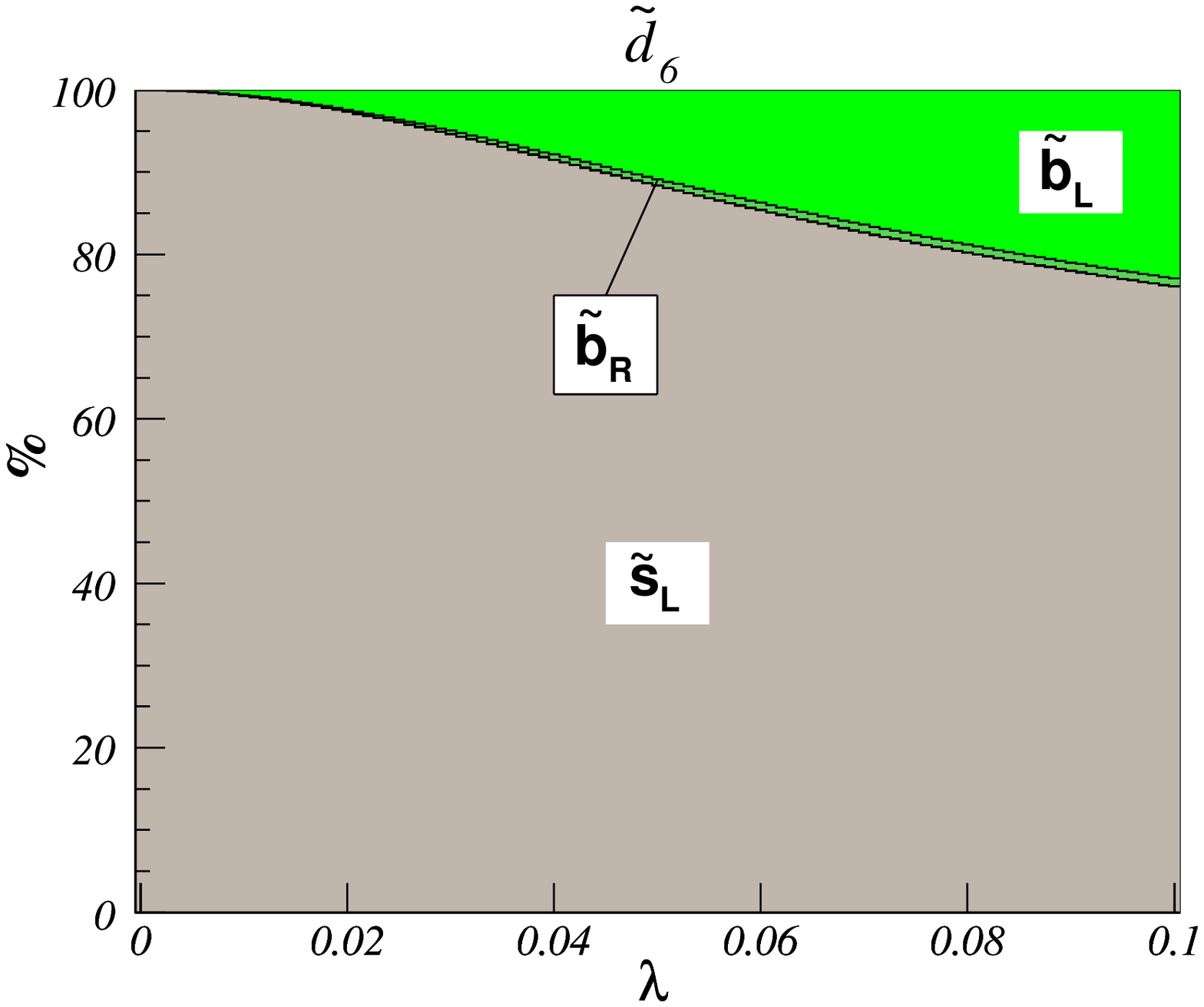}
 \caption{\label{fig:18p}Same as Fig.\ \ref{fig:18} for $\lambda\in
          [0;0.1]$.}
\end{figure}
%
%
\begin{figure}
 \centering
 \includegraphics[width=0.21\columnwidth]{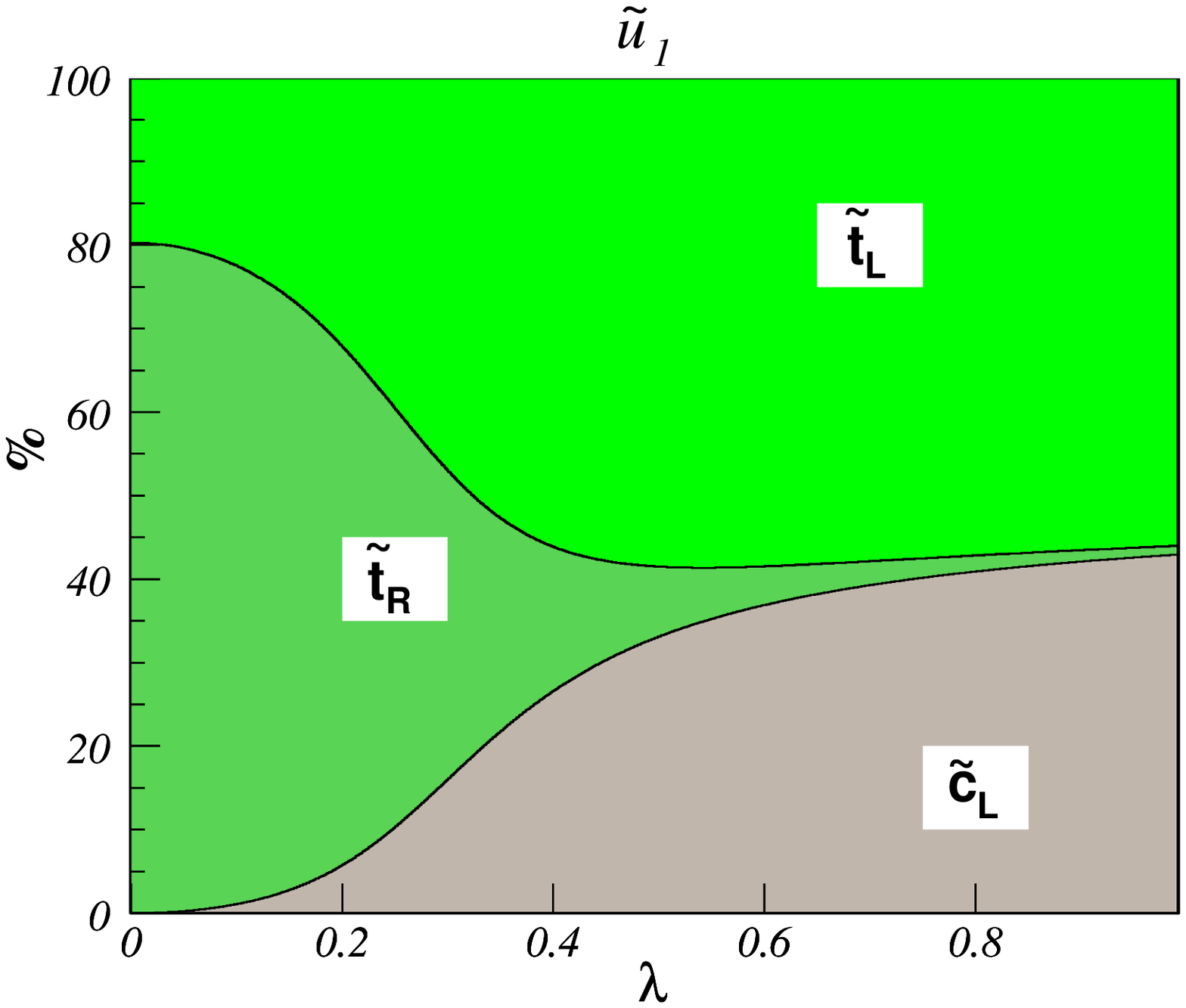}\hspace{2mm}
 \includegraphics[width=0.21\columnwidth]{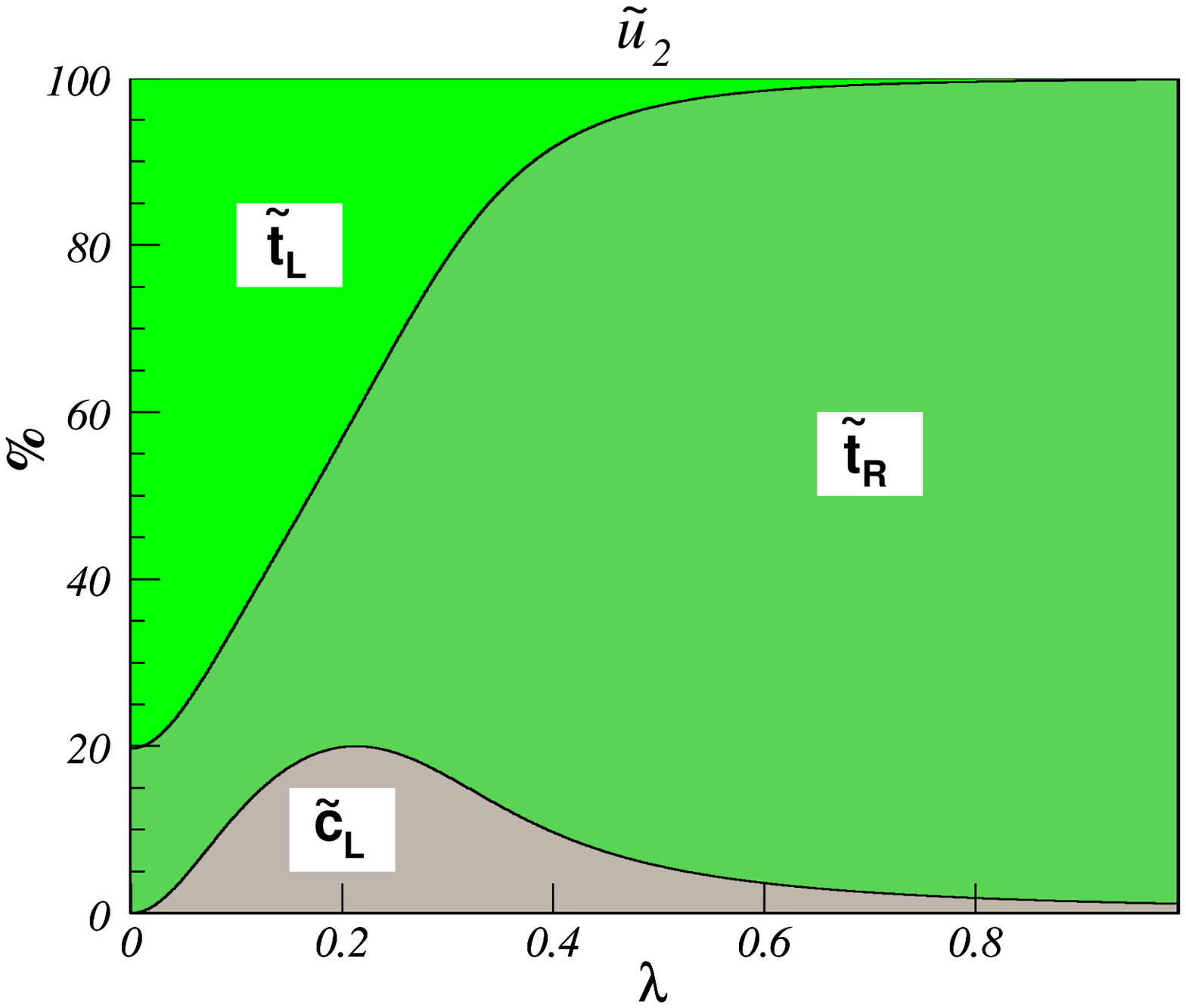}\hspace{2mm}
 \includegraphics[width=0.21\columnwidth]{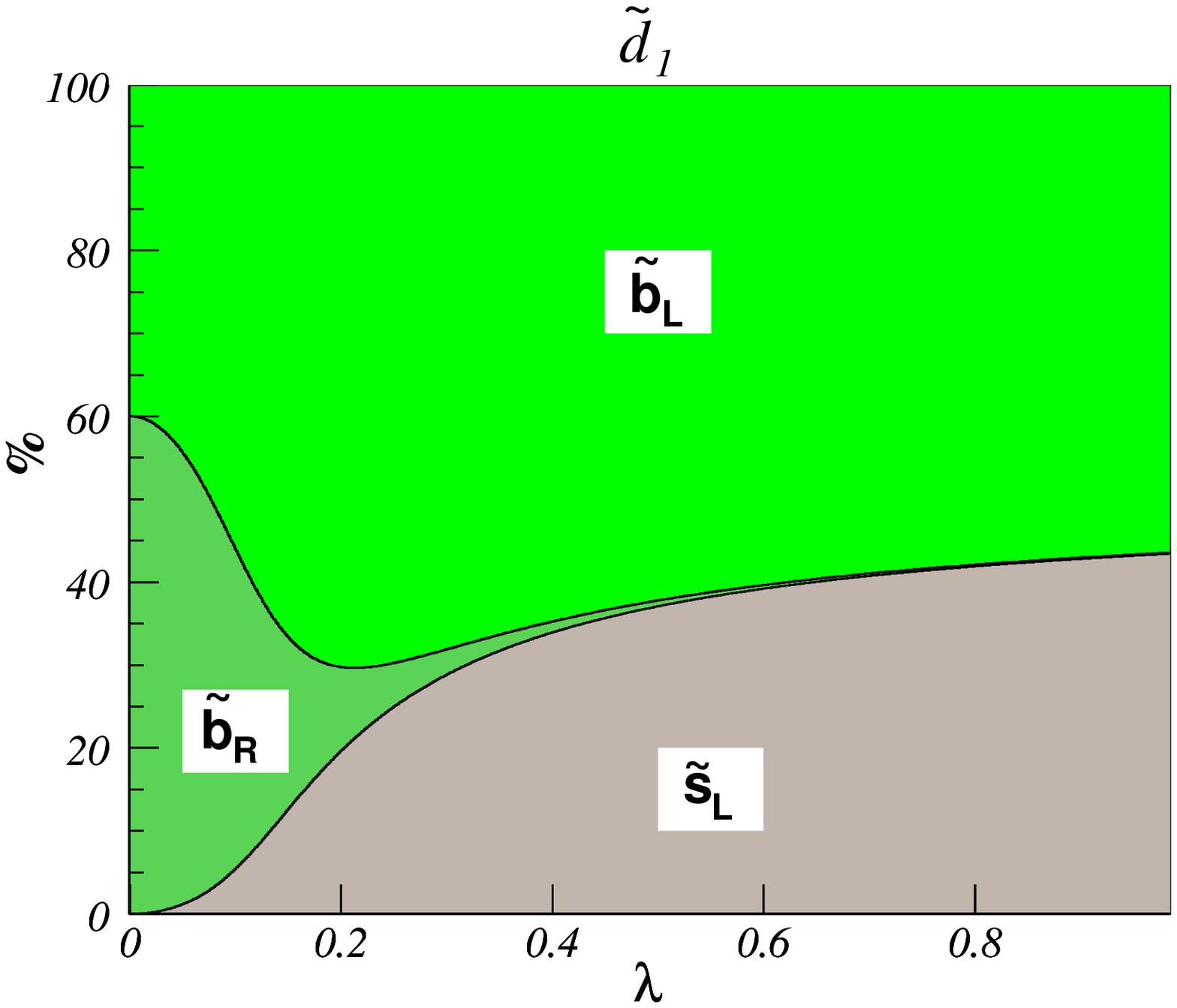}\hspace{2mm}
 \includegraphics[width=0.21\columnwidth]{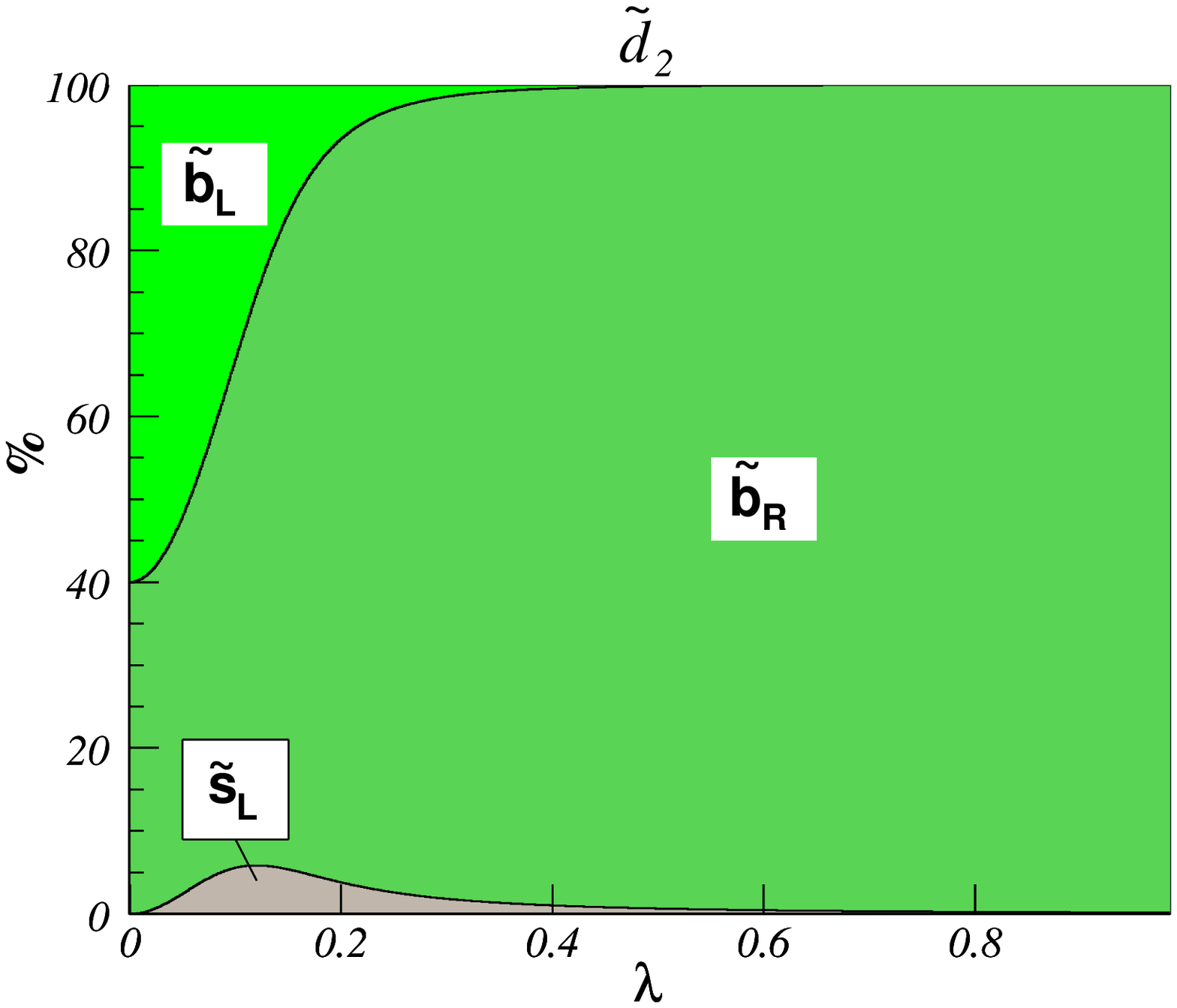}\vspace*{4mm}
 \includegraphics[width=0.21\columnwidth]{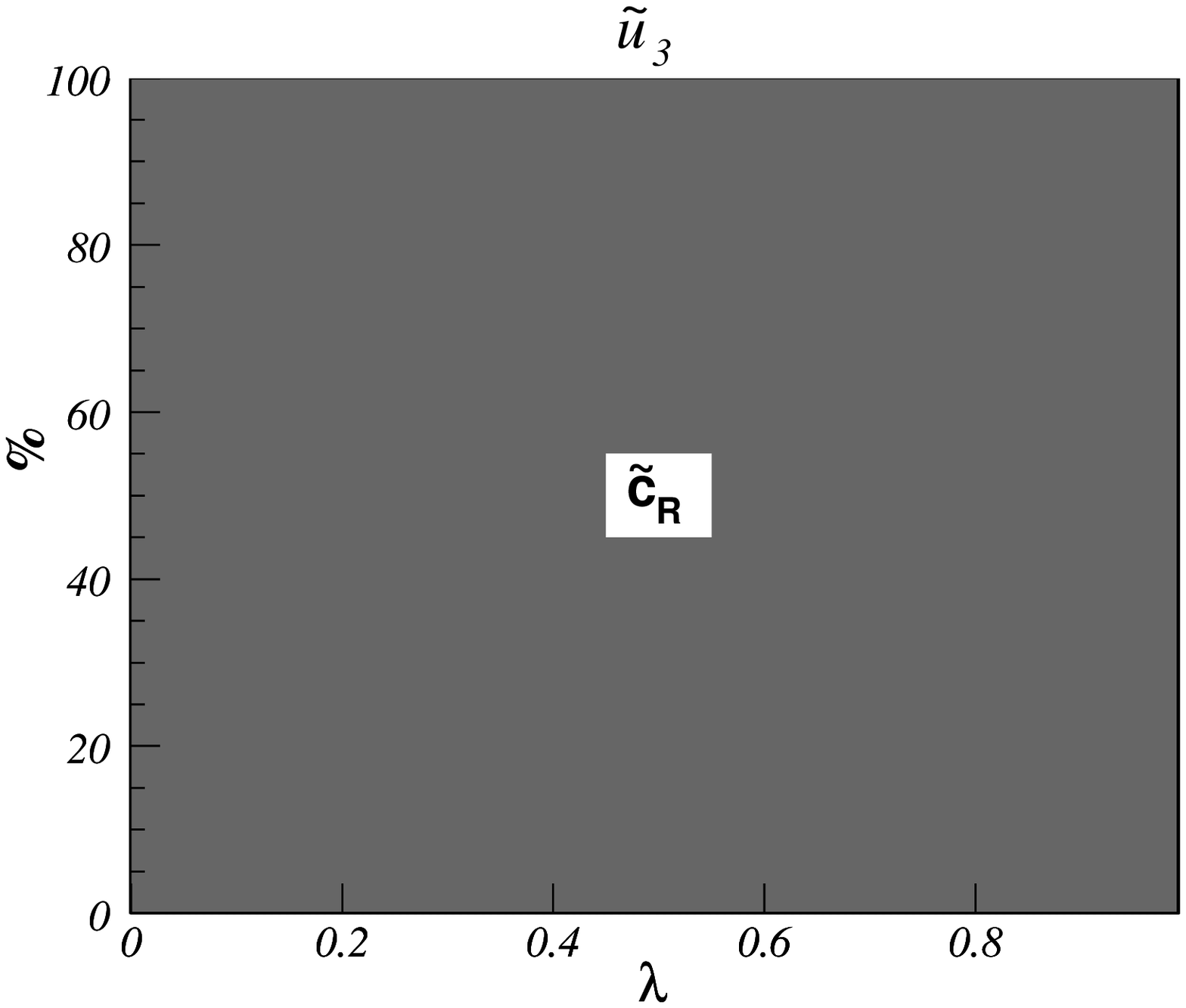}\hspace{2mm}
 \includegraphics[width=0.21\columnwidth]{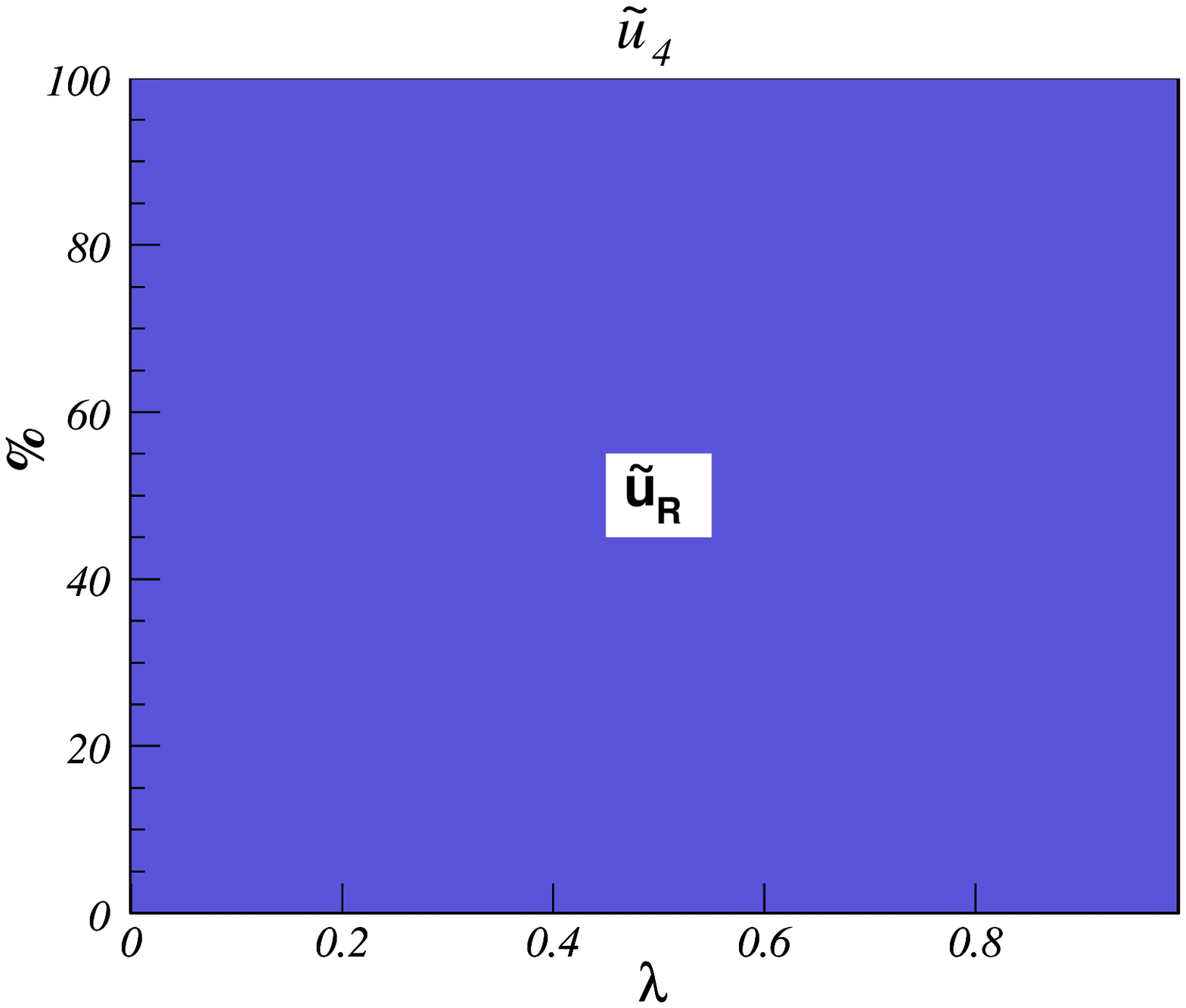}\hspace{2mm}
 \includegraphics[width=0.21\columnwidth]{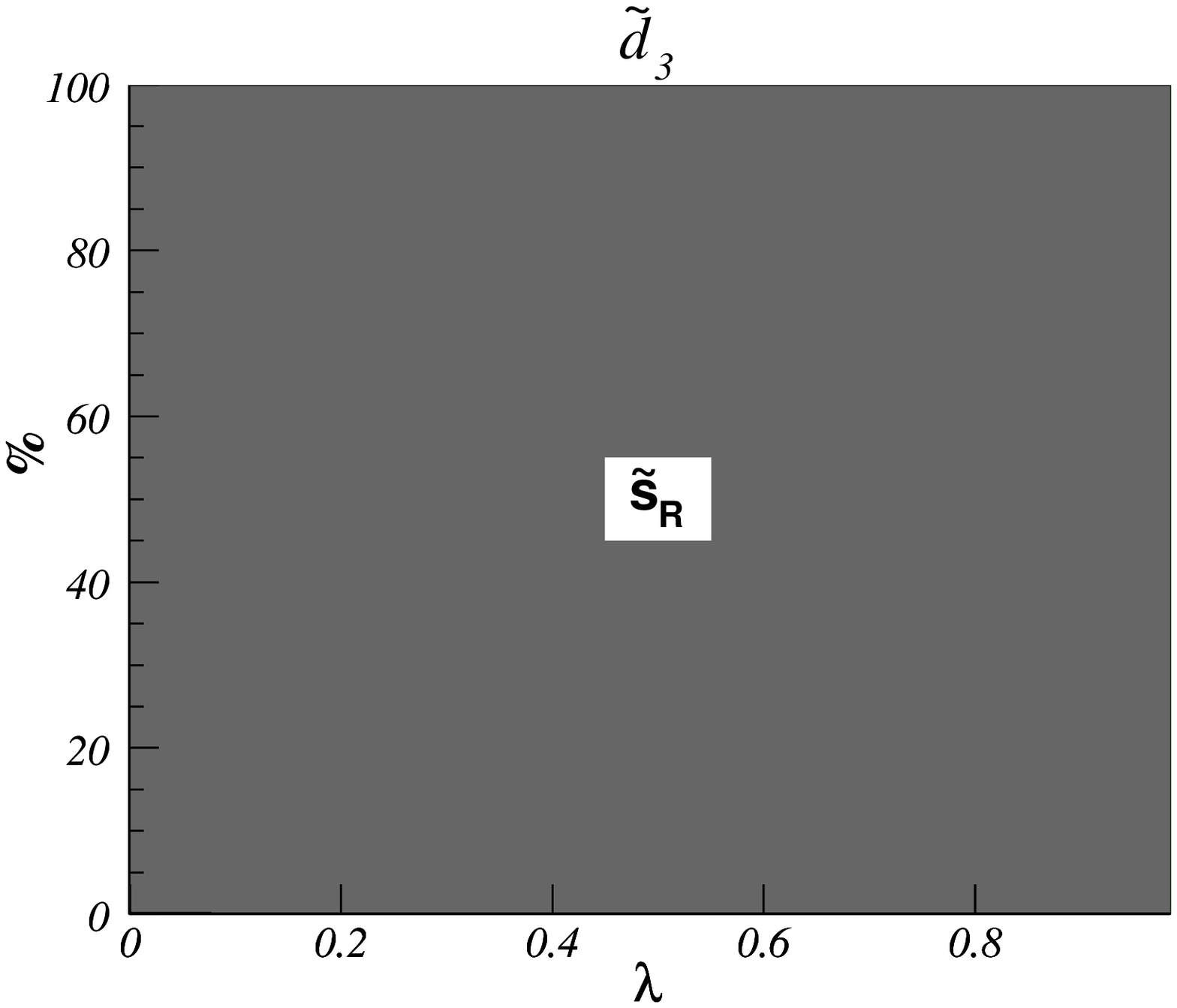}\hspace{2mm}
 \includegraphics[width=0.21\columnwidth]{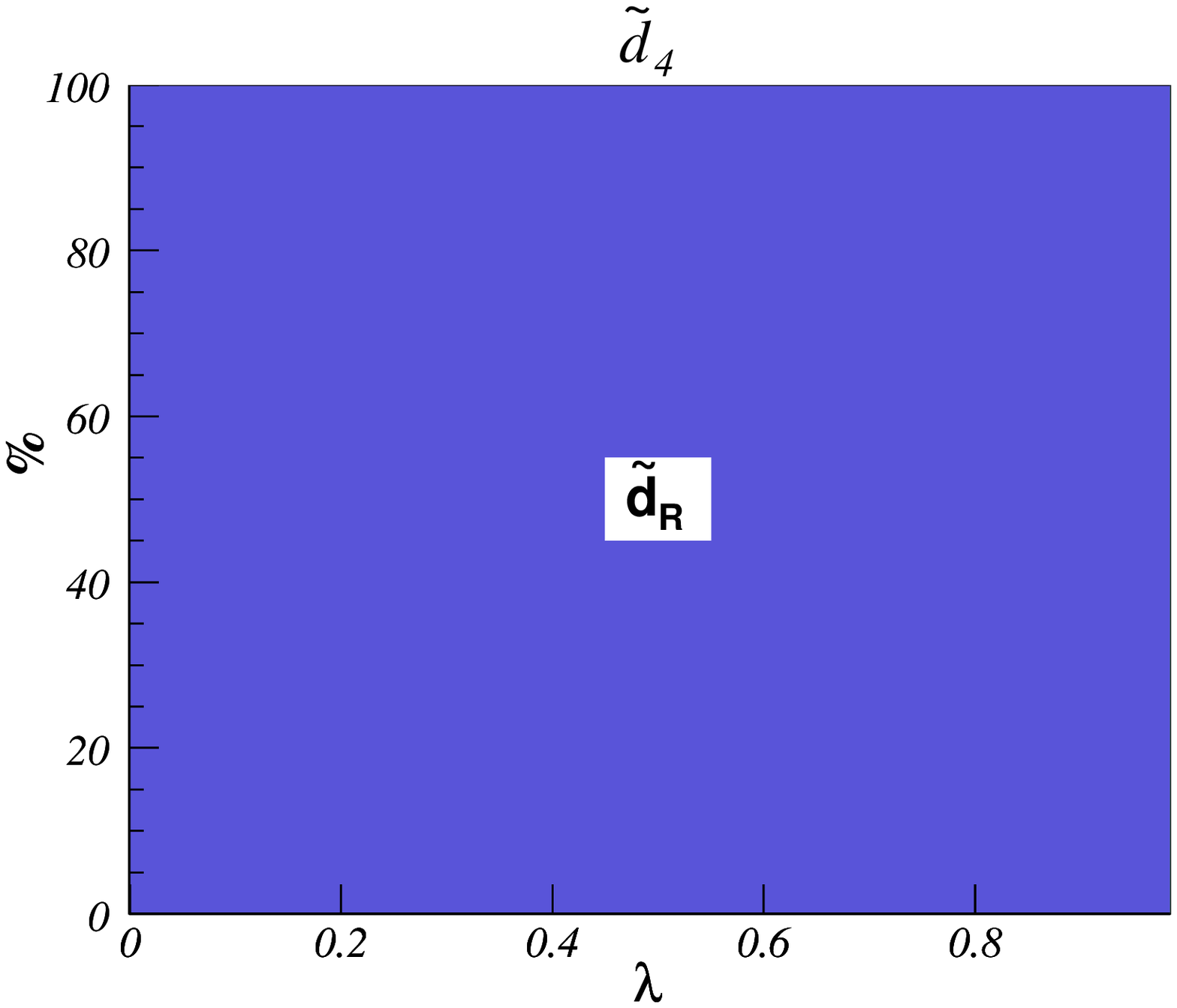}\vspace*{4mm}
 \includegraphics[width=0.21\columnwidth]{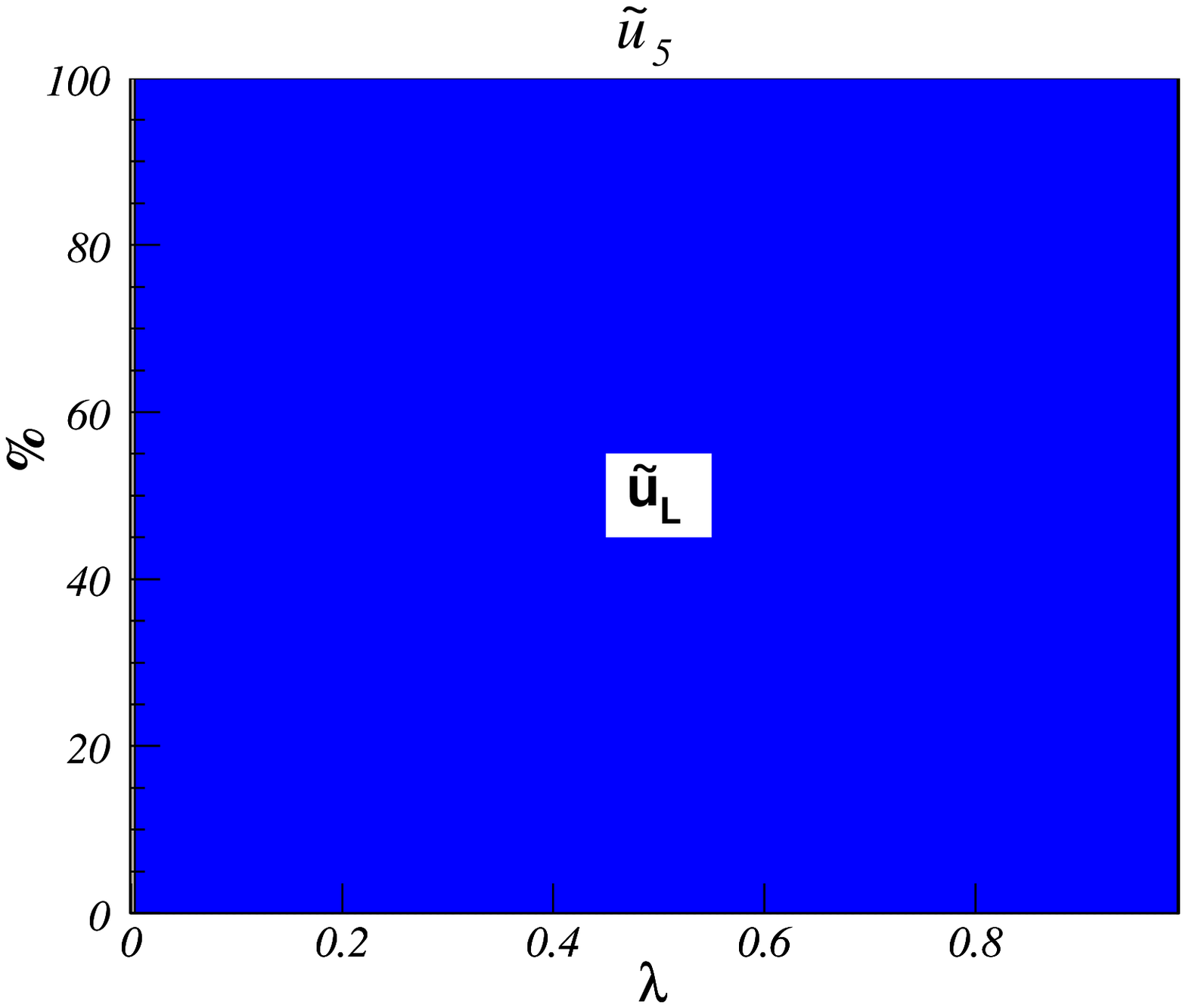}\hspace{2mm}
 \includegraphics[width=0.21\columnwidth]{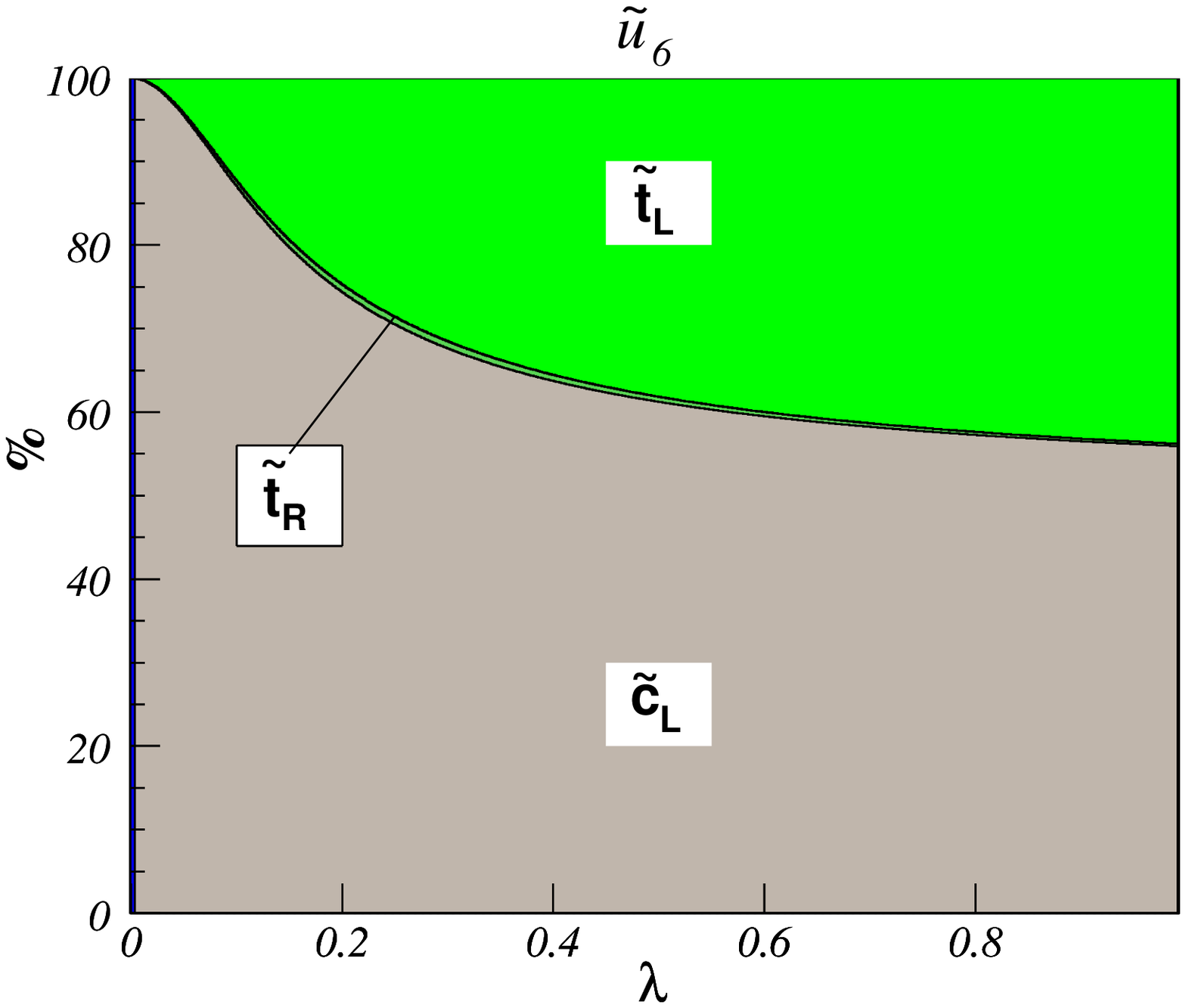}\hspace{2mm}
 \includegraphics[width=0.21\columnwidth]{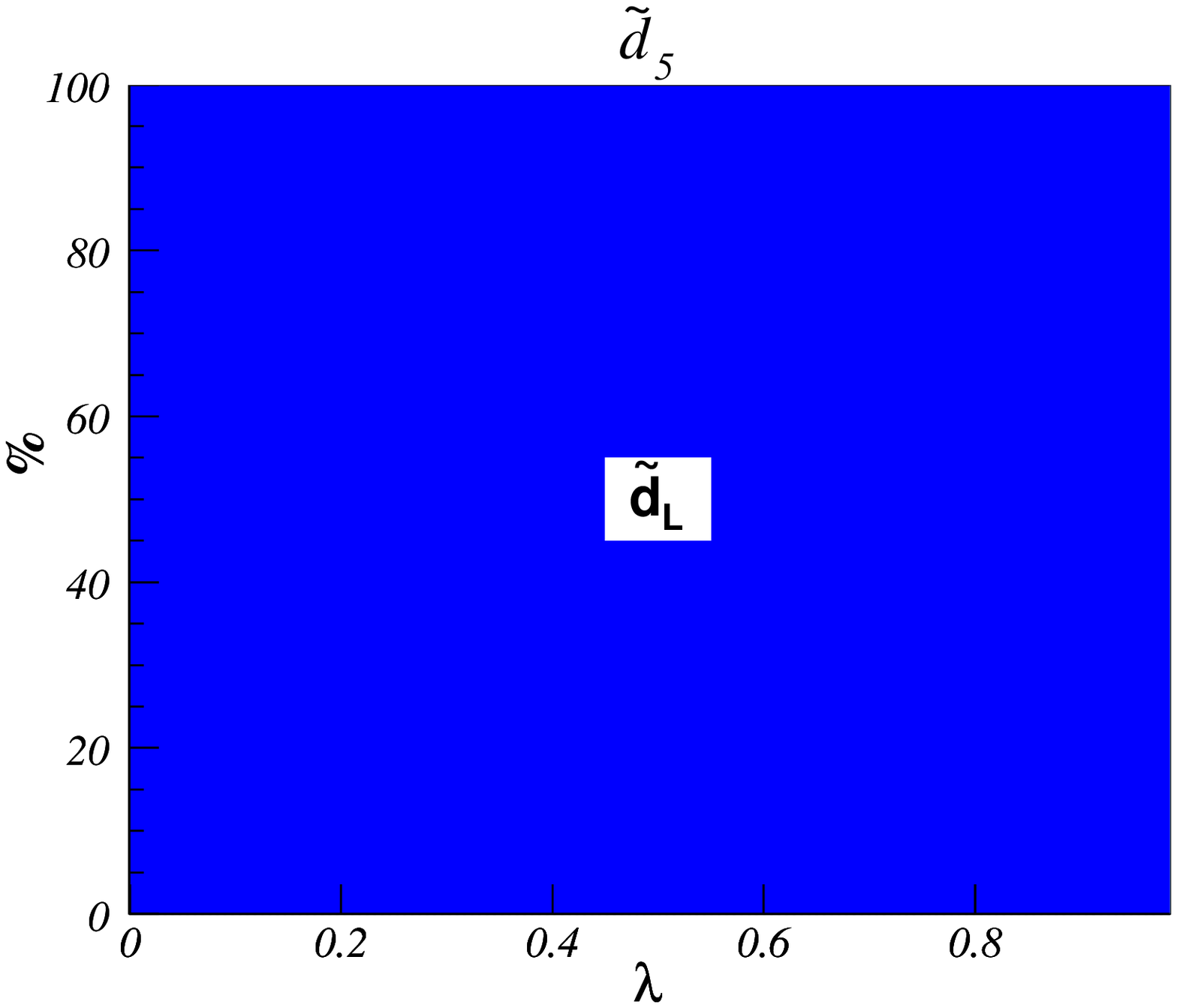}\hspace{2mm}
 \includegraphics[width=0.21\columnwidth]{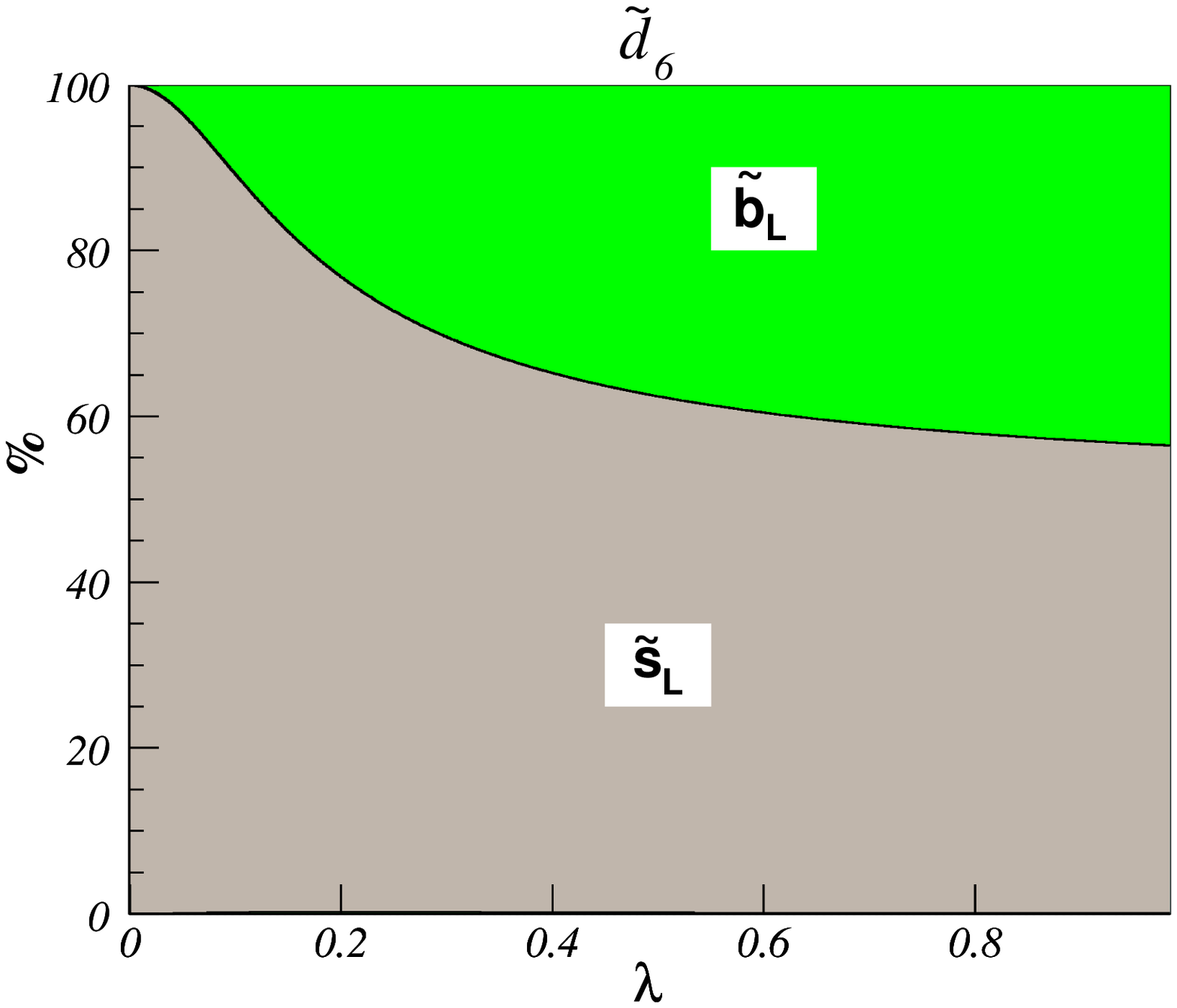}
 \caption{\label{fig:19}Same as Fig.\ \ref{fig:16} for benchmark point D.}
\end{figure}
%
%
\begin{figure}
 \centering
 \includegraphics[width=0.21\columnwidth]{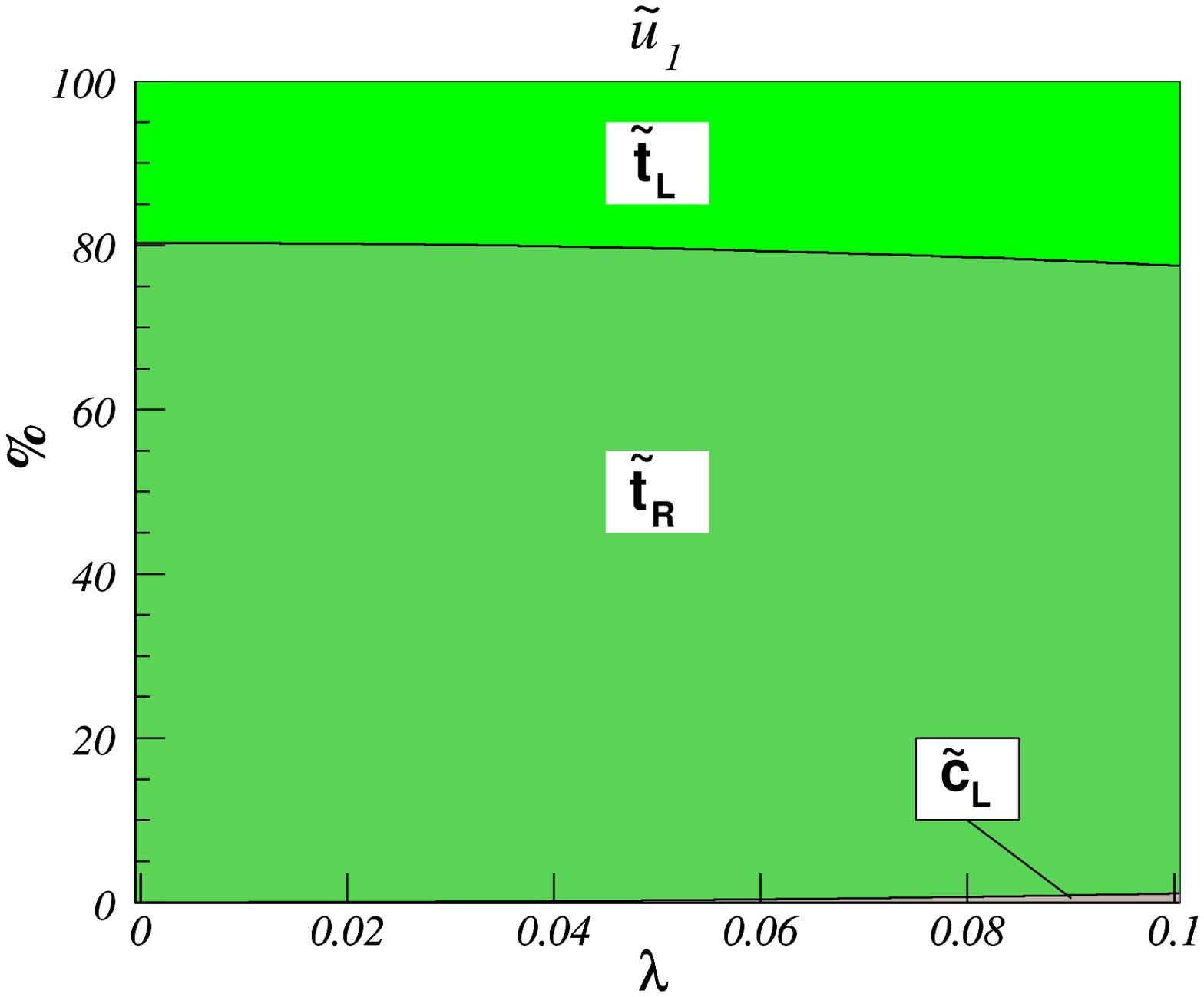}\hspace{2mm}
 \includegraphics[width=0.21\columnwidth]{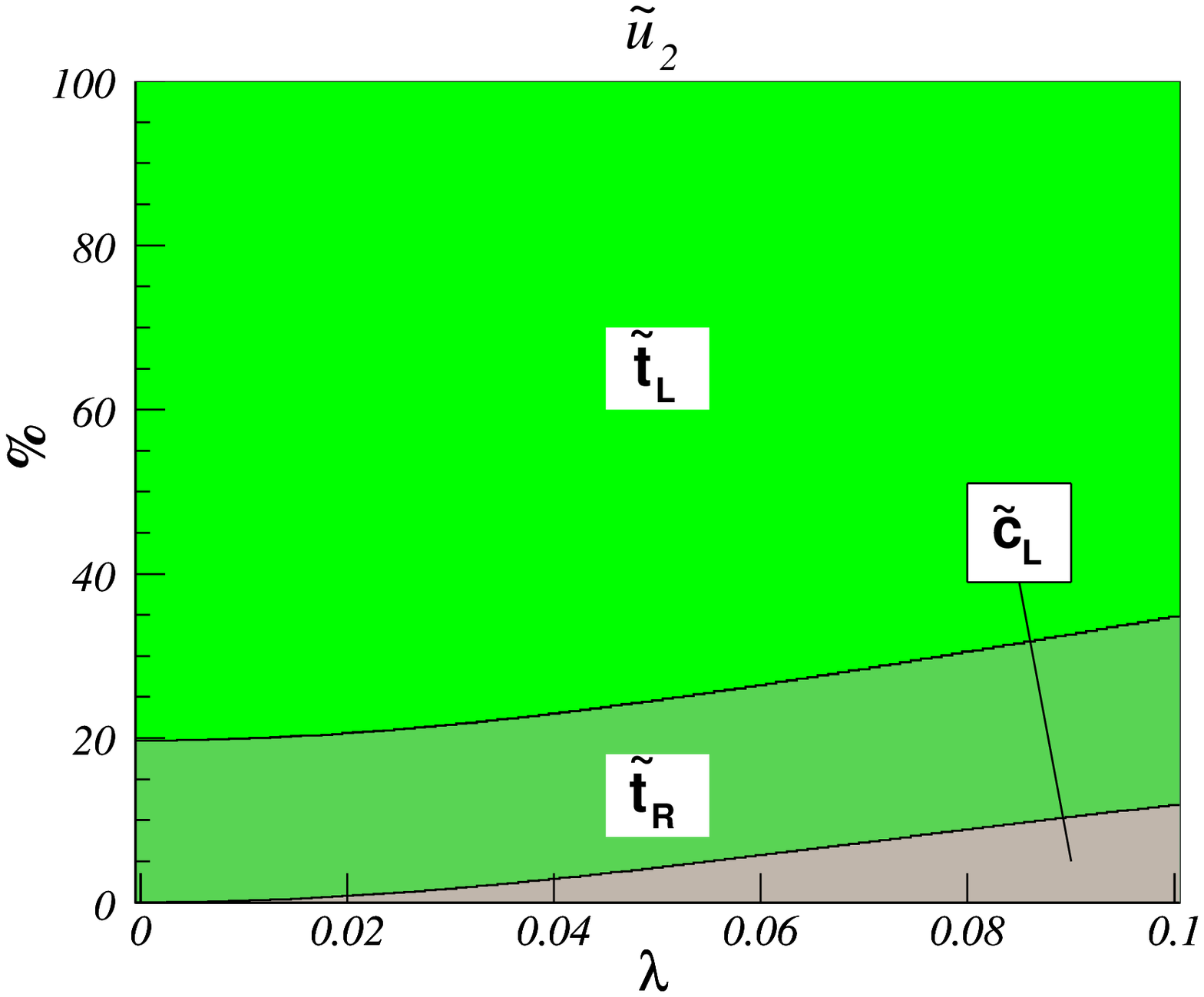}\hspace{2mm}
 \includegraphics[width=0.21\columnwidth]{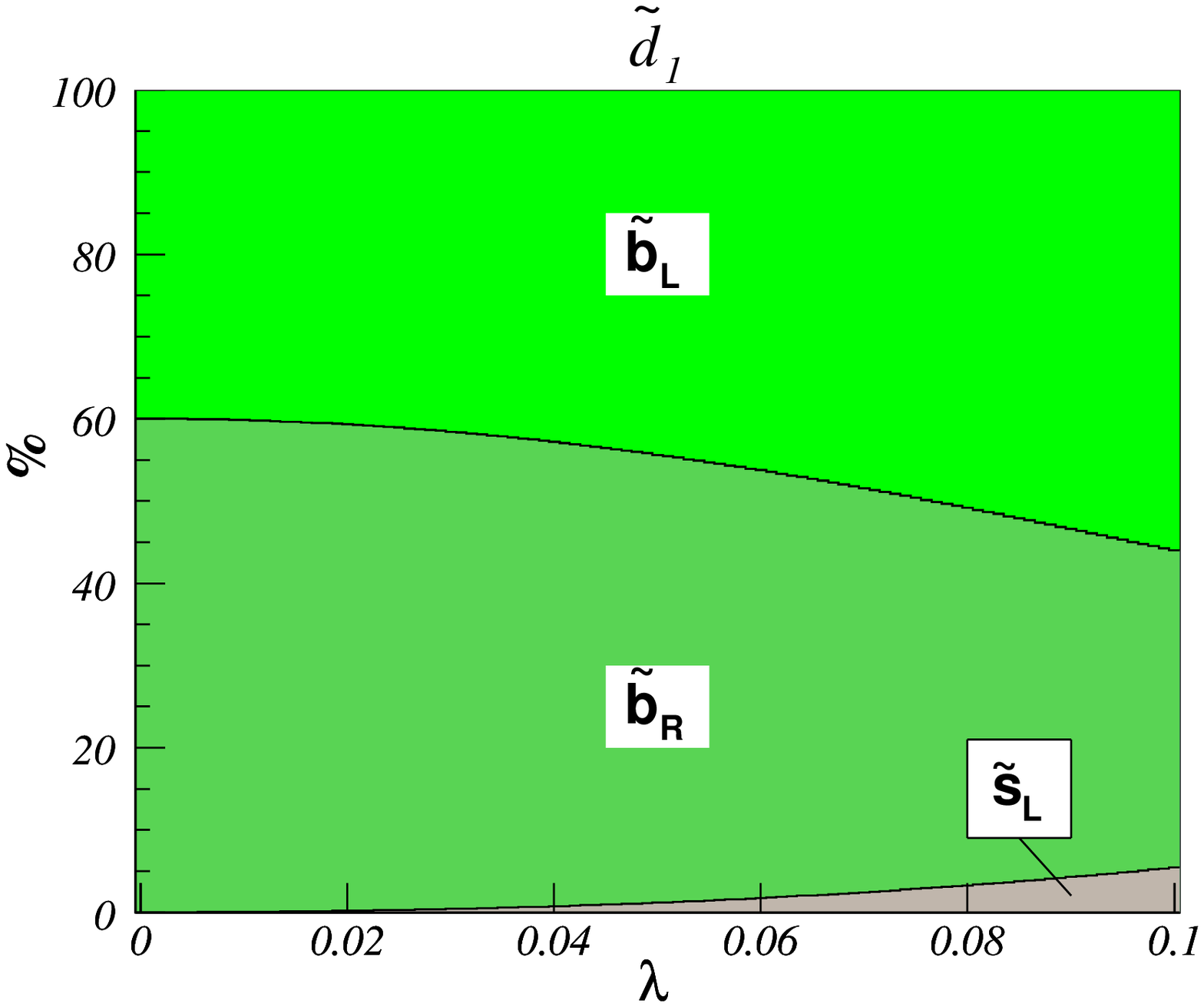}\hspace{2mm}
 \includegraphics[width=0.21\columnwidth]{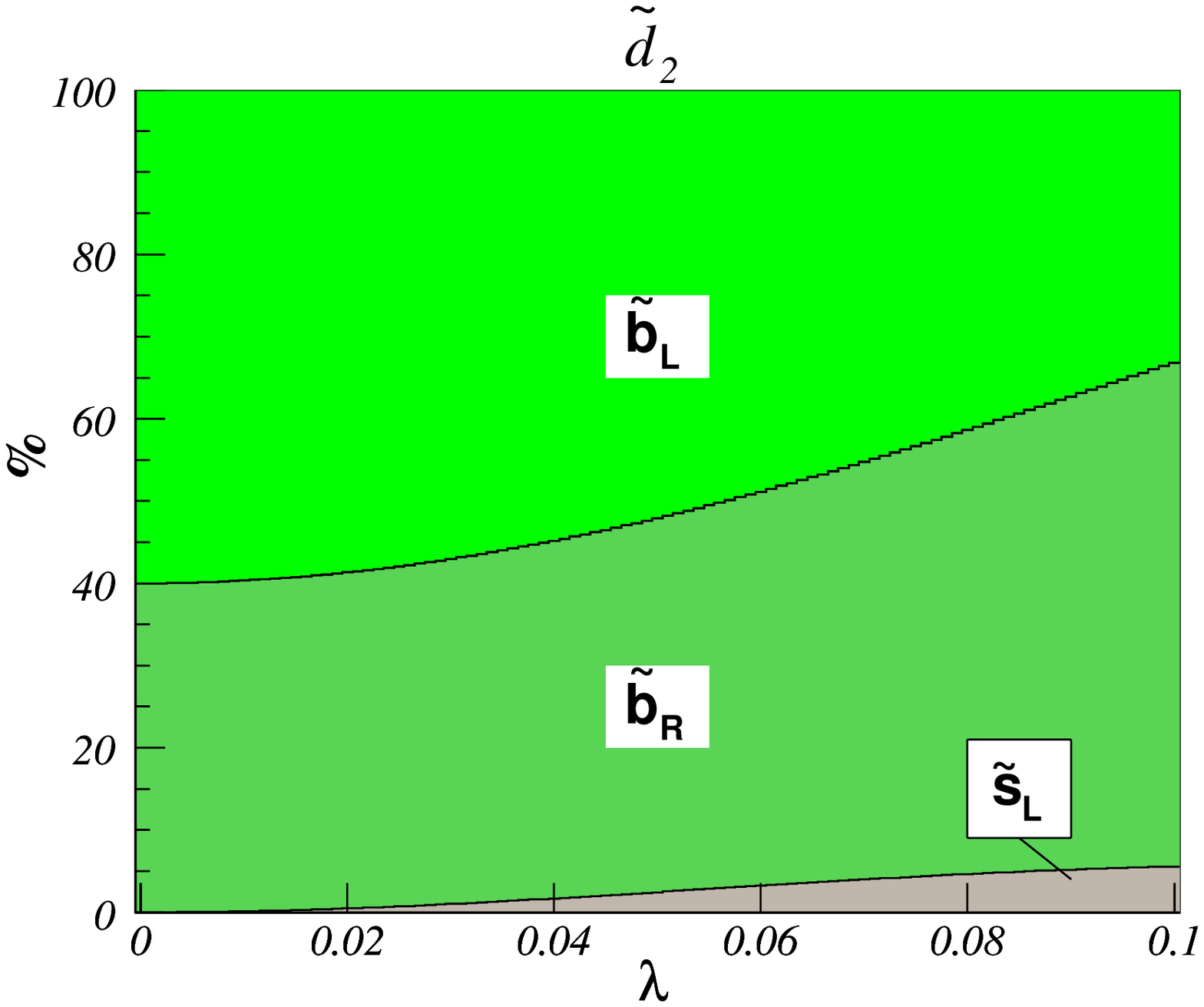}\vspace*{4mm}
 \includegraphics[width=0.21\columnwidth]{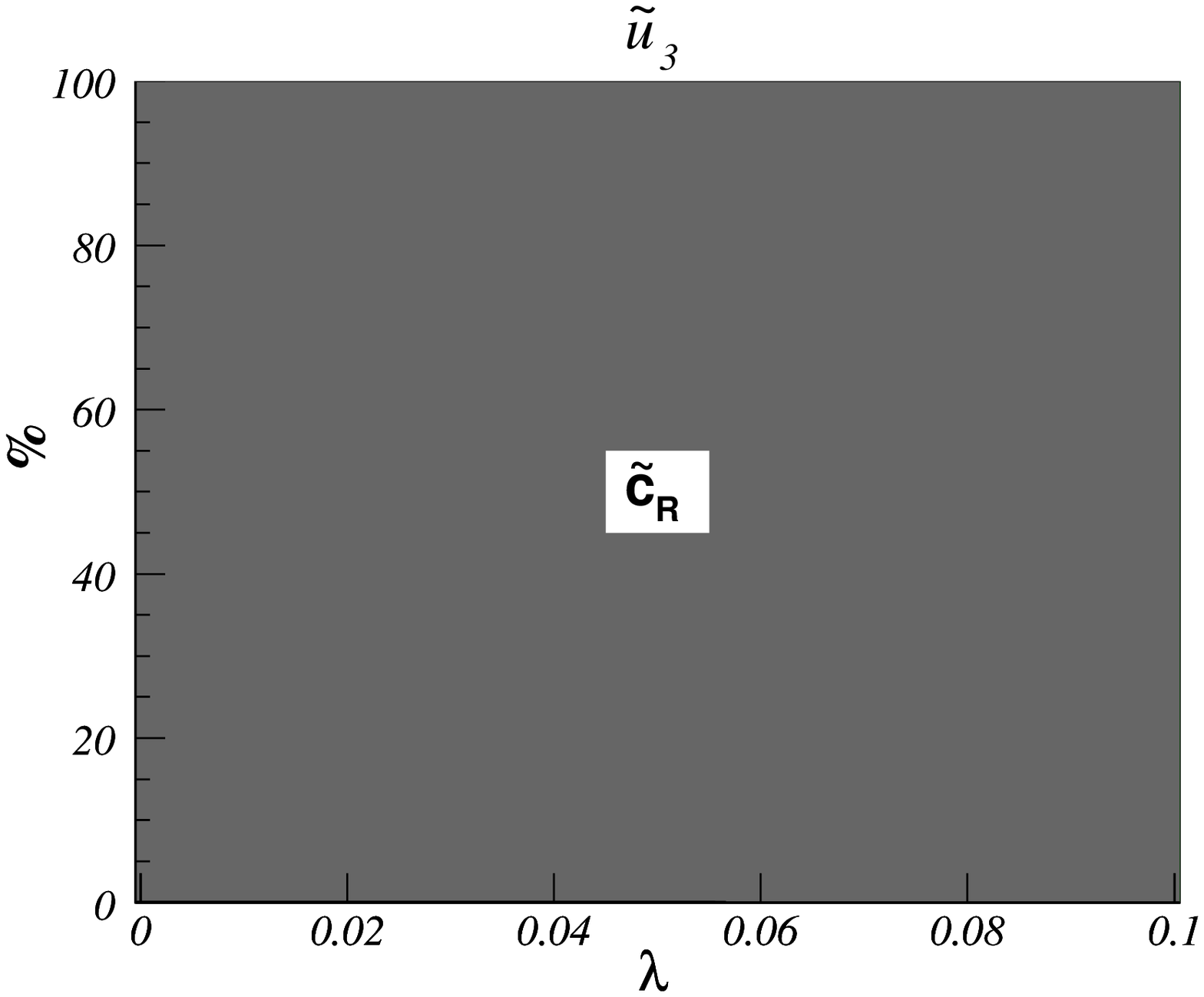}\hspace{2mm}
 \includegraphics[width=0.21\columnwidth]{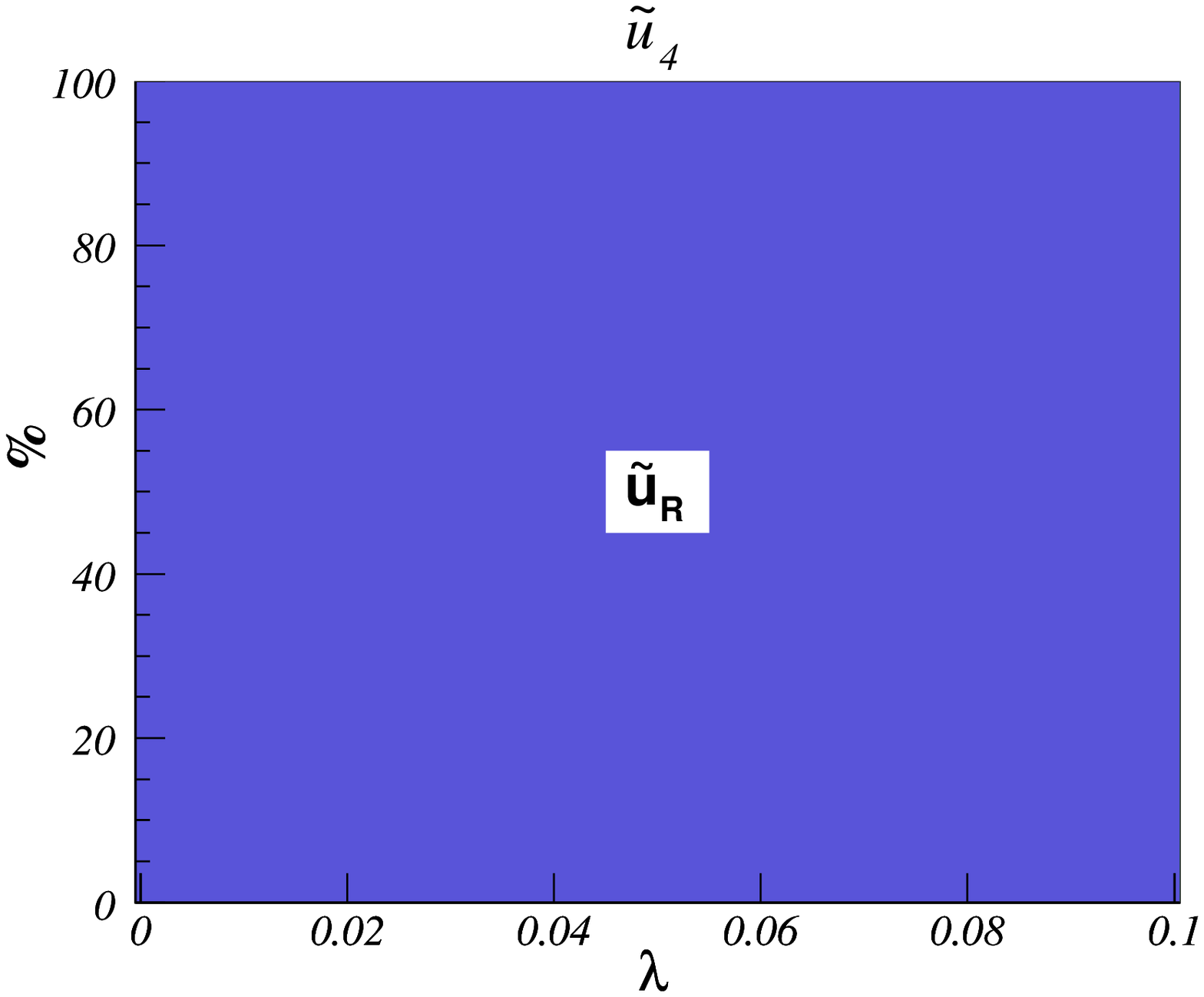}\hspace{2mm}
 \includegraphics[width=0.21\columnwidth]{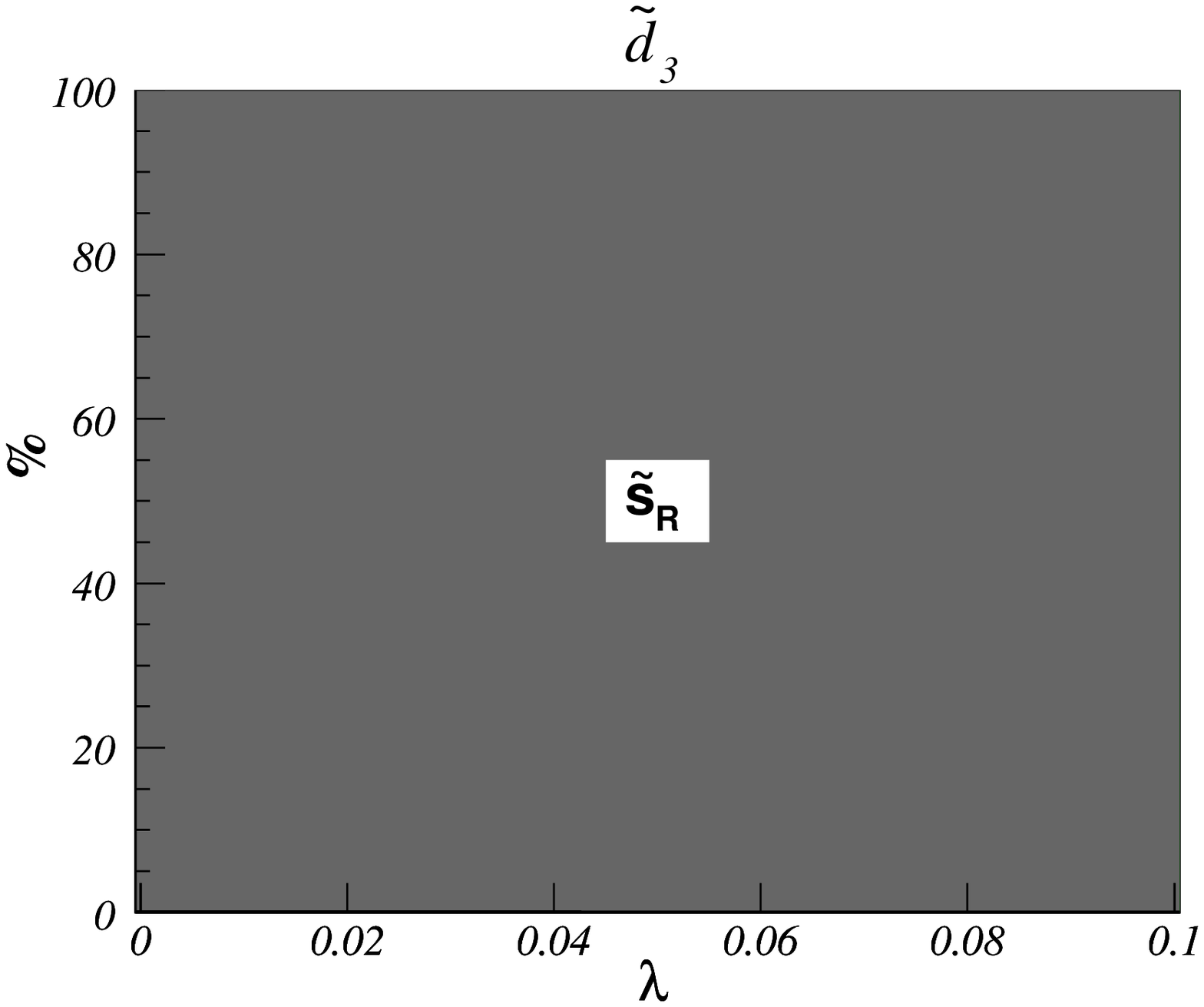}\hspace{2mm}
 \includegraphics[width=0.21\columnwidth]{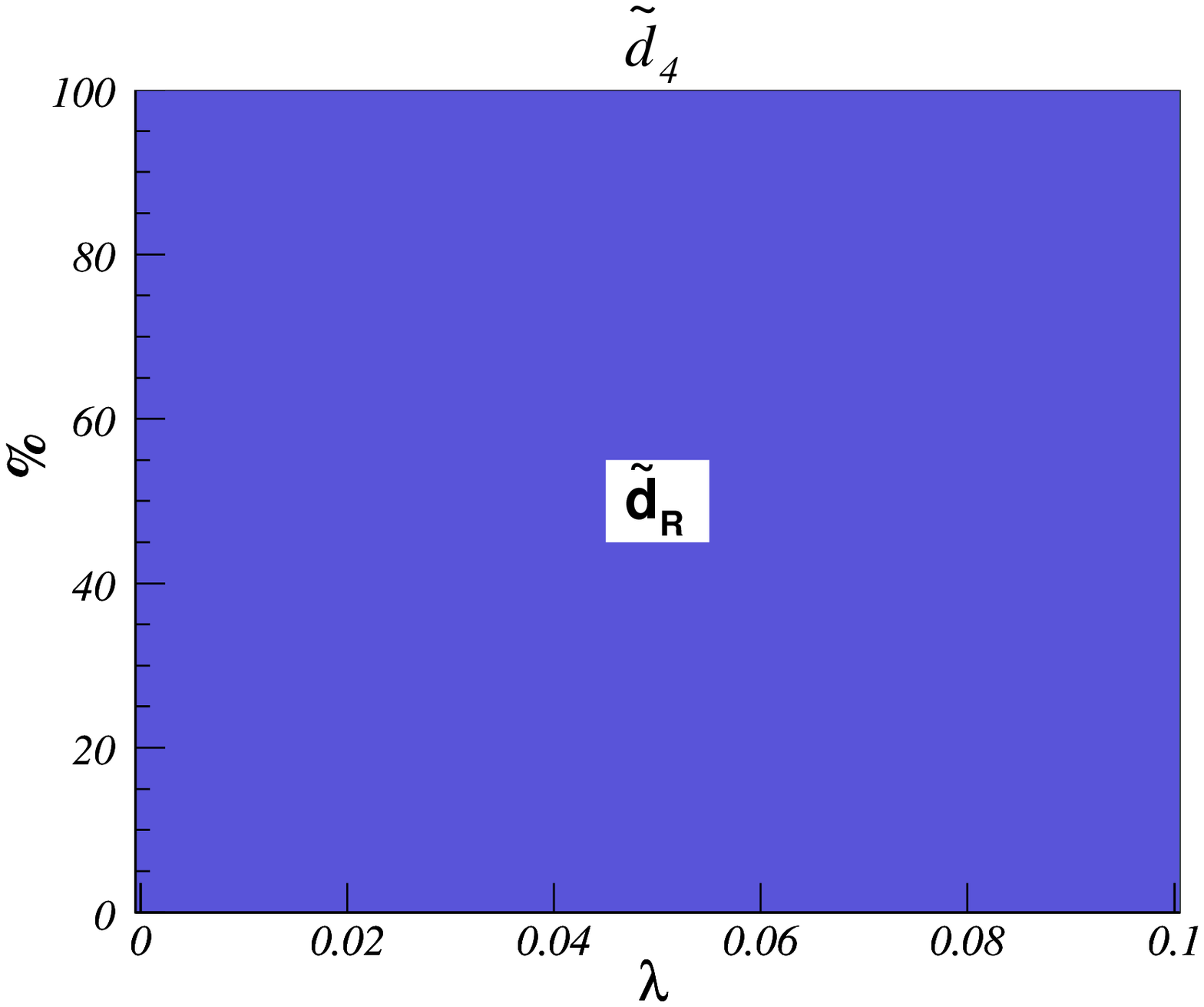}\vspace*{4mm}
 \includegraphics[width=0.21\columnwidth]{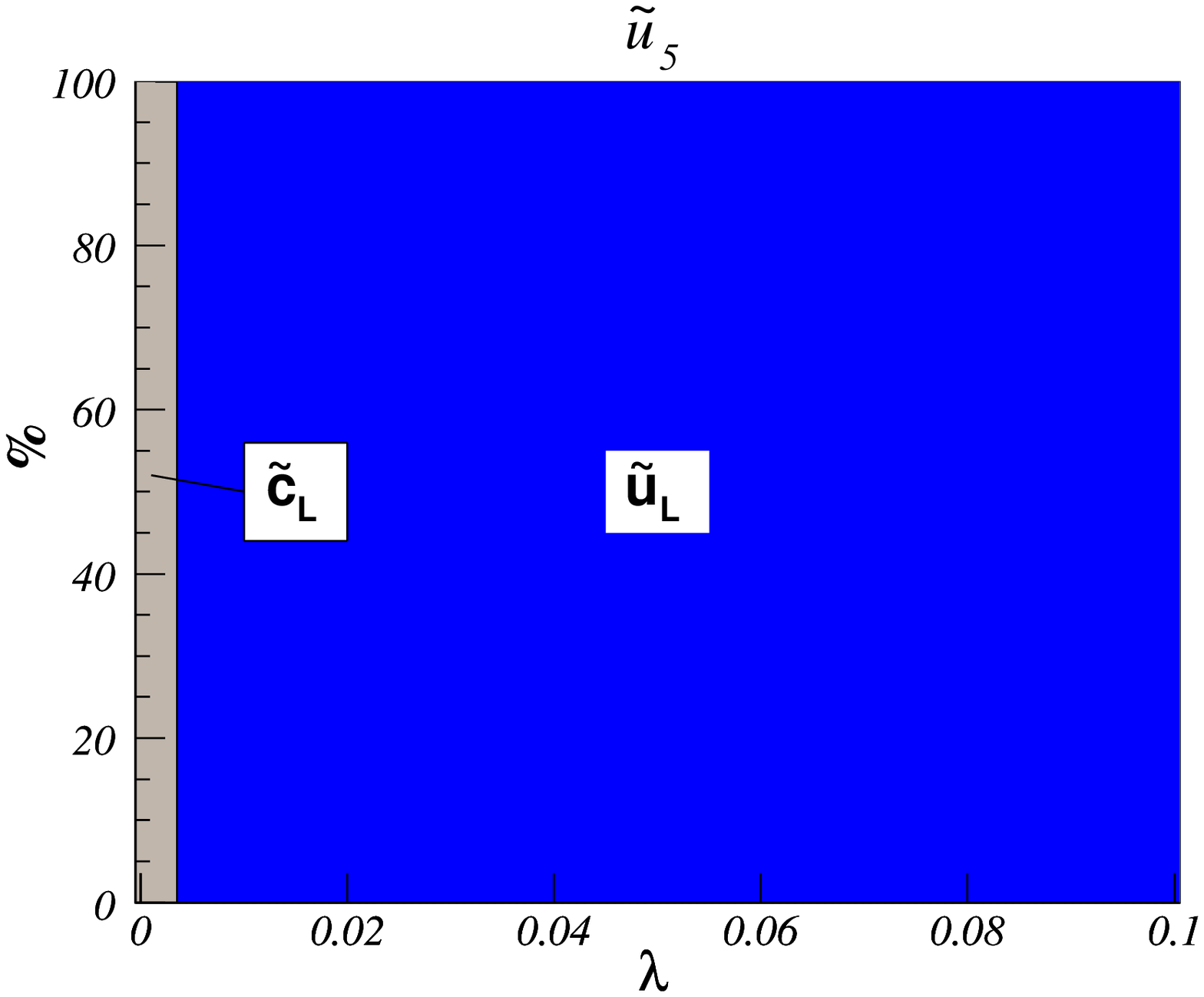}\hspace{2mm}
 \includegraphics[width=0.21\columnwidth]{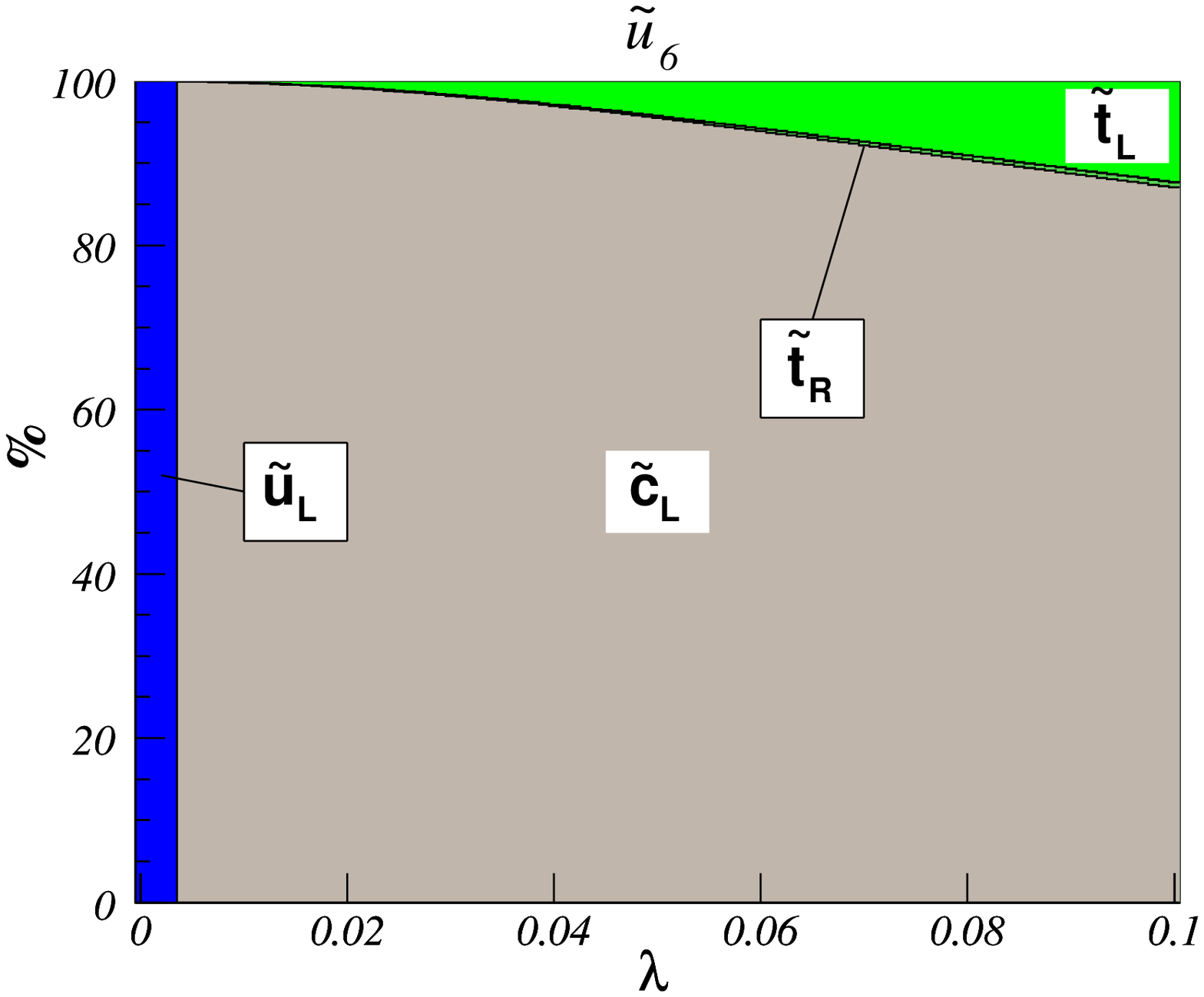}\hspace{2mm}
 \includegraphics[width=0.21\columnwidth]{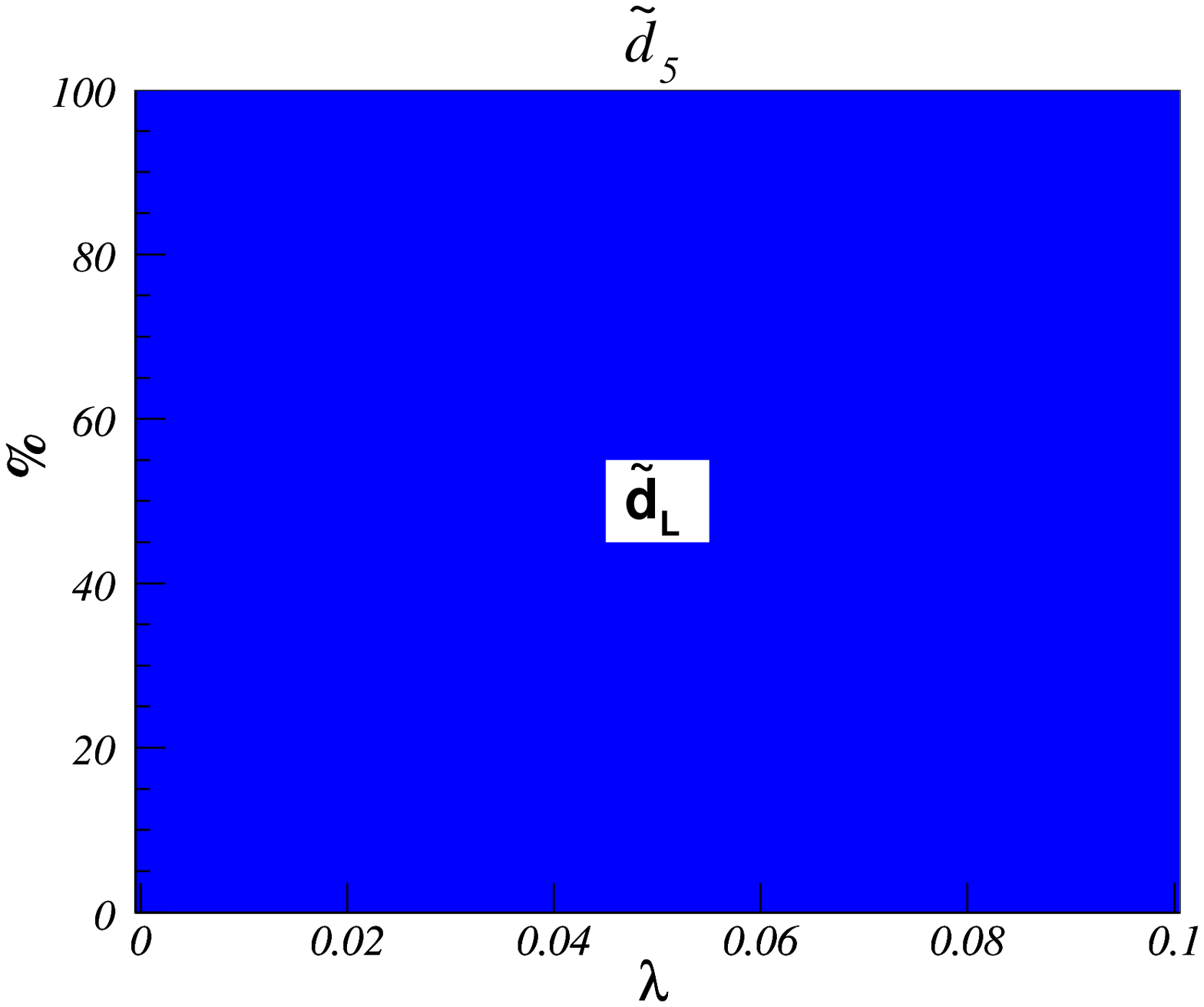}\hspace{2mm}
 \includegraphics[width=0.21\columnwidth]{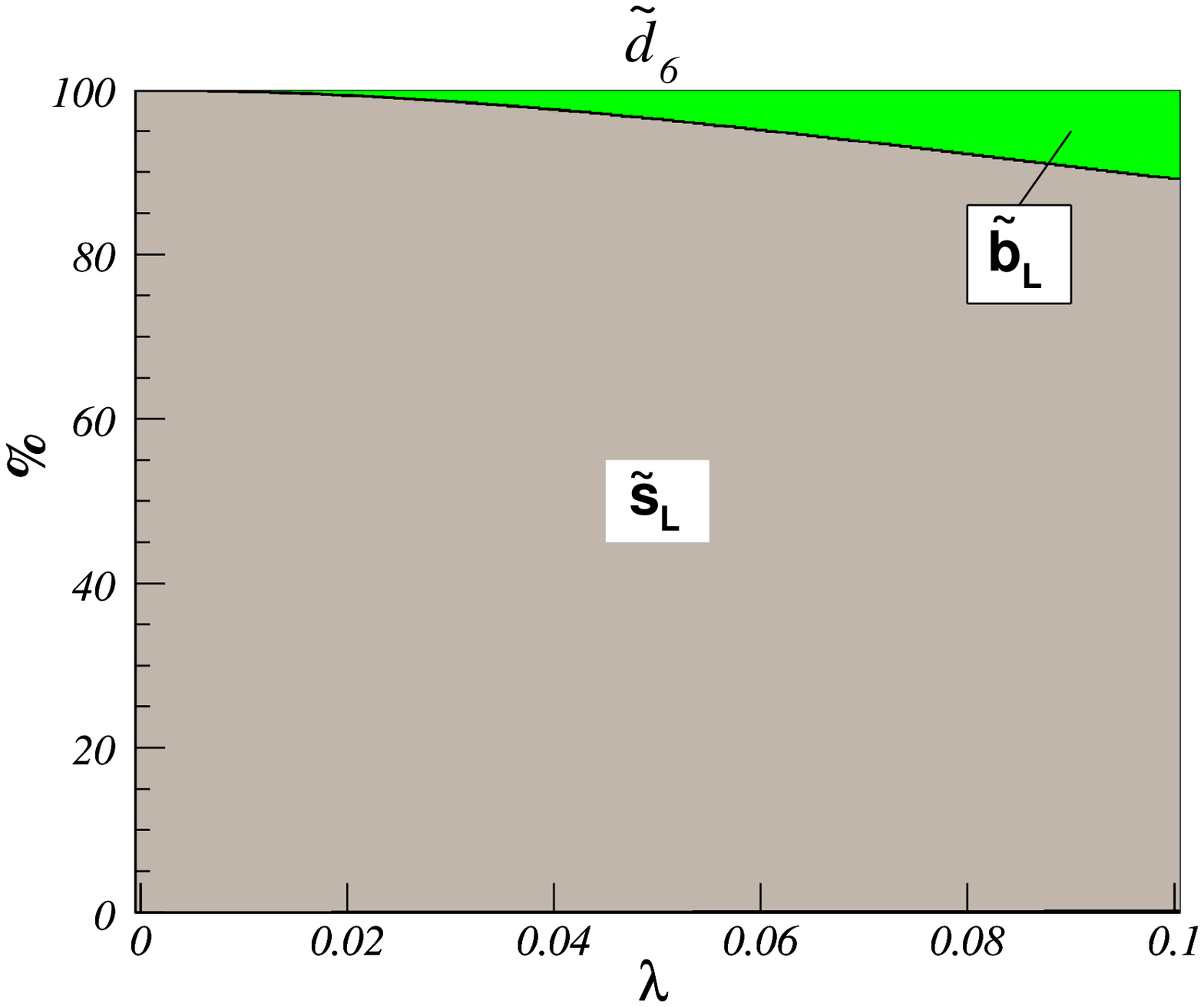}
 \caption{\label{fig:19p}Same as Fig.\ \ref{fig:19} for $\lambda\in
          [0;0.1]$.}
\end{figure}
%
In NMFV, squarks will not only exhibit the traditional mixing of
left- and right-handed helicities of third-generation flavour
eigenstates, but will in addition exhibit generational mixing. As
discussed before, we restrict ourselves here to the simultaneous
mixing of left-handed second- and third-generation up- and
down-type squarks. For our benchmark scenario A, the helicity and
flavour decomposition of the six up-type (left) and down-type
(right) squark mass eigenstates is shown in Fig.\ \ref{fig:16} for
the full range of the parameter $\lambda\in[0;1]$ and in Fig.\
\ref{fig:16p} for the experimentally favoured range in the
vicinity of (c)MFV, $\lambda\in [0;0.1]$. First-generation and
right-handed second-generation squarks remain, of course,
helicity- and flavour-diagonal, \bea
 \tilde{u}_3~=~\tilde{c}_R &,& \tilde{d}_2~=~\tilde{s}_R, \nonumber\\
 \tilde{u}_4~=~\tilde{u}_R &,& \tilde{d}_3~=~\tilde{d}_R, \nonumber\\
 \tilde{u}_5~=~\tilde{u}_L &,& \tilde{d}_5~=~\tilde{d}_L,
\eea
with the left-handed and first-generation squarks being slightly heavier due
their weak isospin coupling (see Eqs.\ (\ref{eq:5})-(\ref{eq:7})) and
different renormalization-group running effects. Production of these states
will benefit from $t$- and $u$-channel contributions of first- and
second-generation quarks with enhanced parton densities in the external
hadrons, but they will not be identified easily with heavy-flavour tagging
and are of little interest for our study of flavour violation. The
lightest up-type squark $\tilde{u}_1$ remains the traditional mixture of
left- and right-handed stops over a large region of $\lambda\leq0.4$, but it
shows at this point an interesting flavour transition, which is in fact
expected from the level reordering phenomenon discussed in the lower central
plot of Fig.\ \ref{fig:12}. The transition happens, however, above the
experimental limit of $\lambda\leq0.1$. Below this limit, it is the states
$\tilde{u}_2$, $\tilde{u}_6$, $\tilde{d}_1$, and in particular $\tilde{d}_4$
and $\tilde{d}_6$ that show, in addition to helicity mixing, the most
interesting and smooth variation of second- and third-generation flavour
content (see Fig.\ \ref{fig:16p}). Note that at very low $\lambda\simeq
0.002$ the states $\tilde{d}_L$ and $\tilde{s}_L$ rapidly switch levels.
This numerically small change was not visible on the linear scale in Fig.\
\ref{fig:12}.

For the benchmark point B, whose helicity and flavour decomposition is
shown in Fig.\ \ref{fig:17}, level reordering occurs at $\lambda\simeq0.4$
for the intermediate-mass up-type squarks,
\bea
 \tilde{u}_{3,4}~=~\tilde{c}_R &,& \tilde{d}_2~=~\tilde{s}_R, \nonumber\\
 \tilde{u}_{4,3}~=~\tilde{u}_R &,& \tilde{d}_3~=~\tilde{d}_R, \nonumber\\
 \tilde{u}_5~=~\tilde{u}_L &,& \tilde{d}_5~=~\tilde{d}_L
\eea
whereas the ordering of down-type squarks is very similar to scenario A.
Close inspection of Fig.\ \ref{fig:17p} shows, however, that also
$\tilde{d}_R$ and $\tilde{s}_R$ switch levels at low values of $\lambda
\simeq0.02$. At $\lambda\simeq0.01$, in addition $\tilde{s}_R$ and
$\tilde{b}_L$ switch levels, and at $\lambda\simeq0.002$ it is the states
$\tilde{u}_L$ and $\tilde{c}_L$. The lightest up-type squark is again
nothing but a mix of left- and right-handed stops up to $\lambda\leq0.4$.
Phenomenologically smooth transitions below $\lambda\leq0.1$ involving
taggable third-generation squarks are observed for $\tilde{u}_4$,
$\tilde{u}_6$, $\tilde{d}_1$, and $\tilde{d}_6$.

The helicity and flavour decomposition for our scenario D, shown in Fig.\
\ref{fig:19}, is rather similar to the one in scenario A, i.e.\
\bea
 \tilde{u}_3~=~\tilde{c}_R &,& \tilde{d}_3~=~\tilde{s}_R, \nonumber\\
 \tilde{u}_4~=~\tilde{u}_R &,& \tilde{d}_4~=~\tilde{d}_R, \nonumber\\
 \tilde{u}_5~=~\tilde{u}_L &,& \tilde{d}_5~=~\tilde{d}_L
\eea
are exactly the same in the up-squark sector, and only the mixed down-type
state $\tilde{d}_4$ is now lighter and becomes $\tilde{d}_2$. The lightest
up-type squark, $\tilde{u}_1$, is again mostly a mix of left- and
right-handed top squarks up to $\lambda \simeq0.4$, where the level
reordering and generation mixing occurs (see lower central part of Fig.\
\ref{fig:15}). At the experimentally favoured lower values of $\lambda\leq
0.1$, the states $\tilde{u}_2$, $\tilde{u}_6$, $\tilde{d}_1$, $\tilde{d}_2$,
and $\tilde{d}_6$ exhibit some smooth variations, shown in detail in Fig.\
\ref{fig:19p}, albeit to a lesser extent than in scenario A. At very low
$\lambda\simeq0.004$, it is now the up-type squarks $\tilde{u}_L$ and
$\tilde{c}_L$ that rapidly switch levels. This numerically small change was
again not visible on a linear scale (see Fig.\ \ref{fig:15}).

For our scenario C, shown in Fig.\ \ref{fig:18}, the assignment of the
intermediate states
\bea
 \tilde{u}_3~=~\tilde{c}_R &,& \tilde{d}_3~=~\tilde{s}_R, \nonumber\\
 \tilde{u}_4~=~\tilde{u}_R &,& \tilde{d}_4~=~\tilde{d}_R, \nonumber\\
 \tilde{u}_5~=~\tilde{u}_L &,& \tilde{d}_5~=~\tilde{d}_L
\eea
is the same as for scenario D above $\lambda\geq0.1$. Just below,
$\tilde{u}_R$ and $\tilde{c}_R$ as well as $\tilde{d}_R$ and $\tilde{s}_R$
rapidly switch levels, whereas $\tilde{u}_L$ and $\tilde{c}_L$ switch levels
at very low $\lambda\simeq0.002$. These changes were already visible upon
close inspection of the lower central and right plots in Fig.\ \ref{fig:14}.
On the other hand, the lightest squarks $\tilde{u}_1$ and $\tilde{d}_1$ only
acquire significant flavour admixtures at relatively large $\lambda\simeq
0.2...0.4$, whereas they are mostly superpositions of left- and right-handed
stops and sbottoms in the experimentally favourable range of $\lambda\leq
0.1$ shown in Fig. \ref{fig:18p}. Here, the heaviest states $\tilde{u}_6$
and $\tilde{d}_6$ show already smooth admixtures of third-generation squarks
as it was the case for the scenarios A and D discussed above. The most
interesting states are, however, $\tilde{u}_2$, $\tilde{u}_4$,
$\tilde{d}_2$, and $\tilde{d}_4$, respectively, since they represent
combinations of up to four different helicity and flavour states and
have a significant, taggable third-generation flavour content.

\section{Numerical Predictions for NMFV SUSY Particle Production at the LHC}
\label{sec:5}

In this section, we present numerical predictions for the production cross
sections of squark-antisquark pairs, squark pairs, the associated production
of squarks and gauginos, and gaugino pairs in NMFV SUSY at the CERN LHC,
i.e.\ for $pp$-collisions at $\sqrt{S}=14$ TeV centre-of-mass energy. Thanks
to the QCD factorization theorem, total unpolarized hadronic cross sections
\bea
 \sigma &~=&
 \int_{4m^2/S}^1\!\d\tau\!\!
 \int_{-1/2\ln\tau}^{1/2\ln\tau}\!\!\d y
 \int_{t_{\min}}^{t_{\max}} \d t \
 f_{a/A}(x_a,M_a^2) \ f_{b/B}(x_b,M_b^2) \ {\d\hat{\sigma}\over\d t}
\eea
can be calculated by convolving the relevant partonic cross sections
d$\hat{\sigma}$/d$t$, computed in Sec.\ \ref{sec:3}, with universal parton
densities $f_{a/A}$ and $f_{b/B}$ of partons $a,b$ in the hadrons $A,B$,
which depend on the longitudinal momentum fractions of the two partons
$x_{a,b} = \sqrt{\tau}e^{\pm y}$ and on the unphysical factorization scales
$M_{a,b}$. For consistency with our leading order (LO) QCD calculation in
the collinear approximation, where all squared quark masses (except for the
top-quark mass) $m_q^2\ll s$, we employ the LO set of the latest CTEQ6
global
parton density fit \cite{Pumplin:2002vw}, which includes $n_f=5$ ``light''
(including the bottom) quark flavours and the gluon, but no top-quark
density. Whenever it occurs, i.e.\ for gluon initial states and gluon or
gluino exchanges, the strong coupling constant $\alpha_s(\mu_R)$ is
calculated with the corresponding LO value of $\Lambda_{\rm LO}^{n_f=5}=165$
MeV. We identify the renormalization scale $\mu_R$ with the factorization
scales $M_a=M_b$ and set the scales to the average mass of the final state
SUSY particles $i$ and $j$, $m=(m_i+m_j)/2$.

%
\begin{figure}
 \centering
 \includegraphics[width=0.32\columnwidth]{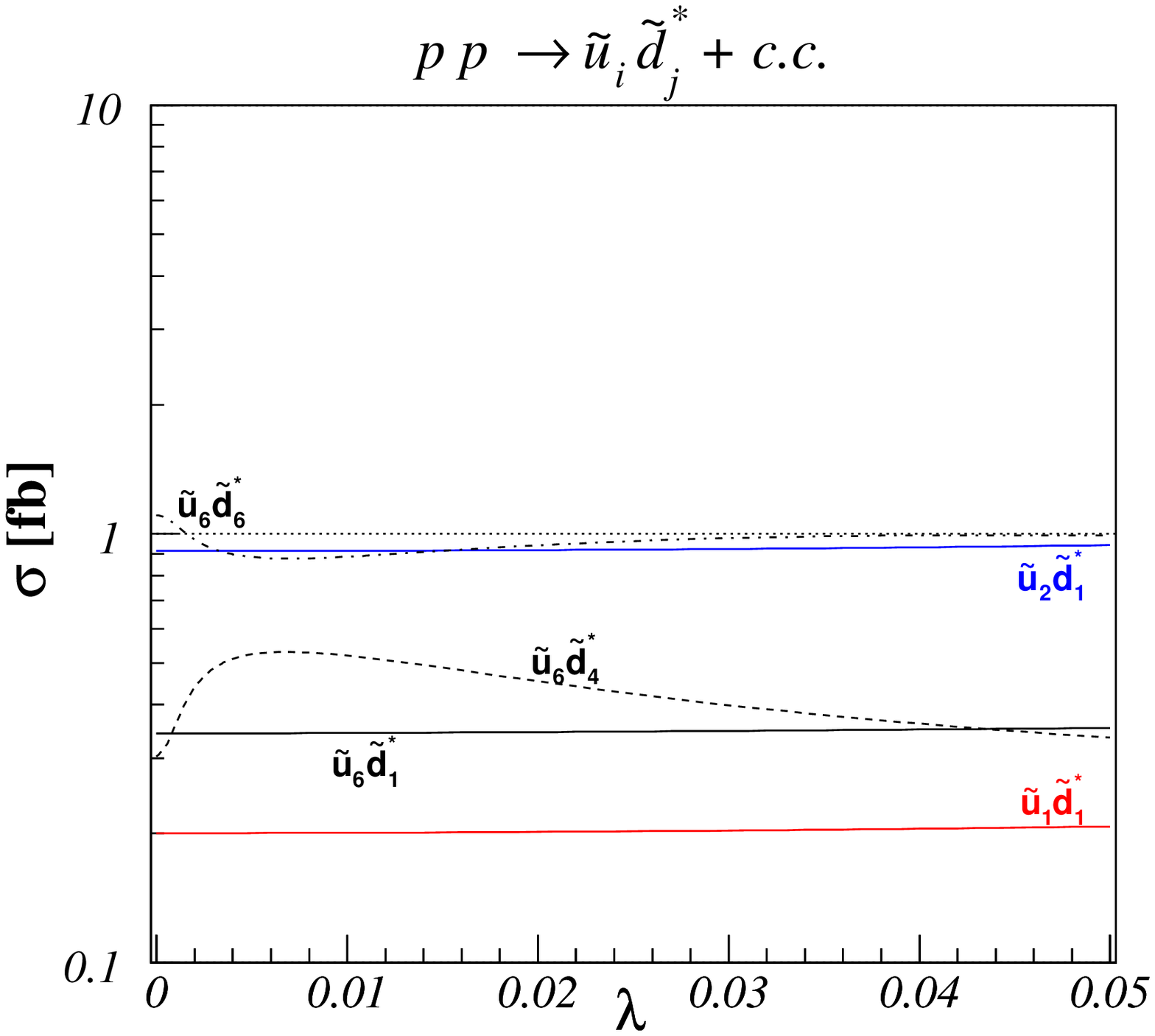}\hspace{2mm}
 \includegraphics[width=0.32\columnwidth]{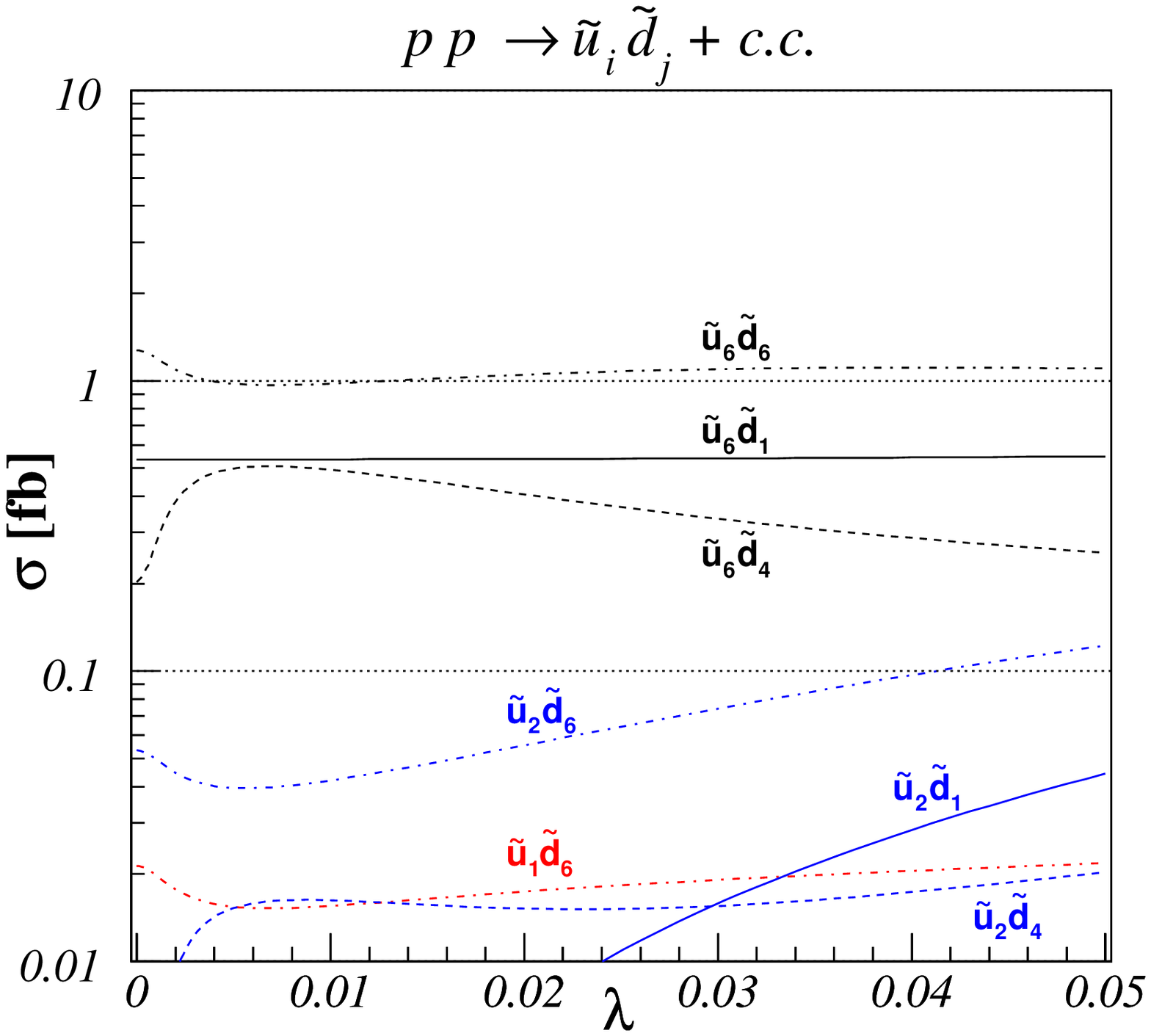}\hspace{2mm}
 \includegraphics[width=0.32\columnwidth]{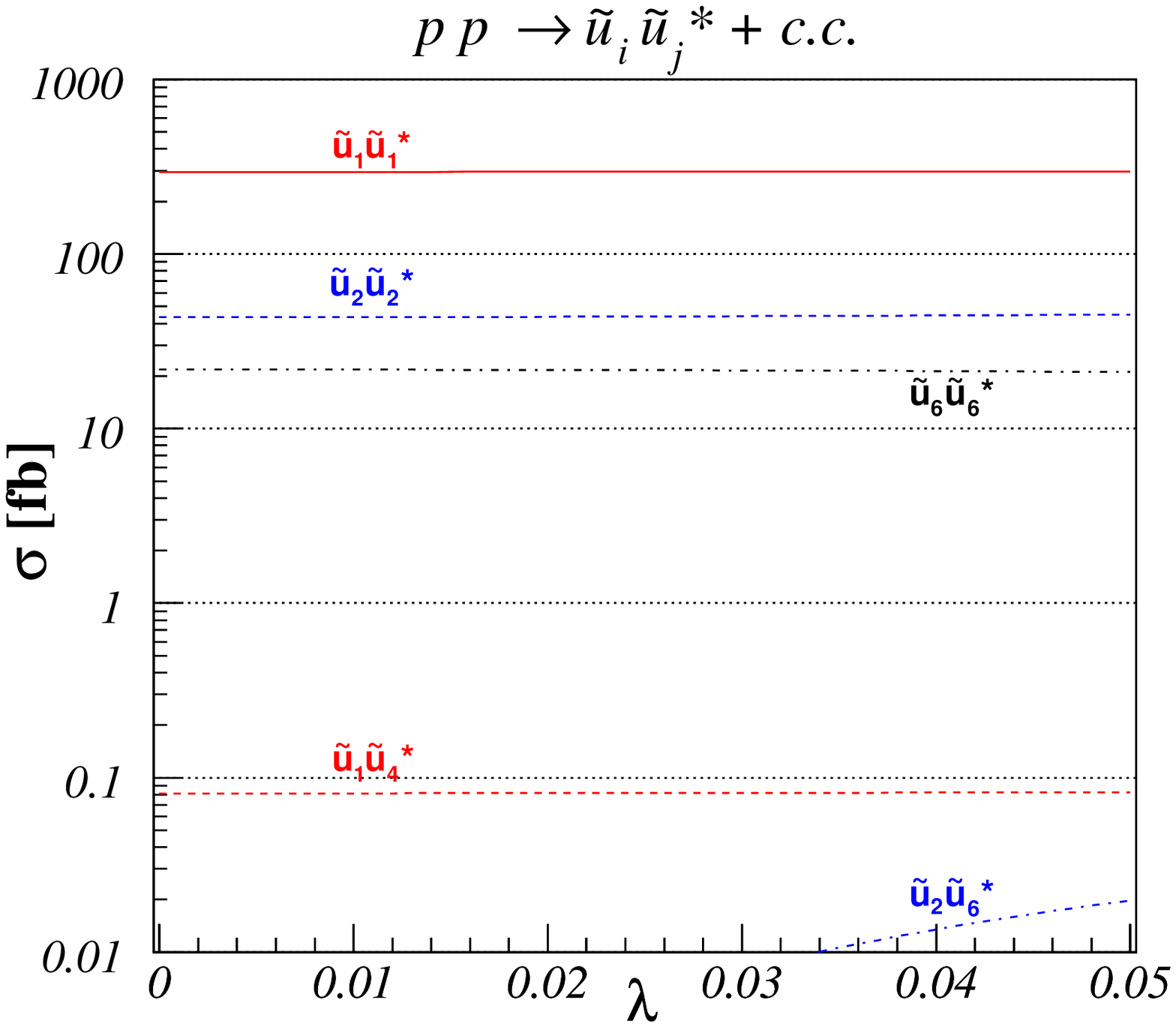}\hspace{2mm}
 \includegraphics[width=0.32\columnwidth]{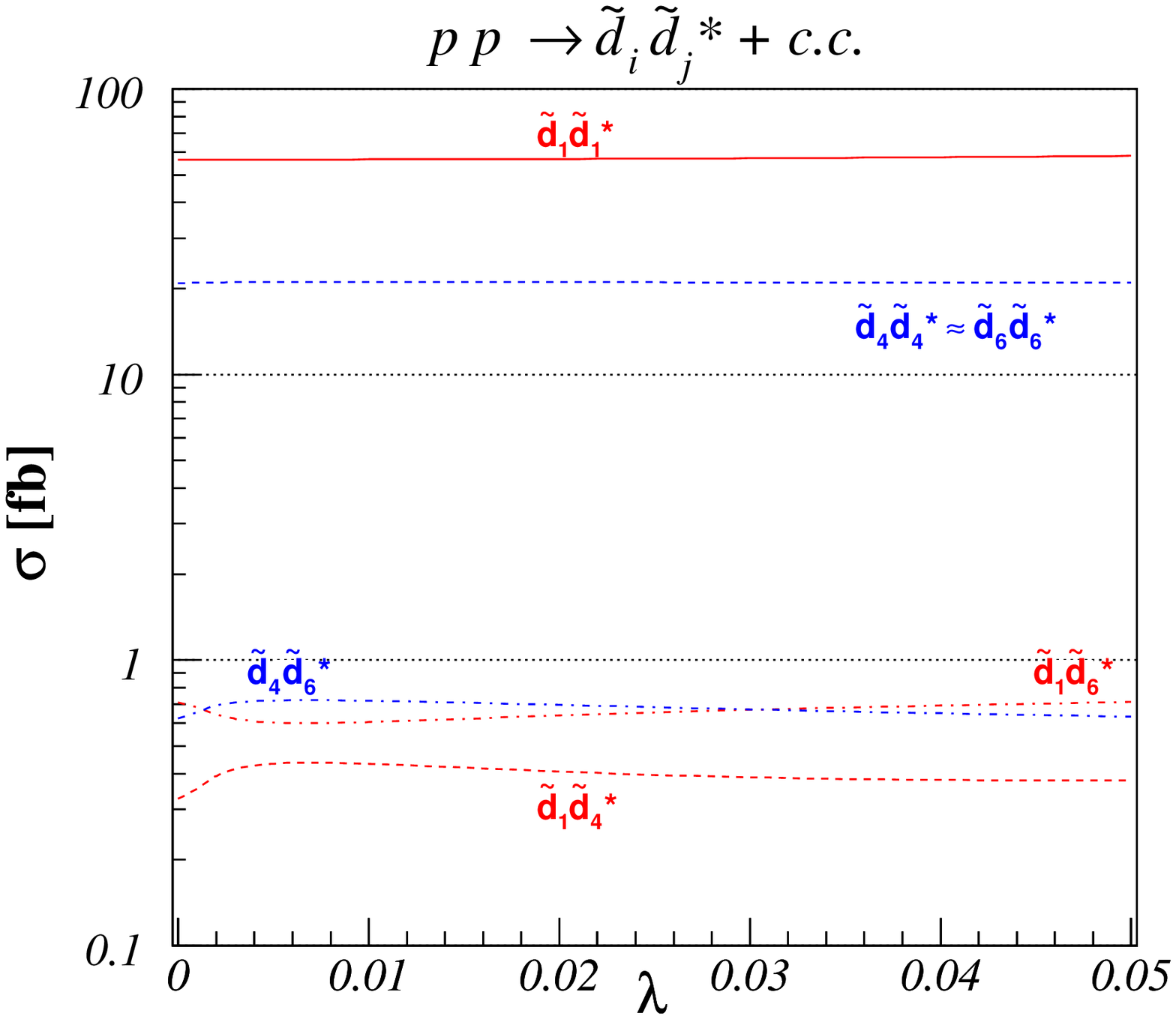}\hspace{2mm}
 \includegraphics[width=0.32\columnwidth]{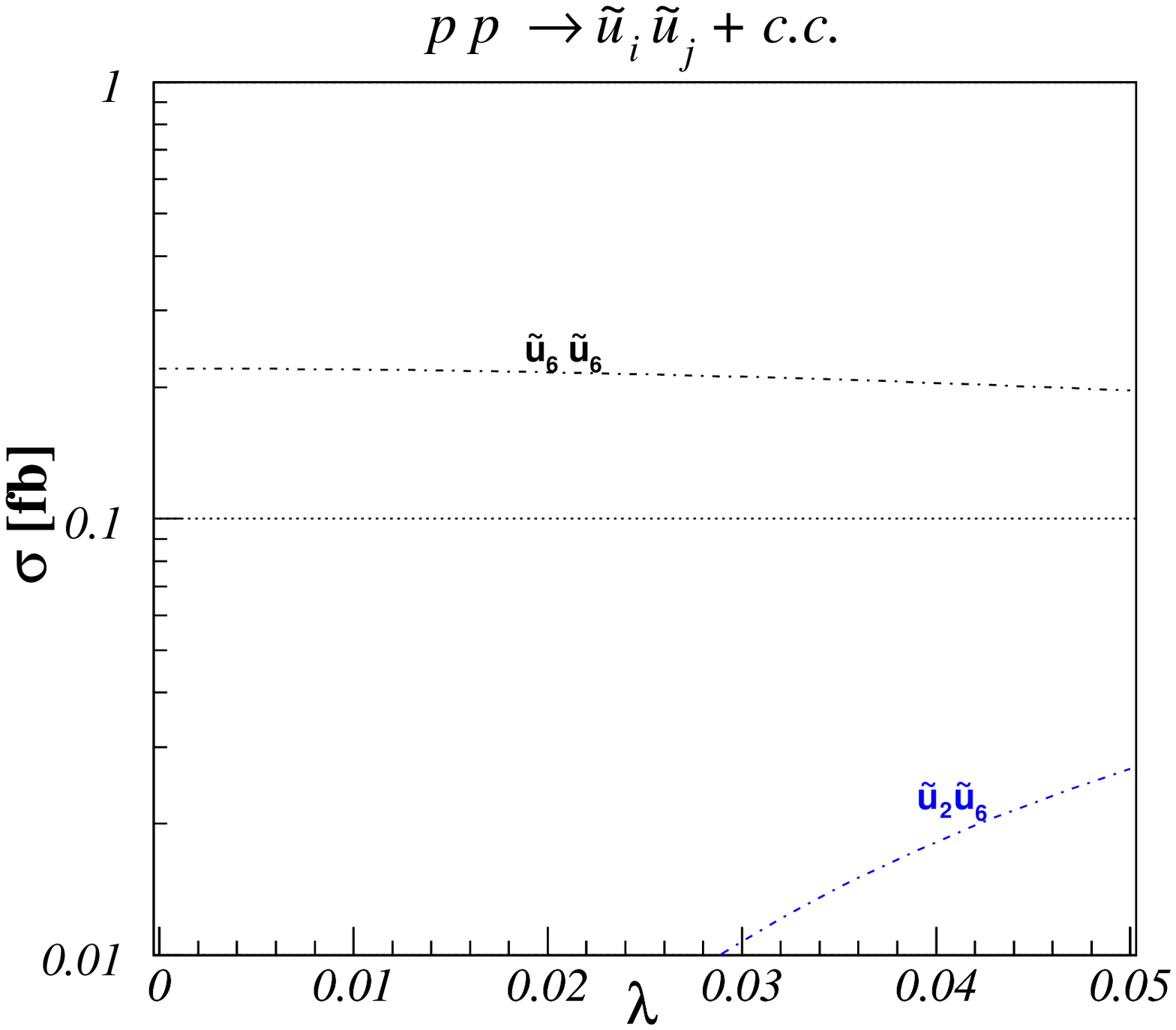}\hspace{2mm}
 \includegraphics[width=0.32\columnwidth]{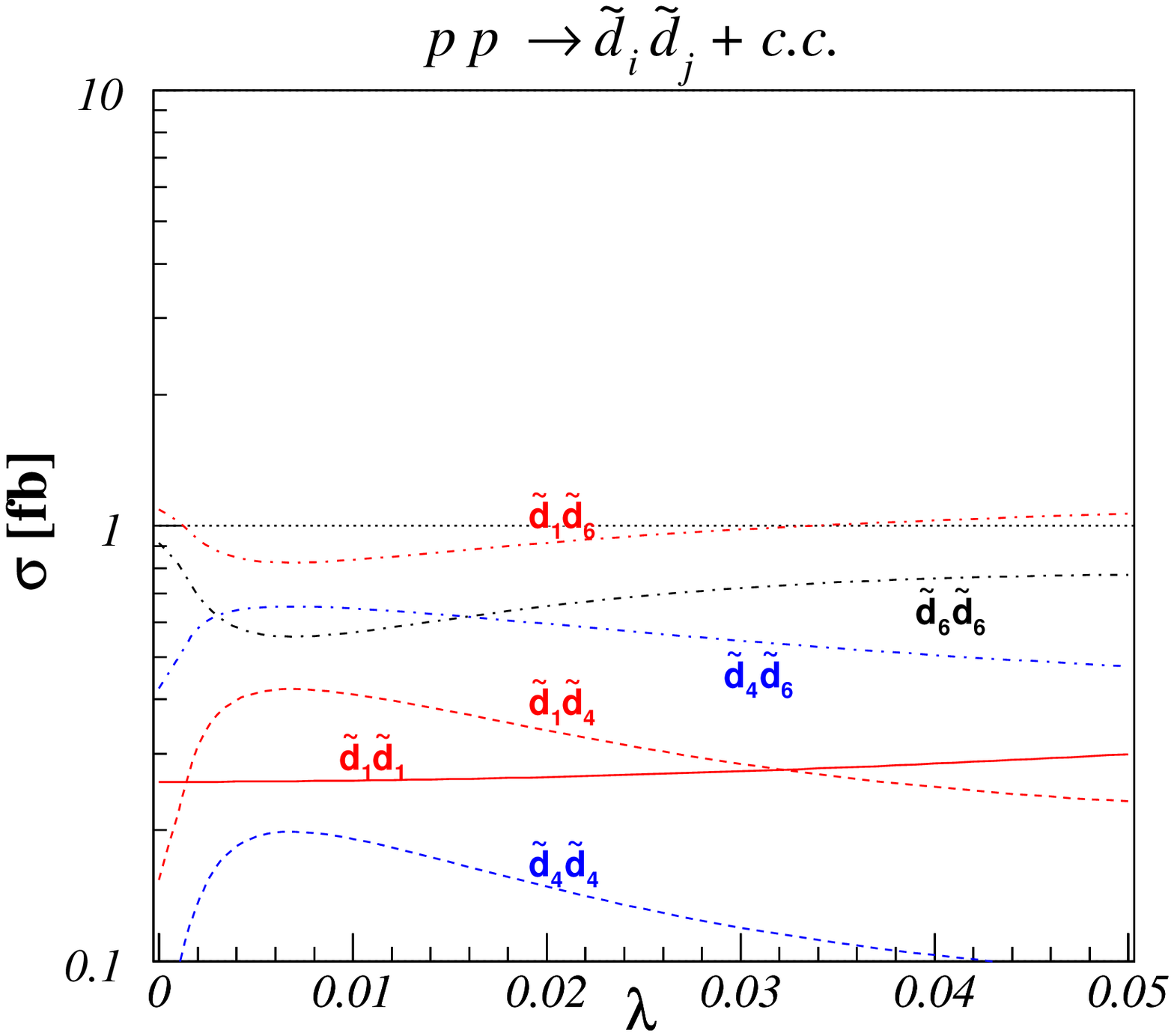}\hspace{2mm}
 \includegraphics[width=0.32\columnwidth]{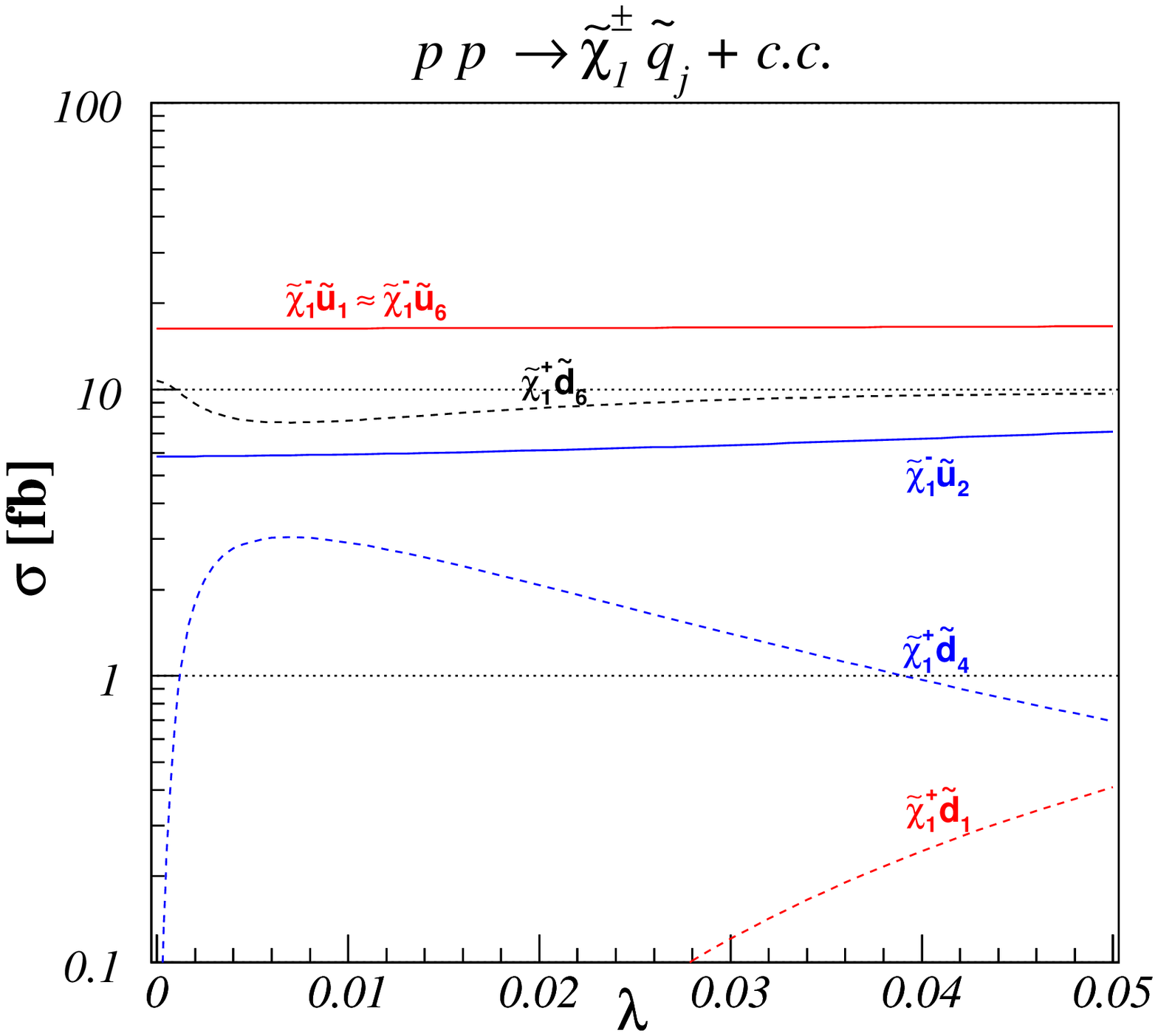}\hspace{2mm}
 \includegraphics[width=0.32\columnwidth]{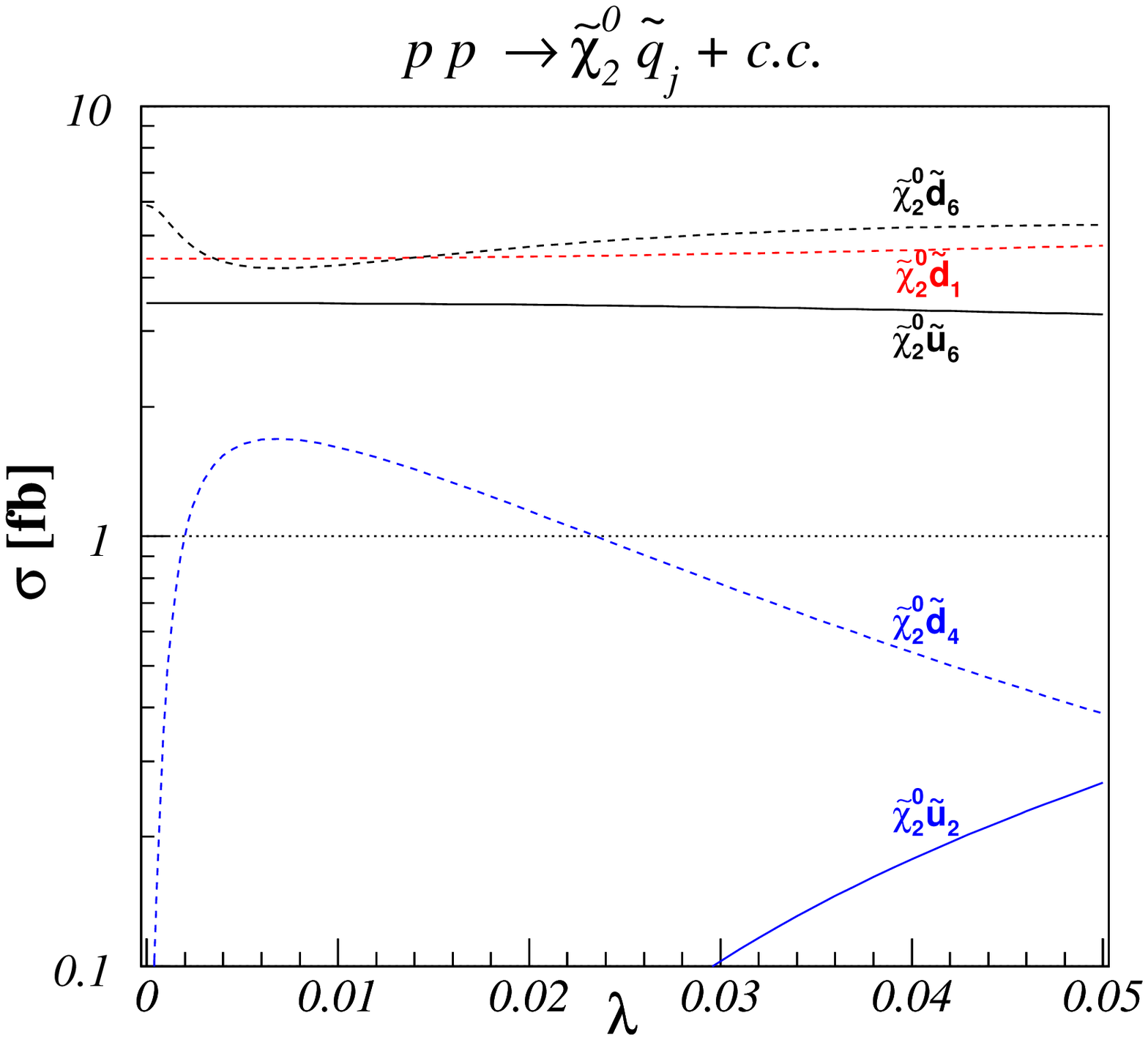}\hspace{2mm}
 \includegraphics[width=0.32\columnwidth]{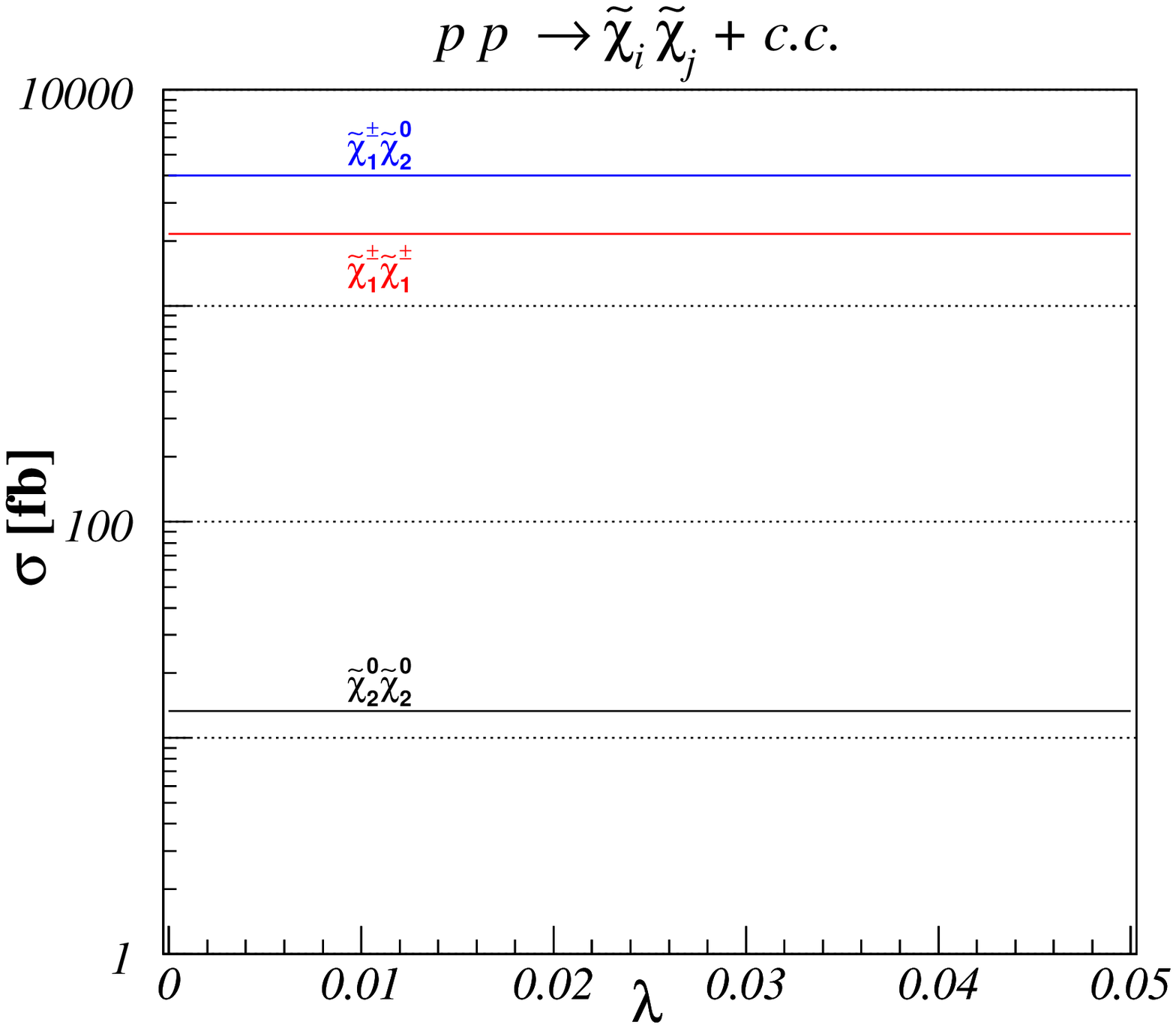}
 \caption{\label{fig:20}Cross sections for charged squark-antisquark (top
          left) and squark-squark (top centre) production, neutral up-type
          (top right) and down-type (centre left) squark-antisquark and
          squark-squark pair (centre and centre right) production,
          associated production of squarks with charginos (bottom left) and
          neutralinos (bottom centre), and gaugino pair production (bottom
          right) at the LHC in our benchmark scenario A.}
\end{figure}
%
%
\begin{figure}
 \centering
 \includegraphics[width=0.32\columnwidth]{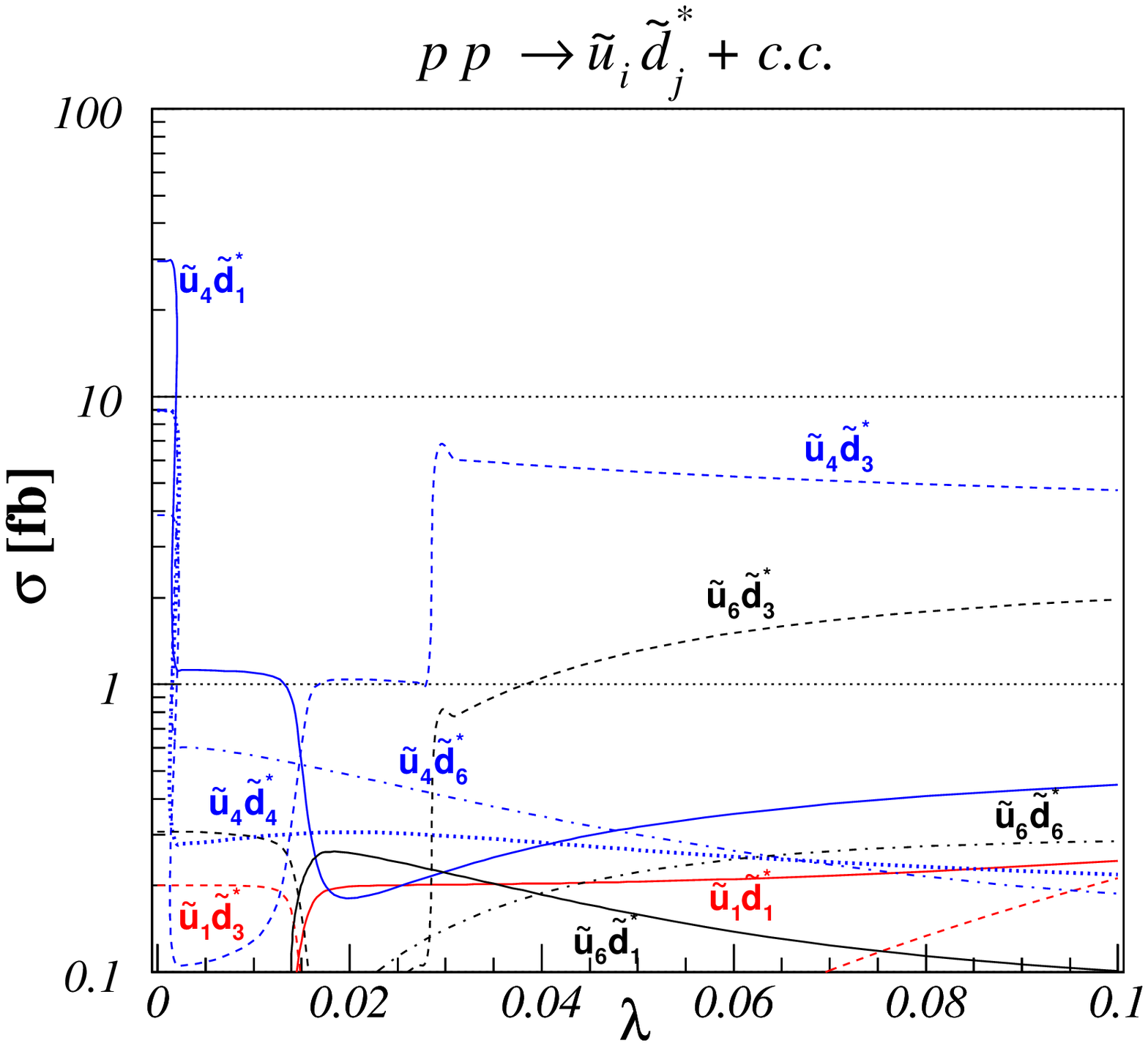}\hspace{2mm}
 \includegraphics[width=0.32\columnwidth]{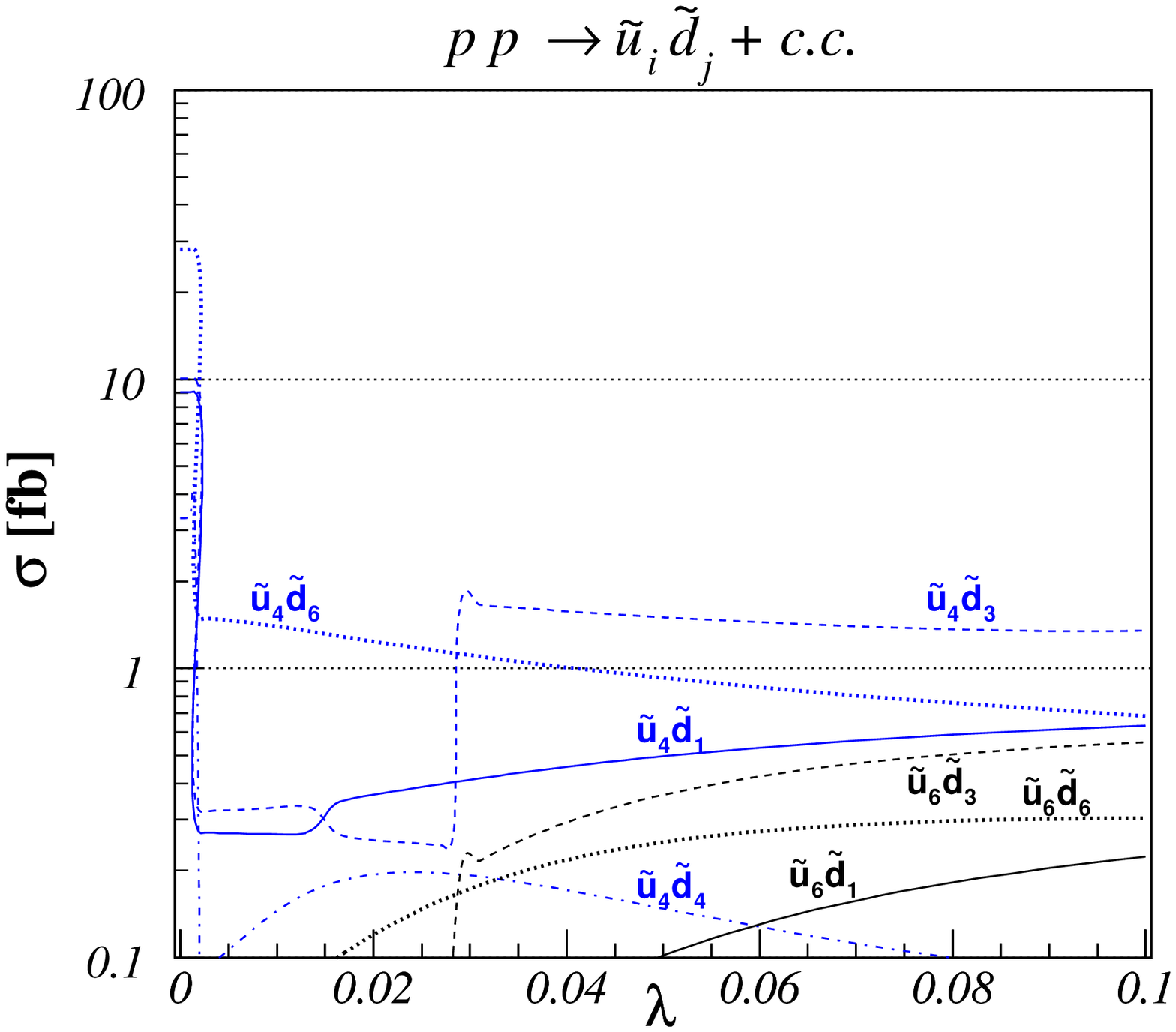}\hspace{2mm}
 \includegraphics[width=0.32\columnwidth]{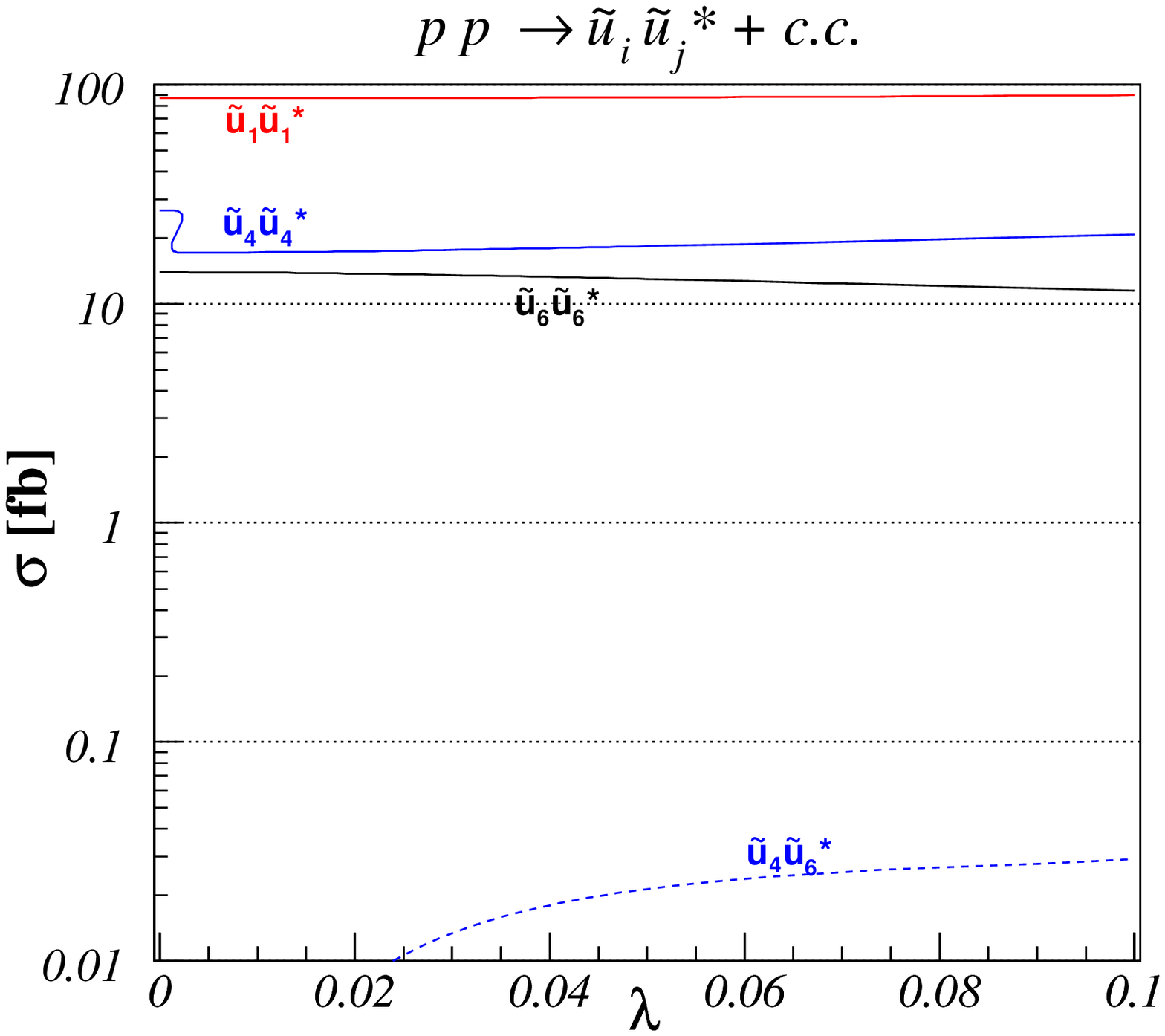}\hspace{2mm}
 \includegraphics[width=0.32\columnwidth]{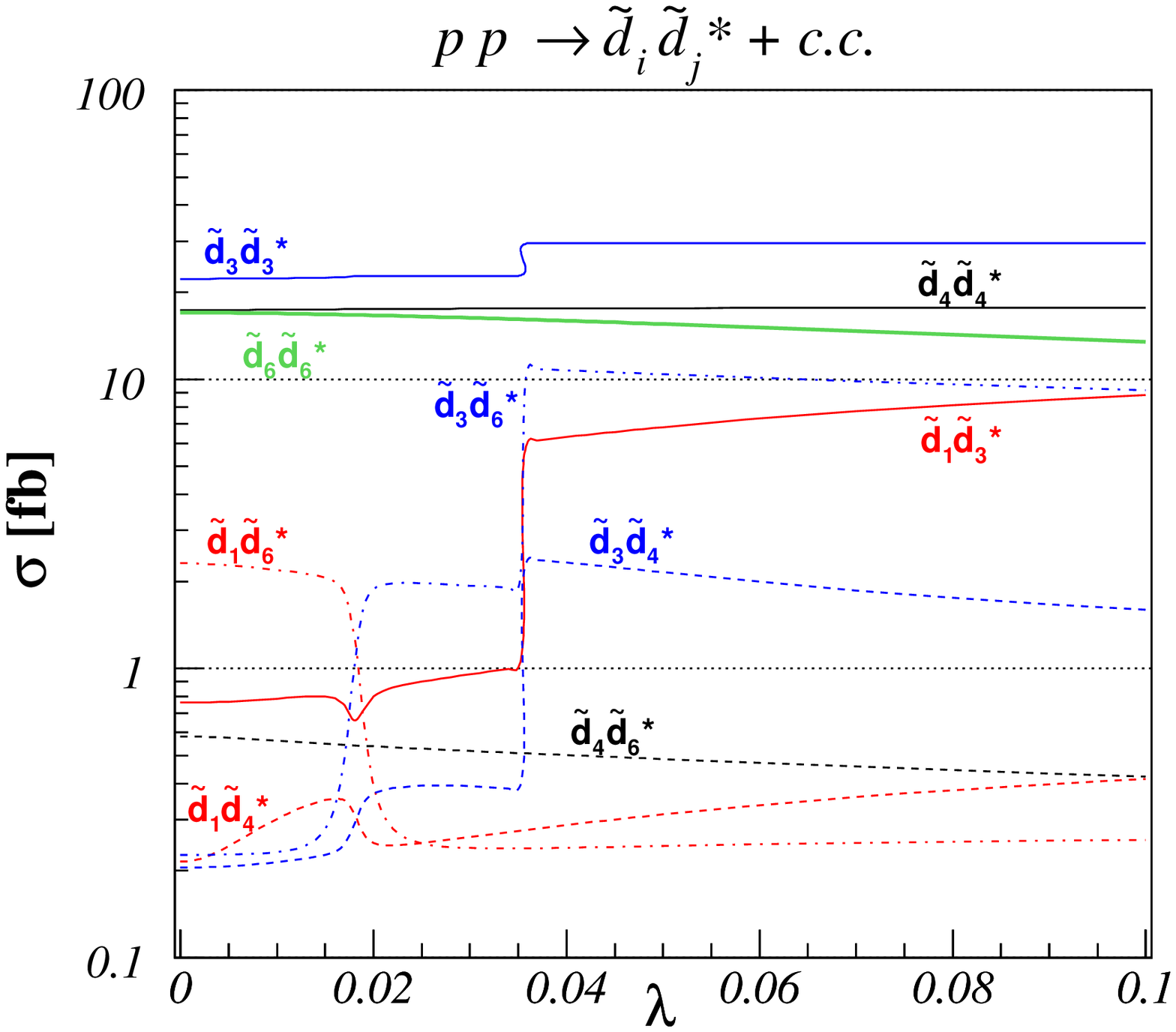}\hspace{2mm}
 \includegraphics[width=0.32\columnwidth]{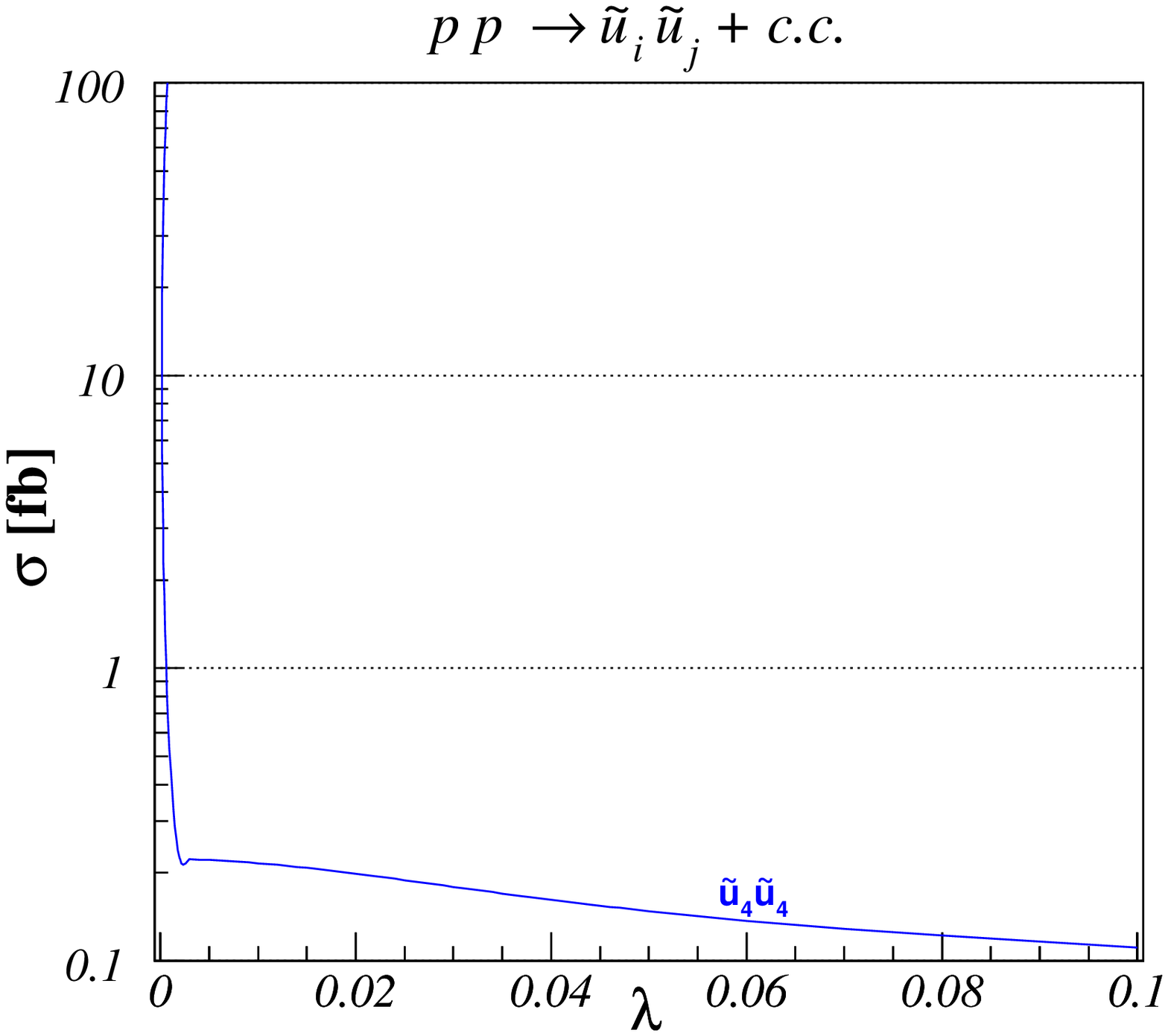}\hspace{2mm}
 \includegraphics[width=0.32\columnwidth]{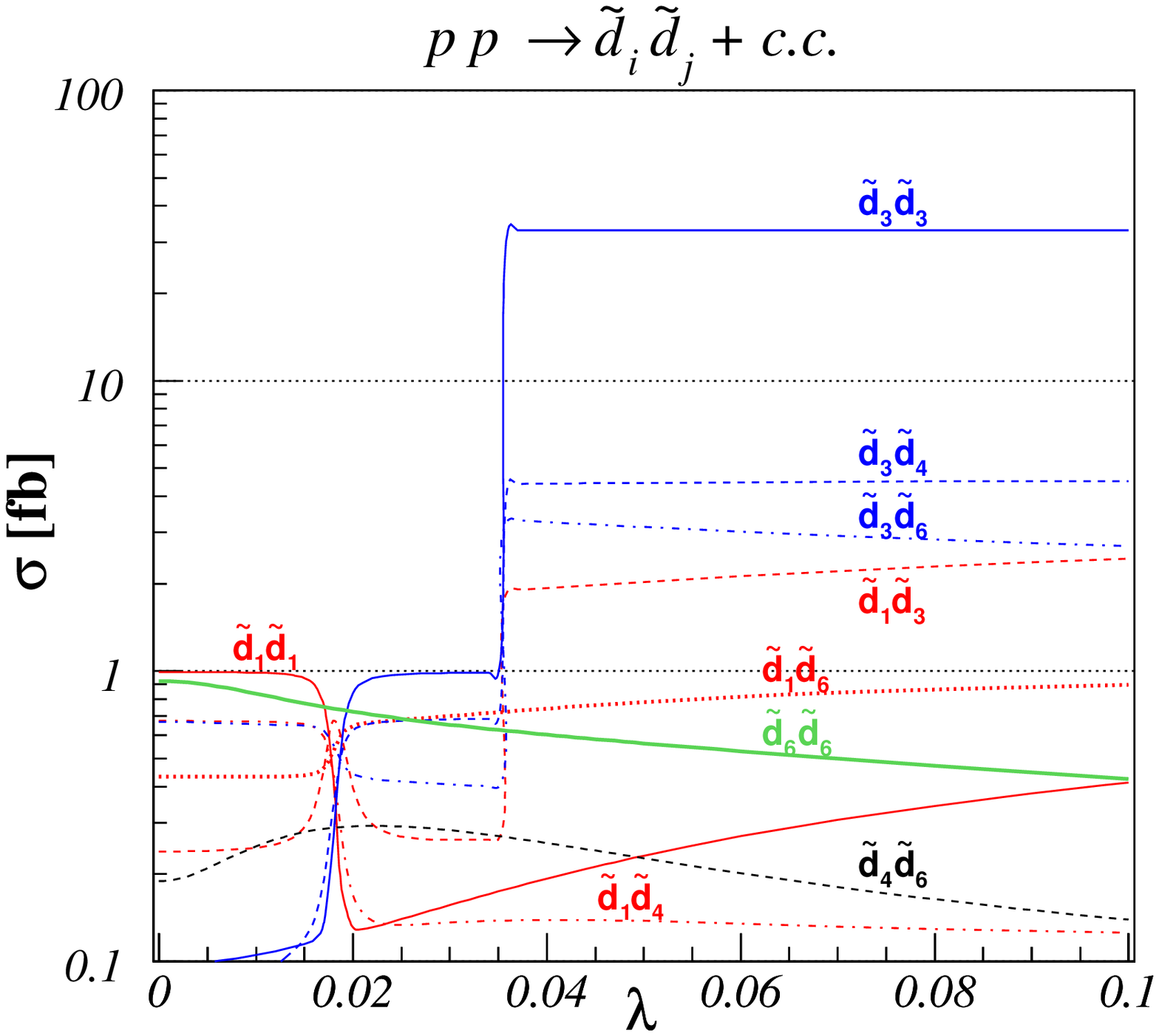}\hspace{2mm}
 \includegraphics[width=0.32\columnwidth]{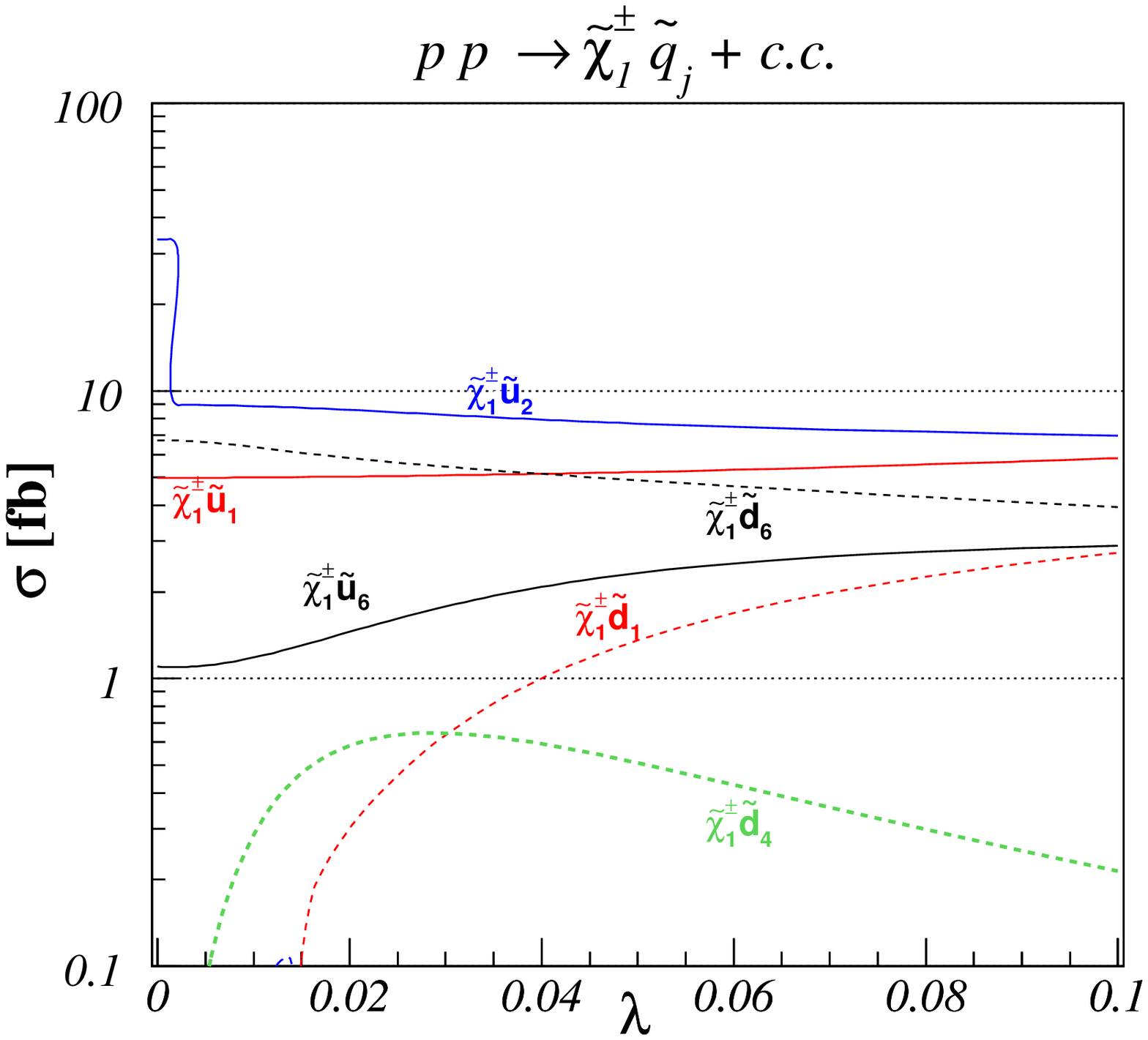}\hspace{2mm}
 \includegraphics[width=0.32\columnwidth]{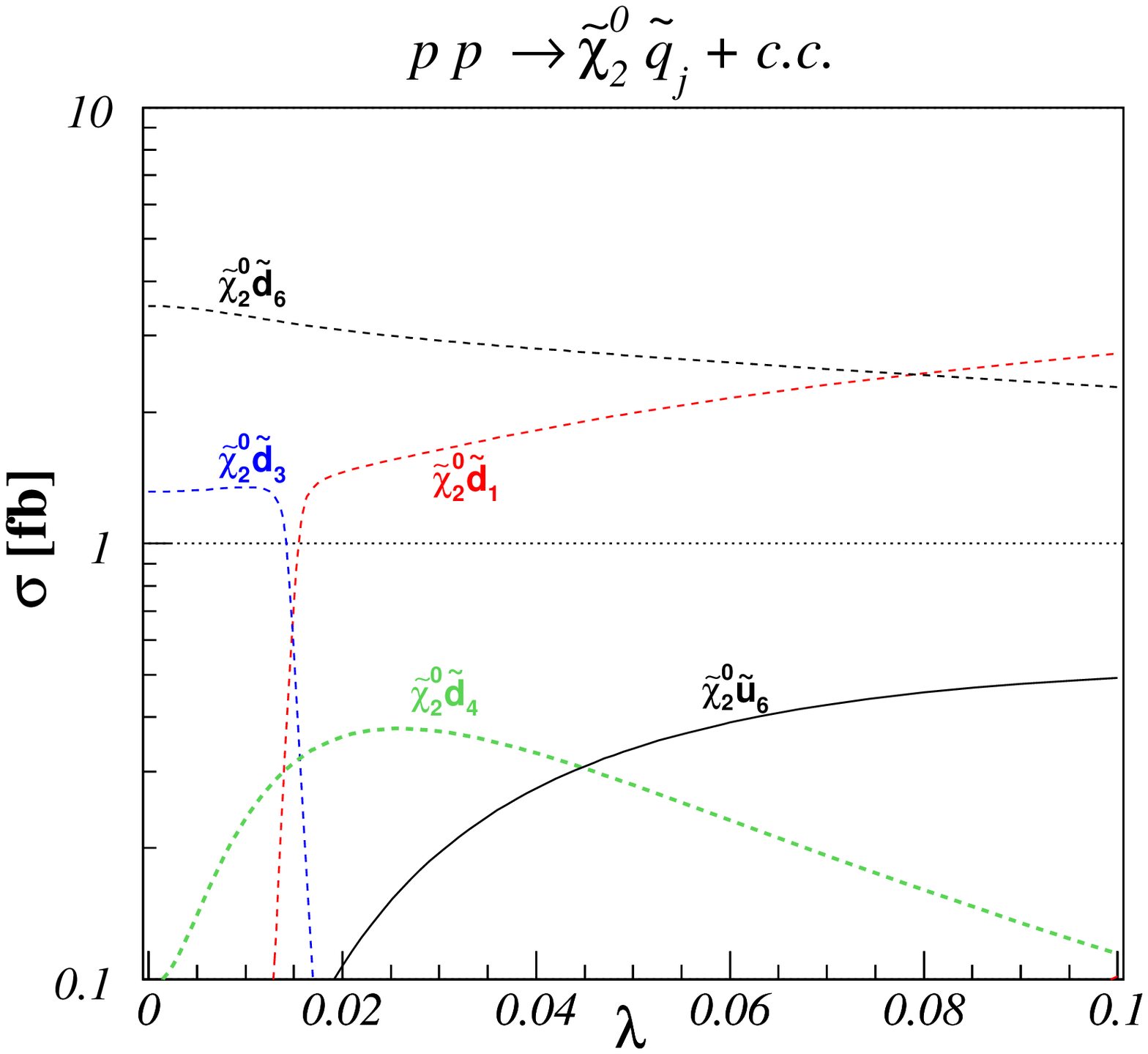}\hspace{2mm}
 \includegraphics[width=0.32\columnwidth]{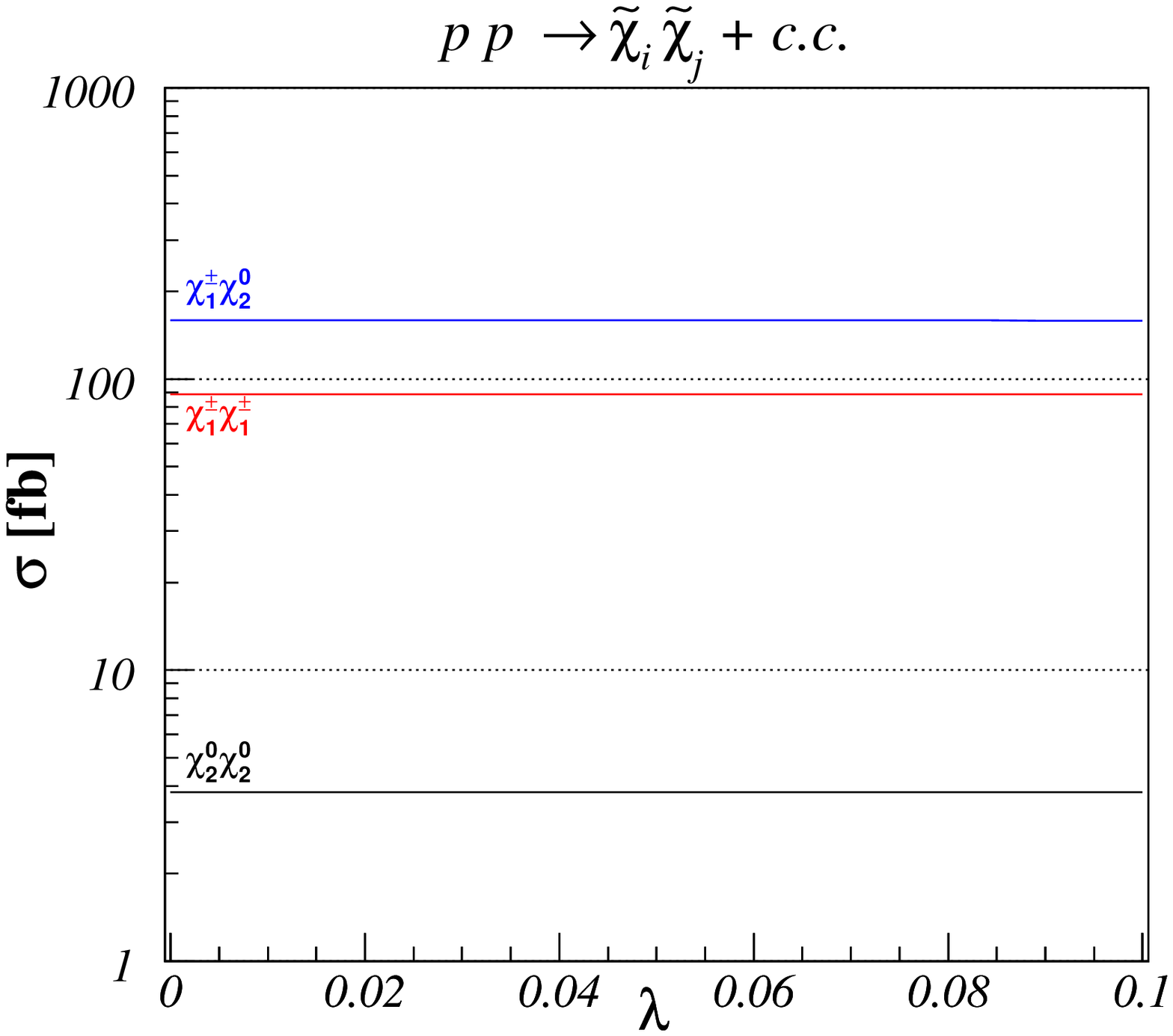}
 \caption{\label{fig:21}Same as Fig.\ \ref{fig:20} for our benchmark
          scenario B.}
\end{figure}
%
%
\begin{figure}
 \centering
 \includegraphics[width=0.32\columnwidth]{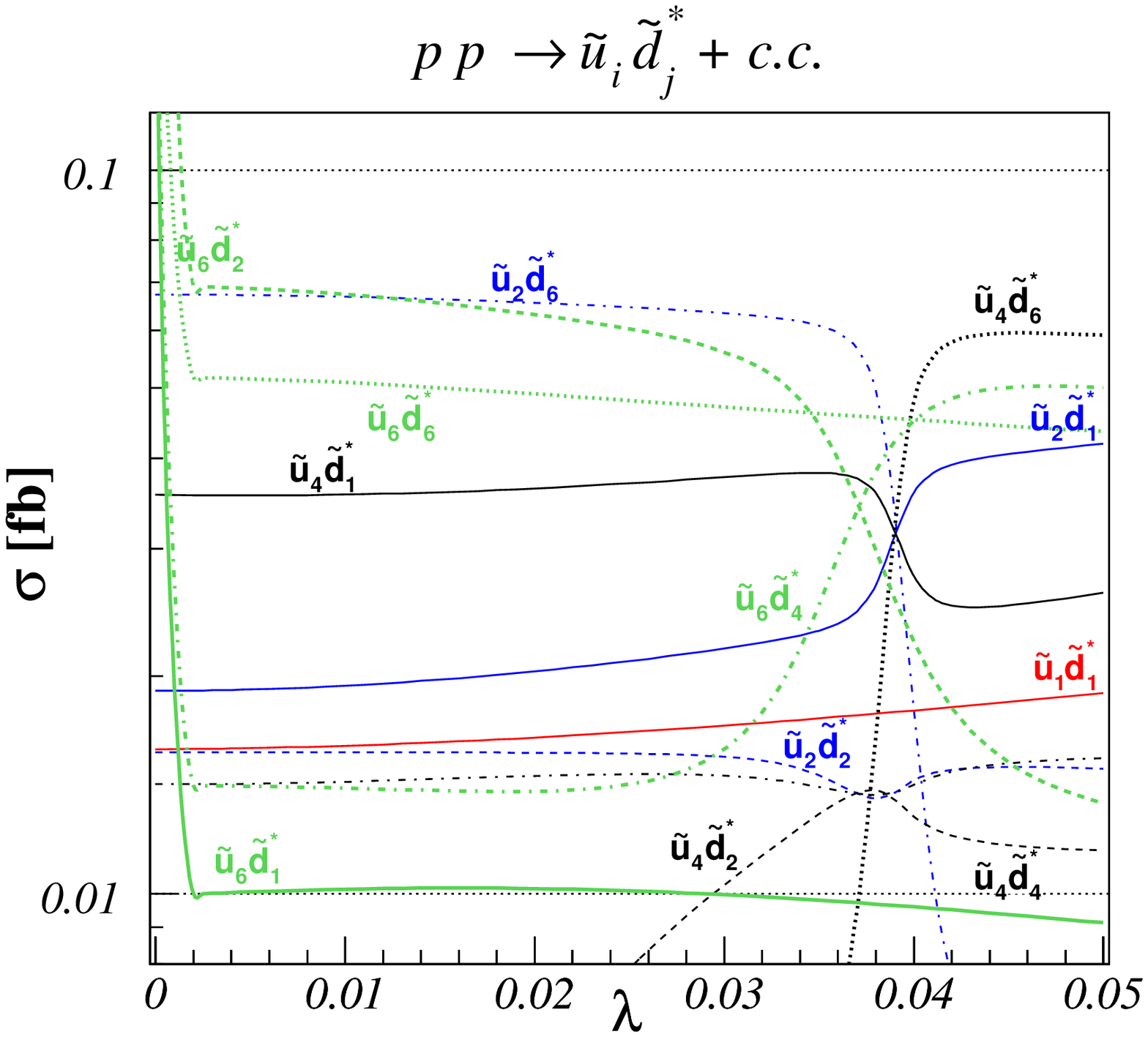}\hspace{2mm}
 \includegraphics[width=0.32\columnwidth]{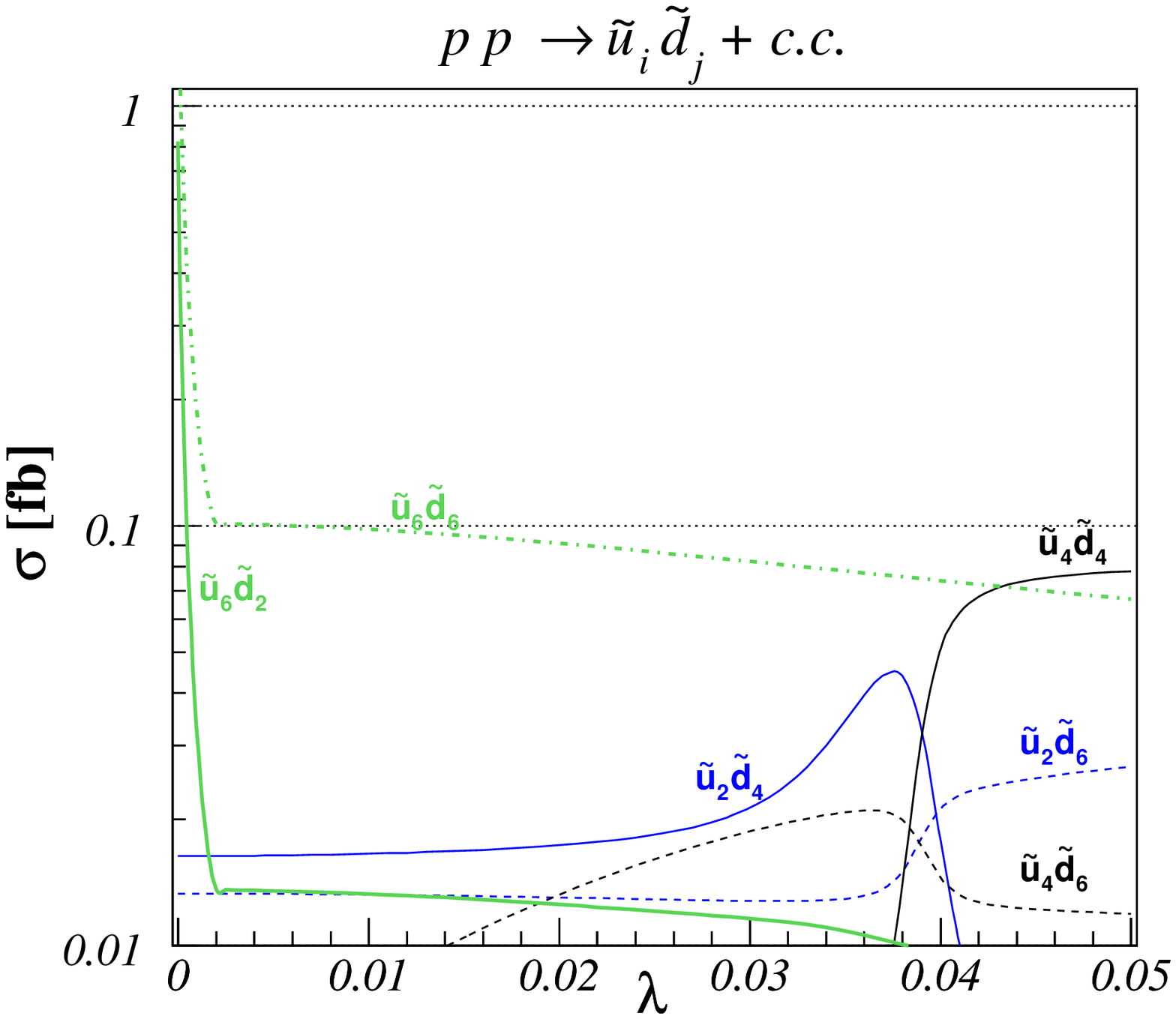}\hspace{2mm}
 \includegraphics[width=0.32\columnwidth]{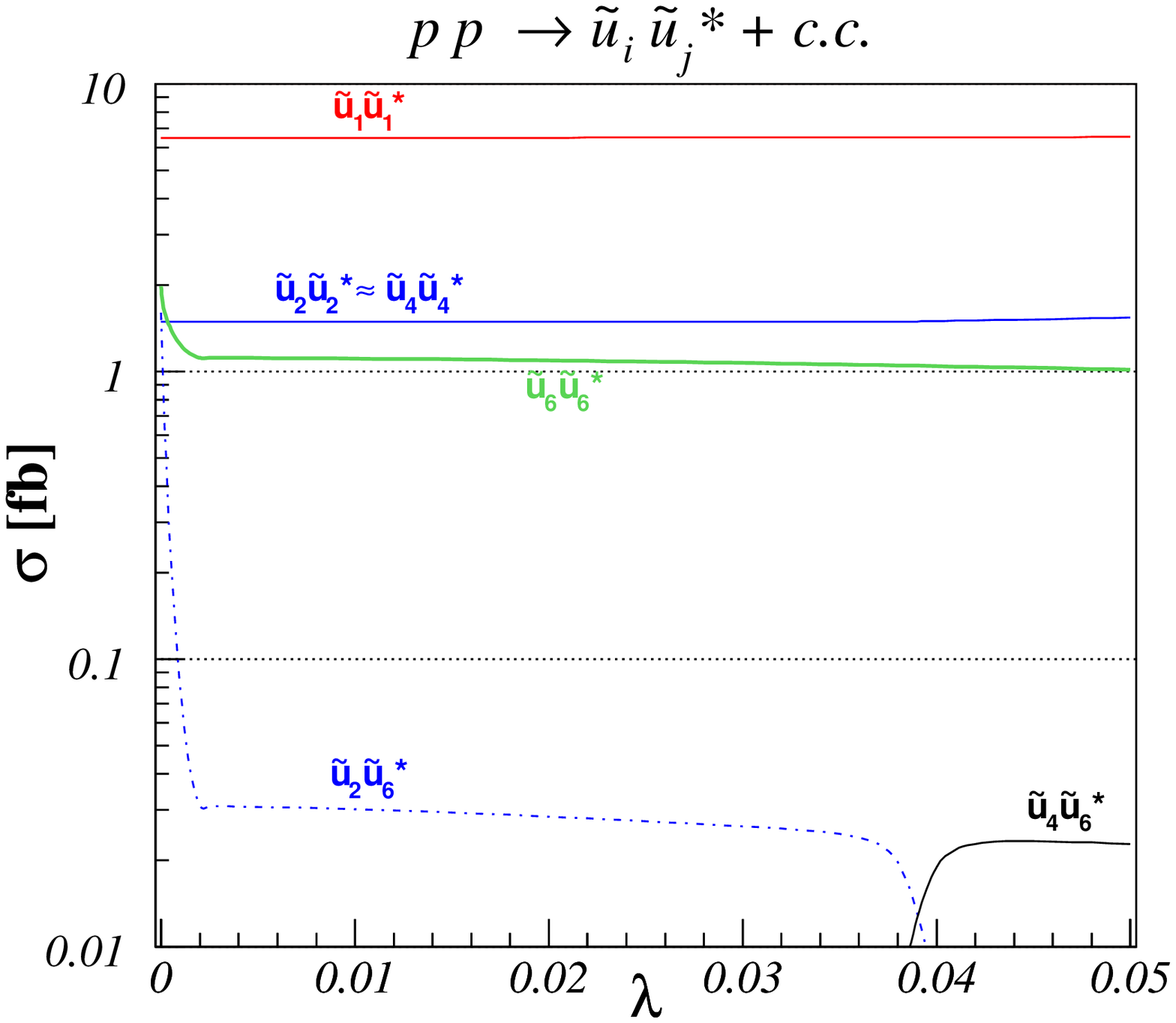}\hspace{2mm}
 \includegraphics[width=0.32\columnwidth]{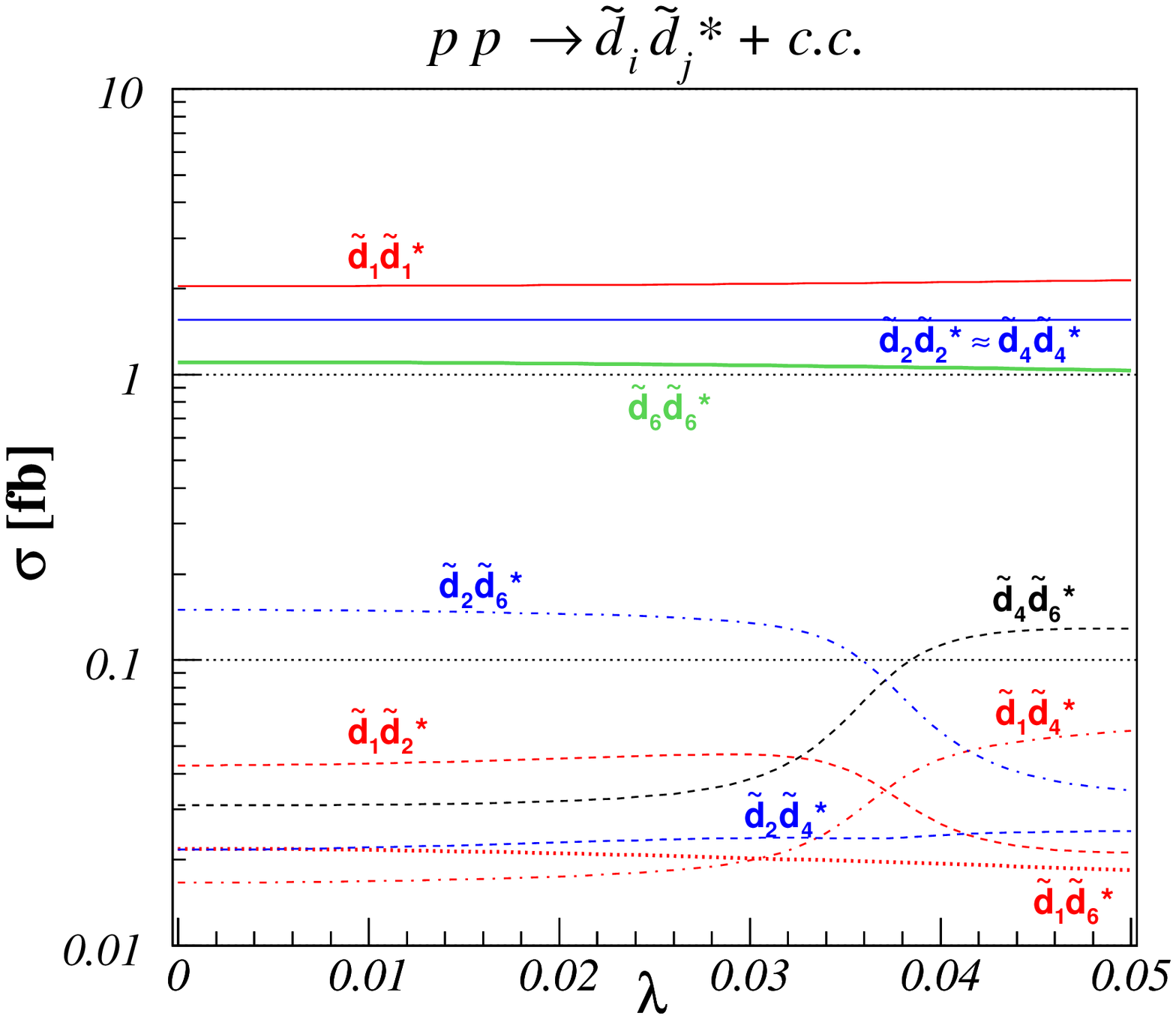}\hspace{2mm}
 \includegraphics[width=0.32\columnwidth]{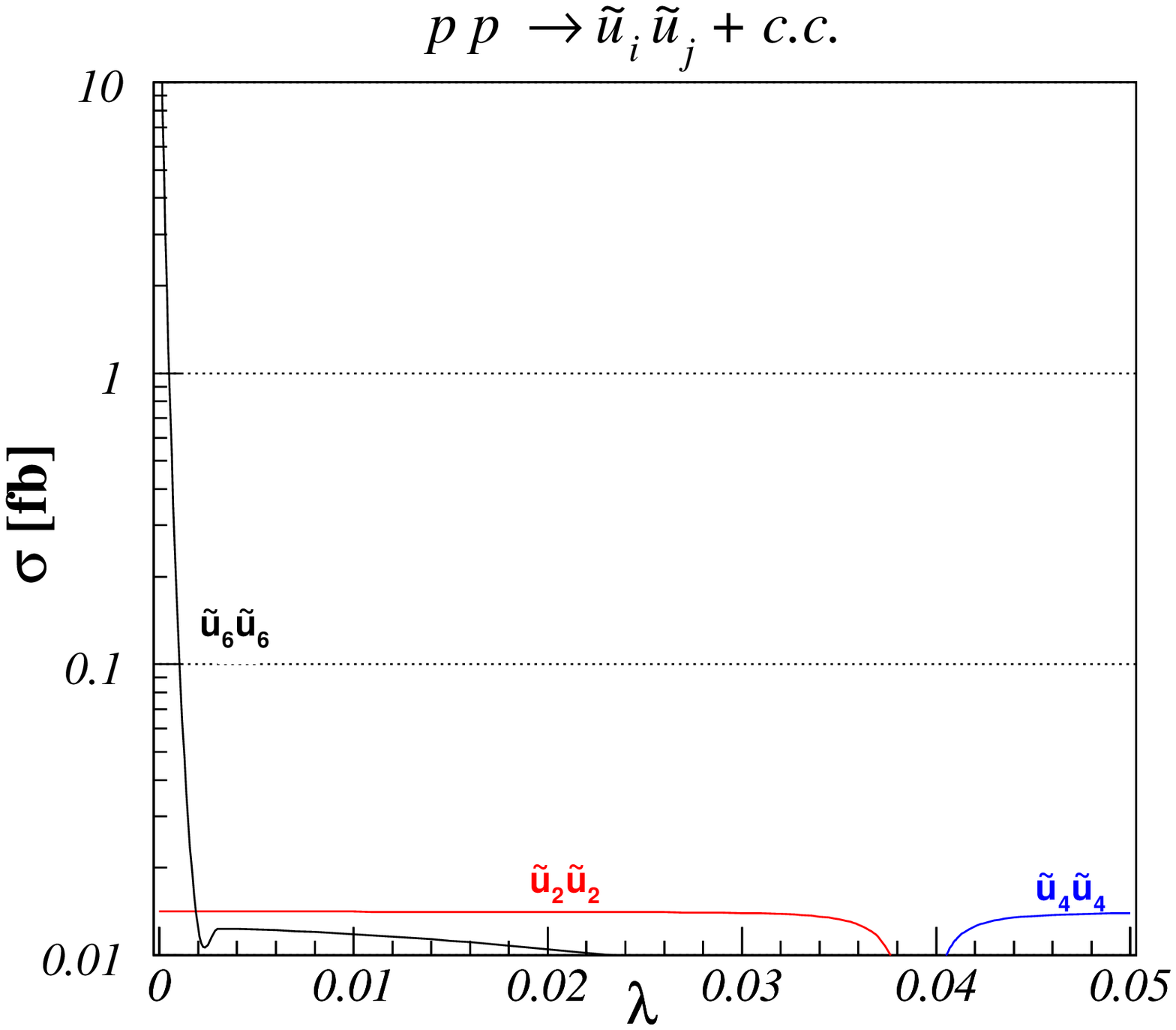}\hspace{2mm}
 \includegraphics[width=0.32\columnwidth]{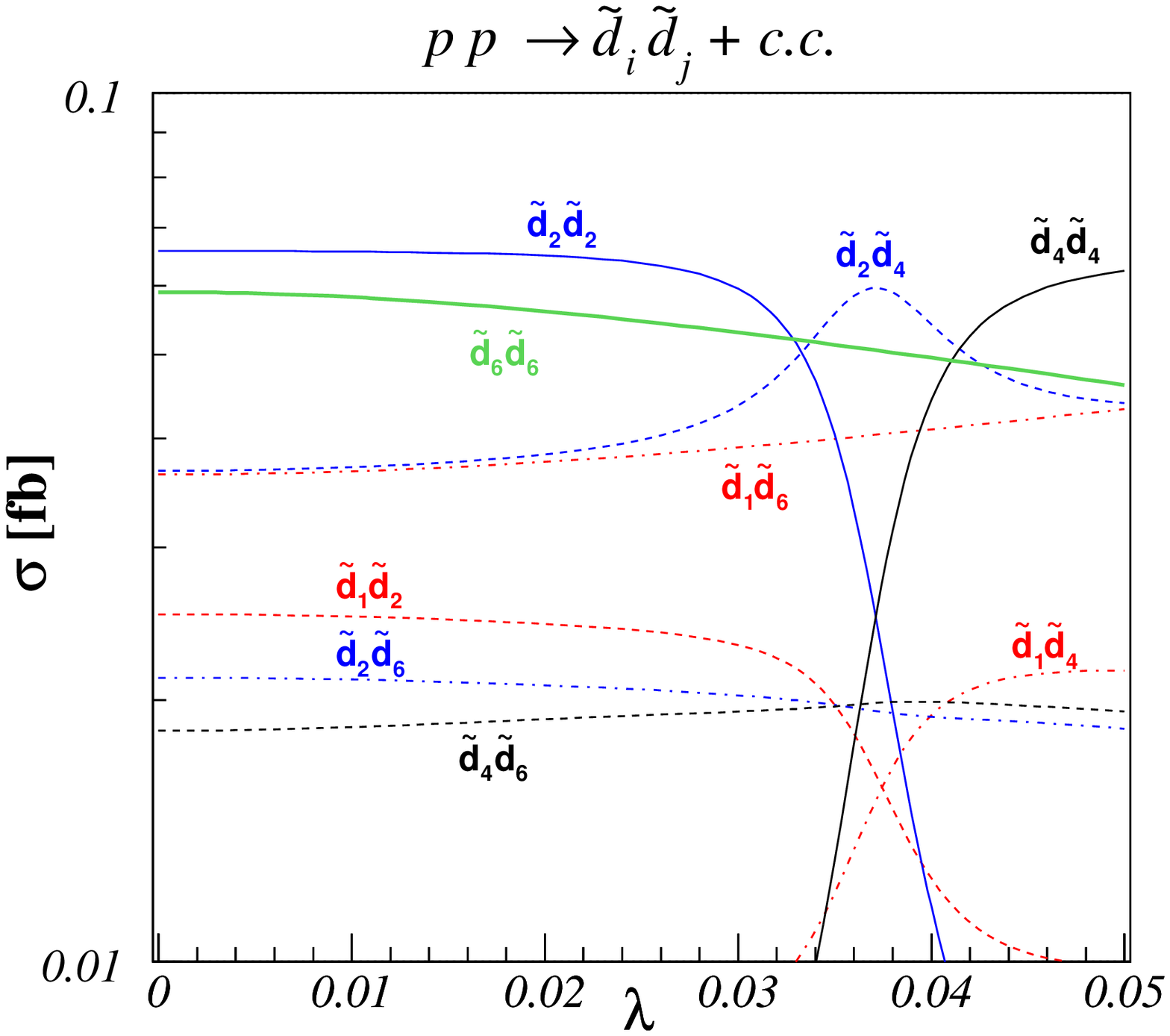}\hspace{2mm}
 \includegraphics[width=0.32\columnwidth]{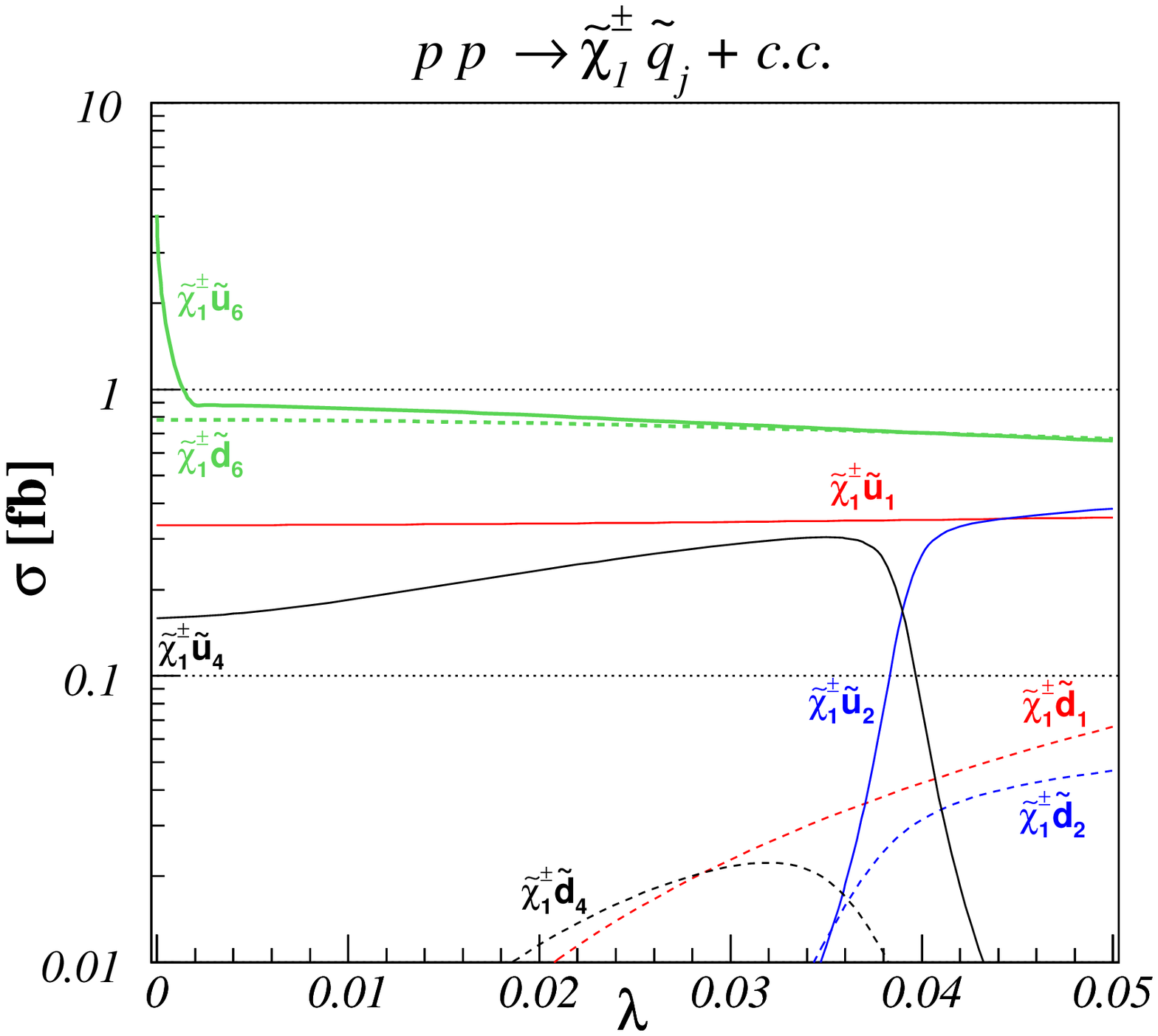}\hspace{2mm}
 \includegraphics[width=0.32\columnwidth]{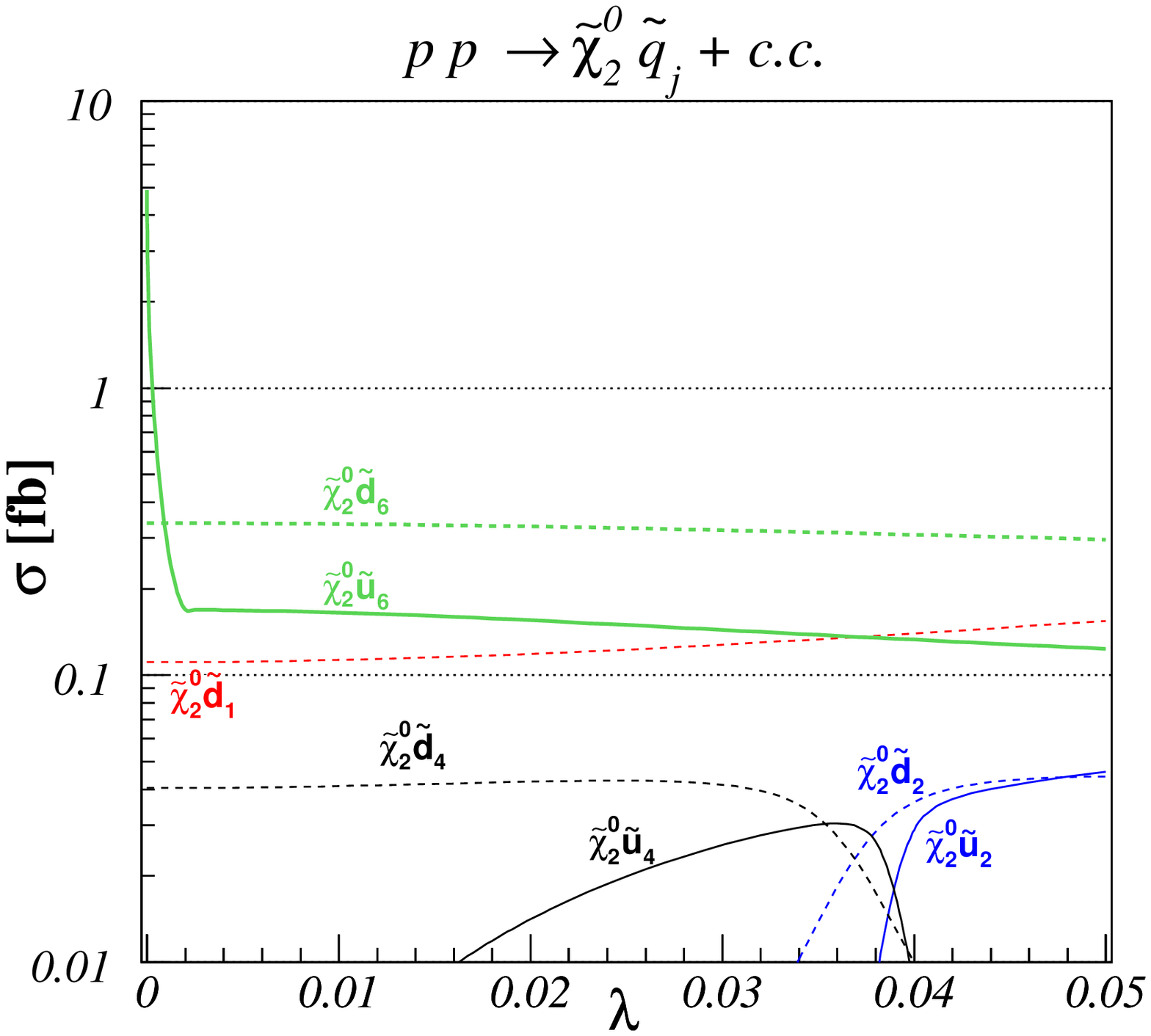}\hspace{2mm}
 \includegraphics[width=0.32\columnwidth]{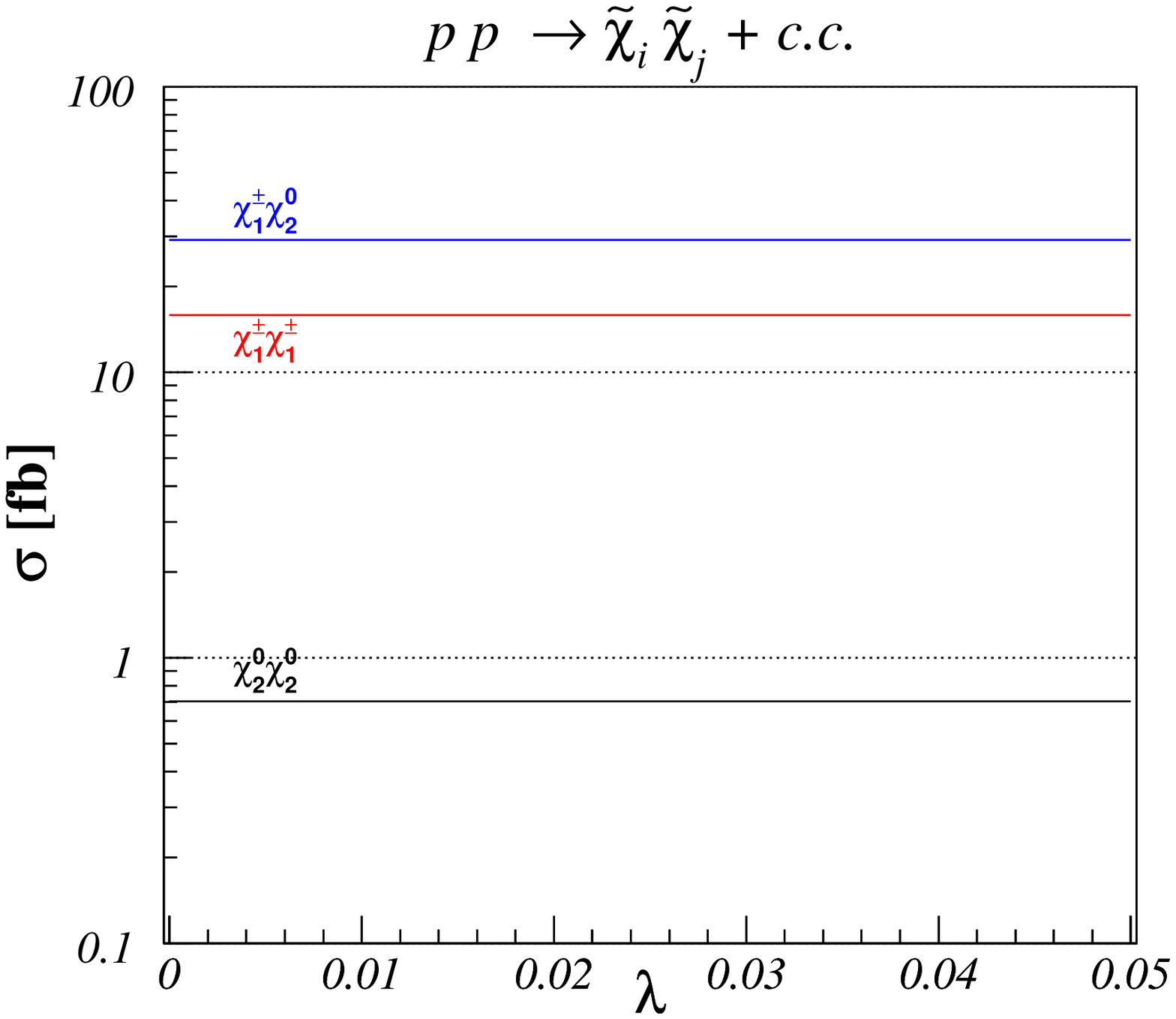}
 \caption{\label{fig:22}Same as Fig.\ \ref{fig:20} for our benchmark
          scenario C.}
\end{figure}
%
%
\begin{figure}
 \centering
 \includegraphics[width=0.32\columnwidth]{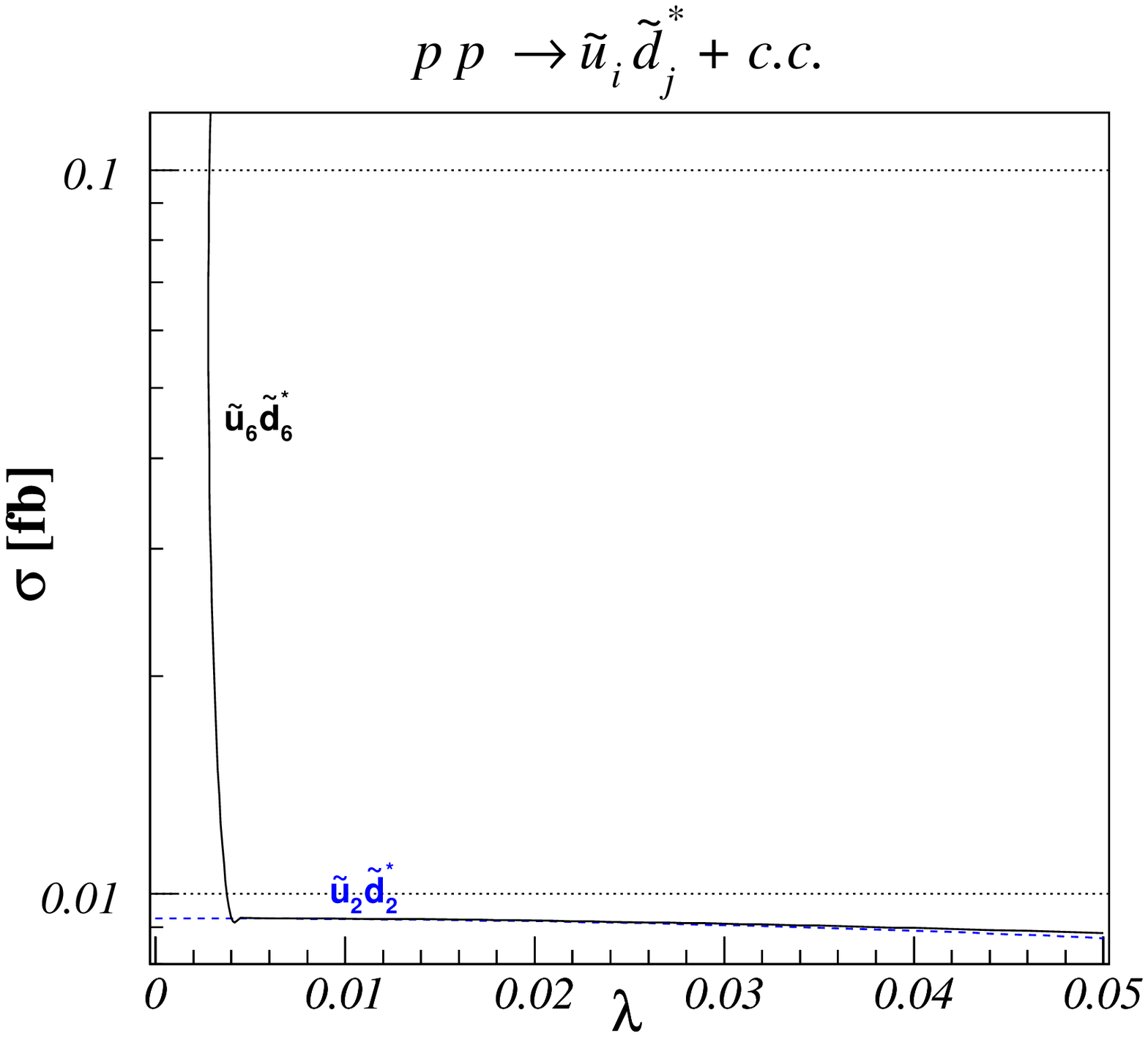}\hspace{2mm}
 \includegraphics[width=0.32\columnwidth]{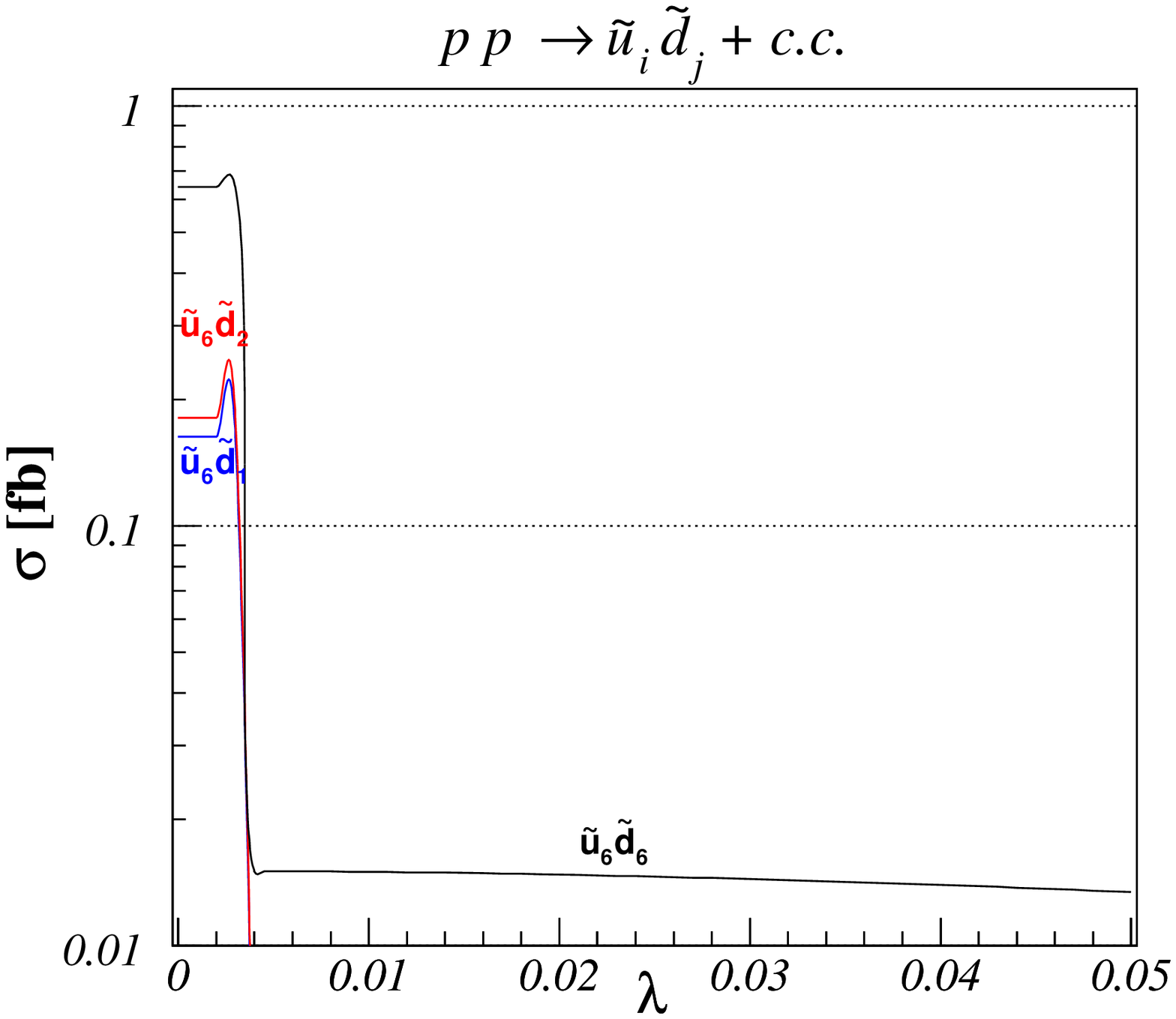}\hspace{2mm}
 \includegraphics[width=0.32\columnwidth]{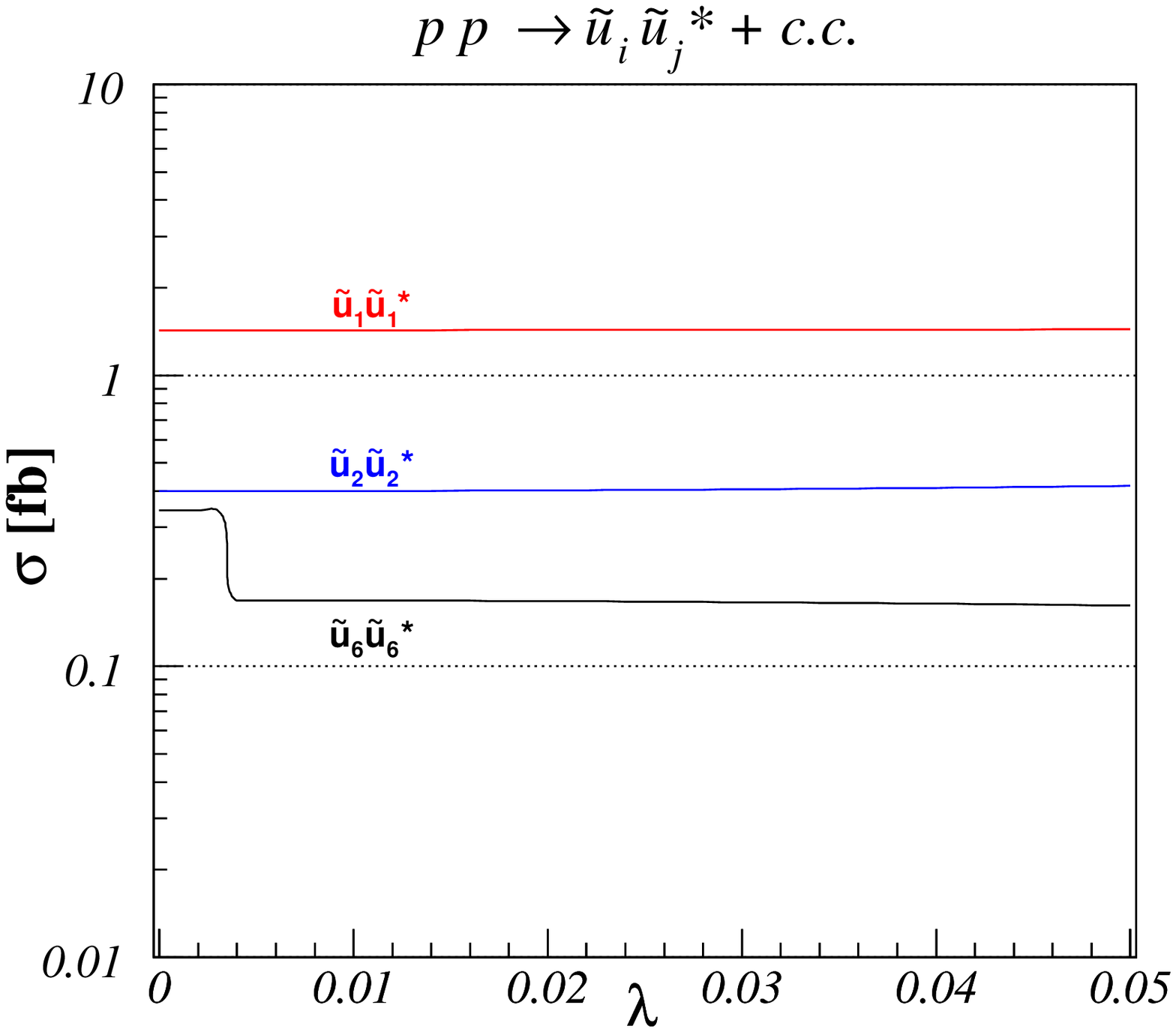}\hspace{2mm}
 \includegraphics[width=0.32\columnwidth]{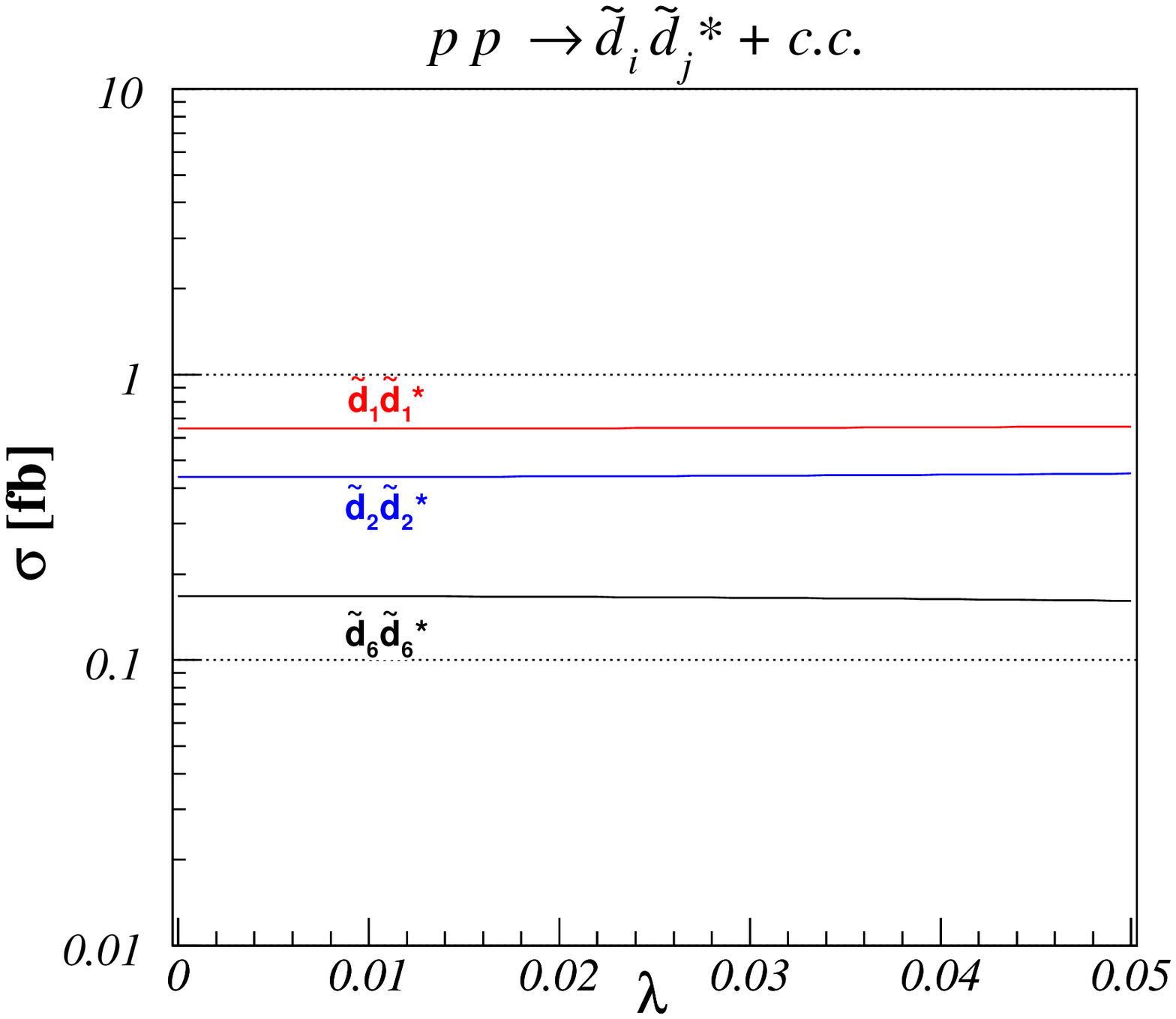}\hspace{2mm}
 \includegraphics[width=0.32\columnwidth]{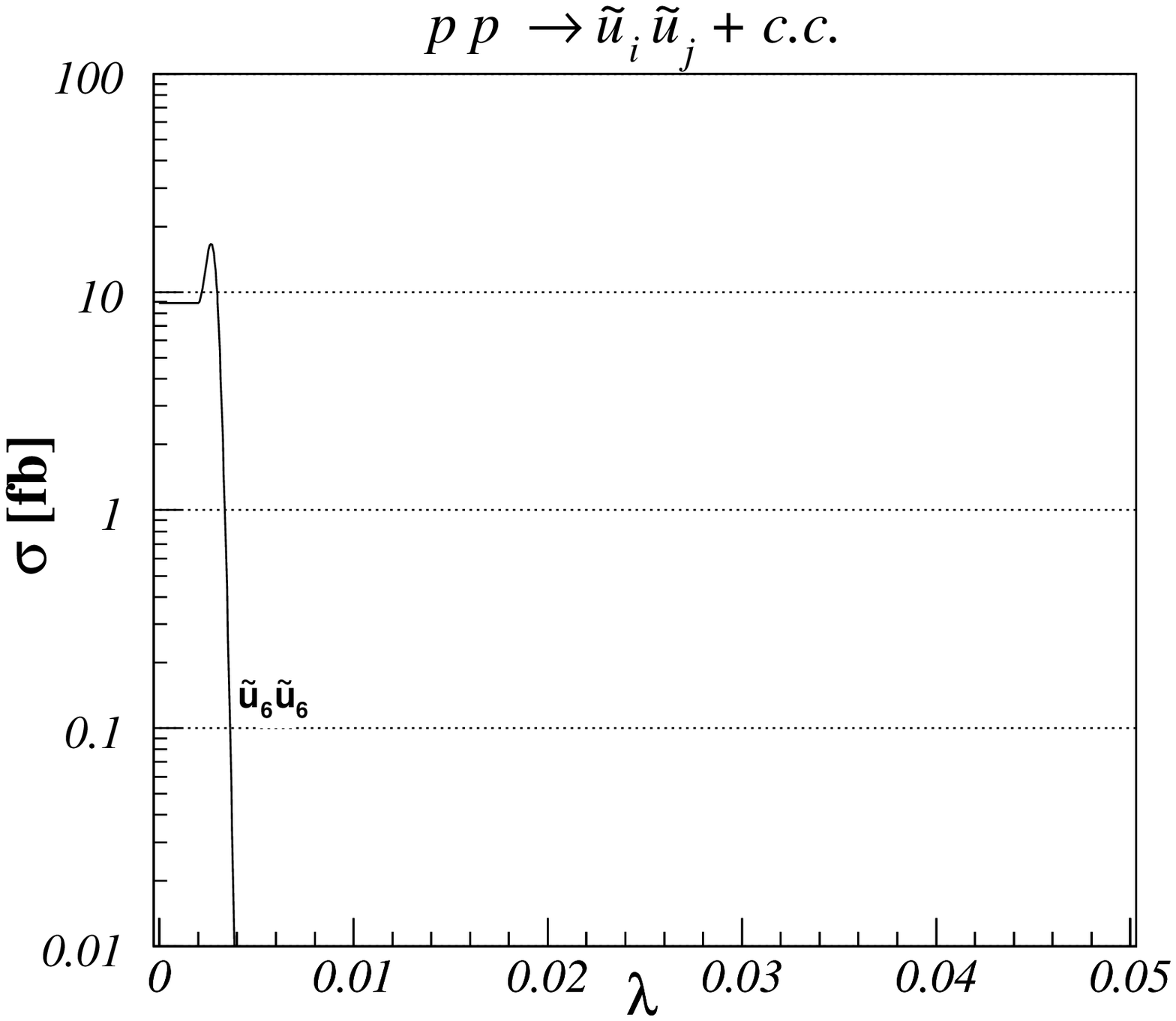}\hspace{2mm}
 \includegraphics[width=0.32\columnwidth]{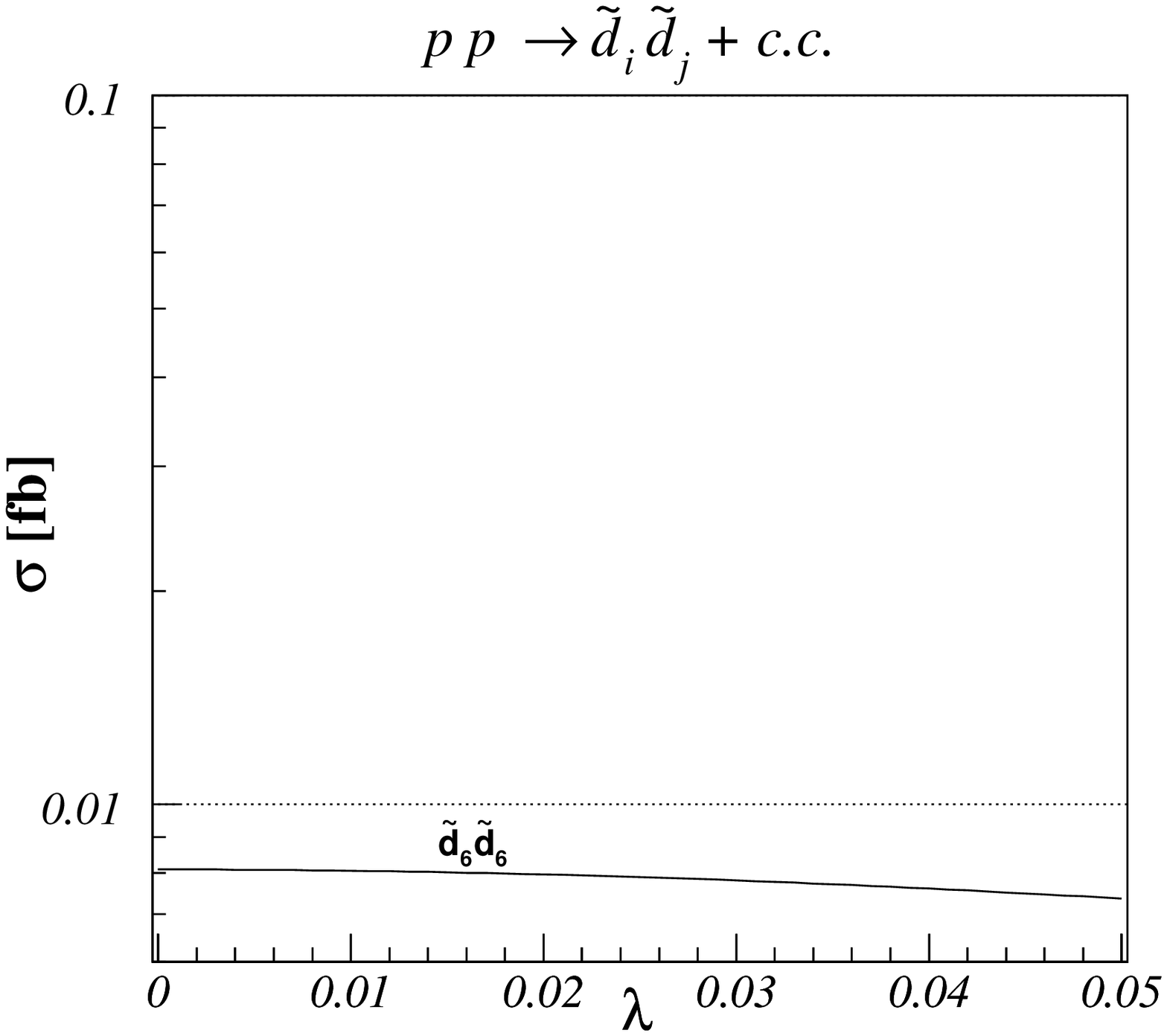}\hspace{2mm}
 \includegraphics[width=0.32\columnwidth]{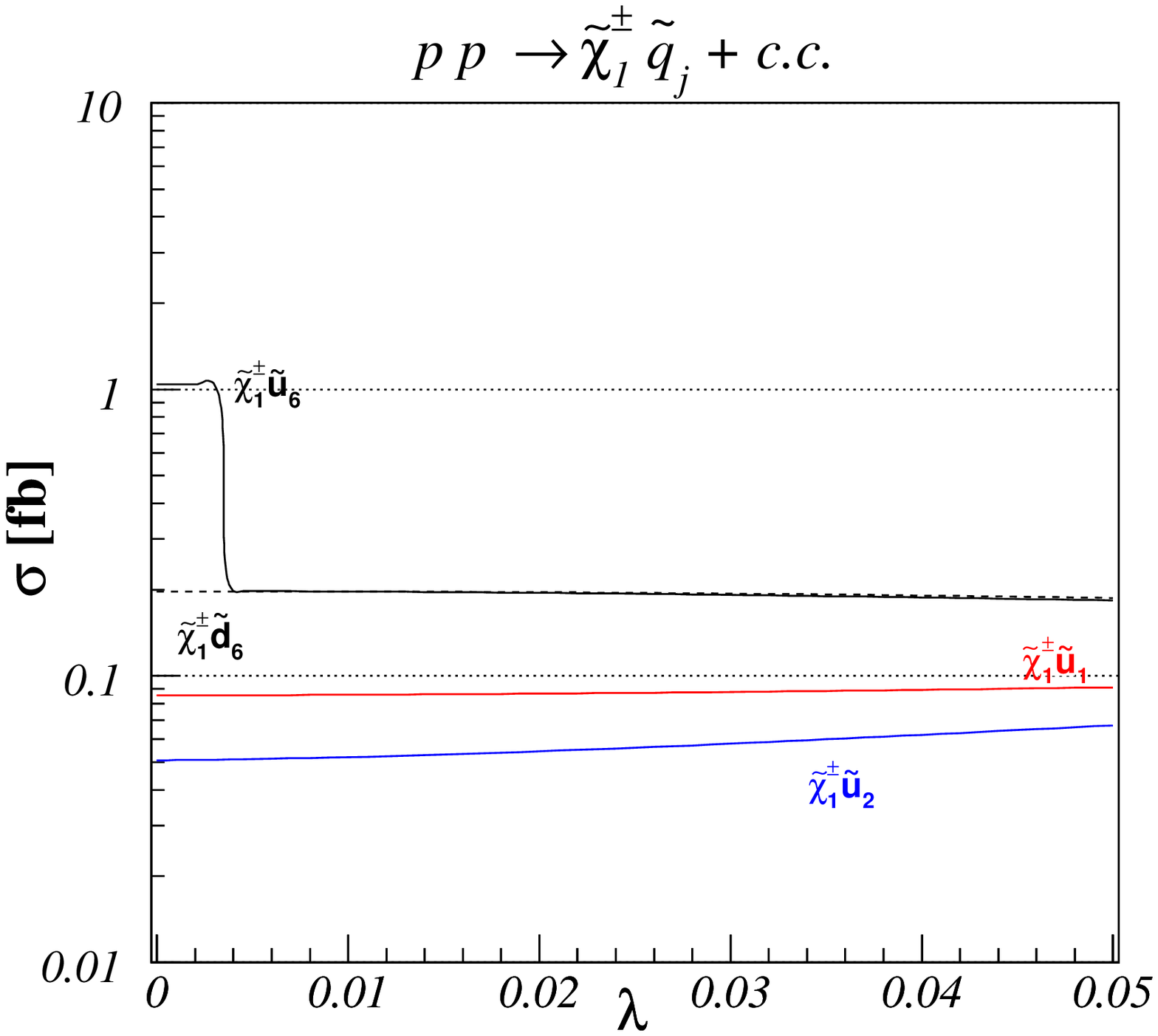}\hspace{2mm}
 \includegraphics[width=0.32\columnwidth]{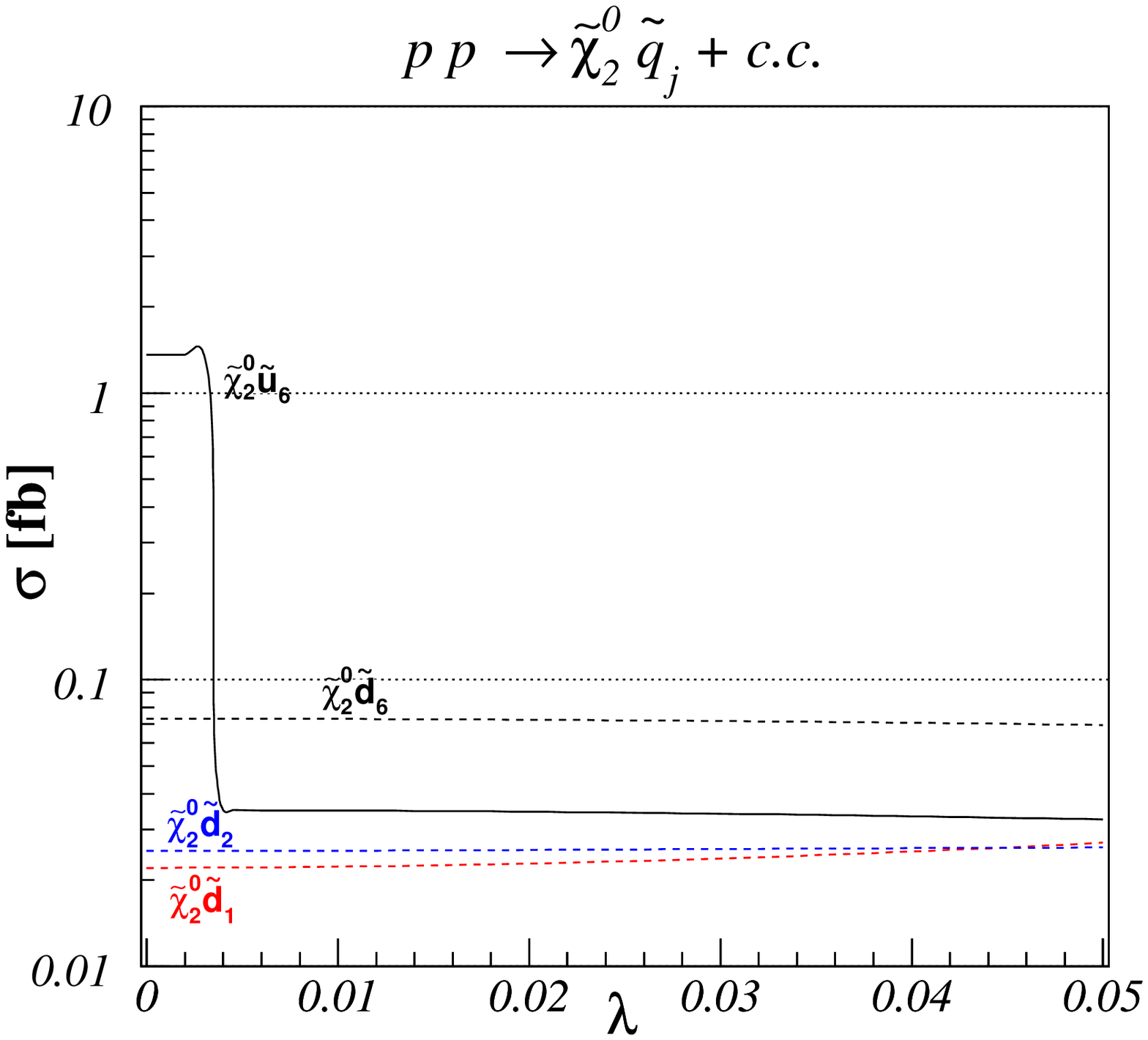}\hspace{2mm}
 \includegraphics[width=0.32\columnwidth]{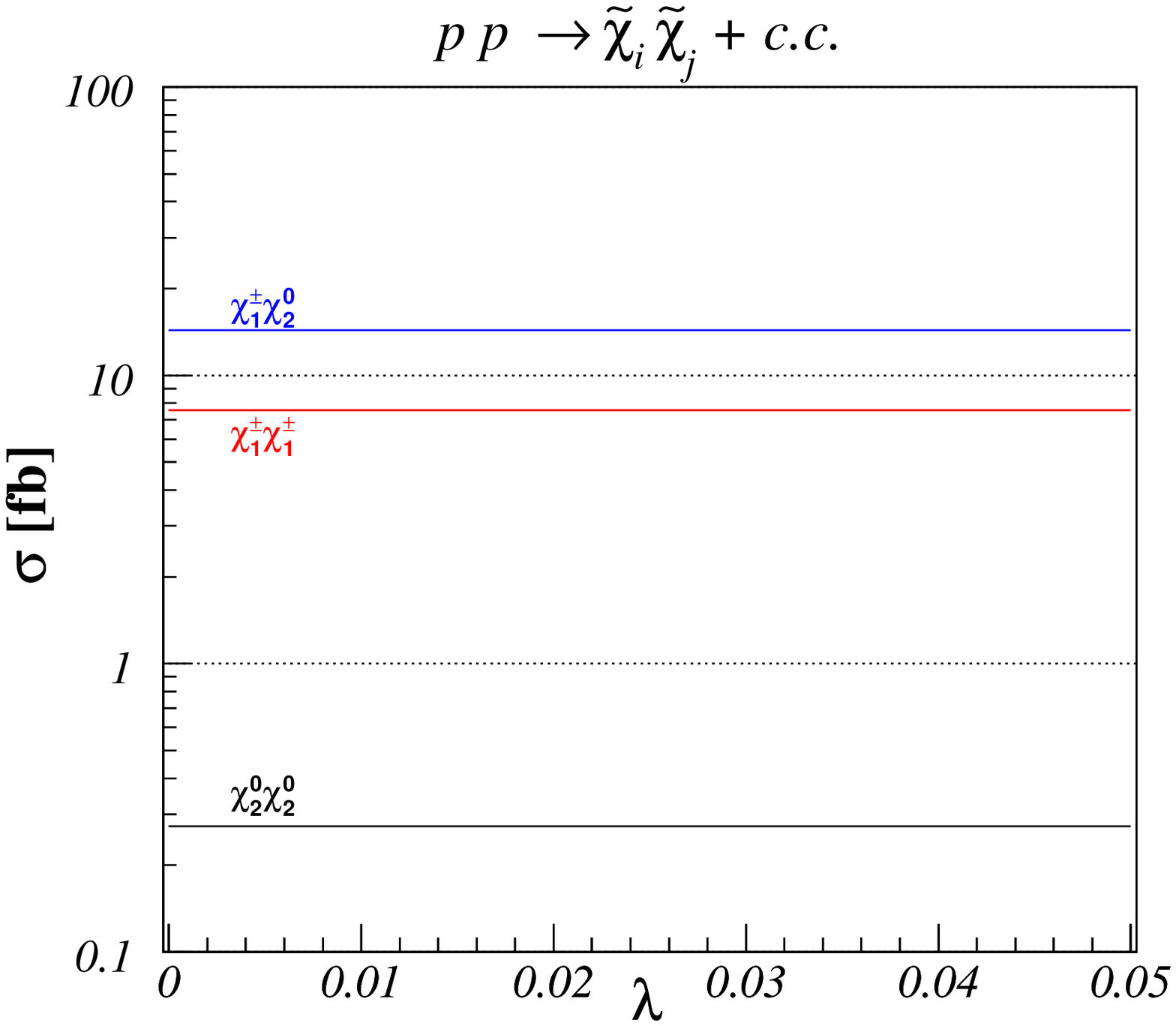}
 \caption{\label{fig:23}Same as Fig.\ \ref{fig:20} for our benchmark
          scenario D.}
\end{figure}
%

The numerical cross sections for charged squark-antisquark and squark-squark
production, neutral up- and down-type squark-antisquark and squark-squark
pair production, associated production of squarks with charginos and
neutralinos, and gaugino pair production are shown in Fig.\ \ref{fig:20}
for our benchmark scenario A, in Fig.\ \ref{fig:21} for scenario B,
in Fig.\ \ref{fig:22} for scenario C, and in Fig.\ \ref{fig:23} for scenario
D. The magnitudes of the cross sections vary from the barely visible level
of $10^{-2}$ fb for weak production of heavy final states over the
semi-strong production of average squarks and gauginos and quark-gluon
initial states to large cross sections of $10^2$ to $10^3$ fb for the strong
production of diagonal squark-(anti)squark pairs or weak production of very
light gaugino pairs. Unfortunately, these processes, whose cross sections
are largest (top right, center left, and lower right parts of Figs.\
\ref{fig:20}-\ref{fig:23}), are practically insensitive to the flavour
violation parameter $\lambda$, as the strong gauge interaction is
insensitive to quark flavours and gaugino pair production cross sections are
summed over exchanged squark flavours.

Some of the subleading, non-diagonal cross sections show, however,
sharp transitions, in particular down-type squark-antisquark
production at the benchmark point B (centre-left part of Fig.\
\ref{fig:21}), but also the other squark-antisquark and
squark-squark production processes. At $\lambda=0.02$, the cross
sections for $\tilde{d}_1 \tilde{d}_6^*$ and
$\tilde{d}_3\tilde{d}_6^*$ switch places. Since the concerned
sstrange and sbottom mass differences are rather small, this is
mainly due to the different strange and bottom quark densities in
the proton. The cross section is mainly due to the exchange of
strongly coupled gluinos despite their larger mass. At
$\lambda=0.035$ the cross sections for $\tilde{d}_3\tilde{d}_6^*$
and $\tilde{d}_1\tilde{d}_3^*$ increase sharply, since
$\tilde{d}_3= \tilde{d}_R$ can then be produced from down-type
valence quarks. The cross section of the latter process increases
with the strange squark content of $\tilde{d}_1$.

At the benchmark point C (Fig.\ \ref{fig:22}), sharp transitions
occur between the $\tilde{u}_2/\tilde{u}_4$ and
$\tilde{d}_2/\tilde{d}_4$ states, which are pure charm/strange
squarks below/above $\lambda=0.035$, for all types of charged and
neutral squark-antisquark and squark-squark production and also
squark-gaugino associated production. As a side-remark we note
that an interesting perspective might be the exploitation of these
$t$-channel contributions to second- and third-generation squark
production for the determination of heavy-quark densities in the
proton. This requires, of course, efficient experimental
techniques for heavy-flavour tagging.

Smooth transitions and semi-strong cross sections of about 1 fb are observed
for the associated production of third-generation squarks with charginos
(lower left diagrams) and neutralinos (lower centre diagrams) and in
particular for the scenarios A and B. For benchmark point A (Fig.\
\ref{fig:20}), the cross section for $\tilde{d}_4$ production decreases with
its strange squark content, while the bottom squark content increases at the
same time. For benchmark point B (Fig.\ \ref{fig:21}), the same (opposite)
happens for $\tilde{d}_6$ ($\tilde{d}_1$), while the cross sections for
$\tilde{u}_6$ increase/decrease with its charm/top squark content.
Even in minimal flavour violation, the associated production of stops and
charginos is a particularly useful channel for SUSY particle spectroscopy,
as can be seen from the fact that cross sections vary over several orders
of magnitude among our four benchmark points (see also Ref.\
\cite{Beccaria:2006wz}).

%
\begin{table}
 \caption{\label{tab:3}Dominant $s$-, $t$-, and $u$-channel contributions to
          the flavour violating hadroproduction of third-generation squarks
          and/or gauginos and the competing dominant flavour-diagonal
          contributions.}
 \begin{tabular}{c|ccc}
  \underline{Exchange} & $s$ & $t$ & $u$ \\
  Final State & & & \\
  \hline
  $\tilde{t}\tilde{b}^*$ & $W$ & NMFV-$\tilde{g}$ & - \\
  $\tilde{b}\tilde{s}^*$ & NMFV-$Z$ & NMFV-$\tilde{g}$ & - \\
  $\tilde{t}\tilde{c}^*$ & NMFV-$Z$ & NMFV-$\tilde{g}$ & - \\
  \hline
  $\tilde{t}\tilde{b}$ & - & - & NMFV-$\tilde{g}$ \\
  $\tilde{b}\tilde{b}$ & - & $\tilde{g}$ & $\tilde{g}$ \\
  $\tilde{t}\tilde{t}$ & - & NMFV-$\tilde{g}$ & NMFV-$\tilde{g}$ \\
  \hline
  $\tilde{\chi}^0\tilde{b}$ & $b$ & $\tilde{b}$ & - \\
  $\tilde{\chi}^\pm\tilde{b}$ & NMFV-$c$ & NMFV-$\tilde{b}$ & - \\
  $\tilde{\chi}^0\tilde{t}$ & NMFV-$c$ & NMFV-$\tilde{t}$ & - \\
  $\tilde{\chi}^\pm\tilde{t}$ & $b$ & $\tilde{t}$ & - \\
  \hline
  $\tilde{\chi}\tilde{\chi}$ & $\gamma,Z,W$ & $\tilde{q}$ & $\tilde{q}$ \\
 \end{tabular}
\end{table}
%
An illustrative summary of flavour violating hadroproduction cross section
contributions for third-generation squarks and/or gauginos is presented in
Tab.\ \ref{tab:3}, together with the competing flavour-diagonal
contributions.

\section{Conclusions}
\label{sec:6}

In conclusion, we have performed an extensive analysis of squark and gaugino
hadroproduction and decays in non-minimal flavour violating supersymmetry.
Within the super-CKM basis, we have taken into account the possible
misalignment of quark and squark rotations and computed all squared
helicity amplitudes for the production and the decay widths of squarks and
gauginos in compact analytic form, verifying that our results agree with the
literature in the case of non-mixing squarks whenever possible.
Flavour violating effects have also been included in our analysis of dark
matter (co-)annihilation processes. We have then analyzed the
NMFV SUSY parameter space for regions allowed by low-energy, electroweak
precision, and cosmological data and defined four new post-WMAP benchmark
points and slopes equally valid in minimal and non-minimal flavour violating
SUSY. We found that left-chiral mixing of second- and third-generation
squarks is slightly stronger constrained than previously believed, mostly
due to smaller experimental errors on the $b\to s\gamma$ branching ratio and
the cold dark matter relic density. For our four benchmark points, we have
presented the dependence of squark mass eigenvalues and the flavour and
helicity decomposition of the squark mass eigenstates on the flavour
violating parameter $\lambda$. We have computed numerically all production
cross sections for the LHC and discussed in detail their dependence on
flavour violation. A full experimental study including heavy-flavour
tagging efficiencies, detector resolutions, and background processes would,
of course, be very interesting in order to establish the experimental
significance of NMFV. While the implementation of our analytical results
in a general-purpose Monte Carlo generator should now be straight-forward,
such a detailed experimental study represents a research project of its own
\cite{inprep} and is beyond the scope of the work presented here.

\acknowledgments
A large part of this work has been performed in the context of the CERN
2006/2007 workshop on ``Flavour in the Era of the LHC''. The authors also
acknowledge interesting discussions with J.\ Debove, A.\ Djouadi, W.\ Porod,
J.M.\ Richard, and P.\ Skands. This work was supported by two
Ph.D.\ fellowships of the French ministry for education and research.

\newpage

\appendix
\section{Gaugino and Higgsino Mixing}
\label{sec:a}

The soft SUSY-breaking terms in the minimally supersymmetric Lagrangian
include a term
\bea
 {\cal L}&\supset&-{1\over2}(\psi^0)^T\,Y\,\psi^0+{\rm h.c.},
\eea
which is bilinear in the (2-component) fermionic partners
\bea
 \psi^0_j&=&(-i\tilde{B},-i\tilde{W}^3,\tilde{H}_1^0,\tilde{H}_2^0)^T
\eea
of the neutral electroweak gauge and Higgs bosons and proportional to the,
generally complex and symmetric, neutralino mass matrix
\bea
 Y &=& \left( \begin{array}{c c c c}
  \hspace{-1mm}M_1 & \hspace{-1mm}0 &
  \hspace{-1mm}-m_Z\,s_W\,c_\beta &
  \hspace{-1mm}m_Z\,s_W\,s_\beta \\
  \hspace{-1mm}0 & \hspace{-1mm}M_2 &
  \hspace{-1mm}m_Z\,c_W\,c_\beta &
  \hspace{-1mm}-m_Z\,c_W\,s_\beta \\
  \hspace{-1mm}-m_Z\,s_W\,c_\beta &
  \hspace{-1mm}m_Z\,c_W\,c_\beta &
  \hspace{-1mm}0 &\hspace{-1mm} -\mu \\
  \hspace{-1mm}m_Z\,s_W\,s_\beta &
  \hspace{-1mm}-m_Z\,c_W\,s_\beta &
  \hspace{-1mm}-\mu & \hspace{-1mm}0
 \end{array} \right).
\eea Here, $M_1$, $M_2$, and $\mu$ are the SUSY-breaking bino,
wino, and off-diagonal higgsino mass parameters with
$\tan\beta=s_\beta/c_\beta=v_u/ v_d$ being the ratio of the vacuum
expectation values $v_{u,d}$ of the two Higgs doublets, while
$m_Z$ is the SM $Z$-boson mass and $s_W$ $(c_W)$ is the sine
(co-sine) of the electroweak mixing angle $\theta_W$. After
electroweak gauge-symmetry breaking and diagonalization of the
mass matrix $Y$, one obtains the neutralino mass eigenstates \bea
 \chi^0_i&=&N_{ij}\,\psi_j^0,~~~i,j=1,\dots,4,
\eea
where $N$ is a unitary matrix satisfying the relation
\bea
 N^*\,Y\,N^{-1}&=&{\rm diag}\, (m_{\tilde{\chi}^0_1},m_{\tilde{\chi}^0_2},
 m_{\tilde{\chi}^0_3},m_{\tilde{\chi}^0_4}).
\eea
In 4-component notation, the Majorana-fermionic neutralino mass eigenstates
can be written as
\bea
 \tilde{\chi}^0_i&=&\lr\begin{array}{c} \chi_i^0\\
 \bar{\chi}_i^0\end{array}\rr,~~~i=1,\dots,4.
\eea
Their mass eigenvalues $m_{\tilde{\chi}^0_i}$ can, e.g., be found in
analytic form in \cite{ElKheishen:1992yv} and can be chosen to be real and
non-negative.

The chargino mass term in the SUSY Lagrangian
\bea
 {\cal L} &\supset& -{1\over2} (\psi^+\psi^-)\lr\begin{array}{cc}
 0&X^T\\X&0\end{array}\rr \lr\begin{array}{c}\psi^+\\ \psi^-\end{array}\rr
 +{\rm h.c.}
\eea
is bilinear in the (2-component) fermionic partners
\bea
 \psi_j^\pm&=&(-i\tilde{W}^\pm,\tilde{H}^\pm_{2,1})^T
\eea
of the charged electroweak gauge and Higgs bosons and proportional to the,
generally complex, chargino mass matrix
\bea
 X &=& \left( \begin{array}{c c} M_{2} & m_{W}\, \sqrt{2}\, s_\beta \\
 m_{W}\, \sqrt{2}\, c_\beta &  \mu \end{array}\right),
\eea
where $m_W$ is the mass of the SM $W$-boson. Its diagonalization leads to
the chargino mass eigenstates
\bea
 \begin{array}{l} \chi_i^+~=~V_{ij}\,\psi_j^+\\
 \chi_i^-~=~U_{ij}\,\psi_j^-\end{array},~~~i,j=1,2,
\eea
where the matrices $U$ and $V$ satisfy the relation
\bea
 U^*\,X\,V^{-1}&=&{\rm diag}\,(m_{\tilde{\chi}^\pm_1},m_{\tilde{\chi}^
 \pm_2}).
\eea
In 4-component notation, the Dirac-fermionic chargino mass eigenstates can
be written as
\bea
 \tilde{\chi}^\pm_i&=&\lr\begin{array}{c} \chi_i^\pm\\
 \bar{\chi}_i^\mp\end{array}\rr.
\eea
The mass eigenvalues can be chosen to be real and non-negative and are given
by \cite{Haber:1984rc}
\bea
 m_{\tilde{\chi}^\pm_{1,2}}^2 & = & \frac{1}{2}\bigg\{ M_{2}^{2} + \mu^{2} +
 2 m_{W}^{2} \mp \Big[(M_{2}^{2}-\mu^2 )^{2} + 4 m_{W}^{4}c^2_{2\beta}+
 4 m_{W}^{2} (M_{2}^{2} + \mu^{2} + 2 M_{2}\,\mu\, s_{2 \beta})\Big]^{1/2}
 \bigg\},
\eea
while the matrices
\bea
 U ~=~ {\cal O}_{-}&~~{\rm and}~~&
 V ~=~ \left\{ \begin{array}{ll} {\cal O}_{+} &,~~ {\rm if~~det}\,X\,
 \geq 0\\ \sigma_{3}\,{\cal O}_{+} &,~~ {\rm if~~ det} \,X\, < 0 \end{array}
 \right.~~~{\rm with}~~~
 {\cal O}_{\pm} ~=~\left(\begin{array}{cc}~~\,\cos\theta_{\pm} & \sin
 \theta_{\pm} \\ -\sin\theta_{\pm} & \cos\theta_{\pm} \end{array}\right)
\eea
are determined by the mixing angles $\theta_\pm$ with $0\leq\theta_\pm\leq
\pi/2$ and
\bea
 \tan\,2\theta_+~=~\frac{2\sqrt{2} m_W \left( M_2 \sin\beta + \mu \cos\beta
 \right)} {M_2^2 -\mu^2 + 2 m_W^2\cos2\beta}&~~{\rm and}~~&
 \tan\,2\theta_-~=~\frac{2\sqrt{2} m_W \left( M_2 \cos\beta + \mu \sin\beta
 \right)} {M_2^2 -\mu^2 - 2 m_W^2\cos2\beta}.
\eea


\end{document}